%% file: thesis.tex
\documentclass[a4paper,11pt,twoside,notitlepage,openright]{report}

\usepackage[spanish,english]{babel}
\usepackage[small,bf]{caption}

\usepackage{color}
\definecolor{chapternumber}{rgb}{0.64,0.16,0.16}
\definecolor{chaptertitle}{rgb}{0.64,0.16,0.16}
\definecolor{boxbackground}{rgb}{1.0,0.25,0.25}

\usepackage[Bjornstrup]{fncychap} 
\ChNumVar{\fontsize{76}{80}\usefont{OT1}{pzc}{m}{n}\selectfont\color{chapternumber}}
\ChTitleVar{\raggedleft\LARGE\bfseries\color{chaptertitle}}

\usepackage{acronym}

\usepackage{times}
\usepackage{latexsym}
\usepackage{tocbibind}

\usepackage{varioref}
\usepackage[pdfstartview=FitV, pdftitle={Harnessing Folksonomies for Resource Classification}, pdfauthor={Arkaitz Zubiaga}, pdfkeywords={folksonomies, social-tagging, classification, social-media, phd, thesis}, pdfsubject={PhD Thesis}, pagebackref, pdfproducer={pdfLaTeX}, colorlinks=true, linkcolor=blue, citecolor=blue, urlcolor=blue]{hyperref}
\usepackage{bookmark}
\usepackage[all]{hypcap}

\usepackage{ucs}
\usepackage[utf8x]{inputenc}
\usepackage[T1]{fontenc}

\usepackage[sc]{mathpazo}
\usepackage{epsfig}
\usepackage{hhline}
\usepackage{rotating}
\usepackage{colortbl}
\usepackage{hyphenat}
\usepackage{textcomp} 

\newcommand{\chaptersummary}[1]{\begin{center}\parbox{10cm}{#1}\end{center}}
\newcommand{\translatedtitle}[1]{\begin{center}\LARGE \textbf{#1}\end{center}}

\newcommand{\rqsvmone}[0]{What kind of SVM classifiers should be used to perform this kind of classification tasks: a native multiclass classifier, or a combination of binary classifiers?}
\newcommand{\rqsvmtwo}[0]{What kind of learning method performs better for this kind of classification tasks: a supervised one, or a semi-supervised one?}
\newcommand{\rqdataone}[0]{How do the settings of social tagging systems affect users' annotations and the resulting folksonomies?}
\newcommand{\rqrepone}[0]{What is the best way of amalgamating users' aggregated annotations on a resource in order to get a single representation for a resource classification task?}
\newcommand{\rqreptwo}[0]{Despite of the usefulness of social tags for these tasks, is it worthwhile considering their combination with other data sources like the content of the resource as an approach to improve the results even more?}
\newcommand{\rqrepthree}[0]{Are social tags also useful and specific enough to classify resources into narrower categories as in deeper levels of hierarchical taxonomies?}
\newcommand{\rqdistone}[0]{Can we further consider the distribution of tags across the collection so that we can measure the overall representativity of each tag to represent resources?}
\newcommand{\rqdisttwo}[0]{What is the best approach to weigh the representativity of tags in the collection for resource classification?}
\newcommand{\rqcatone}[0]{Can we discriminate different user profiles so that we can find a subset of users who provide annotations that better fit a classification scheme?}
\newcommand{\rqcattwo}[0]{What are the features that identify a user as a good contributor to the resource classification?}
\newcommand{\problemstatement}[0]{How can the annotations provided by users on social tagging systems be exploited to yield the most accurate resource classification task?}

\newcommand{\rqsvmonees}[0]{\textquestiondown Qu\'{e} tipo de clasificador SVM deber\'{i}a utilizarse para llevar a cabo este tipo de tareas de clasificaci\'{o}n: un clasificador multiclase nativo, o una combinaci\'{o}n de clasificadores binarios?}
\newcommand{\rqsvmtwoes}[0]{\textquestiondown Qu\'{e} m\'{e}todo de aprendizaje rinde mejor para este tipo de tareas de clasificaci\'{o}n: uno supervisado o uno semisupervisado?}
\newcommand{\rqdataonees}[0]{\textquestiondown C\'{o}mo afecta la configuraci\'{o}n de los sistemas de etiquetado social en las anotaciones de los usuarios y las folksonom\'{i}as resultantes?}
\newcommand{\rqreponees}[0]{\textquestiondown Cu\'{a}l es la mejor manera de acumular las anotaciones de los usuarios sobre un recurso con el fin de obtener una representaci\'{o}n?}
\newcommand{\rqreptwoes}[0]{A pesar de la utilidad de las etiquetas sociales para estas tareas, \textquestiondown merece la pena considerar otras fuentes de datos como el contenido de los recursos para mejorar a\'{u}n m\'{a}s los resultados?}
\newcommand{\rqrepthreees}[0]{\textquestiondown Son las etiquetas sociales tambi\'{e}n \'{u}tiles y suficientemente espec\'{i}ficas para clasificar recursos en categor\'{i}as a nivel m\'{a}s bajo?}
\newcommand{\rqdistonees}[0]{\textquestiondown Podemos tener en cuenta la distribuci\'{o}n de etiquetas a lo largo de la colecci\'{o}n para as\'{i} medir la representatividad general de la etiqueta?}
\newcommand{\rqdisttwoes}[0]{\textquestiondown Cu\'{a}l es la mejor aproximaci\'{o}n para establecer la representatividad de las etiquetas en la colecci\'{o}n?}
\newcommand{\rqcatonees}[0]{\textquestiondown Podemos discriminar diferentes perfiles de usuario de manera que encontremos un subconjunto de usuarios que proporciona anotaciones que se ajustan en mayor medida a la tarea de clasificaci\'{o}n?}
\newcommand{\rqcattwoes}[0]{\textquestiondown Cu\'{a}les son las caracter\'{i}sticas que identifican a un usuario como apropiado para la tarea de clasificaci\'{o}n de recursos?}
\newcommand{\problemstatementes}[0]{\textquestiondown C\'{o}mo se pueden aprovechar las anotaciones provistas por usuarios en sistemas de etiquetado social de forma que se obtenga una clasificaci\'{o}n de recursos m\'{a}s precisa?}

\newcommand{\rqsvmoneeu}[0]{Zein SVM sailkatzaile mota erabili beharko litzateke sailkapen ataza hauek burutzeko: jatorrizko klase-anitzeko sailkatzailea, ala sailkatzaile bitarren konbinazio bat?}
\newcommand{\rqsvmtwoeu}[0]{Zein ikasketa motak ematen du errendimendu hobea sailkapen ataza hauek bu\-ru\-tze\-ko: gainbegiratu batek, ala erdi-gainbegiratu batek?}
\newcommand{\rqdataoneeu}[0]{Nola eragiten dute etiketa sozialen sistemetako ezarpenek bertako erabiltzaileen anotazioetan eta ondorioz sortutako folksonomietan?}
\newcommand{\rqreponeeu}[0]{Zein da baliabide baten gainean erabiltzaileek egindako anotazio guztiak adierazpen bakarrean bateratzeko modurik egokiena?}
\newcommand{\rqreptwoeu}[0]{Etiketa sozialek ataza hauetarako duten balioaz gainera, merezi al du baliabidearen barne edukia bezalako beste datu iturri batzuk kontuan hartzea emaitzak are gehiago hobetzeko?}
\newcommand{\rqrepthreeeu}[0]{Baliabideak maila baxuagoko kategoria zehatzagoetan sailkatu ahal izateko nahikoa erabilgarriak eta zehatzak dira etiketa sozialak?}
\newcommand{\rqdistoneeu}[0]{Etiketen adierazgarritasuna neurtzera bidean, kolekzioan zehar etiketek duten distribuzioa kontuan har al daiteke?}
\newcommand{\rqdisttwoeu}[0]{Zein da etiketek kolekzioan duten adierazgarritasuna neurtzeko hurbilpenik egokiena?}
\newcommand{\rqcatoneeu}[0]{Ba al dago erabiltzaile profilak ezberdintzerik, sailkapen ataza batera ahalik eta gehien hurbiltzen diren anotazioak egiten dituzten erabiltzaileak bilatu asmoz?}
\newcommand{\rqcattwoeu}[0]{Zeintzu dira erabiltzaile bat baliabideen sailkapen on bat egiten ari dela zehazten duten ezaugarriak?}
\newcommand{\problemstatementeu}[0]{Nola egin daiteke erabiltzaileek etiketen sistema sozialetan egindako anotazioak ustiatzeko baliabideen sailkapen ahalik eta zehatzena lortuz?}

\input epsf

\usepackage{amsmath}
\usepackage{url}
\usepackage{epigraph}

\usepackage{fancyhdr}
\fancypagestyle{plain}{
        \fancyhead{} 
        
}

\usepackage{natbib}
\bibpunct{(}{)}{;}{a}{,}{,}

\let\oldmarginpar\marginpar
\renewcommand\marginpar[1]{\-\oldmarginpar[\raggedleft\small\sf#1]{\raggedright
\small\sf#1}}

\usepackage{appendix}

\begin{document}


\include{cover}
\include{inside-cover}
\include{license}
\include{quotes}

\include{acknowledgments}

\include{abstract}


\tableofcontents
\listoffigures
\listoftables

\include{introduction}

\include{state-of-the-art}

\include{svm-classification}

\include{datasets}

\include{tag-representation}

\include{tag-distribution}

\include{tag-categorizers}

\include{conclusions}


\bibliographystyle{plainnat}
\bibliography{thesis}

\include{publications}

\cleardoublepage
\appendix
\include{topx-tags}

\include{key-terms}

\include{acronyms}

\include{sp-summary}

\include{bq-summary}


\end{document}

%% file: cover.tex
\thispagestyle{empty}
\newcommand{\HRule}{\rule{\linewidth}{1mm}}

\begin{center}
   \textsc{Universidad Nacional de Educación a Distancia} \\
 \small Departamento de Lenguajes y Sistemas Informáticos\\  
 \small  \textit{Escuela Técnica Superior de Ingeniería Informática} \\
  
 \end{center}

\vspace*{\stretch{1.5}} 

\begin{center}
\includegraphics[width=1in]{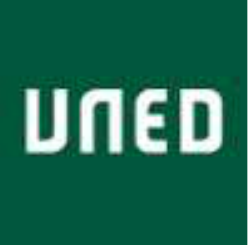}
\end{center}

\vspace*{\stretch{1.5}} 
\HRule
\begin{flushright}
\begin{center}
  \large {\textbf{HARNESSING FOLKSONOMIES FOR RESOURCE CLASSIFICATION}}
\end{center}
\end{flushright}
\HRule 
\vspace*{\stretch{2}}

\begin{center}
  \Large \textbf{PhD THESIS} \\ [45mm]
\end{center}

\begin{center}
 \large {\bfseries Arkaitz Zubiaga Mendialdua} \\ 
\small MSc in Computer Science \\
 2011\\ 
\end{center}

\newpage
\thispagestyle{empty}
\vspace*{1cm}

%% file: inside-cover.tex
\newpage
\thispagestyle{empty}

\begin{center}
   \textsc{Universidad Nacional de Educación a Distancia} \\
   \small Departamento de Lenguajes y Sistemas Informáticos\\  
 \small  \textit{Escuela Técnica Superior de Ingeniería Informática} \\ [1cm]
 \end{center}

\begin{center}
\includegraphics[width=1in]{logo-uned.pdf}
\end{center}

\begin{flushright}
\begin{center}
  \large \textbf{HARNESSING FOLKSONOMIES FOR RESOURCE CLASSIFICATION} \\ [25mm]

\end{center}
\end{flushright}

\begin{center}
  \large \textbf{Arkaitz Zubiaga Mendialdua} \\
\small MSc in Computer Science, Mondragon Unibertsitatea \\ [25mm]
\end{center}

\begin{center}
Advisors:  \\ [5mm]

\large \textbf{V\'ictor Fresno Fern\'andez} \\ [3mm]
\small Assistant Professor in the Lenguajes y Sistemas Inform\'aticos Department at
       Universidad Nacional de Educación a Distancia \\ [5mm]

\large \textbf{Raquel Mart\'inez Unanue} \\ [3mm]
\small Associate Professor in the Lenguajes y Sistemas Inform\'aticos Department at
       Universidad Nacional de Educación a Distancia \\ [5mm]

\end{center}

\newpage
\thispagestyle{empty}
\vspace*{1cm}

%% file: license.tex
\newpage
\thispagestyle{empty}

\vspace*{15cm}
\small
 \copyright 2011 Arkaitz Zubiaga Mendialdua

 This work is licensed under the

 Creative Commons Attribution-ShareAlike 3.0 License.


 To view a copy of this license, visit

 \url{http://creativecommons.org/licenses/by-sa/3.0/}

 or send a letter to

 Creative Commons,

 543 Howard Street, 5th Floor,

 San Francisco, California, 94105, USA.

\newpage
\thispagestyle{empty}
\vspace*{1cm}

%% file: quotes.tex
\newpage
\thispagestyle{empty}

\vspace*{5cm}
\begin{flushright}
  \small
  \textit{"A free culture has been our past, but it will only be our future if we change the path we are on right now."}
  
  ---Lawrence Lessig
  
  \vspace*{2cm}
  
  \textit{"Free software is a matter of liberty, not price. To understand the concept, you should think of free as in free speech, not as in free beer."}

  ---Richard Stallman

  \vspace*{2cm}

  \textit{"The only valid censorship of ideas is the right of people not to listen."}

  ---Tommy Smothers
\end{flushright}

\newpage
\thispagestyle{empty}
\vspace*{1cm}

%% file: acknowledgments.tex
\chapter*{Acknowledgments}
\label{c:acknowledgments}

First and foremost I would like to earnestly thank my advisors, V\'{i}ctor and Raquel, who have been helping me throughout this research. Their hard work guiding me has enabled me to acquire a deep insight into both research and the field of social media mining. Their guidance has been of vital importance for pursuing first the Master Thesis, and the PhD Thesis afterward. Since the moment I decided to work on the novel field of social media, which was also new to them, they responded with a very willing and excellent attitude to assist with this project.

I would also like to thank my colleagues at the University, with whom I have discussed many research matters, that allowed me to grow as a researcher. Furthermore, all the leisure times we enjoyed together were of utmost importance to help me feel welcome and comfortable in the department. I would especially like to thank my academic brother Alberto, who has worked alongside me. With Alberto I have learned lots of new things and together worked on significant research that was presented at an international conference.

I am also very grateful to Markus and Christian, from the Graz University of Technology, with whom I had the pleasure of cooperating. They kindly hosted me in Graz, Austria, during my research stay. The stay was helpful in many aspects, such as getting to know other researchers who work in the same field as I, and sharing our thoughts and knowledge. It was especially useful to produce a sound joint research work which was presented at a consolidated and prestigious international conference.

I deeply appreciate and want to thank Jake for his kindness and effort in reviewing this dissertation. Jake provided a handful of tips and suggestions to improve the English in this document.

I also want to thank all the friends who have been with me all this time. I would like to thank not only those in my hometown Arrasate in the Basque Country, with whom I spent most of my childhood, but also my friends in Madrid, who have helped me settle in the city.

Last, but not least, I would sincerely like to thank my parents, Jos\'{e} Ram\'{o}n and Maribi, and my sister Igone. Without them, I would not be who I am today. Their help and patience throughout so many years has been fundamental in helping me reach this goal. And finally, my most sincere thanks to Nayeli, with whom I have shared most of my time during this work. She has been patient and understanding throughout the process of writing this dissertation and I will always wholeheartedly appreciate her.

\section*{Institutional Acknowledgements}
\label{sec:institutional-acknowledgments}

The research presented in this work has been in part funded by the Regional Government of Madrid under the Research Networks MAVIR (S-0505/TIC-0267) and MA2VICMR (S-2009/TIC-1542), the Regional Ministry of Education of the Community of Madrid, and the Spanish Ministry of Science and Innovation projects QEAVis-Catiex (TIN2007-67581-C02-01) and Holopedia (TIN2010-21128-C02-01).

%% file: abstract.tex
\chapter*{Abstract}
\label{c:abstract}

In our daily lives, organizing resources into a set of categories is a common task. Organizing resources into categories makes searching through those resources easier by limiting the focus to a specific category. Limiting the focus significantly reduces the amount of information one must search. Categorization becomes more useful as the collection of resources increases, when managing resources becomes more and more difficult if they are not organized appropriately. Large collections like those made up by books, movies, and web pages, for instance, are usually cataloged in libraries, organized in databases and classified in directories, respectively. However, the usual largeness of these collections requires a vast endeavor and an outrageous expense to organize manually.

Recent research is moving towards developing automated classifiers that reduce the increasing costs and effort of the task. Most of the research in this field has focused on self-content, where the publisher is the only author, as a data source to discover the aboutness of the resource. Self-content presents the problem that it is not always representative enough, and sometimes it is difficult to access depending on the type of resource. Little work has been done analyzing the appropriateness of and exploring how to harness the annotations provided by users on social tagging systems as a data source. Users on these systems save resources as bookmarks in a social environment by attaching annotations in the form of tags. It has been shown that these tags facilitate retrieval of resources not only for the annotators themselves but also for the whole community. Likewise, these tags provide meaningful metadata that refers to the content of the resources.

In this thesis, we deal with the utilization of these user-provided tags in search of the most accurate classification of resources as compared to expert-driven categorizations. After performing a set of experiments to choose a suitable classifier for this kind of task, we explore social annotations looking for a way to best use them. For this purpose, we have created three large-scale datasets including tagging data for resources from well-known social tagging systems: Delicious, LibraryThing, and GoodReads. Those resources are accompanied by categorization data from sound and consolidated expert-driven taxonomies. From these resources the appropriateness of social tags for predicting categories can be evaluated.

Specifically, we first study several ways of representing the massive number of social tags by amalgamating the contributions of large communities of users. We analyze their suitability for the classification task, upon both broader top level categories and narrower deep level categories. Then, we explore the nature, characteristics, and distributions of tags in folksonomies, in order to determine how the settings of each system affect the tagging behavior and the usefulness of tags for the classification task. We go deeper into tag distributions by analyzing the usefulness of weighting schemes based on inverse frequency values. Finally, using state-of-the-art user behavior detection processes, we identify users on social tagging systems who better fit the classification task.

To the best of our knowledge, this is the first research work performing actual classification experiments utilizing social tags. By exploring the characteristics and nature of these systems and the underlying folksonomies, this thesis sheds new light on the way of getting the most out of social tags for the sake of automated resource classification tasks. Therefore, we believe that the contributions in this work are of utmost interest for future researchers in the field, as well as for the scientific community in order to better understand these systems and further utilize the knowledge garnered from social tags.

%% file: introduction.tex
\chapter{Introduction}
\label{c:introduction}

\textit{``Ideals are like stars; you will not succeed in touching them with your hands. But like the seafaring man on the desert of waters, you choose them as your guides, and following them you will reach your destiny.''}

--- Carl Schurz

\section{Motivation}
\label{sec:motivation}

Organizing resources into predefined categories is a natural idea in our daily lives. Assigning categories to resources helps facilitate the search for resources by reducing the focus to a specific category or categories. Categorization effectively reduces the amount of resources one has to search. For instance, librarians usually organize books into groups of related subjects. Also, movie databases, music catalogs, and file systems, among others, tend to be categorized in a way that eases access to their resources. Likewise, web directories such as the Yahoo! Directory and the Open Directory Project organize web pages into categories. Web page classification can substantially enhance search engines by reducing the scope of results to the category of user's interest \citep{qi_webpage_2009}.

The process of manually categorizing resources becomes expensive as the collection of resources grows. For instance, the Library of Congress reported that the average cost of cataloging each bibliographic record by professionals was \$94.58 in 2002\footnote{http://www.loc.gov/loc/lcib/0302/collections.html}. For the 291,749 records they cataloged that year, the total cost came to more than \$27.5 million. Given the expensiveness of this task, switching to automated classifiers seems to be a good alternative to facilitate the task and keep catalogs updated by reducing manual effort.

Until now, most of the automated classifiers rely on the content of the resources, especially regarding web page classification tasks (\cite{qi_webpage_2009}). Nonetheless, the lack of representative data within many resources makes the classification task more complicated. In some cases, it may not be feasible to obtain enough data for certain kinds of resources such as books or movies. For example, usually the full text of books is not available, and it is not easy to represent movies as text or processable data. Without sufficient data, representing the content becomes more challenging.

As a means to solve these issues, social tagging systems provide an easier and cheaper way to obtain metadata related to resources. Social tagging systems are a means to save, organize, and search resources, by annotating them with tags that the user provides. Systems like Delicious\footnote{http://delicious.com}, LibraryThing\footnote{http://www.librarything.com} and GoodReads\footnote{http://www.goodreads.com} collect user annotations in the form of tags on their respective collections of resources. These user-generated tags give rise to meaningful data describing the content of the resources \citep{heymann_can_social_2008}. User-provided annotations can be useful as a data source by providing meaningful information that can help infer the categorization of the resources. Our hypothesis is that these large collections of annotations can enhance the automated resource classification task in a noticeable manner.

By providing tags, users are creating their own categorization system for the given resource. The aggregation of users in an active community can create many bookmarks, tags, and therefore annotated resources. With more users contributing bookmarks and tags to these systems, the more accurately these resources can be annotated.

\begin{quote}
 \textit{``Each individual categorization scheme is worth less than a professional categorization scheme. But there are many, many more of them''}, Joshua Schachter, founder of Delicious, at the 2006 FOWA summit in London, England\footnote{http://simonwillison.net/2006/Feb/8/summit/}.
\end{quote}

Given that a large number of users are providing their own annotations on each resource, our objective is focused on finding out an approach to amalgamate their contributions in such a way that resembles the categorization by professionals. In this context, where users are providing large amounts of metadata, our challenge lies in making the most of them in order to enhance resource categorization tasks.

\begin{quote}
 \textit{``We've entered an era where data is cheap, but making sense of it is not''}, Danah Boyd, Social Media Researcher at Microsoft Research New England, at the WWW2010 conference in Raleigh, North Carolina, United States\footnote{http://www.danah.org/papers/talks/2010/WWW2010.html}.
\end{quote}

\subsection{Resource Classification}
\label{resource-classification}

Resource classification can be defined as the task of labeling and organizing resources within a set of predefined categories. In this work, we use Support Vector Machines (SVM, \cite{joachims98text}), a state-of-the-art classification approach. This type of classification relies on previously categorized or labeled training sets of resources. The classifier uses these sets of resources to gather knowledge which, in turn, is used to classify new unknown resources.

Different settings can be used for resource different classification problems. The system's learning technique may be \emph{supervised} or \emph{semi-supervised}. \emph{Supervised} learning requires that all training resources are previously categorized where \emph{semi-supervised} learning permits unlabeled resources to be taken into account during the learning phase. Classification may be \textit{binary}, where only two possible categories can be assigned to each resource, or \textit{multiclass}, where three or more categories can be assigned. \emph{Binary} classification systems are commonly used for filtering systems --e.g., an email application that filters out spam messages--, whereas the \emph{multiclass} systems are necessary for thematic classification with larger taxonomies --i.e., classification by topic or subject.

For thematic classification on large collections of resources, like web pages on the Web, or books in libraries, the taxonomies are usually defined by more than two categories, and the subset of previously labeled resources tends to be tiny. Accordingly, we believe that the application of both semi-supervised and multiclass approaches should be considered and analyzed to perform this kind of task.

In this thesis, we propose the analysis of several classification approaches using SVM, with the aim of analyzing their suitability to these tasks. These include different approaches to solving multiclass problems, as well as the study of \emph{supervised} and \emph{semi-supervised} algorithms.

\subsection{Social Annotations}
\label{social-annotations}

Social tagging sites allow users to save and annotate their favorite resources --e.g., web pages, movies, books, photos or music--, socially sharing them with the community. These annotations are usually provided by users in the form of tags. Tagging is an open way to assign tags or keywords to resources, in order to describe and organize them. It enables the later retrieval of resources in an easier way, using tags as metadata describing resources. Usually, there are no predefined tags, and therefore users can freely choose the words they want as tags.

\begin{quote}
 \textit{``Tagging is mostly user interface - a way for people to recall things, what they were thinking about when they saved it. Fairly useful for recall, OK for discovery, terrible for distribution (where publishers add as many tags as possible to get it in lots of boxes).''}, Joshua Schachter, founder of Delicious, at the 2006 FOWA summit in London, England\footnote{http://simonwillison.net/2006/Feb/8/summit/}.
\end{quote}

This tagging process generates a tag structure so-called folksonomy on a social tagging system, i.e., a user-driven organization of resources. Folksonomy is a portmanteau of the words \textit{folk} (people), \textit{taxis} (classification) and \textit{nomos} (management). It is also known as a community-based taxonomy, where the classification scheme is non-hierarchical, as opposed to a classical taxonomy-based categorization scheme. Thus, a folksonomy has to do with expert-driven taxonomies, insofar as resources are labeled and put together into groups.

These annotations are said to belong to a social environment when they are accessible and profitable by any user. This feature enables searching resources by taking advantage of annotations provided by others. This encourages the contribution of large communities of users.

Not all the annotations are shared in the same way, though. The social tagging site itself may establish some constraints, mainly by setting who is able to annotate each resource. In this regard, two kinds of systems can be distinguished \citep{smith_tagging_2008}:

\begin{itemize}
 \item \textbf{Simple tagging systems:} users can describe their own resources, such as photos on Flickr\footnote{http://www.flickr.com}, news on Digg\footnote{http://digg.com} or videos on Youtube\footnote{http://www.youtube.com}, but nobody annotates others' resources. Usually, the author of the resource is who annotates it. This means that no more than one user tags a resource. In a simple tagging system, there is a set of users ($U$), who are annotating resources ($R$) using tags ($T$). A user $u_{i} \in U$ annotates their resource $r_{j} \in R$ with a set of tags $T_{j} = \{t_{j1},...,t_{jp}\}$, with a variable number $p$ of tags. The set of tags assigned to $r_{j}$ will always be limited to $T_{j}$, since nobody else can annotate it.
 
 \item \textbf{Collaborative tagging systems:} many users annotate the same resource, and all of them can tag it with tags in their own vocabulary. The collection of tags assigned by a single user creates a smaller folksonomy, also known as personomy. As a result, several users tend to post the same resource. For instance, CiteULike\footnote{http://www.citeulike.org}, LibraryThing and Delicious are based on collaborative annotations, where each resource (papers, books and URLs, respectively) can be annotated and tagged by all the users who consider it interesting. A collaborative tagging system is more complex than a simple one, where there is a set of users ($U$), who are posting bookmarks ($B$) for resources ($R$) annotated by tags ($T$). Each user $u_{i} \in U$ can post a bookmark $b_{ij} \in B$ of a resource $r_{j} \in R$ with a set of tags $T_{ij} = \{t_{ij1},...,t_{ijp}\}$, with a variable number $p$ of tags. After $k$ users posted $r_{j}$, it is described with a weighted set of tags $T_{j} = \{w_{j1} t_{j1},...,w_{jn} t_{jn}\}$, where $w_{j1},...,w_{jn} \leq k$ represent the number of assignments of a specific tag. Accordingly, each bookmark is a triple of a user, a resource, and a set of tags: $b_{ij}: u_{i} \times r_{j} \times T_{ij}$. Thus, each user saves bookmarks of different resources, and a resource has bookmarks posted by different users. The result of aggregating tags within bookmarks by a user is known as the personomy of the user: $T_{i} = \{w_{i1} t_{i1},...,w_{im} t_{im}\}$, where $m$ is the number of different tags in user's personomy.
\end{itemize}

\begin{figure}[ht]
\begin{center}
 \includegraphics[width=100mm,clip]{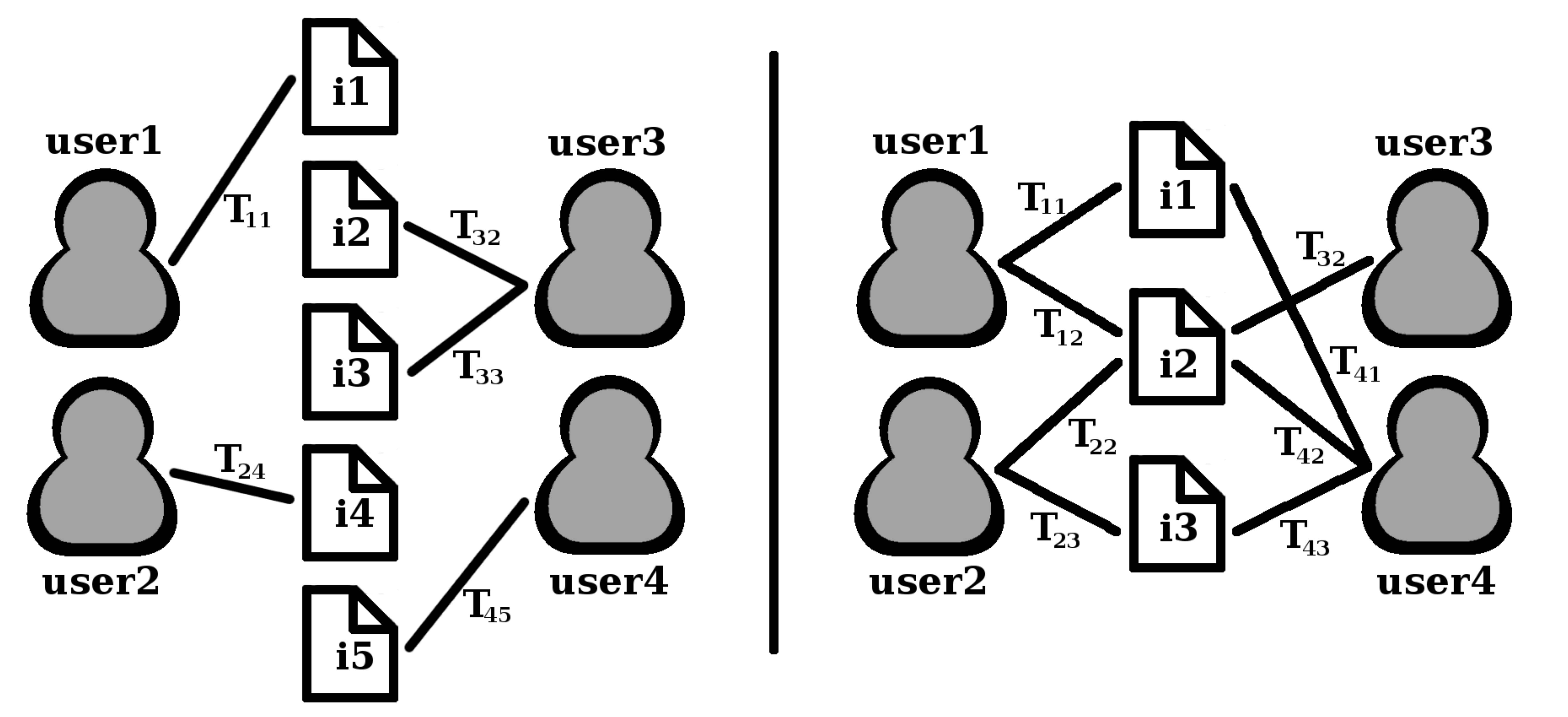}
\end{center}
\caption[Simple tagging vs collaborative tagging]{Comparison of user annotations on simple and collaborative tagging systems.}
\label{fig:simple-vs-collaborative}
\end{figure}

Figure \ref{fig:simple-vs-collaborative} shows an example comparing the behavior of both systems.

In this thesis, we will focus on collaborative tagging systems. Tags present a high likelihood of coincidence across users annotating the same resource, making the aggregated tags of collaborative tagging systems especially strong rather than simple tagging systems, i.e., multiple users annotate the same resource.

In a collaborative tagging system, for instance, a user could tag this work as \texttt{social-tagging}, \texttt{research}, and \texttt{thesis}, whereas another user could use the tags \texttt{social-tagging}, \texttt{social-bookmarking}, \texttt{phd}, and \texttt{thesis} to annotate it. Users' behavior may considerably differ on these systems. Because of this, the aggregation of their annotations is usually considered as the consensus. For instance, the aggregation of the above annotations would be the following: \texttt{thesis} (2), \texttt{social-tagging} (2), \texttt{social-bookmarking} (1), \texttt{phd} (1), and \texttt{research} (1). In this example the values represent the weighted union of all tags.

In this thesis, we analyze and study the annotations provided by end users on social tagging systems. We present different methods to use these annotations to classify resources as accurately as possible. Specifically, we focus on the analysis of the usefulness of tags on user-driven folksonomies as a means to get an organization that resembles the categories on expert-driven taxonomies. In this context, we study several representations of social annotations, in search of an approach that resembles the classification by experts as much as possible. Especially, we focus on getting the most out of social tags, by both looking for the best representation, and measuring the impact of the distribution of tags across the triple of resources, bookmarks and users. Finally, we also study the application of state-of-the-art user behavior analysis approaches for the detection of users who rather provide tags for categorization purposes.

\section{Scope of the Thesis}
\label{sec:scope-of-the-thesis}

In this thesis, we investigate how annotations gathered together on social tagging systems can be harnessed for resource classification. Specifically, this thesis focuses on the study of several resource representation approaches using social tags. We perform the evaluation of such representations by measuring their similarity to classifications by experts. In this context, we consider the classification provided by experts as a ground truth for the evaluation process. We perform the classification experiments by using a state-of-the-art classification method, so-called Support Vector Machines. To choose the appropriate settings for the classifier, we also perform a preliminary study in this regard.

As meaningful metadata to enhance the resource classification task, we explore social tags provided by users from a statistical and distributional point of view, and we do not consider other details such as analyzing their linguistic and semantic meanings. For us, each text string representing a tag is treated as a different token, regardless of its meaning. Thereby, rather than analyzing the meaning of tags, we focus on analyzing the structure of folksonomies, represented by triples of users, bookmarks and resources.

\section{Problem Statement and Research Questions}
\label{sec:problem-statement}

The main goal of this thesis is to shed new light on the appropriate use of the great deal of data gathered on social tagging systems. Given the interest of classifying resources, and the lack of representative data in many cases, we aim at analyzing the extent to which and how social tags can enhance a resource classification task. At the beginning of this work, we found no works dealing with this insofar as no special attention had been paid at how to represent resources using social tags, and no actual classification experiments had been performed. Thus, we were motivated to carry out this research work. To that end, we set forth the following problem statement, which summarizes the main focus of this thesis:

\begin{description}
 \item[Problem Statement] \hfill \\
 \textit{\problemstatement}
\end{description}

Regarding the classification algorithm, we rely on Support Vector Machines (SVM) as a state-of-the-art classification method. Using this method, several approaches have already been proposed to work on binary and multiclass scenarios, as well as supervised and semi-supervised ones. Nonetheless, there is little work comparing different approaches in the multiclass scenario. We assume that these kinds of tasks are usually multiclass, and the number of prior annotated resources tends to be tiny as compared to the whole collection of resources. Accordingly, the first two research questions we formulate in this thesis are:

\begin{description}
 \item[Research Question 1] \hfill \\
 \textit{\rqsvmone}
 \item[Research Question 2] \hfill \\
 \textit{\rqsvmtwo}
\end{description}

Moreover, regarding social annotations, it has been shown that they provide useful metadata for improving resource management. Nevertheless, there is little work analyzing the usefulness of social tags for performing classification tasks. Preliminary analyses have shown encouraging results, and conclude that these annotations may be helpful for classification. However, they did not analyze the annotations in more depth, and it is not clear whether the representation they used was good enough.

We believe that several factors should be taken into account when representing resources using social tags. In contrast to classical document repositories, social annotations rely on a triple of users, resources, and tags, which should be analyzed in more depth for the representation task. In this context, apart from representing the resources, it is worthwhile considering that not all the tags have to be equally representative, and not all the users provide equally good annotations. In this thesis, our main goal is to deepen on the way social annotations can be used to the greatest extent in search of an accurate classification of the resources. Based on these ideas, we formulate the following research questions:

\begin{description}
 \item[Research Question 3] \hfill \\
 \textit{\rqdataone}
 \item[Research Question 4] \hfill \\
 \textit{\rqrepone}
 \item[Research Question 5] \hfill \\
 \textit{\rqreptwo}
 \item[Research Question 6] \hfill \\
 \textit{\rqrepthree}
 \item[Research Question 7] \hfill \\
 \textit{\rqdistone}
 \item[Research Question 8] \hfill \\
 \textit{\rqdisttwo}
 \item[Research Question 9] \hfill \\
 \textit{\rqcatone}
 \item[Research Question 10] \hfill \\
 \textit{\rqcattwo}
\end{description}

\section{Research Methodology}
\label{sec:research-methodology}

The research methodology we followed throughout this work includes 6 parts:

\begin{enumerate}
 \item Review of the literature and understanding of social tagging systems.
 \item Looking for an appropriate SVM classifier to perform the work.
 \item Looking for existing social tagging datasets. Since we did not find any that fulfilled our requirements, we created three large-scale social tagging datasets instead.
 \item Thinking of and proposing approaches to classifying using social tags.
 \item Evaluating the proposed approaches.
 \item Performing a thorough analysis of the results, in order to understand them for drawing conclusions.
 \item Showing and presenting partial results at several national and international conferences and workshops, in order to get useful comments and feedback from other researchers.
 \item Summarizing the research, contributions, and conclusions drawn throughout this work by writing this dissertation.
\end{enumerate}

Step 4 through 6 was an iterative process.

\section{Structure of the Thesis}
\label{sec:structure-of-the-thesis}

This thesis consists of 8 chapters. Below we provide a brief overview summarizing the contents of each of these chapters.

\begin{description}
 \item[Chapter \vref{c:introduction}] \hfill \\
 \textbf{Introduction} \\
  We present the motivation for the study on the use of social annotations for resource classification. We formalize the problem, and motivate the need of such a study.
  
 \item[Chapter \vref{c:related-work}] \hfill \\
 \textbf{Related Work} \\
  We provide a survey of previous works in the field. We summarize the advances in related fields, not only on the use of social annotations, but also on resource classification.
  
 \item[Chapter \vref{c:svm-classification}] \hfill \\
 \textbf{Support Vector Machines for Large-Scale Classification} \\
  We perform a study on different SVM approaches to the problem of classifying large-scale resource collections on multiclass taxonomies. It gives rise to the best SVM approach, which we use to perform the rest of the classification experiments along the work.
  
 \item[Chapter \vref{c:datasets}] \hfill \\
 \textbf{Generation of Social Tagging Datasets} \\
  We describe and analyze in detail the social tagging datasets we created. We detail in depth the process of creation of such datasets, and we analyze the main characteristics of the underlying folksonomies.
  
 \item[Chapter \vref{c:tag-representation}] \hfill \\
 \textbf{Representing the Aggregation of Tags} \\
  We propose and evaluate different representations of resources based on social tags for a resource classification task. We study the usefulness of social tags as compared to other data sources, and propose the best representation approach to get the most out of them. We also deal with the combination of social tags with other data sources to yield a better performance.
  
 \item[Chapter \vref{c:tag-distribution-classification}] \hfill \\
 \textbf{Analyzing the Distribution of Tags for Resource Classification} \\
  We deal with the task of considering the representativity of tags within a collection of social annotations on a social tagging system for resource classification. We study the application of weighting schemes adapted to social tagging systems, and analyze their suitability by taking into account the settings of each system.
  
 \item[Chapter \vref{c:analyzing-appropriateness-users}] \hfill \\
 \textbf{Analyzing the Behavior of Users for Classification} \\
  We explore the effect of user behavior on social tagging systems for the resource classification task. Previous works suggest the existence of two types of users: Categorizers, who use tags to categorize resources, and Describers, who use tags to describe resources. Based on these works, we study whether tags by Categorizers are better than tags by Describers for the resource classification.
  
 \item[Chapter \vref{c:conclusions}] \hfill \\
 \textbf{Conclusions and Future Research} \\
  We discuss and summarize the main conclusions and contributions of the work. We present the answers to the formulated research questions, and the outlook on future directions of the work.
\end{description}

Additionally, the thesis contains the following appendices at the end, with complementary information and summaries in other languages:

\begin{description}
 \item[Appendix \vref{c:topx-tags}] \hfill \\
 \textbf{Additional Results} \\
  We present some additional results, which we did not include in the main content of the thesis, but are also worth including to prove and help understand some conclusions.
  
 \item[Appendix \vref{c:key-terms}] \hfill \\
 \textbf{Key Terms and Definitions} \\
  We list the most relevant terms related to social tagging systems, and provide a detailed definition of them.
  
 \item[Appendix \vref{c:acronyms}] \hfill \\
 \textbf{List of Acronyms} \\
  We provide a list of the acronyms used along the work, and what they stand for.
  
 \item[Appendix \vref{c:sp-summary}] \hfill \\
 \textbf{Resumen (Spanish Summary)} \\
  We summarize the contents of this work in Spanish language.
  
 \item[Appendix \vref{c:bq-summary}] \hfill \\
 \textbf{Laburpena (Basque Summary)} \\
  We summarize the contents of this work in Basque language.
\end{description}

\section{Writing Conventions}
\label{sec:conventions}

Next, we detail some conventions we defined while writing this thesis. These conventions include formatting of text, and some issues regarding English language.

\subsection{Formatting}
\label{ssec:formatting}

In the thesis, we mention names of tags many times, either to show them as examples or to clarify some explanations. When those tags appear in the text, we use a monospaced typeface to differentiate them easily from the rest of the text. For instance: \texttt{reference}.

In the same manner, we emphasize with italic text those inline appearances of math formulas, or terms that for some reason have certain importance in the context.

\subsection{Language Issues}
\label{ssec:language-issues}

This thesis, being focused on social media, deals with users of social tagging systems at some points. When we refer to a single user, but no distinction is made between genders, we use the pronoun \textit{they} instead of either \textit{he} or \textit{she}. For instance:

\textit{When \textbf{a user} saves a bookmark, the tags annotated by \textbf{them} are added to \textbf{their} personomy.}

This is grammatically incorrect in English. However, a person's gender is explicit in the third person singular pronouns, and there is no perfect solution to this issue. Sometimes, the wording \textit{he/she} is used, but using it all along this work would become cumbersome, and would harm its readability. We rely on tips by the Oxford English Dictionary for this decision\footnote{http://www.oxforddictionaries.com/page/heshethey/he-or-she-versus-they}.

%% file: state-of-the-art.tex
\chapter{Related Work}
\label{c:related-work}

\textit{``If you have an apple and I have an apple and we exchange these apples then you and I will still each have one apple.  But if you have an idea and I have an idea and we exchange these ideas, then each of us will have two ideas.''}

--- George Bernard Shaw

\chaptersummary{This chapter introduces the previous work we found in the literature. Specifically, the works in the research areas related to this work are put together, summarized, and contextualized. Next, in Section \vref{sec:resource-classification-sota} we define and provide a background on the resource classification problem. In Section \vref{sec:svm-classification-sota} we summarize the previous efforts towards an SVM approach that enables the classification of resources within an environment where the taxonomy is multiclass (i.e., made up by more than two classes), and the number of labeled resources use to be tiny as compared to the unlabeled ones. In Section \vref{sec:benefiting-from-social-annotations-sota} we summarize the works in which annotations from social tagging systems have been profited to enhance information search, management and access. Specifically, we first summarize in Subsection \vref{ssec:social-annotations-for-information-management} the use of social annotations to enhance information management tasks. Then, we present in detail in Subsection \vref{ssec:social-text-classification} the works regarding the use of social annotations for classification. The latter is the most important topic for this thesis, but due to the novelty of the research field and the lack of work on it, it does not extend as much as the former.}

\section{Resource Classification}
\label{sec:resource-classification-sota}

Resource classification is the task of assigning categories from a predefined taxonomy to a set of resources. Formally, it consists of associating a Boolean value to each pair $\langle r_j, c_i \rangle \in R \times C$, where $R = \{ r_1, ..., r_{\arrowvert R \arrowvert}\}$ is the set of resources, and $C = \{ c_1, ..., c_{\arrowvert C \arrowvert}\}$ is the set of predefined categories. The goal of the task aims at letting the classifier give predictions by means of the function $\phi^* : R \times C \rightarrow \{T, F\}$, in such a way that it resembles as much as possible the function $\phi : R \times C \rightarrow \{T, F\}$, which defines the ideal classification of the resources \citep{sebastiani02machine}. Upon this, several settings can define a different classification approach. Next, we briefly define the main settings.

Usually, a classification task comprises two subsets of resources when it relies on a machine learning approach. Some of the resources are already labeled with corresponding categories, and others are unlabeled. The former are used by the classifier to learn the characteristics of each category, creating a model for each category after the learning process. The latter are the instances to be predicted by the classifier. Relying on the models created during the learning phase, the classifier provides a category for each unlabeled resource as its prediction.


\subsection{Binary and Multiclass Classification}
\label{ssec:binary-vs-multiclass}

As regards to the taxonomies with predefined categories considered in the classification, where the resources are organized, the task is said to be either binary or multiclass. Even though the sole apparent difference is the number of classes making up the taxonomy --2 for the binary, and 3 or more for the multiclass--, the tasks tend to follow a different goal.

A binary classification is usually part of a filtering process, where the classes are the positive and the negative case. These tasks aim at separating the resources that want to be considered from the resources that want to be ruled out. For instance, a common binary classifier used as a filter is an email application that keeps the interesting messages in the inbox, whereas it sends the unwanted stuff to the spam folder.

A multiclass classification involves a larger taxonomy, and is usually used on thematic classification, i.e., where the categories represent the aboutness of the underlying resources. This kind of classification enables to organize resources into groups of related matters, and it has several applications such as creating directories of resources to ease later browsing, providing customized suggestions by users' topics of interest, or allowing to handle resources from different categories in a separate way, among many others.

In this thesis, we deal with thematic classification and, thus, we consider the task to be multiclass.

\subsection{Single-label vs Multilabel Classification}
\label{ssec:single-vs-multilabel}

The number of categories or labels that can be assigned to each resource is another setting that describes a resource classification. The number of labels for a single resource can be constrained to just one, thus becoming a single-label task, or it must be extended to allow more labels or even unlimited, when it is called multilabel. In practice, it refers to whether a resource can be related to several categories, or it can be included into just one. Besides the classification task itself, this feature also modifies the subsequent organization and browsing of categories.

In this thesis, we focus on single-label classification, mainly because the taxonomies we use as the ground truth provide this kind of categorization data.

\subsection{Semi-supervised vs Supervised Classification}
\label{ssec:semi-vs-supervised}

With regard to the learning method used by the classifier, it can vary in the instances considered to learn and create the model. A supervised learning method learns from the instances in the training set, and creates a model from them. A semi-supervised goes further by also considering unlabeled instances in the learning method. After creating the model from labeled instances, it includes its predictions on the unlabeled instances in the learning process enabling an incremental evolution of the model. The latter is especially useful when the training set is small, and the lack of sufficient learning data is worth an upsize of the labeled data.

Based upon these two learning methods, we summarize the related work on the use of SVM classifiers in the next section. We rely on SVM as a state-of-the-art classification algorithm widely used in the field.

\section{Support Vector Machines for Classification}
\label{sec:svm-classification-sota}

\begin{figure}[ht]
\begin{center}
 \includegraphics[width=90mm,clip]{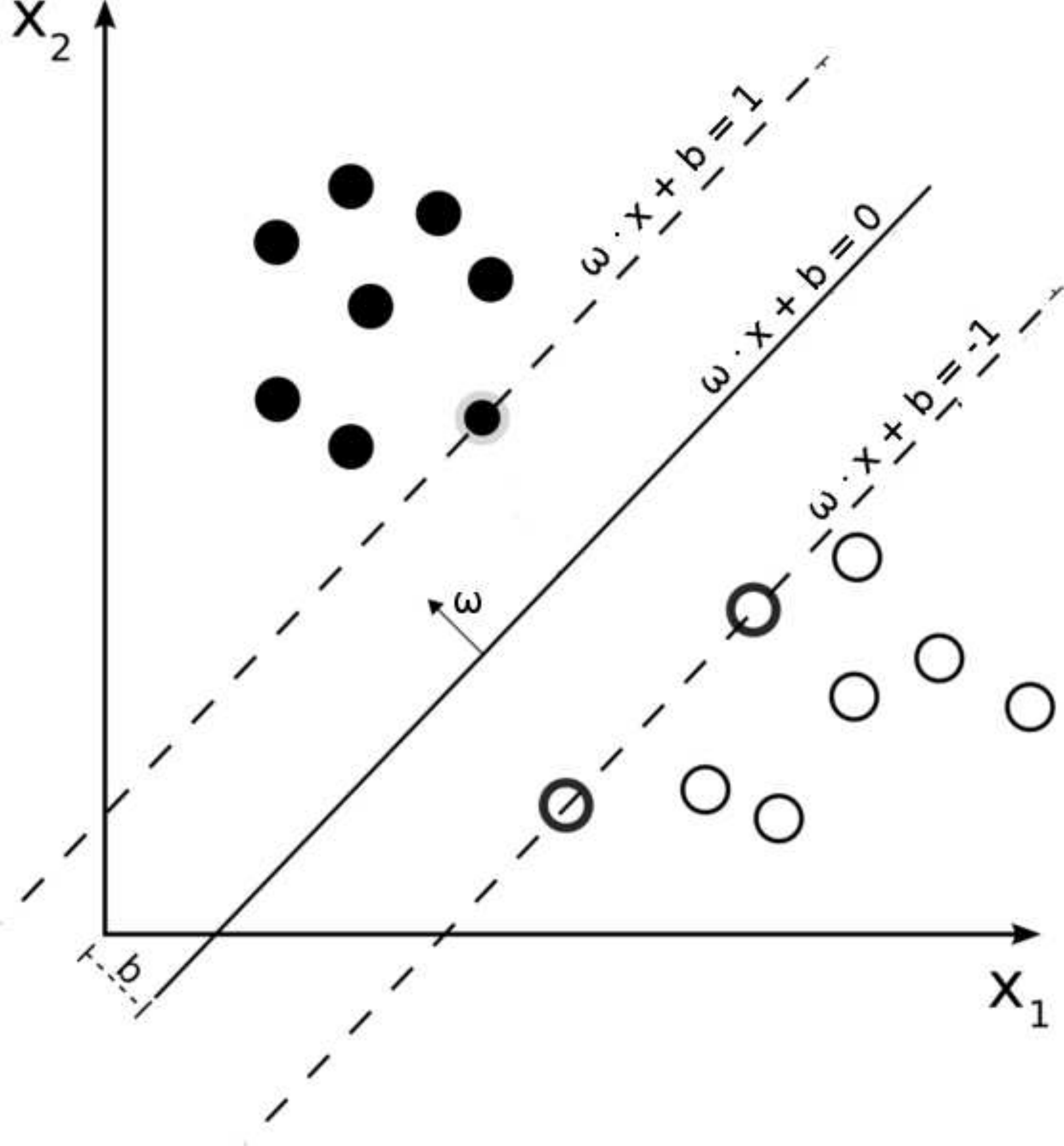}
\end{center}
\caption[Example of binary SVM classification]{An example of binary SVM classification, separating two classes (black dots from white dots). Source: Wikimedia Commons.}
\label{fig:svm-function}
\end{figure}

In the last decade, SVM has become one of the most widely studied techniques for text classification, due to the positive results it has shown. This technique uses the vector space model to represent the resources, and assumes that resources in the same class should fall into separable spaces of the representation. Upon this, it looks for a hyperplane that separates the classes; therefore, this hyperplane should maximize the distance between it and the nearest resources, which is called the margin. Equation \ref{eq:basic-svm} defines such a hyperplane (see Figure \vref{fig:svm-function}).

\begin{equation}
  f(\mathbf{x}) = \mathbf{w} \cdot \mathbf{x} + b
  \label{eq:basic-svm}
\end{equation}

In order to resolve this function, though, all the possible values should be considered and, after that, the values of $w$ and $b$ that maximize the margin should be selected; this would be computationally expensive. The equivalent Equation \ref{eq:svm} is thus used to relax it \citep{boser92training, cortes95supportvector}:

\begin{equation}
  \min \left[ \frac{1}{2} ||\mathbf{w}||^2 + C \sum_{i = 1}^l \xi_i^d \right]
  \label{eq:svm}
\end{equation}

Subject to:

\begin{eqnarray}
  \quad y_{i} (\mathbf{w} \cdot \mathbf{x_{i}} + b) \ge 1 - \xi_{i},\mbox{ }\xi_{i} \ge 0 \nonumber
  \label{eq:svm-subject-to}
\end{eqnarray}

where \emph{C} is the penalty parameter, \emph{$\xi_{i}$} is an stack variable for the \emph{i}$^{th}$ resource, \emph{l} is the number of labeled resources, and \emph{d} is the sigma parameter which defines the non-linear mapping from the input space to some high-dimensional feature space.

When the value of \emph{d} is set to 1, this function can only solve linearly separable problems. The use of a kernel function is sometimes required for the redimension of the space. This redimension creates a new space with higher number of dimensions, which enables a linear separation. After that, the redimension is undone, so the hyperplane will be transformed to the original space, respecting the classification function. Best-known kernel functions include linear, polynomial, radial basis function (RBF) and sigmoid, among others. Different kernel functions' performance has been studied in \cite{scholkopf99advances} and \cite{kivinen02online}. Linear kernel is most widely used for text classification.



Note that the function above can only resolve binary and supervised problems, so different variants are necessary to handle semi-supervised or multiclass tasks.

\subsection[Semi-supervised Learning for SVM (S3VM)]{Semi-supervised Learning for SVM (S$^{3}$VM)}
\label{s3vm}

\begin{figure}[ht]
\begin{center}
 \includegraphics[width=60mm,clip]{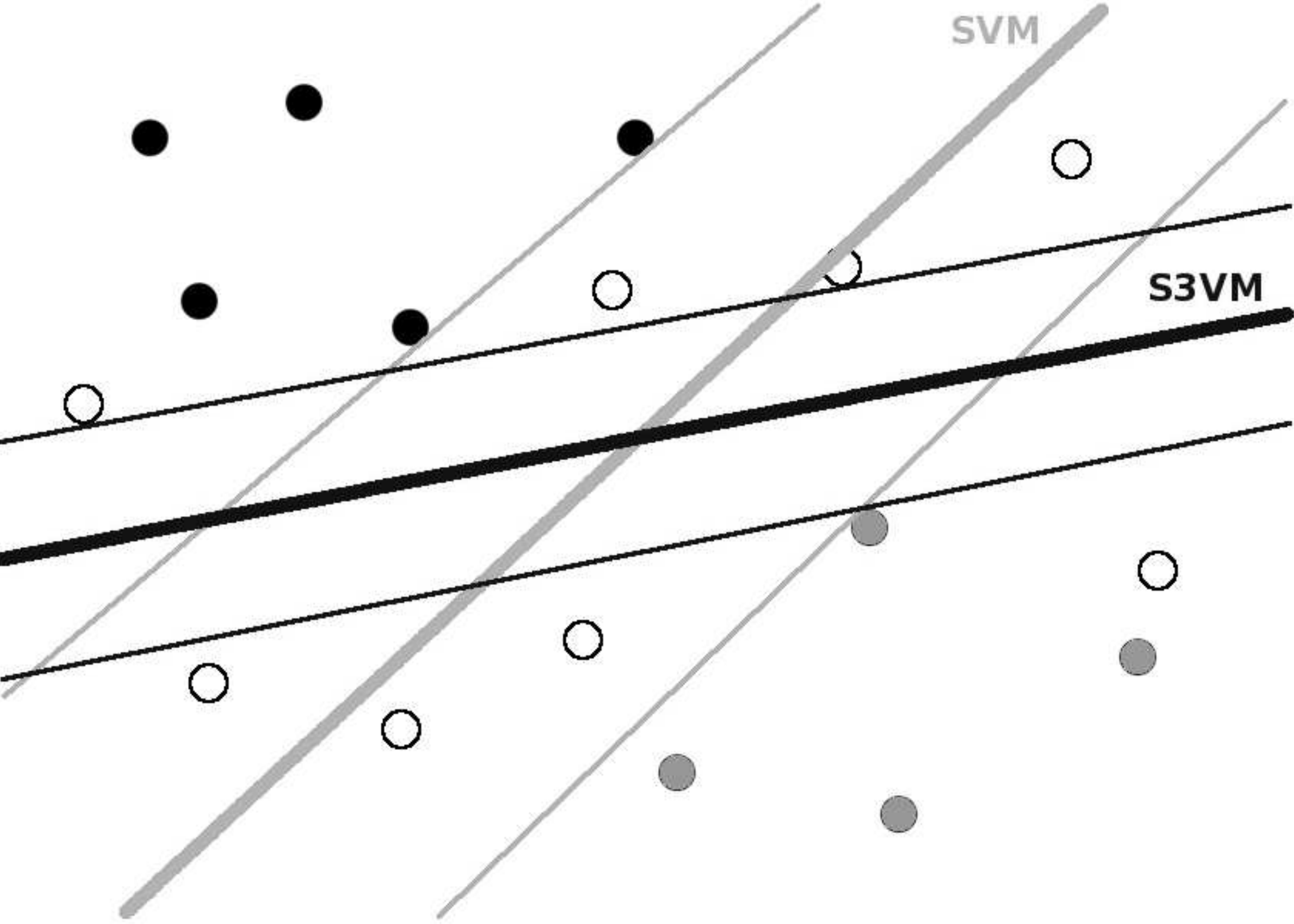}
\end{center}
\caption[Supervised vs Semi-Supervised SVM]{SVM vs S$^{3}$VM, where black and grey dots are labeled resources, and white dots are unlabeled resources. It can be seen that the few labeled resources give rise to a certain separation (grey line, SVM), whereas including unlabeled ones helps infer a more accurate separation (black line, S$^{3}$VM).}
\label{fig:svm-vs-s3vm}
\end{figure}

Semi-supervised learning approaches differ in the learning way of the classifier. As opposed to supervised approaches, unlabeled data is used during the learning phase. Taking into account unlabeled data to learn can help improve the performance of supervised classifiers, especially when its predictions provide new useful information, as shown in Figure \ref{fig:svm-vs-s3vm}. However, the noise added by incorrect predictions can worsen the learned model and, therefore, the performance of the classifier. This makes interesting the study on whether relying on semi-supervised approaches is suitable for a certain kind of task.

Semi-supervised learning for SVM, also known as S$^{3}$VM, was first introduced by \cite{joachims99transductive} in a transductive way, by modifying the original SVM function. To do that, the author proposed to add an additional term to the optimization function (see Equation \ref{eq:s3vm}).

\begin{equation}
  \min \left[ \frac{1}{2} \cdot ||\mathbf{w}||^{2} + C \cdot \sum_{i = 1}^{l} \xi_{i}^{d} + C^{*} \cdot \sum_{j = 1}^{u} \xi_{j}^{*^{d}} \right]
  \label{eq:s3vm}
\end{equation}

where \textit{u} is the number of unlabeled data, and the parameters with an asterisk (*) refer to the unlabeled instances included in the learning phase.

Nevertheless, the adaptation of SVM to semi-supervised learning significantly increases its computational cost, due to the non-convex nature of the resulting function, and so obtaining the minimum value is even more complicated. In order to relax the function, convex optimization techniques such as semi-definite programming are commonly used \citep{xu07efficient}, where minimizing the function gets much easier.

By means of this approach, \cite{joachims99transductive} demonstrated a large performance gap between the original supervised SVM and his proposal for a semi-supervised SVM, in favor of the latter one. He showed that for binary classification tasks, the smaller is the training set size, the larger gets the difference among these two approaches. He used the Reuters-21578, Ohsumed and WebKB datasets for that purpose. Although he worked with multiclass datasets, he split the problems into smaller binary ones, and so he did not demonstrate whether the same performance gap occurs for multiclass classification. More recently, \cite{chapelle08optimization} presented a comprehensive review of advances in binary S$^{3}$VM approaches.

\subsection{Multiclass SVM}
\label{multiclass-svm}

Due to the dichotomic nature of SVM, it came up the need to implement new methods to solve multiclass problems, where more than two classes must be considered. Different approaches have been proposed to achieve this. On the one hand, as a native approach, \cite{weston98multiclass} proposed modifying the optimization function getting into account all the \textit{k} classes at once (see Equation \ref{eq:multiclass-svm}).

\begin{equation}
  \min \left[ \frac{1}{2} \sum_{m=1}^{k} ||\mathbf{w_{m}}||^2 + C \sum_{i = 1}^l \sum_{m \neq y_{i}} \xi_{i}^m \right]
  \label{eq:multiclass-svm}
\end{equation}

Subject to:

\begin{eqnarray}
  \mathbf{w_{y_{i}}} \cdot \mathbf{x_{i}} + b_{y_{i}} \ge \mathbf{w_{m}} \cdot \mathbf{x_{i}} + b_{m} + 2 - \xi_{i}^m, \xi_{i}^m \ge 0 \nonumber
  \label{eq:multiclass-svm-subject-to}
\end{eqnarray}

The main novelty of this approach, as compared to previous ones, is that apart from the choice of a kernel, it is parameterless. Their experiments on benchmark datasets from the UCI repository show results similar to SVMs which have been tuned to have the best choice of parameter.

On the other hand, the original binary SVM classifier has usually been combined to obtain a multiclass solution. As combinations of binary SVM classifiers, two different approaches to \textit{k-class} classifiers can be emphasized \citep{hsu01comparison}:

\begin{itemize}
 \item \textit{one-against-all} constructs \textit{k} classifiers defining that many hyperplanes, each of them separating the class \textit{i} from the rest \textit{k-1}. For instance, for a problem with 4 classes, the following classifiers would be created: \textit{1 vs 2-3-4}, \textit{2 vs 1-3-4}, \textit{3 vs 1-2-4} and \textit{4 vs 1-2-3}. Unlabeled resources will be categorized in the class of the classifier that maximizes the margin: $\hat{C}_{i} = \arg\max_{i = 1,...,k} (w_{i} x + b_{i})$.  As the number of classes increases, the amount of classifiers will increase linearly.

 \item \textit{one-against-one} constructs \textit{$\frac{k (k - 1)}{2}$} classifiers, one for each possible category pair. For instance, for a problem with 4 classes, the following classifiers would be created: \textit{1 vs 2}, \textit{1 vs 3}, \textit{1 vs 4}, \textit{2 vs 3}, \textit{2 vs 4} and \textit{3 vs 4}. After that, it classifies each new document by using all the classifiers, where a vote is added for the winning class over each classifier; the method will propose the class with more votes as the result. As the number of classes increases, the amount of classifiers will increase in an exponential way, and so the problem could became very expensive for large taxonomies.
\end{itemize}

Both \cite{weston98multiclass} and \cite{hsu01comparison} compare the native multiclass approach to the \textit{one-against-one} and \textit{one-against-all} binary classifier combining approaches. They agree concluding that the native approach does not outperform the results by \textit{one-against-one} or \textit{one-against-all}, although it considerably reduces the computational cost because the number of support vector machines it defines is smaller. Among the binary combining approaches, they show the performance of \textit{one-against-one} to be superior to \textit{one-against-all}.

Although these approaches have been widely used in supervised learning environments, they have scarcely been applied to semi-supervised learning. Accordingly, we believe that the study on its applicability and performance for this type of problems is necessary before proceeding with additional experiments.

\subsection[Multiclass S3VM]{Multiclass S$^{3}$VM}
\label{multiclass-s3vm}

When the taxonomy is defined by more than two classes and the number of previously labeled documents is very small, the combination of both multiclass and semi-supervised approaches could be required, i.e., a multiclass S$^{3}$VM approach. A common web page classification problem meets these characteristics, with a taxonomy of more than two categories, and it could be helpful to increase the tiny amount of labeled documents by including predictions on unlabeled data for the learning phase.

However, little work has been done on transforming SVM into both a semi-supervised and multiclass approach, and especially on comparing them to other approaches. As a native approach, \cite{yajima06optimization} modified the original SVM function by fitting it to multiclass semi-supervised tasks (see Equation \ref{eq:multiclass-s3vm}).

\begin{equation}
  \min \frac{1}{2} \sum_{i = 1}^h \beta^{i^{T}} K^{-1} \beta^{i} + C \sum_{j = 1}^l \sum_{i \neq y_{j}} \max \{0, 1 - (\beta_{j}^{y_{j}} - \beta_{j}^i)\}^2
  \label{eq:multiclass-s3vm}
\end{equation}

where $\beta$ represents the product of a vector of variables and a kernel matrix defined by the author.

The authors showed that the proposed approach outperformed other non-SVM algorithms, but they did not show if it was better than other SVM settings. As far as we know, the software was not made publicly available, and no further work has been done using this approach.

\cite{chapelle06continuation} present another native multiclass S$^{3}$VM approach by using the Continuation Method. This is the only work, to the best of our knowledge, where \textit{one-against-all} and \textit{one-against-one} approaches had been tested in a semi-supervised environment. They apply these methods to news datasets, yielding worse performance. Moreover, they show that \textit{one-against-one} is not sufficient for real-world multiclass semi-supervised learning, since the unlabeled data cannot be restricted to the two classes under consideration.

On the other hand, others relied on combining SVM with other algorithms in search of a multiclass semi-supervised SVM approach. \cite{qi04multiclass} use Fuzzy C-Means (FCM) to predict labels on unlabeled resources. After that, multiclass SVM is used to learn with the augmented training set, classifying the test set. \cite{xu05unsupervised} rely on a clustering-based approach to label the unlabeled data. Afterwards, they apply a multiclass SVM classifier to the labeled training set. 

It is worthwhile noting that most of the above works introduced their approaches and only compared them to other semi-supervised classification methods, such as Expectation-Maximization (EM) or Naive Bayes. As an exception, \cite{chapelle06continuation} compared a semi-supervised and a supervised SVM approach, but only over image datasets. In this thesis, we do not aim at proposing new SVM approaches. However, we believe that evaluating and comparing multiclass SVM and multiclass S$^{3}$VM approaches is necessary to conclude with a suitable approach. This would help discover whether learning upon unlabeled resources is helpful for multiclass problems when using SVM as a classifier.

\section{Benefiting from Social Annotations}
\label{sec:benefiting-from-social-annotations-sota}

Since it was introduced along with the Web 2.0 phenomenon in the early 2000s, social annotations have gained popularity and interest with the creation of well-known social tagging sites like Delicious. This section summarizes how social annotations have helped improve information access and management. Among the existing works, it is especially focused on their use for classification tasks.

Social tagging systems arose as an idea of Joshua Schachter, founder of Delicious \citep{smith_tagging_2008}. In late 1990s, after the bookmarks in his web browser had overflowed, he used to save his favorite URLs in a text file, with an entry per line. Each entry was a URL, followed by a set of tags. These tags enabled him to easily refind the URL he was looking for. He just had to filter by keyword to search for a URL. He also published online such a list at Muxway.org (currently discontinued and not accessible). Later, in September 2003, he released Delicious, the first online tool that enabled saving and tagging URLs as he used to, but enhanced by a social environment. This social tool enabled users to search among saved URLs, not only by their own tags, but also taking advantage of others'.

The research on social tagging systems did not arise until 2006. An early work by \cite{golder_structure_2006} performed a study of the characteristics of Delicious, followed by an increasingly interest of researchers that gave rise to large number of research works in the field. Next, we focus on some of the most relevant advances on the use of social tags for information management, and go in more depth for the specific task of resource classification.

\subsection{Social Annotations for Information Management}
\label{ssec:social-annotations-for-information-management}

Social annotations have been widely used for the sake of information management tasks. They have shown to very useful for several tasks in which the availability of data is of utmost importance \citep{gupta2010survey}:

\begin{itemize}
 \item \textbf{Search:} Social tags have been successfully applied to web search. \cite{bao2007optimizing} found that social annotations can enhance web search (1) as a good summaries of corresponding web pages, and (2) as a way to compute the popularity of web pages by considering the number of users who annotate them. \cite{heymann_can_social_2008} analyzed the usefulness of tags from Delicious for web search, and concluded that these metadata can provide additional and meaningful data not available in other sources, though it may currently lack the size to get a significant impact. Also, \cite{dmitriev2006using} showed the usefulness of social annotations for improving the quality of intranet search. As a specific approach for searching on social tagging systems, \cite{hotho2006informationretrieval} proposed FolkRank, a search algorithm that fits the structure of folksonomies. They found this approach useful for providing personalized rankings of the resources in a folksonomy, as well as for finding communities of users within these systems.
 \item \textbf{Recommender Systems:} \cite{shepitsen2008} and \cite{li2008} introduced recommendation algorithms based on user-generated tags. They show that social annotations are effective to discover user interests and, therefore, to recommend them new resources. In \cite{bogers2008recommending}, the authors take advantage of annotations provided by users on a social reference manager for recommending research papers to scientists. In \cite{cantador2011categorising}, the authors present a mechanism to automatically filter and classify social tags in a set of purpose-oriented categories, so that they can rely on suitable tags to recommend resources to users.
 \item \textbf{Enhanced Browsing:} Social annotations can be helpful to improve the navigation of resources as well. \cite{smith_tagging_2008} describes three new navigation ways emerged from folksonomies: pivot browsing (moving through an information space by choosing a reference point to browse), popularity-driven navigation (retrieving the resources that are popular for a given tag), and filtering (social tagging allows to separate the resources you do not want from the resources you do want). In a preliminary study, we integrated tags from Delicious to Wikipedia \citep{zubiaga2009enhancing}. Tags provided new data that was not available in the content of encyclopedia articles, providing a means to enhance the navigation and search on the site.
 \item \textbf{Clustering and Classification:} Social tags have also shown to be useful for resource organization tasks, including clustering and classification. This point will be explained in more depth in the next section.
\end{itemize}

\subsection{Social Annotations for Classification}
\label{ssec:social-text-classification}

Before we began working on this thesis, there was little work dealing with the analysis of the applicability and usefulness of social tags for resource classification tasks. Most of them had focused on classifying web pages, and had just explored the appropriateness of social tags for this kind of tasks. However, none of them performed real classification experiments but just statistical analyses. Accordingly, they did not further explore on how to get the most out social tags in order to improve the performance.

\cite{noll_exploring_2008} presented a study of the characteristics of social annotations provided by end users, in order to determine their usefulness for web page classification. In this work, the authors weight the tags by normalizing the number of users annotating them. The least popular tag is given a value of 0, whereas the most popular is given a value of 1. This way, they remove those least popular tags as they were useless. Moreover, they did not pay attention at whether or not this representation approach was appropriate to carry out the task. The authors matched user-supplied tags of a page against its categorization by the expert editors of the Open Directory Project (ODP). They analyzed at which hierarchy depth matches occurred, concluding that tags may perform better for broad categorization of documents rather than for more specific categorization. The study also points out that since users tend to bookmark and tag top level web documents, this type of metadata will target classification of the entry pages of websites, whereas classification of deeper pages might require more direct content analysis. They observed that in the power law curve formed by the popularity of social tags, not only popular tags, but also the tags in the tail provide helpful data for information retrieval and classification tasks in general. In a previous work, the same authors \citep{noll_authors_2007} suggested that tags provide additional information about a web page, which is not directly contained within its content.

Also, \cite{noll_metadata_2008} studied three types of metadata about web documents: social annotations (tags),
anchor texts of incoming hyperlinks, and search queries to access them. They concluded that tags are better suited for classification purposes than anchor texts or search keywords.


As regards to clustering tasks, \cite{ramage_clusteringtagged_2009} included tagging data for improving the performance of two clustering algorithms when compared to content-based clustering. They found that tagging data was more effective for specific collections than for a collection of general documents. They weighted the tags by both using the number of users annotating them, and reweighting this value considering the Inverse Document Frequency (IDF) value of the tag across the resources in the collection. They showed a superiority for the use of the IDF weighting scheme.

Even though those were the only works published by then, the interest on this research area has increased lately. After the presentation of our earliest work in the field \citep{zubiaga2009getting}, more scientists have shown their interest on it, and have presented new works.

In \cite{aliakbary_webpage_2009}, the authors integrated social annotations as an approach to extending web directories. They relied on the number of users annotating each tag as a weight. Upon that, they created a model vector for each category, and computed the cosine similarity to new web pages to generate predictions. They observed that the annotations provided a multi-faceted summary of the web pages, and that they better represent the aboutness of web pages than the content itself. Also, they conclude that the more users annotate a URL, the better it is classified.

In another work where social tags were exploited for the benefit of web page classification, \cite{godoy2010exploiting} also showed the usefulness of social tags for web page classification, which outperformed classifiers based on full-text of documents. Similar to our previous work \citep{zubiaga2009getting}, they compare tag-based resource representations relying on all the tags and the top 10 of tags for each resource, corroborating our findings that the former performs better. Going further, they concluded that stemming the tags reduces the performance of such classification, even though some operations such as removal of symbols, compound words and reduction of morphological variants have a discrete positive impact on the task.

\cite{xia2010optimizing} studied the usefulness of social tags as a complementary source for improving the classification of academic conferences into corresponding topics. Using tagging data gathered from WikiCFP\footnote{http://www.wikicfp.com}, and weighing the tags according to the number of users annotating them, they compare the classification of conferences by using only the content of the call for papers, and by integrating tagging data along with it. Their experiments yielded slightly better performance for the integration of social tags, with roughly 1\% improvement.

With regard to the classification of resources other than web pages, \cite{lu2010user} present a comparison of
tags annotated on books and their Library of Congress subject headings. Actually, no classification experiments are
performed, but a statistical analysis of the tagging data shows encouraging results. By means of a shallow analysis of the distribution of tags across the subject headings, they conclude that user-generated tags seem to provide an
opportunity for libraries to enhance the access to their resources.

Using a graph-based approach, in \cite{yin2009exploring} the authors present a method to classify products from Amazon into their corresponding categories using social tags. They conclude that social tags can enhance web products classification by representing them in a meaningful feature space, interconnecting them to indicate relationship, and bridging heterogeneous products so that category information can be propagated from one domain to another.

There is also a set of works dealing with user behavior in social tagging systems. Even though they do not perform classification experiments, they suggest the existence of a subset of users in these systems who are rather categorizing the resources when annotating. Specifically, early works such as \cite{hammond2005social} and \cite{marlow2006ht} suggest the existence of two types of users: on one hand, users can be motivated by categorization (so-called \emph{Categorizers}). These users view tagging as a means to categorize resources according to some (shared or personal) high-level conceptualizations. On the other hand, users who are motivated by description (so-called \emph{Describers}) view tagging as a means to accurately and precisely define the content of resources. These proposals have been further studied in \cite{koerner2010stop} and \cite{koerner2010categorizers}. However, they focused on showing that the difference of motivation among those two kinds of users actually exists, and they did not pay attention at whether Categorizers are better suited to the classification task.

Also, there has recently been an increasingly interest in using social tags for the benefit of clustering tasks. In \cite{lu2009exploit} the authors not only cluster the annotated resources, but also users and tags. \cite{zhang2009clustering} found that the effectiveness of clustering blog posts using tags from a simple tagging system was quite limited, and they combined these data with relations in the blogosphere to get better results.

So far, there is little work on the analysis of folksonomies for the classification task. A few works have shown their suitability for this purpose, but no special attention has been paid into further studying these metadata structures. In this thesis, we aim at analyzing these structures to find an approach to amalgamate and represent the tags in order to perform a resource classification task with a high accuracy.

%% file: svm-classification.tex
\chapter{Support Vector Machines for Large-Scale Classification}
\label{c:svm-classification}

\textit{``Science is the systematic classification of experience.''}

--- George Henry Lewes

\chaptersummary{We study the appropriateness of several SVM approaches for large-scale classification in this chapter. We are not going to introduce any new approaches in it, but to perform a preliminary comparison study among different approaches, in search of a suitable approach for large-scale classification tasks. We also evaluate the real contribution of unlabeled data for multiclass SVM-based classification tasks. Specifically, we compare a native multiclass approach to the combination of binary classifiers, as well as a supervised to a semi-supervised approach. We carry out such an experimentation by using three web page datasets. The Web is a good example of the problem we are dealing with, where the number of resources is very large, and the number of labeled ones tends to be tiny as compared to the whole collection. The chapter is organized as follows. Next, in Section \vref{sec:definition-large-scale-classification} we define and present the features of a large-scale classification task. We enumerate and describe the SVM approaches compared in our study in Section \vref{sec:compared-svm-approaches}. Then, in Section \vref{sec:svm-experiment-settings} we detail the settings of the experiments, showing their results in Section \vref{sec:svm-results}. Finally, we discuss the results in Section \vref{sec:svm-discussion}, and conclude and answer the following research questions in Section \vref{sec:svm-conclusions}:}

\chaptersummary{
\begin{description}
 \item[Research Question 1] \hfill \\
 \textit{\rqsvmone}
\end{description}
}
\chaptersummary{
\begin{description}
 \item[Research Question 2] \hfill \\
 \textit{\rqsvmtwo}
\end{description}
}

\section{Definition of Large-Scale Classification}
\label{sec:definition-large-scale-classification}

The Web comprises lots of collections of web documents and other resources that scale up constantly. The increasingly amount of resources on the Web has in part influenced the recent upsize of research on large-scale datasets. Accordingly, extending earlier studies on SVM classification to large-scale collections rises in importance.

In this regard, we believe that a thematic classification task commonly meets the following conditions when it comes to large-scale collections of resources:

\begin{itemize}
 \item \textbf{Tininess of the training set:} getting a manual classification of a subset as a training set is very expensive and entails a lot of time and effort. Thus, the previously categorized subset will be tiny as compared to the uncategorized subset. This suggests considering semi-supervised approaches besides supervised ones, as a way of extending the training set.
 \item \textbf{Multiclass taxonomy:} a taxonomy is usually composed of more than two categories, and thereby it is considered as a multiclass task instead of a binary one.
\end{itemize}

We assume that the large-scale thematic classification tasks we are undertaking in this thesis will fulfill those two features.

\section{Compared SVM Approaches}
\label{sec:compared-svm-approaches}

The two features showed above encourage the study of different SVM settings to conclude with a suitable one. On the one hand, the tininess of the training set requires analyzing whether or not it is worthwhile relying on a semi-supervised approach extending it instead of a supervised one. An early work by \cite{joachims99transductive} showed the outperformance of the former for binary classification, but it is not clear whether or not the same happens for multiclass scenarios. Especially because labeling instances on a larger taxonomy is harder, and it seems much likelier to introduce noise in the extension of the training set. On the other hand, the classification on a multiclass taxonomy can be faced up in two different ways: as a single multiclass task, or as several smaller binary tasks. Little work has been done comparing these two settings, though. The lack of analyses on the appropriateness of the aforementioned methods for the task conveyed us to perform such a study prior to getting to work on large-scale classification tasks.

Our study involves both supervised SVM and semi-supervised SVM (S$^{3}$VM) approaches on different multiclass settings. Specifically, we rely on three multiclass settings which were introduced in Section \vref{multiclass-svm}: a native multiclass algorithm, and \textit{one-against-one} and \textit{one-against-all} based on binary classifiers. When naming the approaches, we add a suffix \textit{-mSVM}, \textit{-SVM} or \textit{-S$^{3}$VM} for clarity, depending if they are multiclass, supervised or semi-supervised, respectively.

\begin{figure}[htb]
 \centering
 \includegraphics[width=350px]{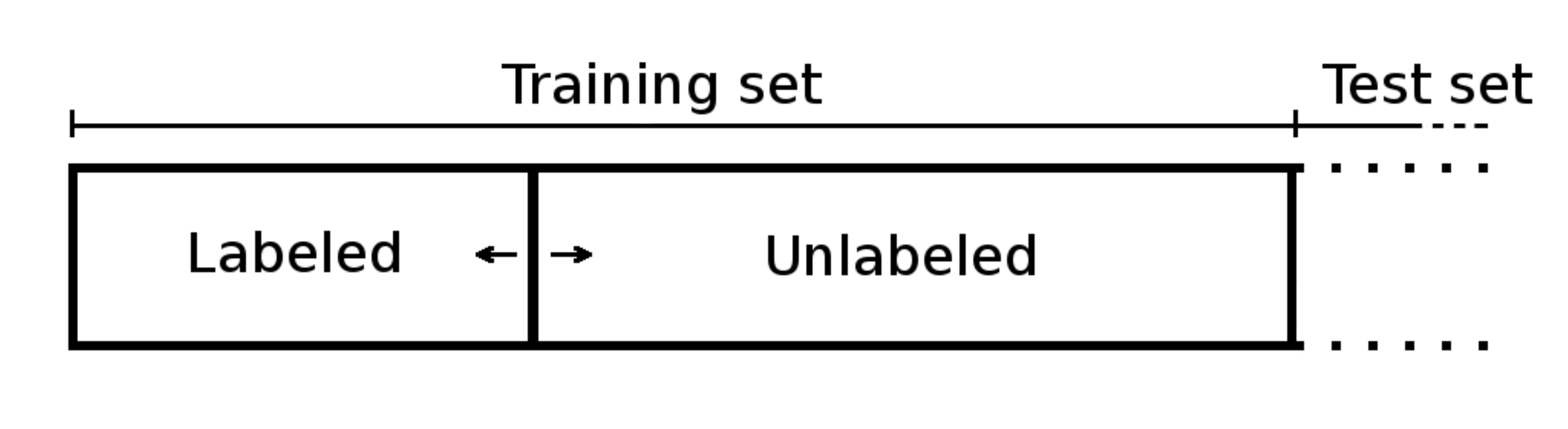}
 \caption[Splitting a training set into labeled and unlabeled subsets]{\small Example of splitting a training set into labeled and unlabeled subsets. The former remains fixed, whereas the size of the latter two changes.}
 \label{fig:training-set}
\end{figure}

In order to compare a supervised approach and different levels of semi-supervision, we created several subsets of labeled and unlabeled instances within the training sets. This enables to analyze different levels of semi-supervision. While the size of the training set remains fixed, smaller subsets of labeled instances in it yield a rather semi-supervised approach. The size of the labeled subset ranges from 50 instances to the whole training set. Figure \vref{fig:training-set} shows how we split training sets into labeled and unlabeled subsets. Supervised approaches learn from the labeled subset, and ignore the unlabeled one, whereas semi-supervised approaches make predictions on the latter to increase the learning base. For each training set size, we perform 6 different selections of labeled subsets. We show the average accuracy of all 6 runs on the results.

\subsection{Native Multiclass Approaches}
\label{ssec:native-multiclass-approaches}

Native multiclass approaches consider the classification as a single task performed by only one classifier. They can be implemented either in a supervised or a semi-supervised basis. However, little work has been done on developing native semi-supervised approaches. The only algorithm was presented by \cite{yajima06optimization}, which as far as we know has not been used in later works. The algorithm is not available for the community, and its implementation does not seem feasible because its description does not provide enough details to reproduce it. Thus, we propose the implementation of a semi-supervised method by following an approach similar to those by \cite{qi04multiclass} and \cite{xu05unsupervised}. They perform a two-step classification task, by extending the training set using a clustering algorithm in the first step. Afterward, they run a supervised SVM on the extended training set. Our approach differs from those two in that we use the same algorithm in both steps. With an approach we call \emph{2-steps-mSVM}, we extend the training set using a supervised SVM, i.e., learning from the labeled subset, and labeling the unlabeled subset relying on classifier's decisions. We run the same algorithm on the extended set after that. As a supervised method, we use a native multiclass SVM, which we call \emph{1-step-mSVM}.

\subsection{One-Against-All Approaches}
\label{ssec:one-against-all-approaches}

\emph{One-against-all} is a method to split the multiclass task into smaller binary problems. Specifically, it creates \emph{k} classifiers defining that many hyperplanes; each of them separates the class \emph{i} from the remainder \emph{k-1}. Thus, the number of classifiers is the same as the number of classes. In the test phase, all the classifiers will provide a margin for each instance, defining whether it belongs to the positive class (class \emph{i}) or the negative class (the remainder \emph{k-1}). Putting together the outputs of all classifiers for an instance (i.e., margins provided by classifiers), the one with the largest positive value will be selected as the system's decision (see Equation \vref{eq:one-against-all-solution}).

\begin{equation}
 \hat{C}_{i} = \arg\max_{i = 1,...,k} (w_{i} x + b_{i})
 \label{eq:one-against-all-solution}
\end{equation}

We implemented this approach with a supervised binary SVM (\emph{one-against-all-SVM}) and a semi-supervised binary SVM (\emph{one-against-all-S$^{3}$VM}).

\subsection{One-Against-One Approaches}
\label{ssec:one-against-one-approaches}

\emph{One-against-one} is another method that divides a multiclass problem into smaller binary ones. Different from the above method, it creates a binary classifier for each possible pair among the \emph{k} categories, what produces \emph{$\frac{k (k - 1)}{2}$} 1-vs-1 classifiers. Again, a margin for all the instances is given by all the classifiers in the test phase, but the way of amalgamating the outputs changes in this case. Considering as a positive vote each time that a class beats the other in the binary classifiers, the class with most positive votes will be predicted by the system.

This method has two major problems, though:
\begin{enumerate}
 \item As the number of classes increases, the amount of classifiers will increase in an exponential way, and so the problem could become very expensive for large taxonomies.
 \item During the test phase, a 1-vs-1 classifier is unable to ignore those instances that actually belong to none of the considered pair of classes. Thus, including all the instances in the test phase is the only solution, and given that binary classifiers will provide a margin for every instance, it seems that the test phase can become noisy. This issue was also pointed out by \cite{chapelle06continuation}.
\end{enumerate}

As for the one-against-all approaches, both a supervised binary SVM and a semi-supervised binary SVM were used to implemented two different settings of this approach: \emph{one-against-one-SVM} and \emph{one-against-one-S$^{3}$VM}.

\section{Experiment Settings}
\label{sec:svm-experiment-settings}

This section introduces the datasets we have used to compare the different SVM approaches, as well as other settings.

\subsection{Datasets}
\label{ssec:datasets}

In order to perform the experimentation on a multiclass scenario, we looked for suitable datasets. As benchmark datasets that have been used several times for research on classification, we chose the following:

\begin{itemize}
 \item \textit{BankSearch} \citep{sinka02large}, a collection of 11,000 web pages over 11 classes, with very different topics: commercial banks, building societies, insurance agencies, java, c, visual basic, astronomy, biology, soccer, motorsports and sports. We removed the category sports, since it includes both soccer and motorsports in it, and it is not at the same level as the rest of categories. This results in 10,000 web pages over 10 categories. 4,000 instances were assigned to the training set, while the other 6,000 were left on the test set.
 \item \textit{WebKB}\footnote{http://www.cs.cmu.edu/afs/cs.cmu.edu/project/theo-20/www/data/}, with a total of 4,518 documents from 4 universities, and classified into 7 classes (student, faculty, personal, department, course, project and other). The class named \textit{other} was removed due to its ambiguity, and so we finally got 6 classes. 2,000 instances fell into the training set, and 2,518 into the test set.
 \item \textit{Yahoo! Science} \citep{tan02use}, with 788 scientific documents classified into 6 classes (agriculture, biology, earth science, math, chemistry and others). We selected 200 documents for the training set, and 588 for the test set.
\end{itemize}

Even though these cannot quite be considered as large-scale datasets, the fact that the selected training sets are small as compared to whole collections makes the problem more similar. The selection of number of instances on the training sets above depend on the number of classes and the size of each dataset.

\subsection{Document Representation}
\label{ssec:document-representation}

SVM relies on a Vector Space Model (VSM) and thereby it requires a vectorial representation of the documents as an input for the classifier, for both train and test phases. To obtain this vectorial representation, we use the textual content of the web pages. To this end, we first converted the original HTML codes into plain text strings, removing all the HTML tags. After that, we removed a set of useless tokens, such as URLs, email addresses and stopwords from a public list\footnote{http://www.textfixer.com/resources/common-english-words.txt}. The vectors representing the documents are composed of the remaining terms, where each dimension corresponds to a term. The weights of these terms in the vectors are defined by the TF-IDF (Term Frequency - Inverse Document Frequency) term weighting function \citep{salton88termweighting}. In order to relax the computational cost of the task, we then removed the least-frequent terms by its document frequency; terms appearing in fewer than 0.5\% of the documents were removed for the representation\footnote{0.5\% was a reasonable value for the number of resources we were dealing with. However, this reduction applies to all the algorithms we compare in this work, and keeping the same reduction for all of them makes their results equally comparable while reducing the computational cost.}. This process yielded term vectors with 8285 dimensions for \textit{BankSearch} dataset, 3115 for \textit{WebKB} and 8437 for \textit{Yahoo! Science}.

\subsection{Algorithmic Implementation}
\label{ssec:implementation}

The 6 SVM approaches presented above require 3 different classifiers to construct them: a supervised multiclass one, and two binaries, one supervised and one semi-supervised. Taking into account that some SVM implementations are freely available for research, we looked for experimented and tested software. Among the studied alternatives, we opted to use \emph{svm-light}\footnote{http://svmlight.joachims.org} and its variants, by Thorsten Joachims \citep{joachims98text}. We used supervised \emph{svm-light} for \emph{one-against-one-SVM} and \emph{one-against-all-SVM} approaches, whereas \emph{one-against-one-S$^{3}$VM} and \emph{one-against-all-S$^{3}$VM} were implemented by using semi-supervised \emph{svm-light}. Finally, we used \emph{svm-multiclass}\footnote{http://svmlight.joachims.org/svm\_multiclass.html} to implement \emph{1-step-mSVM} and \emph{2-steps-mSVM} approaches.

\subsection{Evaluation Measures}
\label{ssec:evaluation-measures}

Most of the results, not only in this chapter but also in the following chapters of this thesis, have to do with classification experiments. In order to evaluate their performance along this work, we use the accuracy as an evaluation measure. It has been widely used for text classification tasks, especially when it comes to multiclass problems. The value computed as the accuracy gives the percentage of correct predictions within the whole test set. We consider all the classes in the taxonomies to be equally relevant to the final performance, so that we do not consider any weightings in the evaluation process. Accordingly, a correct guess adds the same positive value on the accuracy, regardless of the class it belongs to.

Tables presenting accuracy values in this thesis show different training set sizes on each column, and different approaches or representation methods on each row. These accuracy values are emphasized in bold for outscoring performances within each table.

\section{Results}
\label{sec:svm-results}

This section presents the results of the experiments comparing the SVM approaches. We show the results organized by dataset, and analyze the different approaches, studying the appropriateness of multiclass or binary classifiers, as well as a supervised or a semi-supervised learning. Table \vref{tab:banksearch-results}, Table \vref{tab:webkb-results} and Table \vref{tab:yahooscience-results} show those results for BankSearch, WebKB and Yahoo! Science datasets respectively.

\begin{table}[p]
\begin{center}
 \tiny
 \begin{tabular}{|l|c|c|c|c|c|c|c|c|}
  \hline
  \multicolumn{9}{|c|}{\scriptsize\strut \textbf{BankSearch}} \\
  \hline
  \hline
  \scriptsize\strut \textbf{\# of labeled instances} & \scriptsize\strut \textbf{50} & \scriptsize\strut \textbf{100} & \scriptsize\strut \textbf{200} & \scriptsize\strut \textbf{500} & \scriptsize\strut \textbf{1000} & \scriptsize\strut \textbf{2000} & \scriptsize\strut \textbf{3000} & \scriptsize\strut \textbf{4000} \\
  \hline
  \scriptsize\strut \textbf{1-step-mSVM} & \scriptsize\strut .579 & \scriptsize\strut .706 & \scriptsize\strut .792 & \scriptsize\strut .869 & \scriptsize\strut .897 & \scriptsize\strut \textbf{.919} & \scriptsize\strut \textbf{.925} & \scriptsize\strut \textbf{.930} \\
  \hline
  \scriptsize\strut \textbf{2-steps-mSVM} & \scriptsize\strut \textbf{.628} & \scriptsize\strut \textbf{.753} & \scriptsize\strut \textbf{.826} & \scriptsize\strut \textbf{.879} & \scriptsize\strut \textbf{.898} & \scriptsize\strut .916 & \scriptsize\strut .923 & \scriptsize\strut \textbf{.930} \\
  \hline
  \scriptsize\strut \textbf{one-against-all-SVM} & \scriptsize\strut .372 & \scriptsize\strut .485 & \scriptsize\strut .575 & \scriptsize\strut .697 & \scriptsize\strut .759 & \scriptsize\strut .816 & \scriptsize\strut .843 & \scriptsize\strut .855 \\
  \hline
  \scriptsize\strut \textbf{one-against-all-S$^{3}$VM} & \scriptsize\strut .506 & \scriptsize\strut .566 & \scriptsize\strut .621 & \scriptsize\strut .709 & \scriptsize\strut .763 & \scriptsize\strut .814 & \scriptsize\strut .842 & \scriptsize\strut .855 \\
  \hline
  \scriptsize\strut \textbf{one-against-one-SVM} & \scriptsize\strut .311 & \scriptsize\strut .443 & \scriptsize\strut .549 & \scriptsize\strut .679 & \scriptsize\strut .744 & \scriptsize\strut .803 & \scriptsize\strut .826 & \scriptsize\strut .840 \\
  \hline
  \scriptsize\strut \textbf{one-against-one-S$^{3}$VM} & \scriptsize\strut .443 & \scriptsize\strut .513 & \scriptsize\strut .567 & \scriptsize\strut .668 & \scriptsize\strut .724 & \scriptsize\strut .782 & \scriptsize\strut .811 & \scriptsize\strut .840 \\
  \hline
 \end{tabular}
\end{center}
\caption[Accuracy results for the BankSearch dataset]{Accuracy results for the BankSearch dataset.}
\label{tab:banksearch-results}
\end{table}

\begin{table}[p]
\begin{center}
 \tiny
 \begin{tabular}{|l|c|c|c|c|c|c|}
  \hline
  \multicolumn{7}{|c|}{\scriptsize\strut \textbf{WebKB}} \\
  \hline
  \hline
  \scriptsize\strut \textbf{\# of labeled instances} & \scriptsize\strut \textbf{50} & \scriptsize\strut \textbf{100} & \scriptsize\strut \textbf{200} & \scriptsize\strut \textbf{500} & \scriptsize\strut \textbf{1000} & \scriptsize\strut \textbf{2000} \\
  \hline
  \scriptsize\strut \textbf{1-step-mSVM} & \scriptsize\strut \textbf{.600} & \scriptsize\strut \textbf{.677} & \scriptsize\strut \textbf{.739} & \scriptsize\strut \textbf{.787} & \scriptsize\strut \textbf{.810} & \scriptsize\strut \textbf{.822} \\
  \hline
  \scriptsize\strut \textbf{2-steps-mSVM} & \scriptsize\strut .582 & \scriptsize\strut .667 & \scriptsize\strut .715 & \scriptsize\strut .750 & \scriptsize\strut .778 & \scriptsize\strut \textbf{.822} \\
  \hline
  \scriptsize\strut \textbf{one-against-all-SVM} & \scriptsize\strut .513 & \scriptsize\strut .587 & \scriptsize\strut .673 & \scriptsize\strut .744 & \scriptsize\strut .776 & \scriptsize\strut .783 \\
  \hline
  \scriptsize\strut \textbf{one-against-all-S$^{3}$VM} & \scriptsize\strut .592 & \scriptsize\strut .642 & \scriptsize\strut .691 & \scriptsize\strut .740 & \scriptsize\strut .773 & \scriptsize\strut .783 \\
  \hline
  \scriptsize\strut \textbf{one-against-one-SVM} & \scriptsize\strut .488 & \scriptsize\strut .554 & \scriptsize\strut .648 & \scriptsize\strut .736 & \scriptsize\strut .775 & \scriptsize\strut .791 \\
  \hline
  \scriptsize\strut \textbf{one-against-one-S$^{3}$VM} & \scriptsize\strut .494 & \scriptsize\strut .579 & \scriptsize\strut .651 & \scriptsize\strut .718 & \scriptsize\strut .754 & \scriptsize\strut .791 \\
  \hline
 \end{tabular}
\end{center}
\caption[Accuracy results for the WebKB dataset]{Accuracy results for the WebKB dataset.}
\label{tab:webkb-results}
\end{table}

\begin{table}[p]
\begin{center}
 \tiny
 \begin{tabular}{|l|c|c|c|}
  \hline
  \multicolumn{4}{|c|}{\scriptsize\strut \textbf{Yahoo! Science}} \\
  \hline
  \hline
  \scriptsize\strut \textbf{\# of labeled instances} & \scriptsize\strut \textbf{50} & \scriptsize\strut \textbf{100} & \scriptsize\strut \textbf{200} \\
  \hline
  \scriptsize\strut \textbf{1-step-mSVM} & \scriptsize\strut .682 & \scriptsize\strut .825 & \scriptsize\strut \textbf{.908} \\
  \hline
  \scriptsize\strut \textbf{2-steps-mSVM} & \scriptsize\strut \textbf{.687} & \scriptsize\strut \textbf{.836} & \scriptsize\strut \textbf{.908} \\
  \hline
  \scriptsize\strut \textbf{one-against-all-SVM} & \scriptsize\strut .506 & \scriptsize\strut .536 & \scriptsize\strut .630 \\
  \hline
  \scriptsize\strut \textbf{one-against-all-S$^{3}$VM} & \scriptsize\strut .570 & \scriptsize\strut .565 & \scriptsize\strut .630 \\
  \hline
  \scriptsize\strut \textbf{one-against-one-SVM} & \scriptsize\strut .436 & \scriptsize\strut .483 & \scriptsize\strut .586 \\
  \hline
  \scriptsize\strut \textbf{one-against-one-S$^{3}$VM} & \scriptsize\strut .467 & \scriptsize\strut .514 & \scriptsize\strut .586 \\
  \hline
 \end{tabular}
\end{center}
\caption[Accuracy results for the Yahoo! Science dataset]{Accuracy results for the Yahoo! Science dataset.}
\label{tab:yahooscience-results}
\end{table}

\subsection{Native Multiclass vs Combining Binary Classifiers}
\label{ssec:native-vs-binary-results}

Our experiments compare two native multiclass approaches to four methods combining binary classifiers. First of all, the results clearly show that those relying on the \emph{one-against-one} setting (i.e., \emph{one-against-one-SVM} and \emph{one-against-one-S$^{3}$VM}) perform much worse than the rest for all the datasets. This outperformance confirms the issue we pointed out above in Section \vref{ssec:one-against-one-approaches}, i.e., the inability of discriminating the instances that do not belong to the considered pair of classes adds noise into the decisions.

Among the other two settings, the native multiclass SVMs and the \emph{one-against-all}, there is a clear outperformance for the former. The performance gap between those two approaches differs depending on the dataset. Even though it is much smaller for the \emph{WebKB} dataset than for the other two, it is undoubtedly clear that the native multiclass approach seems a better setting to face this kind of tasks. Moreover, regardless of the size of the labeled subset, the multiclass settings always outperform the others.

\subsection{Supervised vs Semi-Supervised Learning}
\label{ssec:supervised-vs-semisupervised-results}

Besides these three settings, we have also compared a supervised and a semi-supervised learning for each of them in our experiments. When comparing the two analogous approaches for each setting, it can be seen that the semi-supervised ones (\emph{2-steps-mSVM}, \emph{one-against-all-S$^{3}$VM} and \emph{one-against-one-S$^{3}$VM}) perform better than the supervised ones (\emph{1-step-mSVM}, \emph{one-against-all-SVM} and \emph{one-against-one-SVM}) in most cases when it comes to the smallest labeled subsets. However, the contrary happens for larger labeled subsets, where the supervised approaches perform better. Looking at these results, it seems that the success of semi-supervised learning for multiclass classification is limited to very small labeled sets, where more instances are required in order to get a sufficient base to learn from.

Going in more depth in the native multiclass approaches, which perform the best, a similar conclusion can be drawn, especially for the largest dataset, \emph{BankSearch}. Even though the semi-supervised \emph{2-steps-mSVM} performs better than the supervised \emph{1-step-mSVM} for the smallest labeled subsets, there is a slight outperformance for the latter when the labeled subset increases. In the case of \emph{WebKB}, \emph{1-step-mSVM} is always the best, probably because it is harder to predict correctly the unlabeled instance in the semi-supervised scenario when the taxonomy is made by closely related categories, and it adds noise in the learning phase. Finally, for \emph{Yahoo! Science}, \emph{2-steps-mSVM} performs slightly better, but since this dataset is quite small, it does not let us see whether \emph{1-step-mSVM} would outperform for larger labeled subsets.

\section{Discussion}
\label{sec:svm-discussion}

In this study, we have compared the required approaches to help us determine (a) if we should use a native multiclass classifier or combine binary classifiers, and (b) whether or not including the predictions on unlabeled instances improves the performance of the classifier. This is not an exhaustive comparison study between SVM approaches for large-scale classification on multiclass taxonomies. An example of this is that we did not consider any native multiclass and semi-supervised approaches like that by \cite{yajima06optimization}, which we did not have access to --reasonwhy it has not been used subsequently. We have compared a set of approaches available for research purposes instead.

\section{Conclusion}
\label{sec:svm-conclusions}

In this chapter, we have analyzed a set of approaches to face a large-scale topical classification task, considering that it fulfills the conditions that (a) it is multiclass with more than two classes in the taxonomy, and (b) the labeled subset tends to be tiny as compared to the whole set to classify. Looking at these two aspects, we have compared 6 different SVM approaches, including (a) semi-supervised and supervised learning, and (b) 3 different settings, a native multiclass and 2 binary settings, \emph{one-against-one} and \emph{one-against-all}. With experiments over 3 different datasets, we have performed a comparison study between the different SVM approaches.

Parts of the research in this chapter have been published in \cite{zubiaga2009unlabeled} and \cite{zubiaga2009comparativa}.

We have also answered the following research questions in this chapter:

\begin{description}
 \item[Research Question 1] \hfill \\
 \textit{\rqsvmone}
\end{description}

We have shown the clear superiority of the native multiclass SVM classifiers over the other approaches combining binary classifiers. Our results show that relying on a set of binary classifiers is not a good option when it comes to multiclass taxonomies. Accordingly, native multiclass classifiers, which consider all the classes at the same time and have more knowledge of the whole task, perform much better.

\begin{description}
 \item[Research Question 2] \hfill \\
 \textit{\rqsvmtwo}
\end{description}

Semi-supervised approaches may perform better when the labeled subset is really small, but supervised approaches, which are computationally less expensive, perform similarly with more labeled documents. Therefore, we have also shown that, unlike binary tasks as shown by \cite{joachims99transductive}, a supervised approach performs very similar to a semi-supervised approach on these environments. It seems reasonable that predicting the class of uncategorized documents is much more difficult when the number of classes increases, and so the miscategorized documents are harmful for classifier's learning.

Thereby, according to these conclusions, we decided to use a supervised multiclass SVM in this thesis, i.e., \emph{svm-multiclass} by \cite{joachims98text}. We use the \emph{1-step-mSVM} approach in Chapter \vref{c:tag-representation}, Chapter \vref{c:tag-distribution-classification} and Chapter \vref{c:analyzing-appropriateness-users}.

%% file: datasets.tex
\chapter{Generation of Social Tagging Datasets}
\label{c:datasets}

\textit{``As a general principle, the more users share about themselves, the more others in the community will learn about them and identify with them.''}

--- Matt Rhodes

\chaptersummary{This chapter describes and analyzes in detail the social tagging datasets we have created to use throughout this work. After looking for existing datasets, we found no one that fulfilled our requirements. Hence, we introduce the process we followed for generating suitable datasets, and we analyze their main characteristics.}

\chaptersummary{The chapter is organized as follows. First, in Section \vref{sec:selection-of-sts} we describe the requirements and criteria that led us to the selection of the appropriate social tagging systems. In Section \vref{sec:characteristics-of-sts} we comprehensively analyze the features of the selected social tagging systems. Next, we present the process we carried out for gathering the datasets from the Web in Section \vref{sec:datasets-generation-process}. Then, we analyze the folksonomies of such datasets and present a set of statistics in Section \vref{sec:dataset-stats}. In Section \vref{sec:gathering-additional-metadata} we introduce the additional data, besides tagging data, we retrieved and included in the datasets. Finally, we conclude and answer the following research question in Section \vref{sec:datasets-conclusion}:

\begin{description}
 \item[Research Question 3] \hfill \\
 \textit{\rqdataone}
\end{description}
}

\section{Selection of Social Tagging Systems}
\label{sec:selection-of-sts}

First of all, we defined a set of conditions that the selected social tagging systems should fulfill according to our requirements:

\begin{enumerate}
 \item They must have a large community of users involved. This enables to further analyze the aggregation of annotations. The fact of considering whether or not a community is large can obviously be subjective, though. We consider it large enough when there is an active community and resources tend to be annotated by many users.
 \item In order to gather the required data, they must provide an accessible API, or an alternative way to access the data by HTML scraping instead. The required data include full access to the triple involved in each bookmark, i.e., the user annotating it, the resource being annotated, and the tags. This is extremely relevant to analyze the nature and structure of folksonomies, and how they are created.
 \item Regarding the ground truth we will assume for the classification tasks, the considered resources must somewhere be classified on consolidated taxonomies by experts. These categorization data will provide a way to quantitatively evaluate the classification tasks.
\end{enumerate}

We thought it would be wise to analyze the existence of social tagging datasets that fulfilled our requirements. Even though we looked for social tagging datasets created and made publicly available by others, we just found a few of them by then, and none of them matched our needs\footnote{By then, the only dataset with categorization data for tagged resources was CABS120k08 by \cite{noll_metadata_2008}, but it did not fulfill our requirements: \\ http://www.michael-noll.com/cabs120k08/}. Therefore, we decided to create new datasets. Before creating the datasets, though, it is of utmost importance to select the appropriate social tagging sites to collect them from. Since we wanted to analyze in depth the tagging structure of folksonomies, we were required to get data as detailed as possible. However, not all the social tagging sites provide all these data.

We analyzed a large set of social tagging sites, and studied whether or not they matched the above requirements. We found that most of them were in the long tail according to the size of the community, with small and almost inactive groups of contributors\footnote{e.g., CiteULike (http://www.citeulike.org) is a bookmarking site for publications where usually there is no enough aggregation of annotations on a resource.}. We ruled them out, and considered those in the head with large and active communities. Not all of them provide all the required data, though. Some social tagging sites show the aggregated list of tags for each resource, but there is no way to extract bookmark data, and thus the exhaustive list of users who contributed and tags assigned by them when saving the resource. Moreover, some social tagging sites only show a list of tags for each resource, without the number of users annotating them\footnote{e.g., GiveALink (http://www.givealink.org) only shows an unweighted list of popular tags for each bookmarked web page.}. Also, there are sites where the annotated resources have no consolidated category data\footnote{e.g., Last.fm (http://www.last.fm) provides large amounts of annotations for musical groups, but there is no standard taxonomy organizing them by musical genres.}. Hence, even though there are lots of social tagging sites available online, most of them restrict the access to data, or do not fulfill all the requirements. Thereby we finally got a smaller list of social tagging sites, since lots of them had to be discarded: (i) Delicious\footnote{http://delicious.com}, where users save and annotate web pages, (ii) LibraryThing\footnote{http://www.librarything.com}, a social tagging site for books, and (iii) GoodReads\footnote{http://www.goodreads.com}, also for books. In fact, all of them consist of bookmarks of resources which are regularly classified by experts. Web pages have been organized into web directories since 1990s, and librarians have been cataloging books into categories for centuries.

\section{Characteristics of the Selected Social Tagging Systems}
\label{sec:characteristics-of-sts}

Even though all the tagging systems have the same end of enabling users to bookmark and annotate the resources of their interest, there are several features that make each of them different from the rest. The design of the interface, constraints on the inputted tags, and other features could influence users' annotations. Thus, it is worthwhile studying the nature of each of the social tagging sites we rely on, in order to understand their underlying folksonomies.

\textbf{Delicious} is a social bookmarking site that allows users to save and tag their favorite web pages, in order to ease the subsequent navigation and retrieval on large collections of annotated bookmarks. Being a social bookmarking site, every web page can be saved, so that the range of covered topics can become as wide as the Web is. It is known that the site is biased to some computer and design related topics though. Tagging web pages is one of the main features of the site, and that is the first thing the system asks for when a user saves a URL as a bookmark. The system suggests tags used earlier for that URL if some users had annotated it before. Thus, new annotators can easily select tags used by earlier users without typing them. This could encourage users to reuse others' tags, reducing the number of new tags assigned to a resource.

\textbf{LibraryThing} and \textbf{GoodReads} are social cataloging\footnote{Both social bookmarking and social cataloging refer to social tagging systems. The sole difference is on the resources, i.e., URLs are bookmarked, whereas books are cataloged.} sites where users save and annotate books. Commonly, users annotate the books they own, they have read, or they are planning to read. We believe that users contributing to this kind of sites are more knowledgeable of the resources than those contributing to social bookmarking systems. Moreover, there are also writers and libraries contributing as users, who have a deep background on the field. This could yield annotations providing further and more detailed knowledge. The main difference among these two systems is that LibraryThing does not suggest tags when saving a book, whereas GoodReads lets the user select from tags within their personomy, that is, tags they previously assigned to other books. The latter makes it easier to reuse users' favorite tags, without re-typing them. This could encourage users to keep a smaller tag vocabulary, where they barely use new tags they did not used previously. Moreover, LibraryThing brings the user to a new page when saving a book, where they can attach tags to it; GoodReads, though, requires the user to click again on the saved book to open the form to add tags. Another remarkable difference is that LibraryThing allows some users to group tags with the same meaning, linking thus typos, misspellings, synonyms and translations to a single tag, e.g., \texttt{science-fiction}, \texttt{sf} and \texttt{ciencia ficci\'on} are grouped into \texttt{science fiction}.

Despite of the aforementioned differences, all of them have some characteristics in common: users save resources as bookmarks, a bookmark can be annotated by a variable number of tags ranging from zero to unlimited, and the vocabulary of the tags is open and unrestricted. Table \vref{tab:st-features} summarizes the main features of the three social tagging sites we study in this thesis.

\begin{table}[htb]
\begin{center}
 \tiny
 \begin{tabular}{|m{2.6cm}|m{2.6cm}|m{2.6cm}|m{2.6cm}|}
  \hline
  & \scriptsize\strut \textbf{Delicious} & \scriptsize\strut \textbf{LibraryThing} & \scriptsize\strut \textbf{GoodReads} \\
  \hline
  \scriptsize\strut \textbf{Resources} & \scriptsize\strut web documents & \scriptsize\strut books & \scriptsize\strut books \\
  \hline
  \scriptsize\strut \nohyphens{\textbf{Tag suggestions}} & \scriptsize\strut \nohyphens{based on earlier bookmarks on the resource} & \scriptsize\strut no & \scriptsize\strut based on user's personomy \\
  \hline
  \scriptsize\strut \textbf{Users} & \scriptsize\strut general & \scriptsize\strut \nohyphens{readers, writers \& libraries} & \scriptsize\strut readers, writers \& libraries \\
  \hline
  \scriptsize\strut \textbf{Tag grouping} & \scriptsize\strut no & \scriptsize\strut selected users suggest merging tags & \scriptsize\strut no \\
  \hline
  \scriptsize\strut \textbf{Vocabulary} & \scriptsize\strut open & \scriptsize\strut open & \scriptsize\strut open \\
  \hline
  \scriptsize\strut \textbf{Tag insertion} & \scriptsize\strut space-separated & \scriptsize\strut comma-separated & \scriptsize\strut one by one text-box \\
  \hline
  \scriptsize\strut \textbf{When saving a resource} & \scriptsize\strut prompts user to add tags & \scriptsize\strut prompts user to add tags at second step & \scriptsize\strut user needs to click again to add tags \\
  \hline
 \end{tabular}
\end{center}
\caption[Characteristics of the studied social tagging systems]{Characteristics of the studied social tagging systems.}
\label{tab:st-features}
\end{table}

\section{Generation Process of Datasets}
\label{sec:datasets-generation-process}

Even though the three chosen social tagging sites provide public access to the full bookmarking activity, getting large collections of data from them turns into a complicated task. All of them have an API for accessing the data, but none of the APIs provides the required data, so crawling the sites and scraping the HTML code instead seems to be the only way to achieve the goal. Moreover, each site sets its own limit on the number of requests, and lots of them must be done in order to obtain large-scale datasets. Hence, we set a crawling policy for each site, and applied it with extra care in order to not get banned while getting as much data as possible.

\subsection{Getting Popular Resources}
\label{ssec:getting-popular-resources}

As a starting point, we focused on getting a set of popular resources from each site. This provided an initial list of popular resources which represented a good seed to start the gathering process from. Those resources were also more likely to have been categorized by experts rather than resources in the tail with fewer annotations. We could also start the process by looking for popular tags or active users, but starting from resources sounds reasonable when those are what we aim to classify. Next, we will focus on the process of gathering the data in such a way that those resources are well represented insofar as involved users and their annotations are taken into account. Apart from representing those resources, we were also interested in gathering additional data, in order to represent involved users and tags to a great extent.

First of all thus we queried the three sites for popular resources. We consider a resource to be popular if at least 100 users have bookmarked it\footnote{It was shown that the tag set of a resource tends to converge when 100 users contribute to it \citep{golder_structure_2006}. Thereby we consider it as a threshold for a resource to be popular.}. In the case of Delicious, we found a set of 87,096 unique URLs fulfilling this requirement. As regards to LibraryThing and GoodReads, we found an intersection of 65,929 popular books. Since the latter two rely on the same resources, we created parallel datasets for them, where the same books have categorization data attached.

\subsection{Looking for Classification Data}
\label{ssec:looking-for-classification-data}

In the next step, we looked for classification labels assigned by experts for both kinds of resources. For the URLs gathered from Delicious, we used the Open Directory Project\footnote{http://www.dmoz.org} (ODP) as a classification scheme. ODP is an open web directory, constructed and maintained by a community of volunteer editors, and it includes categorization data on a hierarchical structure for more than 4 million URLs. A matching between popular URLs on Delicious and those in the ODP returned a set of 12,616 URLs with a category assigned. For the set of books, we fetched their classification for both the Dewey Decimal Classification (DDC) and the Library of Congress Classification (LCC) systems. The former is a classical taxonomy that is still widely used in libraries, whereas the latter is used by most research and academic libraries. We found that 27,299 books were categorized on DDC, and 24,861 books had an LCC category assigned. In total, there are 38,148 books with category data from either one or both category schemes.

In this thesis, we will focus on both the top level and the second level of the taxonomies. This enables to evaluate the usefulness of social tags for classification on both broader and narrower categories. Even though taxonomies are made up by more than 2 levels of categorization, going into deeper levels would lack of enough number of resources for each category, and would not enable an appropriate experimentation. Table \vref{tab:classes-docs} summarizes the number of classes in each taxonomy and level, as well as the number of resources with categorization data for each of them. We kept the structure of all the taxonomies as they were, but made a little change for LCC: we merged E (\emph{History of America}) and F (\emph{History of the United States and British, Dutch, French, and Latin America}) categories into a single one, as it is not clear that they are disjoint categories. Also, note that the number of resources is slightly smaller for second levels. This is because we removed second-level categories and their underlying resources when there were fewer than 5 resources in them, due to the low representativity\footnote{The threshold of 5 resources is arbitrary. It is reasonable from our point of view, because it increases the likelihood of having more than one learning instance for each category, and the reduction of the dataset is minimal.}.

\begin{table}[htb]
 \begin{center}
  \tiny
  \begin{tabular}{ | l | c | c | c | c | }
   \hline
   & \multicolumn{2}{ c | }{\scriptsize\strut \textbf{Top level}} & \multicolumn{2}{ c | }{\scriptsize\strut \textbf{Second level}} \\
   \hline
   & \scriptsize\strut \textbf{Resources} & \scriptsize\strut \textbf{Classes} & \scriptsize\strut \textbf{Resources} & \scriptsize\strut \textbf{Classes} \\
   \hline
   \scriptsize\strut \textbf{ODP} & \scriptsize\strut 12,616 & \scriptsize\strut 17 & \scriptsize\strut 12,286 & \scriptsize\strut 243 \\
   \hline
   \scriptsize\strut \textbf{DDC} & \scriptsize\strut 27,299 & \scriptsize\strut 10 & \scriptsize\strut 27,040 & \scriptsize\strut 99 \\
   \hline
   \scriptsize\strut \textbf{LCC} & \scriptsize\strut 24,861 & \scriptsize\strut 20 & \scriptsize\strut 23,565 & \scriptsize\strut 204 \\
   \hline
  \end{tabular}
 \end{center}
 \caption[Number of resources and classes for the classification experiments]{Number of resources and classes for the classification experiments.}
 \label{tab:classes-docs}
\end{table}

\subsection{Gathering Tagging Data}
\label{ssec:gathering-tagging-data}

Finally, we queried (a) Delicious for gathering all the personomies involved in the set of categorized URLs, and (b) LibraryThing and GoodReads for gathering all the personomies involved in the set of categorized books. By personomy, we consider the whole list of bookmarks posted by a user, including an identifier of the resources and the tags attached by them. All three sites present no restrictions on the bookmarks shown in personomies, so that they return all available public bookmarks for the queried users.

The process above results in a large collection of bookmarks for each dataset. Within the gathered data, we focus on the following information for each bookmark:

\begin{itemize}
 \item \textbf{User (U):} an identifier of the user who annotated the resource.
 \item \textbf{Resource (R):} the resource annotated by the bookmark. It is a URL in the case of Delicious, and the ISBN identifier of a book in the case of LibraryThing and GoodReads.
 \item \textbf{Tags (T):} the set of tags, in case it is available, annotated by the user to the resource.
\end{itemize}

That is, the triple of $U \times R \times T$ involved in a bookmark. In this process, we consider all the tags attached to each bookmark, except for GoodReads. In this case, a tag is automatically attached to each bookmark depending on the reading state of the book: \texttt{read}, \texttt{currently-reading} or \texttt{to-read}. We do not consider this to be part of the tagging process, but just an automated step that does not provide useful information for classification, and we removed all their appearances in our dataset.

\section{Statistics and Analysis of the Datasets}
\label{sec:dataset-stats}

In order to understand the nature and characteristics of each dataset, and to analyze how the settings of each social tagging system affect the folksonomies, we study and present statistics of the created datasets.

It is worthwhile noting that, as we stated above, attaching tags to a bookmark is an optional step, so that depending on the social tagging site, a number of bookmarks may remain without tags. Table \vref{tab:dataset-stats} presents the number of users, bookmarks and resources we gathered for each of the datasets, as well as the percent with attached annotations. In this work, as we rely on tagging data, we only consider annotated data, ruling out bookmarks without tags. Thus, from now on, all the results and statistics presented are based on annotated bookmarks. From these statistics, it stands out that most users (above 87\%) provide tags for bookmarks on Delicious, whereas there are fewer users who tend to assign tags to resources on LibraryThing and GoodReads (roughly 38\% and 17\%, respectively). This shows the importance of Delicious' encouragement to adding tags, and GoodReads' disencouragement to this end, requiring the user to click twice on the book in order to add tags. The latter makes the tagging process cumbersome, and yields a large number of untagged bookmarks. LibraryThing is halfway between those two, which automatically conveys the user to the tagging form, but at a skippable second step after saving the book.

\begin{table}[htb]
\begin{center}
 \tiny
 \begin{tabular}{|l|c|c|c|}
  \hline
  \multicolumn{4}{|c|}{\scriptsize\strut \textbf{Delicious}} \\
  \hline
  & \scriptsize\strut Annotated & \scriptsize\strut Total & \scriptsize\strut Ratio \\
  \hline
  \scriptsize\strut Users & \scriptsize\strut 1,618,635 & \scriptsize\strut 1,855,792 & \scriptsize\strut 87.22\% \\
  \hline
  \scriptsize\strut Bookmarks & \scriptsize\strut 273,478,137 & \scriptsize\strut 300,571,231 & \scriptsize\strut 91.00\% \\
  \hline
  \scriptsize\strut Resources & \scriptsize\strut 92,432,071 & \scriptsize\strut 102,828,761 & \scriptsize\strut 89.89\% \\
  \hline
  \multicolumn{2}{|l|}{\scriptsize\strut Tags} & \scriptsize\strut 11,541,977 & \scriptsize\strut - \\
  \hline
  \multicolumn{4}{|c|}{\scriptsize\strut \textbf{LibraryThing}} \\
  \hline
  & \scriptsize\strut Annotated & \scriptsize\strut Total & \scriptsize\strut Ratio \\
  \hline
  \scriptsize\strut Users & \scriptsize\strut 153,606 & \scriptsize\strut 400,336 & \scriptsize\strut 38.37\% \\
  \hline
  \scriptsize\strut Bookmarks & \scriptsize\strut 22,343,427 & \scriptsize\strut 44,612,784 & \scriptsize\strut 50.08\% \\
  \hline
  \scriptsize\strut Resources & \scriptsize\strut 3,776,320 & \scriptsize\strut 5,002,790 & \scriptsize\strut 75.48\% \\
  \hline
  \multicolumn{2}{|l|}{\scriptsize\strut Tags} & \scriptsize\strut 2,140,734 & \scriptsize\strut - \\
  \hline
  \multicolumn{4}{|c|}{\scriptsize\strut \textbf{GoodReads}} \\
  \hline
  & \scriptsize\strut Annotated & \scriptsize\strut Total & \scriptsize\strut Ratio \\
  \hline
  \scriptsize\strut Users & \scriptsize\strut 110,344 & \scriptsize\strut 649,689 & \scriptsize\strut 16.98\% \\
  \hline
  \scriptsize\strut Bookmarks & \scriptsize\strut 9,323,539 & \scriptsize\strut 47,302,861 & \scriptsize\strut 19.71\% \\
  \hline
  \scriptsize\strut Resources & \scriptsize\strut 1,101,067 & \scriptsize\strut 1,890,443 & \scriptsize\strut 58.24\% \\
  \hline
  \multicolumn{2}{|l|}{\scriptsize\strut Tags} & \scriptsize\strut 179,429 & \scriptsize\strut - \\
  \hline
 \end{tabular}
\end{center}
\caption[Statistics on availability of tags in users, bookmarks, and resources]{Statistics on availability of tags in users, bookmarks, and resources for the three datasets.}
\label{tab:dataset-stats}
\end{table}

The crawling process enabled us to gather large amounts of bookmarks. Not all of them correspond to the resources with categorization data from experts, though. When gathering personomies, we also gathered lots of bookmarks for resources without categorization data. Table \vref{tab:dataset-categorized-stats} shows the statistics on resources' and bookmarks' belonging to the categorized or uncategorized subset of resources, according to the categorization data we gathered from expert-driven taxonomies. It can be seen that the number of categorized bookmarks or resources is always much lower than the number of uncategorized ones. This enables to analyze a larger folksonomy as a whole for finding out tagging patterns on each site, in order to experiment afterward on the categorized subset.

\begin{table}[htb]
\begin{center}
 \tiny
 \setlength{\tabcolsep}{4pt}
 \begin{tabular}{|l|c|c|c|c|c|c|}
  \hline
  \multicolumn{7}{|c|}{\scriptsize\strut \textbf{Resources}} \\
  \hline
  & \multicolumn{3}{c|}{\scriptsize\strut \textbf{Top level}} & \multicolumn{3}{c|}{\scriptsize\strut \textbf{Second level}} \\
  \hline
  & \scriptsize\strut Categ. & \scriptsize\strut Uncateg. & \scriptsize\strut Ratio & \scriptsize\strut Categ. & \scriptsize\strut Uncateg. & \scriptsize\strut Ratio \\
  \hline
  \scriptsize\strut Delicious (ODP) & \scriptsize\strut 12,616 & \scriptsize\strut 92,419,455 & \scriptsize\strut 0.014\% & \scriptsize\strut 12,286 & \scriptsize\strut 92,419,785 & \scriptsize\strut 0.013\% \\
  \scriptsize\strut LibraryThing (DDC) & \scriptsize\strut 23,617 & \scriptsize\strut 3,752,703 & \scriptsize\strut 0.629\% & \scriptsize\strut 22,409 & \scriptsize\strut 3,753,911 & \scriptsize\strut 0.597\% \\
  \scriptsize\strut LibraryThing (LCC) & \scriptsize\strut 24,861 & \scriptsize\strut 3,751,459 & \scriptsize\strut 0.636\% & \scriptsize\strut 23,566 & \scriptsize\strut 3,752,754 & \scriptsize\strut 0.628\% \\
  \scriptsize\strut GoodReads (DDC) & \scriptsize\strut 23,617 & \scriptsize\strut 1,077,450 & \scriptsize\strut 2.192\% & \scriptsize\strut 22,409 & \scriptsize\strut 1,078,658 & \scriptsize\strut 2.077\% \\
  \scriptsize\strut GoodReads (LCC) & \scriptsize\strut 24,861 & \scriptsize\strut 1,076,206 & \scriptsize\strut 2.310\% & \scriptsize\strut 23,566 & \scriptsize\strut 1,077,501 & \scriptsize\strut 2.187\% \\
  \hline
  \multicolumn{7}{|c|}{\scriptsize\strut \textbf{Bookmarks}} \\
  \hline
  & \multicolumn{3}{c|}{\scriptsize\strut \textbf{Top level}} & \multicolumn{3}{c|}{\scriptsize\strut \textbf{Second level}} \\
  \hline
  & \scriptsize\strut Categ. & \scriptsize\strut Uncateg. & \scriptsize\strut Ratio & \scriptsize\strut Categ. & \scriptsize\strut Uncateg. & \scriptsize\strut Ratio \\
  \hline
  \scriptsize\strut Delicious (ODP) & \scriptsize\strut 10,984,426 & \scriptsize\strut 262,493,711 & \scriptsize\strut 4.185\% & \scriptsize\strut 10,773,505 & \scriptsize\strut 262,704,632 & \scriptsize\strut 4.101\% \\
  \scriptsize\strut LibraryThing (DDC) & \scriptsize\strut 4,266,445 & \scriptsize\strut 18,076,982 & \scriptsize\strut 23.602\% & \scriptsize\strut 4,238,774 & \scriptsize\strut 18,104,653 & \scriptsize\strut 23.413\% \\
  \scriptsize\strut LibraryThing (LCC) & \scriptsize\strut 3,777,353 & \scriptsize\strut 18,566,074 & \scriptsize\strut 20.345\% & \scriptsize\strut 3,607,935 & \scriptsize\strut 18,735,492 & \scriptsize\strut 19.257\% \\
  \scriptsize\strut GoodReads (DDC) & \scriptsize\strut 1,615,235 & \scriptsize\strut 7,708,304 & \scriptsize\strut 20.954\% & \scriptsize\strut 1,611,833 & \scriptsize\strut 7,711,706 & \scriptsize\strut 20.901\% \\
  \scriptsize\strut GoodReads (LCC) & \scriptsize\strut 1,465,740 & \scriptsize\strut 7,857,799 & \scriptsize\strut 18.653\% & \scriptsize\strut 1,432,073 & \scriptsize\strut 7,891,466 & \scriptsize\strut 18.147\% \\
  \hline
 \end{tabular}
\end{center}
\caption[Ratio of resources and bookmarks belonging to categorized or uncategorized data]{Ratio of resources and bookmarks belonging to categorized or uncategorized data. The ratio value represents the percent of categorized bookmarks as compared to the uncategorized ones.}
\label{tab:dataset-categorized-stats}
\end{table}

A first glance at the vocabulary employed in each folksonomy can be performed by looking at the top tags on each site. The top 10 of tags set by users for each of the datasets is listed in Table \vref{tab:dataset-pop-tags}. On one hand, top tags on Delicious include tags like \texttt{design}, \texttt{software} and \texttt{blog}, showing its computer and design related bias. On the other hand, top tags on LibraryThing and GoodReads share some similarities, where tags related to literary genres stand out. Moreover, the latter shows that \texttt{non-fiction} and \texttt{nonfiction} are two of the most popular tags, whereas they appear grouped for the former.

\begin{table}[htb]
\begin{center}
 \tiny
 \begin{tabular}{|c|c|c|}
  \hline
  \scriptsize\strut \textbf{Delicious} & \scriptsize\strut \textbf{LibraryThing} & \scriptsize\strut \textbf{GoodReads} \\
  \hline
  \scriptsize\strut \texttt{design} & \scriptsize\strut \texttt{fiction} & \scriptsize\strut \texttt{fiction} \\
  \scriptsize\strut \texttt{blog} & \scriptsize\strut \texttt{non-fiction} & \scriptsize\strut \texttt{fantasy} \\
  \scriptsize\strut \texttt{tools} & \scriptsize\strut \texttt{fantasy} & \scriptsize\strut \texttt{non-fiction} \\
  \scriptsize\strut \texttt{software} & \scriptsize\strut \texttt{history} & \scriptsize\strut \texttt{own} \\
  \scriptsize\strut \texttt{webdesign} & \scriptsize\strut \texttt{mystery} & \scriptsize\strut \texttt{young-adult} \\
  \scriptsize\strut \texttt{web} & \scriptsize\strut \texttt{science fiction} & \scriptsize\strut \texttt{classics} \\
  \scriptsize\strut \texttt{reference} & \scriptsize\strut \texttt{read} & \scriptsize\strut \texttt{mystery} \\
  \scriptsize\strut \texttt{programming} & \scriptsize\strut \texttt{biography} & \scriptsize\strut \texttt{romance} \\
  \scriptsize\strut \texttt{music} & \scriptsize\strut \texttt{poetry} & \scriptsize\strut \texttt{wishlist} \\
  \scriptsize\strut \texttt{web2.0} & \scriptsize\strut \texttt{novel} & \scriptsize\strut \texttt{nonfiction} \\
  \hline
 \end{tabular}
\end{center}
\caption[Top 10 most popular tags]{Top 10 most popular tags on the datasets.}
\label{tab:dataset-pop-tags}
\end{table}

\begin{figure*}[htb]
\centering
\includegraphics[width=350px]{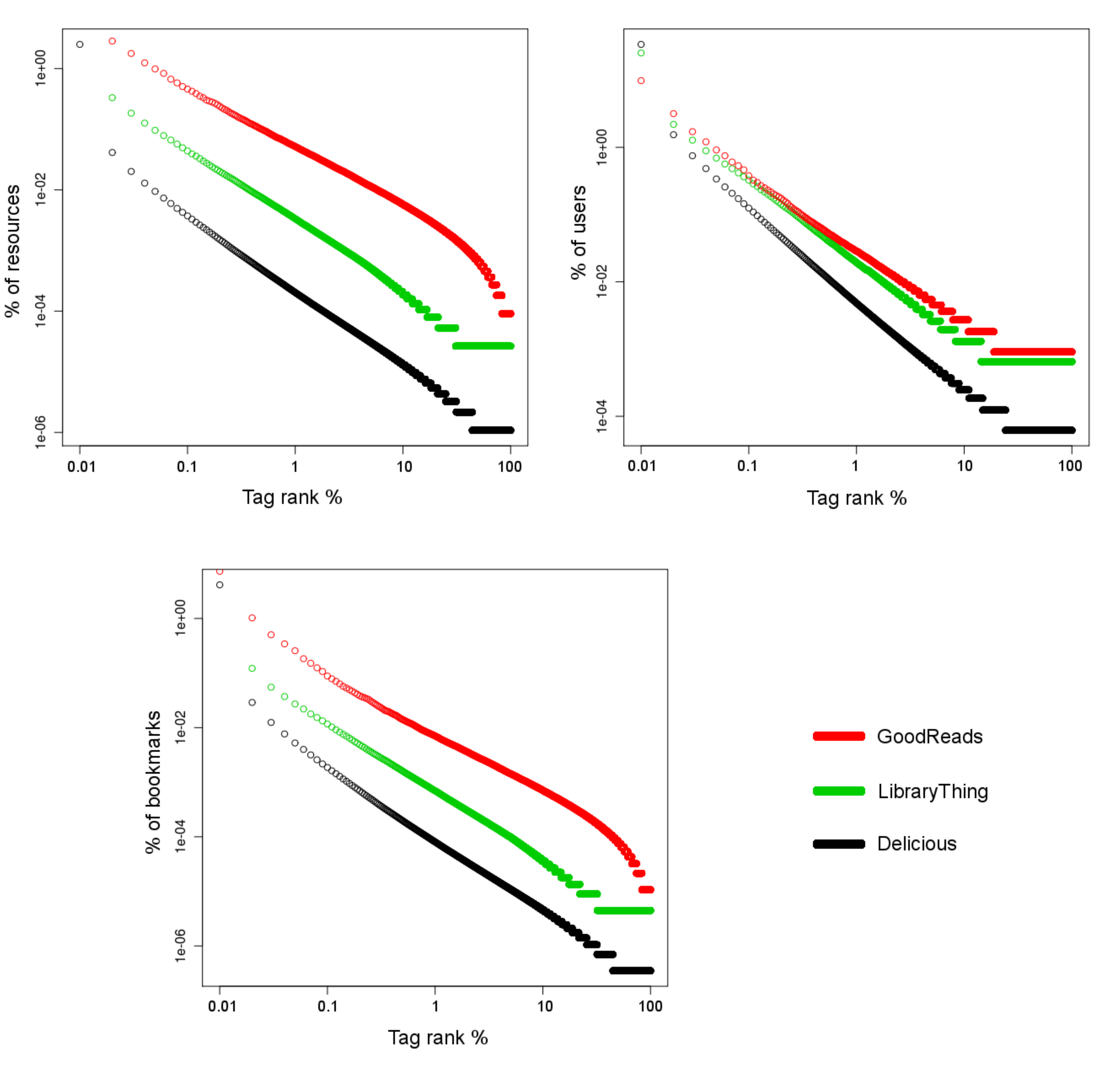}
\caption[Tag usage percentages in the collection]{Tag usage percentages in the collection. These 3 graphs represent, on a logarithmic scale for both \textit{x} and \textit{y} axes, the percent of annotations to resources, users, and bookmarks per tag rank.}
\label{fig:tag-usage}
\end{figure*}

\begin{figure*}[htb]
\centering
\includegraphics[width=350px]{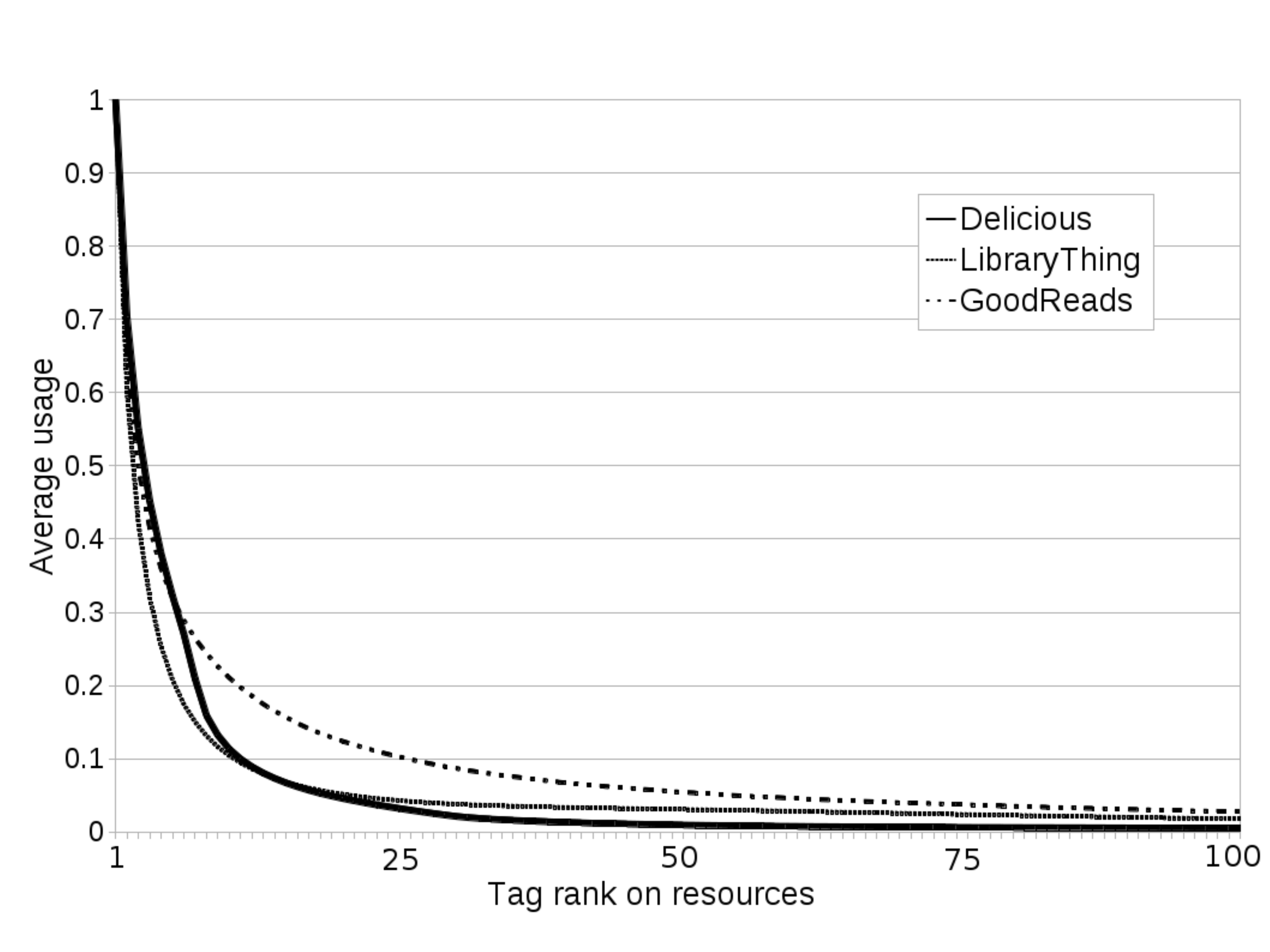}
\caption[Tag usage percentages on resources]{Tag usage percentages on resources. Each tag rank represents the average usage of tags appearing in that position on resources as compared to the top ranked tag.}
\label{fig:tag-popularity-resources}
\end{figure*}

Regarding the distribution of tags across all the resources, users and bookmarks in the datasets, there is a clear difference of behavior among the three collections. Figure \vref{fig:tag-usage} shows, on a logarithmic scale, the percent of resources, users and bookmarks on which tags are annotated according to their rank on the system. That is, the X axis refers to the percent of the tag rank, whereas the Y axis represents the percent of appearances in resources, users and bookmarks. For instance, if the tag ranked first had been annotated on the half of the resources, the value for the top ranked tag on resources would be 50\%. Thus, these graphs enable to analyze how popular are the tags in the top as compared to the tags in the tail on each site. Figure \vref{fig:tag-popularity-resources} shows the average usage of tags in a given rank for resources for each dataset. That is, we give a value of 1 to the tag annotated the most on a resource, hence ranked first for that resource. The second tag is given the value according to the fraction of users annotating it as compared to the first one. And so on for tags ranked third, fourth,... on resources. Finally, we compute the average of tags ranked on each position, which is shown in the graph. It helps infer the popularity gap between top tags on resources and tags ranked lower. Looking at those two figures, and combining their meanings, it stands out that GoodReads has the highest usage of tags in the tail, but Delicious presents the highest usage of tags in the top. Delicious is the site with highest diversity of tags, where a few tags become really popular (both in the whole collection and on resources), and many tags are seldom-used. We believe that the reasons for these differences on tag distributions are:

\begin{itemize}
 \item Since Delicious suggests tags that have been annotated by previous users to a resource, it is obvious that those tags on the top are likely to happen more frequently, whereas others may barely be used.
 \item LibraryThing and GoodReads do not suggest tags used by earlier users and, therefore, tags other than those in the top tend to be used more frequently than on Delicious.
 \item GoodReads suggests tags from previous bookmarks of the same user, instead of tags that others assigned to the resource being tagged. Thus, this encourages reusing tags in their personomy, making it remain with a smaller number of tags (see Table \vref{tab:tag-per-item}). In addition, users tend to assign fewer tags to a bookmark on average, probably due to the one-by-one tag insertion method of site's interface.
\end{itemize}

\begin{table}[htb]
\begin{center}
 \tiny
 \begin{tabular}{|l|c|c|c|}
  \hline
  \scriptsize\strut \textbf{\# of tags} & \scriptsize\strut \textbf{Delicious} & \scriptsize\strut \textbf{LibraryThing} & \scriptsize\strut \textbf{GoodReads} \\
  \hline
  \scriptsize\strut Per resource & \scriptsize\strut 33.35 & \scriptsize\strut 14.53 & \scriptsize\strut 13.33 \\
  \hline
  \scriptsize\strut Per user & \scriptsize\strut 632.714 & \scriptsize\strut 357.15 & \scriptsize\strut 131.03 \\
  \hline
  \scriptsize\strut Per bookmark & \scriptsize\strut 3.75 & \scriptsize\strut 2.46 & \scriptsize\strut 1.55 \\
  \hline
 \end{tabular}
\end{center}
\caption[Average counts of different tags]{Average counts of different tags.}
\label{tab:tag-per-item}
\end{table}

\begin{figure}[htb]
 \centering
 \includegraphics[width=350px]{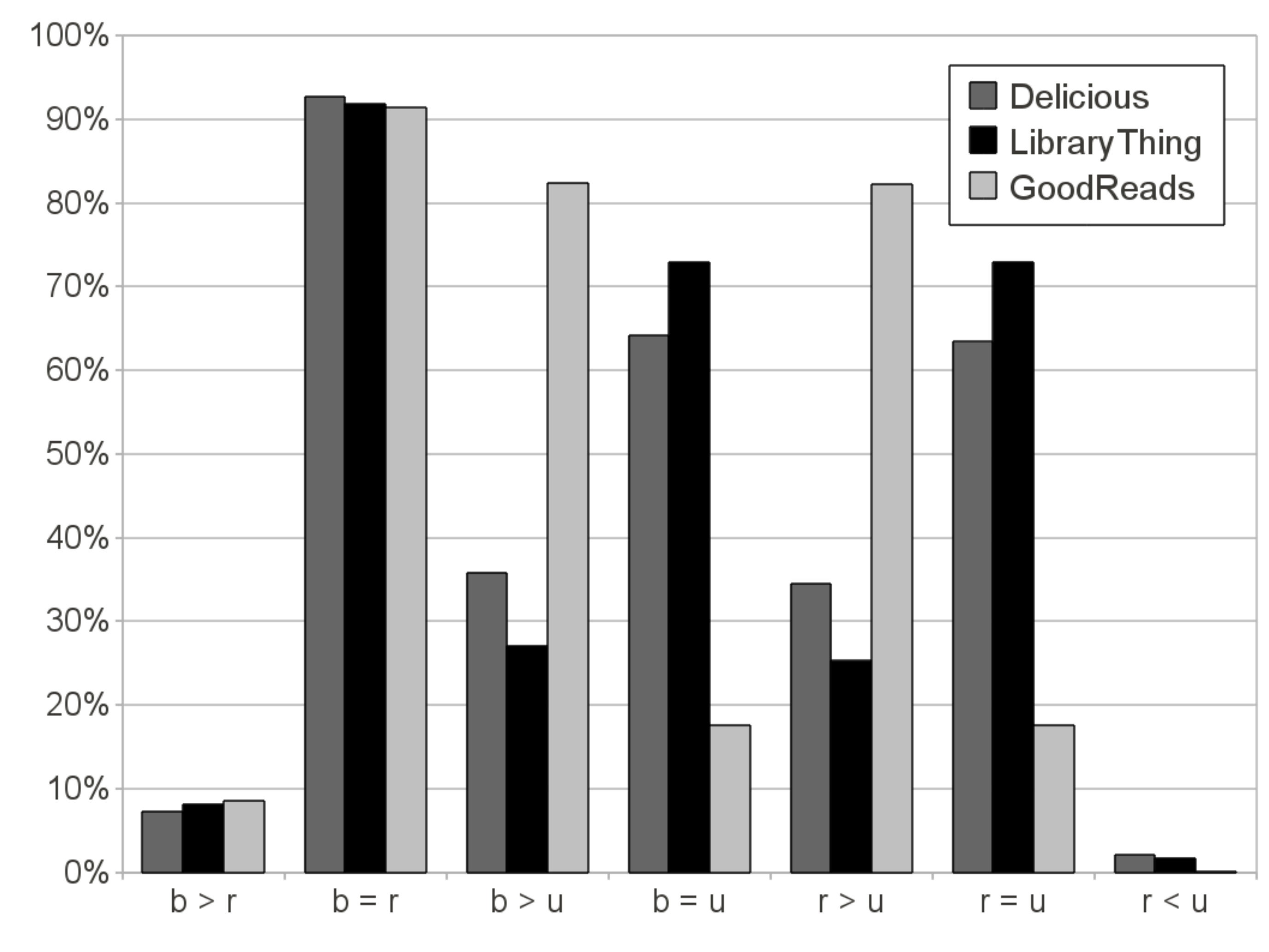}
 \caption[Tag distribution across resources, users and bookmarks]{Tag distribution across resources (r), users (u) and bookmarks (b). Each bar represents the percent of tags that match the condition on X axis.}
 \label{fig:tag-distribution}
\end{figure}

Regarding the distribution of tags across resources, users, and bookmarks, Figure \vref{fig:tag-distribution} shows percents of tags appearing more, equal or less frequently in an item (i.e., resources, users or bookmarks) than in another. It is obvious that a tag cannot appear in a smaller number of bookmarks than users or resources, by definition. Looking at the rest of data, it stands out that tags tend to appear in more bookmarks than users ($b > u$) and more resources than users ($r > u$) for GoodReads, due to the same feature that allows users to select among tags in their personomy. However, LibraryThing and Delicious have many tags present in the same number of bookmarks and users ($b = u$), and resources and users ($r = u$), even though the difference is more marked for the former site. This reflects the large number of tags that users utilize just once on these sites. All three sites have two features in common: there are a few exceptions of tags utilized by more users than the number of resources it appears in ($r < u$), and almost all the tags are present in the same number of bookmarks and resources ($b = r$). The latter, combined with the lower ($b = u$) values, means there is a large number of users spreading personal tags across resources that only have a bookmark with that tag, especially on GoodReads, but also for the other two sites.

\begin{figure}[htb]
 \centering
 \includegraphics[width=350px]{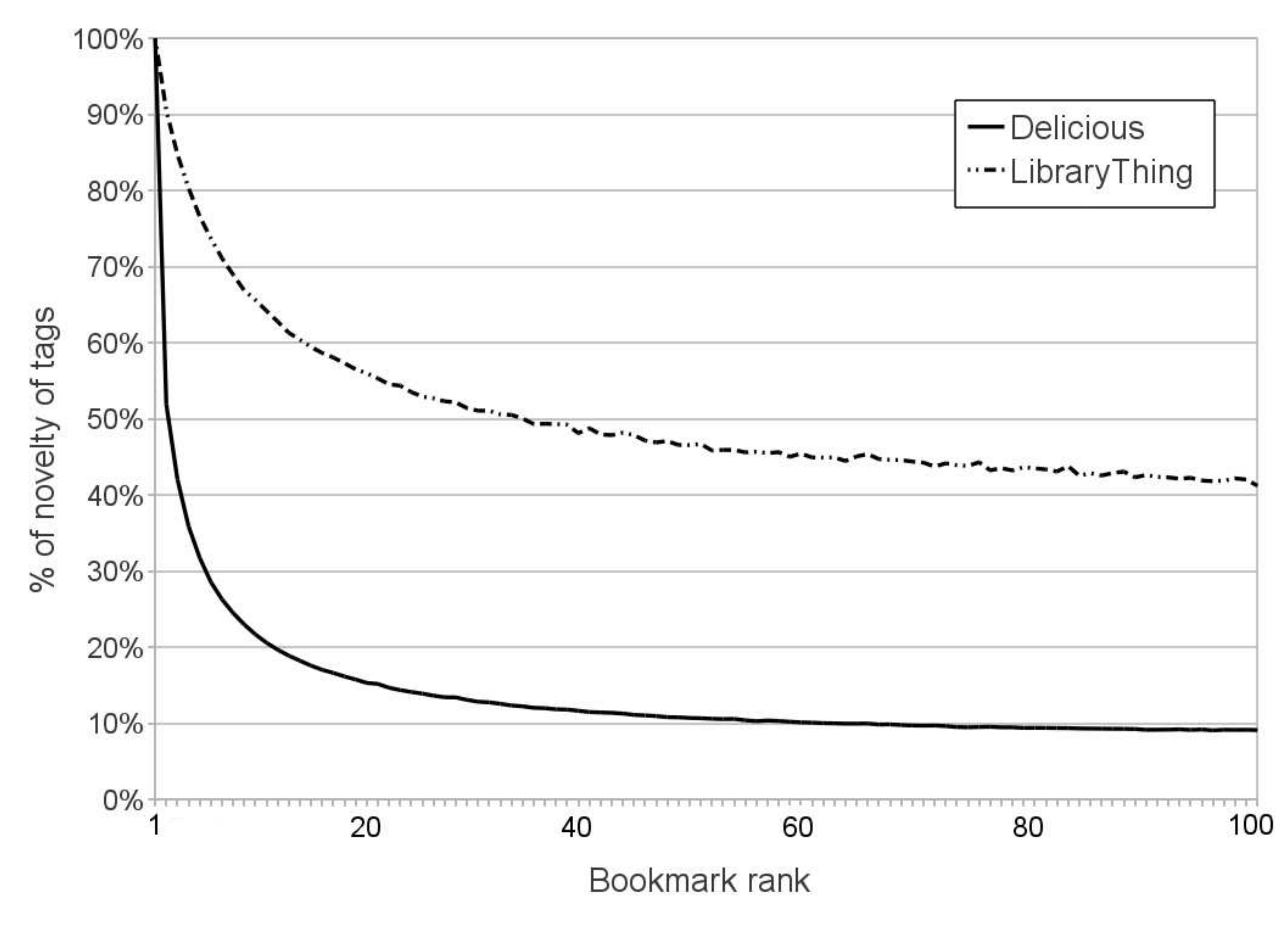}
 \caption[Novelty ratio of tags per rank of bookmark]{Novelty ratio of tags per rank of bookmark.}
 \label{fig:tag-novelty}
\end{figure}

Finally, we analyze to what extent a bookmark introduces new tags into a resource that were not present in earlier bookmarks. Figure \vref{fig:tag-novelty} shows these statistics for Delicious and LibraryThing. The same graph for GoodReads is not shown because neither the timestamp nor the ordering of the bookmarks is available in our dataset. The graph shows, on average, the ratio of new tags, not present in earlier bookmarks of a resource, assigned in bookmarks that rank from first to 100th bookmark, i.e., if \texttt{tag$_1$} and \texttt{tag$_2$} were annotated in the first bookmark of a resource, and \texttt{tag$_2$} and \texttt{tag$_3$} in the second bookmark for the same resource, the ratio of novelty for the second bookmark is of 50\%. It stands out the marked inferiority of tag novelty on Delicious as against to LibraryThing. This is, again, due to the tag suggestion policy of Delicious, what brings about a higher likelihood of reusing previously existing tags.

\section{Gathering Additional Data}
\label{sec:gathering-additional-metadata}

Besides all the aforementioned tagging data, we also gathered some more data about the categorized resources. We needed other data sources in order to perform comparisons with tagging data along the experimentation. Specifically, we compare the usefulness of tagging data as against to other sources for the classification task in Chapter \vref{c:tag-representation}, and also require additional data to analyze the descriptiveness of tags in Chapter \vref{c:analyzing-appropriateness-users}.

On one hand, we got the following data for the categorized URLs:
\begin{itemize}
 \item \textbf{Self-content:} it is the content of the web page itself, i.e., the HTML code fetched from the original URL.
 \item \textbf{Notes:} a note can be defined as a free text describing the content of a web page. It is available on Delicious, and it is intended to provide a means to briefly summarize the aboutness of a web page.
 \item \textbf{Reviews:} a review may be considered to be fairly similar to notes. However, reviews as they were collected from StumbleUpon\footnote{http://www.stumbleupon.com}, usually have a subjective bias, where users tend to valuate how they like the content of a web page.
\end{itemize}

On the other hand, with regard to the categorized books, there is no easy way to get the content of the book. We did not have access to the books, since most of them are not freely available. Thus, we got the following metadata associated to the books:
\begin{itemize}
 \item \textbf{Synopses:} a synopsis is a brief summary of the content of a book, which is usually printed on the back cover. We fetched synopses from the book retailer Barnes\&Noble\footnote{http://www.barnesandnoble.com}.
 \item \textbf{Editorial reviews:} summaries written by the publisher, or other professionals, are considered as editorial reviews. We gathered them from Amazon\footnote{http://www.amazon.com}.
 \item \textbf{User reviews:} we also collected reviews written by users on LibraryThing, GoodReads and Amazon, where they comment on the books with their summaries and thoughts.
\end{itemize}

Since we do not have access to the self-content of the books, we will consider both synopses and editorial reviews as a summary of their contents.

\section{Conclusion}
\label{sec:datasets-conclusion}

We have studied the characteristics of several social tagging systems, and concluded with three sites that fulfill our requirements: Delicious, LibraryThing and GoodReads. We have created three large-scale social tagging datasets from these sites including millions of bookmarks, not only for web pages, but also for books, which enables further analyzing other kinds of resources. To the best of our knowledge, these are the largest social tagging datasets used for research so far. Also, we have analyzed the statistics of the datasets and the features of the underlying folksonomies.

Even after we created these social tagging datasets, and made publicly available parts of them\footnote{http://nlp.uned.es/social-tagging/datasets/}, little work has been done on creating more datasets and especially on releasing them. In \cite{koerner2010call}, the authors present a list of publicly available social tagging datasets, among which our datasets are also included. However, the authors set out the problem of the unavailability of more datasets, and encourage researchers to create and release new ones.

In this chapter, we have answered the following research question:

\begin{description}
 \item[Research Question 3] \hfill \\
 \textit{\rqdataone}
\end{description}

To this end, we have analyzed several features that can be found in different settings of social tagging systems. Among the analyzed features, we have shown the impact of tag suggestions, which considerably alters the resulting folksonomy. In the studied social tagging sites, all of them differ on the settings regarding suggestions:

\begin{itemize}
 \item \textbf{Resource-based suggestions (Delicious):} when the system suggests tags assigned by other users to the resource at the time of bookmarking it, the likelihood of using new tags to further describe such a resource descreases. In this case, users provide less originality and tend to rely on system suggestions.
 \item \textbf{Personomy-based suggestions (GoodReads):} when the system suggests tags previously used by the user, the vocabulary in their personomy tends to be much smaller. However, users do not know how others annotated a resource, and thus they are likely to provide new tags to the resource.
 \item \textbf{Without suggestions (LibraryThing):} when the system does not suggest any tags to the user, the vocabulary in their personomy increases, as well as the diversity of tags in each resource.
\end{itemize}

From now on, in Chapter \vref{c:tag-representation}, Chapter \vref{c:tag-distribution-classification} and Chapter \vref{c:analyzing-appropriateness-users}, we will use these three datasets for experimentation, and we will analyze in more depth their features and how they affect the performance of a resource classification task.

%% file: tag-representation.tex
\chapter{Representing the Aggregation of Tags}
\label{c:tag-representation}

\textit{``If we wish to discuss knowledge in the most highly developed contemporary society, we must answer the preliminary question of what methodological representation to apply to that society.''}

--- Jean-Francois Lyotard

\chaptersummary{In this chapter, we set out to propose and evaluate different representations of resources based on social tags for a resource classification task. Each user contributing to the annotations on a resource provides their own tags, which commonly differ from others'. We explore different ways of representing large amounts of annotations provided by users and aggregated on resources on social tagging systems. We also measure the potential of social tags as compared to other data sources including self-content and user reviews, and analyze the suitability of combining them in search of a better performance of the classifier.}

\chaptersummary{This chapter is organized as follows. Next, in Section \vref{sec:aggregation-of-user-annotations} we describe the way user annotations are aggregated on a resource to go into the problem. In Section \vref{sec:representing-resources-using-tags} we propose several representation approaches for social tags. Then, we present the results of the tag-based classification in Section \vref{sec:tag-based-classification}, and compare them to the results by other data sources in Section \vref{sec:comparing-tags-to-other-sources}. We describe the experiments on combining data sources in Section \vref{sec:getting-the-most-out-of-all-data-sources}, and conclude the chapter in Section \vref{sec:tag-representation-conclusion}.}

\chaptersummary{The following research questions are addressed in this chapter:}

\chaptersummary{
 \begin{description}
  \item[Research Question 4] \hfill \\
  \textit{\rqrepone}
 \end{description}
}

\chaptersummary{
 \begin{description}
  \item[Research Question 5] \hfill \\
  \textit{\rqreptwo}
 \end{description}
}

\chaptersummary{
 \begin{description}
  \item[Research Question 6] \hfill \\
  \textit{\rqrepthree}
 \end{description}
}

\section{Aggregation of User Annotations}
\label{sec:aggregation-of-user-annotations}

Social tagging systems allow users to annotate on resources that others have previously annotated. This enables the aggregation of annotations provided by many users on the same resource. Obviously, each user provides their own annotations, so that tags tend to be different from user to user. These annotations are listed all together in a detailed manner (see Table \vref{tab:flickr-user-bookmarks}), and merged into a single list of top tags which summarizes the Full Tagging Activity (in the following, FTA) on a resource (see Table \vref{tab:flickr-top-tags}).

\begin{table}[htb]
\begin{center}
 \tiny
 \begin{tabular}{|l|l|}
  \hline
  \multicolumn{2}{|c|}{\scriptsize\strut \textbf{User annotations: Flickr.com}} \\
  \hline
  \scriptsize\strut \textbf{User 1:} & \scriptsize\strut \texttt{photo}, \scriptsize\strut \texttt{photography}, \scriptsize\strut \texttt{images}, \scriptsize\strut \texttt{pictures} \\
  \hline
  \scriptsize\strut \textbf{User 2:} & \scriptsize\strut \texttt{photo}, \scriptsize\strut \texttt{web2.0}, \scriptsize\strut \texttt{social}, \scriptsize\strut \texttt{tools}, \scriptsize\strut \texttt{blog} \\
  \hline
  \scriptsize\strut \textbf{User 3:} & \scriptsize\strut \texttt{cloud}, \scriptsize\strut \texttt{pictures}, \scriptsize\strut \texttt{sharing} \\
  \hline
  \scriptsize\strut \textbf{User 4:} & \scriptsize\strut \texttt{flickr}, \scriptsize\strut \texttt{photos} \\
  \hline
  \scriptsize\strut \textbf{User 5:} & \scriptsize\strut \texttt{photo}, \scriptsize\strut \texttt{sharing}, \scriptsize\strut \texttt{tool} \\
  \hline
 \end{tabular}
\end{center}
\caption[Example of annotations for Flickr.com on Delicious]{Example of annotations for the URL Flickr.com on the social bookmarking site Delicious.}
\label{tab:flickr-user-bookmarks}
\end{table}

\begin{table}[htb]
\begin{center}
 \tiny
 \begin{tabular}{|l|r|}
  \hline
  \multicolumn{2}{|c|}{\scriptsize\strut \textbf{Top tags: Flickr.com}} \\
  \multicolumn{2}{|c|}{\scriptsize\strut \textbf{(79,681 users)}} \\
  \hline
  \scriptsize\strut \texttt{photos} & \scriptsize\strut 22,712 \\
  \hline
  \scriptsize\strut \texttt{flickr} & \scriptsize\strut 19,046 \\
  \hline
  \scriptsize\strut \texttt{photography} & \scriptsize\strut 15,968 \\
  \hline
  \scriptsize\strut \texttt{photo} & \scriptsize\strut 15,225 \\
  \hline
  \scriptsize\strut \texttt{sharing} & \scriptsize\strut 10,648 \\
  \hline
  \scriptsize\strut \texttt{images} & \scriptsize\strut 9,637 \\
  \hline
  \scriptsize\strut \texttt{web2.0} & \scriptsize\strut 9,528 \\
  \hline
  \scriptsize\strut \texttt{community} & \scriptsize\strut 4,571 \\
  \hline
  \scriptsize\strut \texttt{social} & \scriptsize\strut 3,798 \\
  \hline
  \scriptsize\strut \texttt{pictures} & \scriptsize\strut 3,115 \\
  \hline
 \end{tabular}
\end{center}
\caption[Example of top 10 tags for Flickr.com on Delicious]{Example of top tags for the URL Flickr.com on the social bookmarking site Delicious: the number associated to each tag represents the number of users annotating it.}
\label{tab:flickr-top-tags}
\end{table}

The tagging activity of a community of users on a resource creates an aggregated list of tags. A resource annotated by $p$ users will have a list of $n$ different tags, where each tag could have been assigned by $p$ or fewer users. The number of users who used a certain tag, $w_t$, defines a weight that allows to infer an ordered list of tags.

This aggregation of social tags was suggested as a means to feed the classification of resources \citep{noll_exploring_2008}. However, to the best of our knowledge, no research work has been conducted on their application to an automated classifier. Moreover, it is not clear what is a good way to represent the aggregation of tags. In this chapter, we will focus on these issues by proposing, analyzing and evaluating different representations for social tags so as to classifying resources, and also comparing their performance to other data sources including self-content and user reviews. We perform such a study on two different levels of the taxonomies, exploring thereby the suitability of social tags for broader and narrower categories.

\section{Representing Resources Using Tags}
\label{sec:representing-resources-using-tags}

We believe there are two major factors that should be considered for the representation of resources using social tags provided by users: (1) the selection of the tags that should be taken into account for the representation, and (2) the weights that should be assigned to those tags.

On the one hand, as regards to the selection of tags, one could think that not all the tags are useful for the representation, but just those in the top that most users have chosen. An important feature of social tagging systems is the ability of users to coincide on some tags provided by others. Thus, the coincidence of user annotations, which is reflected on the top tags, can be considered as a consensus of the main tags that better fit the description of the resource. However, the diversity on the annotations can also give users the opportunity to assign seldom-used tags that further detail the resource. The latter could encourage considering even tags in the tail. We will thus explore both tags in the top and the whole set.

On the other hand, the weight of each tag must be defined appropriately. We propose 4 different ways of assigning those weights:

\begin{itemize}
 \item \textbf{Tag ranks:} the weight is assigned according to the position of a tag in the ranked top of tags. Tags corresponding to the top 10 list of a resource are assigned a value in a rank-based way. The first-ranked tag is always set the value 1, 0.9 for the second, 0.8 for the third, and so on. This approach respects the position of each tag in the top 10, but the different gaps among tag weights are ignored.
 \item \textbf{Tag fractions:} the weight is computed according to the fraction of users who annotate a tag, $w_t / p$, i.e., the number of users annotating a tag on a resource, divided by the total number of users who annotated the resource. Taking into account both the number of users who bookmarked a resource $r$ and the weight of each tag $w_t$, it is possible to define the fraction of users assigning each tag. A tag would have been annotated by the totality of the users when its weight matches the user count of a resource, getting a value of 1 as the fraction. According to this, the value set to each tag is higher than 0 (since the considered tags have annotated by at least one user), and can be up to 1. This representation approach is similar to that by \cite{noll_exploring_2008} for their analysis of the similarity between social tags and the classification by experts. However, they ignore the least popular tags by giving a value of 0, what may give rise to the removal of several tags from the representation.
 \item \textbf{Unweighted:} in a binary way, the presence of a tag represents a value of 1, and its absence a value of 0. The only feature considered for this representation is the occurrence or non-occurrence of a tag in the annotations of a resource. This approach thereby ignores the weights of tags, and assigns a binary value to each feature in the vector.
 \item \textbf{Weighted according to user counts:} it considers the number of users assigning the tag ($w_t$) as a weight. The weight for each of the tags of a resource (${w_1,...,w_n}$) is considered as it is in this approach. Now, by definition, the weights of the tags are fully respected, although the amount of users bookmarking a resource is ignored. Note that different orders of magnitude are mixed up now, since the count of bookmarking users range within very different values. For instance, \cite{ramage_clusteringtagged_2009} used this approach in their work for clustering web pages, but they assumed it without comparing it to other representations.
\end{itemize}

Table \vref{tab:tag-weighting} shows an example of annotations on a resource, and how each of the 4 weighting measures would look like for the example.

\begin{table}[htb]
\begin{center}
 \tiny
 \begin{tabular}{|l|c|c|c|c|c|c|c|c|c|}
  \hline
  & \multicolumn{9}{c|}{\scriptsize\strut \textbf{FTA}} \\
  \hline
  & \multicolumn{6}{c|}{\scriptsize\strut \textbf{Top 10}} & \multicolumn{3}{c|}{} \\
  \hline
  \textbf{} & \scriptsize\strut \textbf{$t_{1}$} & \scriptsize\strut \textbf{$t_{2}$} & \scriptsize\strut \textbf{$t_{3}$} & \scriptsize\strut \textbf{...} & \scriptsize\strut \textbf{$t_{9}$} & \scriptsize\strut \textbf{$t_{10}$} & \scriptsize\strut \textbf{$t_{11}$} & \scriptsize\strut \textbf{...} & \scriptsize\strut \textbf{$t_{n}$} \\
  \hline
  \scriptsize\strut \textbf{Ranks} & \scriptsize\strut 1 & \scriptsize\strut 0.9 & \scriptsize\strut 0.8 & \scriptsize\strut ... & \scriptsize\strut 0.2 & \scriptsize\strut 0.1 & \scriptsize\strut 0 & \scriptsize\strut ... & \scriptsize\strut 0 \\
  \hline
  \scriptsize\strut \textbf{Fractions} & \scriptsize\strut 0.5 & \scriptsize\strut 0.3 & \scriptsize\strut 0.2 & \scriptsize\strut ... & \scriptsize\strut 0.02 & \scriptsize\strut 0.01 & \scriptsize\strut 0.01 & \scriptsize\strut ... & \scriptsize\strut 0.01 \\
  \hline
  \scriptsize\strut \textbf{Unweighted} & \scriptsize\strut 1 & \scriptsize\strut 1 & \scriptsize\strut 1 & \scriptsize\strut ... & \scriptsize\strut 1 & \scriptsize\strut 1 & \scriptsize\strut 1 & \scriptsize\strut ... & \scriptsize\strut 1 \\
  \hline
  \scriptsize\strut \textbf{Weighted} & \scriptsize\strut 50 & \scriptsize\strut 30 & \scriptsize\strut 20 & \scriptsize\strut ... & \scriptsize\strut 2 & \scriptsize\strut 1 & \scriptsize\strut 1 & \scriptsize\strut ... & \scriptsize\strut 1 \\
  \hline
 \end{tabular}
\end{center}
\caption[Example of the 4 representations of social tags]{Example of the 4 representations of social tags on a resource annotated by 100 users, and tags ranked 1st, 2nd and 3rd were annotated by 50, 30 and 20 users, respectively.}
\label{tab:tag-weighting}
\end{table}

Taking into account the factors above, we propose and analyze the 7 representation approaches summarized in Table \vref{tab:tag-representations}. All 4 weighting measures are included, as well as two selections of tags: the FTA including the whole set of tags, and the top 10 tags of each resource including the best-weighted ones\footnote{We selected the top 10 because it is usual to find that number of tags on social tagging systems. However, we could have chosen another value instead, yielding comparable conclusions. We provide additional results and information on this in Appendix \vref{c:topx-tags}.}. In the case of the rank-based weighting, we only apply it to the top 10 of tags, because it is defined to give a weight for only 10 tags.

\begin{table}[htb]
\begin{center}
 \tiny
 \begin{tabular}{|l|c|c|}
  \hline
  & \scriptsize\strut \textbf{Top 10} & \scriptsize\strut \textbf{FTA} \\
  \hline
  \scriptsize\strut \textbf{Tag ranks} & \scriptsize\strut x & \scriptsize\strut \\
  \hline
  \scriptsize\strut \textbf{Tag fractions} & \scriptsize\strut x & \scriptsize\strut x \\
  \hline
  \scriptsize\strut \textbf{Unweighted} & \scriptsize\strut x & \scriptsize\strut x \\
  \hline
  \scriptsize\strut \textbf{Weighted} & \scriptsize\strut x & \scriptsize\strut x \\
  \hline
 \end{tabular}
\end{center}
\caption[Summary of tag representations]{Summary of tag representations.}
\label{tab:tag-representations}
\end{table}

\section{Tag-based Classification}
\label{sec:tag-based-classification}

According to the experimental results in Chapter \vref{c:svm-classification}, we have used a multiclass SVM algorithm to perform the classification tasks, feeding the classifier with social tags from the three datasets introduced in Chapter \vref{c:datasets}. We got different sizes of training sets for each dataset, and generated 6 different random selections for each size. We present the accuracy results corresponding to the average of those 6 runs. Results are split into separate tables, with a table corresponding to each dataset (Delicious, LibraryThing, GoodReads). Each table includes results for all 7 representations introduced above, and both top and second levels of the corresponding taxonomy.

\begin{table}[htb]
\begin{center}
 \tiny
 \begin{tabular}{|l|c|c|c|c|c|c|c|}
  \hline
  \multicolumn{8}{|c|}{\scriptsize\strut \textbf{Delicious - ODP}} \\
  \hline
  \hline
  \multicolumn{8}{|c|}{\scriptsize\strut \textbf{Top level}} \\
  \hline
  \textbf{} & \scriptsize\strut \textbf{600} & \scriptsize\strut \textbf{1400} & \scriptsize\strut \textbf{2200} & \scriptsize\strut \textbf{3000} & \scriptsize\strut \textbf{4000} & \scriptsize\strut \textbf{5000} & \scriptsize\strut \textbf{6000} \\
  \hline
  \scriptsize\strut \textbf{Tag Ranks} & \scriptsize\strut .462 & \scriptsize\strut .501 & \scriptsize\strut .493 & \scriptsize\strut .501 & \scriptsize\strut .498 & \scriptsize\strut .501 & \scriptsize\strut .484 \\
  \hline
  \scriptsize\strut \textbf{Tag Fractions (Top 10)} & \scriptsize\strut .430 & \scriptsize\strut .447 & \scriptsize\strut .456 & \scriptsize\strut .467 & \scriptsize\strut .466 & \scriptsize\strut .462 & \scriptsize\strut .464 \\
  \hline
  \scriptsize\strut \textbf{Tag Fractions (FTA)} & \scriptsize\strut .442 & \scriptsize\strut .463 & \scriptsize\strut .457 & \scriptsize\strut .460 & \scriptsize\strut .461 & \scriptsize\strut .461 & \scriptsize\strut .461 \\
  \hline
  \scriptsize\strut \textbf{Unweighted Tags (Top 10)} & \scriptsize\strut .505 & \scriptsize\strut .510 & \scriptsize\strut .512 & \scriptsize\strut .517 & \scriptsize\strut .520 & \scriptsize\strut .522 & \scriptsize\strut .531 \\
  \hline
  \scriptsize\strut \textbf{Unweighted Tags (FTA)} & \scriptsize\strut .530 & \scriptsize\strut .556 & \scriptsize\strut .566 & \scriptsize\strut .572 & \scriptsize\strut .569 & \scriptsize\strut .571 & \scriptsize\strut .572 \\
  \hline
  \scriptsize\strut \textbf{Weighted Tags (Top 10)} & \scriptsize\strut .509 & \scriptsize\strut .576 & \scriptsize\strut .606 & \scriptsize\strut .625 & \scriptsize\strut .638 & \scriptsize\strut .645 & \scriptsize\strut .654 \\
  \hline
  \scriptsize\strut \textbf{Weighted Tags (FTA)} & \scriptsize\strut \textbf{.533} & \scriptsize\strut \textbf{.600} & \scriptsize\strut \textbf{.629} & \scriptsize\strut \textbf{.647} & \scriptsize\strut \textbf{.660} & \scriptsize\strut \textbf{.669} & \scriptsize\strut \textbf{.680} \\
  \hline
  \hline
  \multicolumn{8}{|c|}{\scriptsize\strut \textbf{Second level}} \\
  \hline
  \textbf{} & \scriptsize\strut \textbf{600} & \scriptsize\strut \textbf{1400} & \scriptsize\strut \textbf{2200} & \scriptsize\strut \textbf{3000} & \scriptsize\strut \textbf{4000} & \scriptsize\strut \textbf{5000} & \scriptsize\strut \textbf{6000} \\
  \hline
  \scriptsize\strut \textbf{Tag Ranks} & \scriptsize\strut .292 & \scriptsize\strut .332 & \scriptsize\strut .345 & \scriptsize\strut .349 & \scriptsize\strut .351 & \scriptsize\strut .349 & \scriptsize\strut .360 \\
  \hline
  \scriptsize\strut \textbf{Tag Fractions (Top 10)} & \scriptsize\strut .262 & \scriptsize\strut .280 & \scriptsize\strut .297 & \scriptsize\strut .304 & \scriptsize\strut .315 & \scriptsize\strut .317 & \scriptsize\strut .349 \\
  \hline
  \scriptsize\strut \textbf{Tag Fractions (FTA)} & \scriptsize\strut .249 & \scriptsize\strut .279 & \scriptsize\strut .294 & \scriptsize\strut .308 & \scriptsize\strut .302 & \scriptsize\strut .302 & \scriptsize\strut .336 \\
  \hline
  \scriptsize\strut \textbf{Unweighted Tags (Top 10)} & \scriptsize\strut .315 & \scriptsize\strut .340 & \scriptsize\strut .354 & \scriptsize\strut .351 & \scriptsize\strut .348 & \scriptsize\strut .365 & \scriptsize\strut .361 \\
  \hline
  \scriptsize\strut \textbf{Unweighted Tags (FTA)} & \scriptsize\strut \textbf{.411} & \scriptsize\strut \textbf{.480} & \scriptsize\strut \textbf{.502} & \scriptsize\strut .509 & \scriptsize\strut .519 & \scriptsize\strut .509 & \scriptsize\strut .529 \\
  \hline
  \scriptsize\strut \textbf{Weighted Tags (Top 10)} & \scriptsize\strut .342 & \scriptsize\strut .432 & \scriptsize\strut .475 & \scriptsize\strut .497 & \scriptsize\strut .517 & \scriptsize\strut .532 & \scriptsize\strut .545 \\
  \hline
  \scriptsize\strut \textbf{Weighted Tags (FTA)} & \scriptsize\strut .359 & \scriptsize\strut .453 & \scriptsize\strut .498 & \scriptsize\strut \textbf{.522} & \scriptsize\strut \textbf{.541} & \scriptsize\strut \textbf{.556} & \scriptsize\strut \textbf{.568} \\
  \hline
 \end{tabular}
\end{center}
\caption[Accuracy results for tag-based web page classification]{Accuracy results for tag-based web page classification.}
\label{tab:delicious-tag-results}
\end{table}

Table \vref{tab:delicious-tag-results} shows the results on the Delicious dataset. At a first glance, it is clear that rank-based and fraction-based approaches perform much worse than the rest. Among the others, the weighted approach performs better than the unweighted one, so that considering the number of users assigning each tag seems to be the best option. Accordingly, considering the total number of users annotating a resource does not seem helpful, as shown by the underperformance of the fraction-based approach.

The unweighted approach may perform better than the weighted one for small training sets when it comes to the second level classification. It seems reasonable that the weighted approach requires more training instances to correctly represent the large diversity of possible values, and especially when the number of categories increases, as it happens on the second level. This is reflected in the underperformance of the weighted approach for the smaller training sets upon the second level of the taxonomy. However, the outperformance of the weighted approach becomes clear when the size of the training set increases.

In most cases, FTA outperforms the top 10, even though the gap is not very large. This shows that top tags are the most useful, but the rest may also be helpful to a lesser extent. Accordingly, tags in the tail chosen by fewer users provide useful data that should not be discarded. The weighted approach on all the tags performs the best for Delicious.

\begin{table}[p]
\begin{center}
 \tiny
 \begin{tabular}{|l|c|c|c|c|c|c|c|}
  \hline
  \multicolumn{8}{|c|}{\scriptsize\strut \textbf{LibraryThing - DDC}} \\
  \hline
  \hline
  \multicolumn{8}{|c|}{\scriptsize\strut \textbf{Top level}} \\
  \hline
  \textbf{} & \scriptsize\strut \textbf{3000} & \scriptsize\strut \textbf{6000} & \scriptsize\strut \textbf{9000} & \scriptsize\strut \textbf{12000} & \scriptsize\strut \textbf{15000} & \scriptsize\strut \textbf{18000} & \scriptsize\strut \textbf{21000} \\
  \hline
  \scriptsize\strut \textbf{Tag Ranks} & \scriptsize\strut .791 & \scriptsize\strut .783 & \scriptsize\strut .778 & \scriptsize\strut .782 & \scriptsize\strut .788 & \scriptsize\strut .787 & \scriptsize\strut .797 \\
  \hline
  \scriptsize\strut \textbf{Tag Fractions (Top 10)} & \scriptsize\strut .719 & \scriptsize\strut .717 & \scriptsize\strut .720 & \scriptsize\strut .721 & \scriptsize\strut .727 & \scriptsize\strut .721 & \scriptsize\strut .724 \\
  \hline
  \scriptsize\strut \textbf{Tag Fractions (FTA)} & \scriptsize\strut .700 & \scriptsize\strut .696 & \scriptsize\strut .701 & \scriptsize\strut .702 & \scriptsize\strut .706 & \scriptsize\strut .701 & \scriptsize\strut .706 \\
  \hline
  \scriptsize\strut \textbf{Unweighted Tags (Top 10)} & \scriptsize\strut .756 & \scriptsize\strut .763 & \scriptsize\strut .753 & \scriptsize\strut .766 & \scriptsize\strut .759 & \scriptsize\strut .759 & \scriptsize\strut .758 \\
  \hline
  \scriptsize\strut \textbf{Unweighted Tags (FTA)} & \scriptsize\strut .624 & \scriptsize\strut .622 & \scriptsize\strut .628 & \scriptsize\strut .629 & \scriptsize\strut .629 & \scriptsize\strut .628 & \scriptsize\strut .624 \\
  \hline
  \scriptsize\strut \textbf{Weighted Tags (Top 10)} & \scriptsize\strut .858 & \scriptsize\strut .861 & \scriptsize\strut .862 & \scriptsize\strut .865 & \scriptsize\strut .866 & \scriptsize\strut .866 & \scriptsize\strut .864 \\
  \hline
  \scriptsize\strut \textbf{Weighted Tags (FTA)} & \scriptsize\strut \textbf{.861} & \scriptsize\strut \textbf{.864} & \scriptsize\strut \textbf{.864} & \scriptsize\strut \textbf{.867} & \scriptsize\strut \textbf{.869} & \scriptsize\strut \textbf{.869} & \scriptsize\strut \textbf{.868} \\
  \hline
  \hline
  \multicolumn{8}{|c|}{\scriptsize\strut \textbf{Second level}} \\
  \hline
  \textbf{} & \scriptsize\strut \textbf{3000} & \scriptsize\strut \textbf{6000} & \scriptsize\strut \textbf{9000} & \scriptsize\strut \textbf{12000} & \scriptsize\strut \textbf{15000} & \scriptsize\strut \textbf{18000} & \scriptsize\strut \textbf{21000} \\
  \hline
  \scriptsize\strut \textbf{Tag Ranks} & \scriptsize\strut .520 & \scriptsize\strut .520 & \scriptsize\strut .526 & \scriptsize\strut .530 & \scriptsize\strut .527 & \scriptsize\strut .525 & \scriptsize\strut .532 \\
  \hline
  \scriptsize\strut \textbf{Tag Fractions (Top 10)} & \scriptsize\strut .511 & \scriptsize\strut .513 & \scriptsize\strut .511 & \scriptsize\strut .513 & \scriptsize\strut .513 & \scriptsize\strut .517 & \scriptsize\strut .521 \\
  \hline
  \scriptsize\strut \textbf{Tag Fractions (FTA)} & \scriptsize\strut .465 & \scriptsize\strut .474 & \scriptsize\strut .469 & \scriptsize\strut .470 & \scriptsize\strut .470 & \scriptsize\strut .472 & \scriptsize\strut .477 \\
  \hline
  \scriptsize\strut \textbf{Unweighted Tags (Top 10)} & \scriptsize\strut .507 & \scriptsize\strut .538 & \scriptsize\strut .538 & \scriptsize\strut .532 & \scriptsize\strut .543 & \scriptsize\strut .528 & \scriptsize\strut .539 \\
  \hline
  \scriptsize\strut \textbf{Unweighted Tags (FTA)} & \scriptsize\strut .515 & \scriptsize\strut .533 & \scriptsize\strut .530 & \scriptsize\strut .533 & \scriptsize\strut .538 & \scriptsize\strut .536 & \scriptsize\strut .538 \\
  \hline
  \scriptsize\strut \textbf{Weighted Tags (Top 10)} & \scriptsize\strut .679 & \scriptsize\strut .687 & \scriptsize\strut .696 & \scriptsize\strut .696 & \scriptsize\strut .701 & \scriptsize\strut .701 & \scriptsize\strut .704 \\
  \hline
  \scriptsize\strut \textbf{Weighted Tags (FTA)} & \scriptsize\strut \textbf{.690} & \scriptsize\strut \textbf{.700} & \scriptsize\strut \textbf{.707} & \scriptsize\strut \textbf{.709} & \scriptsize\strut \textbf{.715} & \scriptsize\strut \textbf{.712} & \scriptsize\strut \textbf{.715} \\
  \hline
  \hline
  \multicolumn{8}{|c|}{\scriptsize\strut \textbf{LibraryThing - LCC}} \\
  \hline
  \hline
  \multicolumn{8}{|c|}{\scriptsize\strut \textbf{Top level}} \\
  \hline
  \textbf{} & \scriptsize\strut \textbf{3000} & \scriptsize\strut \textbf{6000} & \scriptsize\strut \textbf{9000} & \scriptsize\strut \textbf{12000} & \scriptsize\strut \textbf{15000} & \scriptsize\strut \textbf{18000} & \scriptsize\strut \textbf{21000} \\
  \hline
  \scriptsize\strut \textbf{Tag Ranks} & \scriptsize\strut .783 & \scriptsize\strut .790 & \scriptsize\strut .788 & \scriptsize\strut .783 & \scriptsize\strut .789 & \scriptsize\strut .795 & \scriptsize\strut .790 \\
  \hline
  \scriptsize\strut \textbf{Tag Fractions (Top 10)} & \scriptsize\strut .739 & \scriptsize\strut .740 & \scriptsize\strut .741 & \scriptsize\strut .743 & \scriptsize\strut .741 & \scriptsize\strut .738 & \scriptsize\strut .746 \\
  \hline
  \scriptsize\strut \textbf{Tag Fractions (FTA)} & \scriptsize\strut .711 & \scriptsize\strut .715 & \scriptsize\strut .715 & \scriptsize\strut .717 & \scriptsize\strut .714 & \scriptsize\strut .712 & \scriptsize\strut .719 \\
  \hline
  \scriptsize\strut \textbf{Unweighted Tags (Top 10)} & \scriptsize\strut .759 & \scriptsize\strut .772 & \scriptsize\strut .764 & \scriptsize\strut .771 & \scriptsize\strut .763 & \scriptsize\strut .770 & \scriptsize\strut .763 \\
  \hline
  \scriptsize\strut \textbf{Unweighted Tags (FTA)} & \scriptsize\strut .654 & \scriptsize\strut .660 & \scriptsize\strut .661 & \scriptsize\strut .661 & \scriptsize\strut .658 & \scriptsize\strut .655 & \scriptsize\strut .661 \\
  \hline
  \scriptsize\strut \textbf{Weighted Tags (Top 10)} & \scriptsize\strut .852 & \scriptsize\strut .854 & \scriptsize\strut \textbf{.856} & \scriptsize\strut .858 & \scriptsize\strut .858 & \scriptsize\strut .855 & \scriptsize\strut .858 \\
  \hline
  \scriptsize\strut \textbf{Weighted Tags (FTA)} & \scriptsize\strut \textbf{.853} & \scriptsize\strut \textbf{.857} & \scriptsize\strut \textbf{.856} & \scriptsize\strut \textbf{.861} & \scriptsize\strut \textbf{.861} & \scriptsize\strut \textbf{.857} & \scriptsize\strut \textbf{.861} \\
  \hline
  \hline
  \multicolumn{8}{|c|}{\scriptsize\strut \textbf{Second level}} \\
  \hline
  \textbf{} & \scriptsize\strut \textbf{3000} & \scriptsize\strut \textbf{6000} & \scriptsize\strut \textbf{9000} & \scriptsize\strut \textbf{12000} & \scriptsize\strut \textbf{15000} & \scriptsize\strut \textbf{18000} & \scriptsize\strut \textbf{21000} \\
  \hline
  \scriptsize\strut \textbf{Tag Ranks} & \scriptsize\strut .519 & \scriptsize\strut .511 & \scriptsize\strut .515 & \scriptsize\strut .518 & \scriptsize\strut .512 & \scriptsize\strut .511 & \scriptsize\strut .520 \\
  \hline
  \scriptsize\strut \textbf{Tag Fractions (Top 10)} & \scriptsize\strut .414 & \scriptsize\strut .413 & \scriptsize\strut .413 & \scriptsize\strut .415 & \scriptsize\strut .417 & \scriptsize\strut .411 & \scriptsize\strut .417 \\
  \hline
  \scriptsize\strut \textbf{Tag Fractions (FTA)} & \scriptsize\strut .408 & \scriptsize\strut .409 & \scriptsize\strut .408 & \scriptsize\strut .410 & \scriptsize\strut .410 & \scriptsize\strut .409 & \scriptsize\strut .410 \\
  \hline
  \scriptsize\strut \textbf{Unweighted Tags (Top 10)} & \scriptsize\strut .542 & \scriptsize\strut .568 & \scriptsize\strut .564 & \scriptsize\strut .565 & \scriptsize\strut .579 & \scriptsize\strut .550 & \scriptsize\strut .576 \\
  \hline
  \scriptsize\strut \textbf{Unweighted Tags (FTA)} & \scriptsize\strut .596 & \scriptsize\strut .612 & \scriptsize\strut .608 & \scriptsize\strut .616 & \scriptsize\strut .615 & \scriptsize\strut .606 & \scriptsize\strut .614 \\
  \hline
  \scriptsize\strut \textbf{Weighted Tags (Top 10)} & \scriptsize\strut .687 & \scriptsize\strut .710 & \scriptsize\strut .716 & \scriptsize\strut .720 & \scriptsize\strut .721 & \scriptsize\strut .722 & \scriptsize\strut .727 \\
  \hline
  \scriptsize\strut \textbf{Weighted Tags (FTA)} & \scriptsize\strut \textbf{.703} & \scriptsize\strut \textbf{.725} & \scriptsize\strut \textbf{.729} & \scriptsize\strut \textbf{.734} & \scriptsize\strut \textbf{.734} & \scriptsize\strut \textbf{.736} & \scriptsize\strut \textbf{.739} \\
  \hline
 \end{tabular}
\end{center}
\caption[Accuracy results for tag-based book classification on LibraryThing]{Accuracy results for tag-based book classification (LibraryThing).}
\label{tab:librarything-tag-results}
\end{table}

Table \vref{tab:librarything-tag-results} shows the results for the LibraryThing dataset, for DDC and LCC taxonomies. Similar to Delicious, the weighted one relying on the FTA is the outperforming approach, for both the top and second levels, and for both schemes.

Different from Delicious, the FTA-based approaches are not always better than those based on the top 10. This difference happens when using the unweighted approach for the top level classification. Many LibraryThing users tend to use personal tags describing whether or not they own the book (e.g., \texttt{own}), and the physical location of it (e.g., \texttt{a1}). These tags are barely used within a book, as personal tags that spread across books by a single user. Thus, ignoring tag weights and giving all of them the same weight overrates those personal and low-ranked tags. This may increase the likelihood of books containing a certain personal tag to be mispredicted for the same category by the classifier. Accordingly, this is the main reason for the slight gap between the top 10 and FTA based representations for the weighted approach. Tags below the top 10 are not as useful as on Delicious. Fortunately, the weighted approach underrates such personal tags, and the classifier is able to discriminate them, profiting from some low-ranked tags to slightly improve the performance. This outperformance is larger on the second level, suggesting that low-ranked tags provide more detailed description, and rather help for deeper classification.

Even though the weighted approach is the best in this case, using rank-based weights performs better than the unweighted approach, different from Delicious. After all, the weighted approach performs the best also for this dataset.

\begin{table}[p]
\begin{center}
 \tiny
 \begin{tabular}{|l|c|c|c|c|c|c|c|}
  \hline
  \multicolumn{8}{|c|}{\scriptsize\strut \textbf{GoodReads - DDC}} \\
  \hline
  \hline
  \multicolumn{8}{|c|}{\scriptsize\strut \textbf{Top level}} \\
  \hline
  \textbf{} & \scriptsize\strut \textbf{3000} & \scriptsize\strut \textbf{6000} & \scriptsize\strut \textbf{9000} & \scriptsize\strut \textbf{12000} & \scriptsize\strut \textbf{15000} & \scriptsize\strut \textbf{18000} & \scriptsize\strut \textbf{21000} \\
  \hline
  \scriptsize\strut \textbf{Tag Ranks} & \scriptsize\strut .652 & \scriptsize\strut .656 & \scriptsize\strut .659 & \scriptsize\strut .654 & \scriptsize\strut .650 & \scriptsize\strut .655 & \scriptsize\strut .668 \\
  \hline
  \scriptsize\strut \textbf{Tag Fractions (Top 10)} & \scriptsize\strut .660 & \scriptsize\strut .658 & \scriptsize\strut .662 & \scriptsize\strut .663 & \scriptsize\strut .671 & \scriptsize\strut .659 & \scriptsize\strut .664 \\
  \hline
  \scriptsize\strut \textbf{Tag Fractions (FTA)} & \scriptsize\strut .654 & \scriptsize\strut .653 & \scriptsize\strut .657 & \scriptsize\strut .658 & \scriptsize\strut .665 & \scriptsize\strut .655 & \scriptsize\strut .659 \\
  \hline
  \scriptsize\strut \textbf{Unweighted Tags (Top 10)} & \scriptsize\strut .647 & \scriptsize\strut .645 & \scriptsize\strut .643 & \scriptsize\strut .650 & \scriptsize\strut .639 & \scriptsize\strut .657 & \scriptsize\strut .647 \\
  \hline
  \scriptsize\strut \textbf{Unweighted Tags (FTA)} & \scriptsize\strut .635 & \scriptsize\strut .638 & \scriptsize\strut .637 & \scriptsize\strut .639 & \scriptsize\strut .639 & \scriptsize\strut .642 & \scriptsize\strut .640 \\
  \hline
  \scriptsize\strut \textbf{Weighted Tags (Top 10)} & \scriptsize\strut .728 & \scriptsize\strut .730 & \scriptsize\strut .736 & \scriptsize\strut .742 & \scriptsize\strut .739 & \scriptsize\strut .740 & \scriptsize\strut .740 \\
  \hline
  \scriptsize\strut \textbf{Weighted Tags (FTA)} & \scriptsize\strut \textbf{.745} & \scriptsize\strut \textbf{.747} & \scriptsize\strut \textbf{.754} & \scriptsize\strut \textbf{.757} & \scriptsize\strut \textbf{.757} & \scriptsize\strut \textbf{.757} & \scriptsize\strut \textbf{.756} \\
  \hline
  \hline
  \multicolumn{8}{|c|}{\scriptsize\strut \textbf{Second level}} \\
  \hline
  \textbf{} & \scriptsize\strut \textbf{3000} & \scriptsize\strut \textbf{6000} & \scriptsize\strut \textbf{9000} & \scriptsize\strut \textbf{12000} & \scriptsize\strut \textbf{15000} & \scriptsize\strut \textbf{18000} & \scriptsize\strut \textbf{21000} \\
  \hline
  \scriptsize\strut \textbf{Tag Ranks} & \scriptsize\strut .435 & \scriptsize\strut .439 & \scriptsize\strut .434 & \scriptsize\strut .447 & \scriptsize\strut .445 & \scriptsize\strut .443 & \scriptsize\strut .447 \\
  \hline
  \scriptsize\strut \textbf{Tag Fractions (Top 10)} & \scriptsize\strut .445 & \scriptsize\strut .450 & \scriptsize\strut .450 & \scriptsize\strut .452 & \scriptsize\strut .452 & \scriptsize\strut .453 & \scriptsize\strut .458 \\
  \hline
  \scriptsize\strut \textbf{Tag Fractions (FTA)} & \scriptsize\strut .432 & \scriptsize\strut .440 & \scriptsize\strut .439 & \scriptsize\strut .440 & \scriptsize\strut .440 & \scriptsize\strut .441 & \scriptsize\strut .445 \\
  \hline
  \scriptsize\strut \textbf{Unweighted Tags (Top 10)} & \scriptsize\strut .430 & \scriptsize\strut .440 & \scriptsize\strut .441 & \scriptsize\strut .443 & \scriptsize\strut .435 & \scriptsize\strut .440 & \scriptsize\strut .449 \\
  \hline
  \scriptsize\strut \textbf{Unweighted Tags (FTA)} & \scriptsize\strut .450 & \scriptsize\strut .460 & \scriptsize\strut .447 & \scriptsize\strut .454 & \scriptsize\strut .453 & \scriptsize\strut .458 & \scriptsize\strut .452 \\
  \hline
  \scriptsize\strut \textbf{Weighted Tags (Top 10)} & \scriptsize\strut .487 & \scriptsize\strut .500 & \scriptsize\strut .503 & \scriptsize\strut .505 & \scriptsize\strut .507 & \scriptsize\strut .508 & \scriptsize\strut .510 \\
  \hline
  \scriptsize\strut \textbf{Weighted Tags (FTA)} & \scriptsize\strut \textbf{.509} & \scriptsize\strut \textbf{.520} & \scriptsize\strut \textbf{.528} & \scriptsize\strut \textbf{.528} & \scriptsize\strut \textbf{.530} & \scriptsize\strut \textbf{.529} & \scriptsize\strut \textbf{.530} \\
  \hline
  \hline
  \multicolumn{8}{|c|}{\scriptsize\strut \textbf{GoodReads - LCC}} \\
  \hline
  \hline
  \multicolumn{8}{|c|}{\scriptsize\strut \textbf{Top level}} \\
  \hline
  \textbf{} & \scriptsize\strut \textbf{3000} & \scriptsize\strut \textbf{6000} & \scriptsize\strut \textbf{9000} & \scriptsize\strut \textbf{12000} & \scriptsize\strut \textbf{15000} & \scriptsize\strut \textbf{18000} & \scriptsize\strut \textbf{21000} \\
  \hline
  \scriptsize\strut \textbf{Tag Ranks} & \scriptsize\strut .625 & \scriptsize\strut .636 & \scriptsize\strut .629 & \scriptsize\strut .630 & \scriptsize\strut .632 & \scriptsize\strut .630 & \scriptsize\strut .631 \\
  \hline
  \scriptsize\strut \textbf{Tag Fractions (Top 10)} & \scriptsize\strut .657 & \scriptsize\strut .664 & \scriptsize\strut .665 & \scriptsize\strut .667 & \scriptsize\strut .667 & \scriptsize\strut .663 & \scriptsize\strut .674 \\
  \hline
  \scriptsize\strut \textbf{Tag Fractions (FTA)} & \scriptsize\strut .650 & \scriptsize\strut .658 & \scriptsize\strut .656 & \scriptsize\strut .658 & \scriptsize\strut .659 & \scriptsize\strut .654 & \scriptsize\strut .663 \\
  \hline
  \scriptsize\strut \textbf{Unweighted Tags (Top 10)} & \scriptsize\strut .625 & \scriptsize\strut .626 & \scriptsize\strut .633 & \scriptsize\strut .633 & \scriptsize\strut .634 & \scriptsize\strut .623 & \scriptsize\strut .629 \\
  \hline
  \scriptsize\strut \textbf{Unweighted Tags (FTA)} & \scriptsize\strut .642 & \scriptsize\strut .648 & \scriptsize\strut .653 & \scriptsize\strut .651 & \scriptsize\strut .647 & \scriptsize\strut .639 & \scriptsize\strut .653 \\
  \hline
  \scriptsize\strut \textbf{Weighted Tags (Top 10)} & \scriptsize\strut .700 & \scriptsize\strut .711 & \scriptsize\strut .711 & \scriptsize\strut .714 & \scriptsize\strut .713 & \scriptsize\strut .713 & \scriptsize\strut .721 \\
  \hline
  \scriptsize\strut \textbf{Weighted Tags (FTA)} & \scriptsize\strut \textbf{.725} & \scriptsize\strut \textbf{.731} & \scriptsize\strut \textbf{.737} & \scriptsize\strut \textbf{.738} & \scriptsize\strut \textbf{.734} & \scriptsize\strut \textbf{.731} & \scriptsize\strut \textbf{.743} \\
  \hline
  \hline
  \multicolumn{8}{|c|}{\scriptsize\strut \textbf{Second level}} \\
  \hline
  \textbf{} & \scriptsize\strut \textbf{3000} & \scriptsize\strut \textbf{6000} & \scriptsize\strut \textbf{9000} & \scriptsize\strut \textbf{12000} & \scriptsize\strut \textbf{15000} & \scriptsize\strut \textbf{18000} & \scriptsize\strut \textbf{21000} \\
  \hline
  \scriptsize\strut \textbf{Tag Ranks} & \scriptsize\strut .404 & \scriptsize\strut .411 & \scriptsize\strut .410 & \scriptsize\strut .403 & \scriptsize\strut .404 & \scriptsize\strut .405 & \scriptsize\strut .407 \\
  \hline
  \scriptsize\strut \textbf{Tag Fractions (Top 10)} & \scriptsize\strut .412 & \scriptsize\strut .421 & \scriptsize\strut .426 & \scriptsize\strut .427 & \scriptsize\strut .430 & \scriptsize\strut .427 & \scriptsize\strut .427 \\
  \hline
  \scriptsize\strut \textbf{Tag Fractions (FTA)} & \scriptsize\strut .418 & \scriptsize\strut .427 & \scriptsize\strut .431 & \scriptsize\strut .432 & \scriptsize\strut .433 & \scriptsize\strut .432 & \scriptsize\strut .433 \\
  \hline
  \scriptsize\strut \textbf{Unweighted Tags (Top 10)} & \scriptsize\strut .414 & \scriptsize\strut .419 & \scriptsize\strut .420 & \scriptsize\strut .415 & \scriptsize\strut .414 & \scriptsize\strut .422 & \scriptsize\strut .435 \\
  \hline
  \scriptsize\strut \textbf{Unweighted Tags (FTA)} & \scriptsize\strut .462 & \scriptsize\strut .475 & \scriptsize\strut .467 & \scriptsize\strut .478 & \scriptsize\strut .477 & \scriptsize\strut .481 & \scriptsize\strut .484 \\
  \hline
  \scriptsize\strut \textbf{Weighted Tags (Top 10)} & \scriptsize\strut .467 & \scriptsize\strut .479 & \scriptsize\strut .487 & \scriptsize\strut .486 & \scriptsize\strut .491 & \scriptsize\strut .491 & \scriptsize\strut .493 \\
  \hline
  \scriptsize\strut \textbf{Weighted Tags (FTA)} & \scriptsize\strut \textbf{.494} & \scriptsize\strut \textbf{.507} & \scriptsize\strut \textbf{.510} & \scriptsize\strut \textbf{.514} & \scriptsize\strut \textbf{.513} & \scriptsize\strut \textbf{.517} & \scriptsize\strut \textbf{.519} \\
  \hline
 \end{tabular}
\end{center}
\caption[Accuracy results for tag-based book classification on GoodReads]{Accuracy results for tag-based book classification (GoodReads).}
\label{tab:goodreads-tag-results}
\end{table}

Table \vref{tab:goodreads-tag-results} shows the results for the GoodReads dataset, for DDC and LCC taxonomies. Again, the FTA-based weighted approach is the best one, with clearly outperforming results for both top and second levels on both schemes, DDC and LCC. Different from LibraryThing, though, FTA-based approaches perform better than top 10 based ones in most cases. This shows that GoodReads users tend to use fewer personal biased tags, making low-ranked tags much more useful than for LibraryThing. Despite these differences, the weighted approach is clearly the best approach for this dataset as well.

For both LibraryThing and GoodReads, the results look very similar for both taxonomies, DDC and LCC. Even though the results are slightly better for the former, both yield similar conclusions when comparing the gaps between representation approaches. This strengthens the usefulness of the weighted approach regardless of the taxonomy being considered.

Summarizing the results for the three datasets, the FTA-based weighted approach has shown to be the best. Even though not every low-ranked tag seems useful for the classification task, the weighted approach is able to establish their representativity to the resource, getting the best results by using all the tags. Thereby, the aggregation of user annotations has shown to be crucial to define the representativity of a tag with respect to the annotated resource.

\section{Comparing Social Tags to Other Data Sources}
\label{sec:comparing-tags-to-other-sources}

After we got the best representation approach to perform the classification experiments using social tags, we aimed at comparing their performance to that by other data sources. As we introduced previously in Chapter \vref{c:datasets}, we gathered additional data for the resources we are working on, i.e., web pages and books. In both cases, we tried to gather two more types of data: content and reviews. Regarding web pages, we rely on the textual content contained in the HTML source and user reviews fetched from social networks. In the case of books, we consider synopses and editorial reviews as a summary of their content, and user-generated reviews on the other hand.

With those content and user reviews, we created a representation based on the bag-of-words model \citep{zellig70distributional}. We merged all the texts available for each source, and created a single text with them. In order to clean up those texts, we stripped HTML tags, removed stop-words and stemmed the remaining words \citep{porter80algorithm}. Then, we weighted the words according to the TF-IDF scheme. The final representation of a resource, either based on content or user reviews, is a vector composed by words weighted by their TF-IDF values.

We use the same method as above for the creation of different training set sizes with 6 runs. As both LibraryThing and GoodReads work on the same books, the content and user reviews are the same in these cases, so that we group their results into a single table. For the three datasets we work with, we show the results of using content and comments, and compare them to the best tag-based approach, that is, the FTA-based weighted approach.

\begin{table}[htb]
\begin{center}
 \tiny
 \begin{tabular}{|l|c|c|c|c|c|c|c|}
  \hline
  \multicolumn{8}{|c|}{\scriptsize\strut \textbf{Delicious - ODP}} \\
  \hline
  \hline
  \multicolumn{8}{|c|}{\scriptsize\strut \textbf{Top level}} \\
  \hline
  \textbf{} & \scriptsize\strut \textbf{600} & \scriptsize\strut \textbf{1400} & \scriptsize\strut \textbf{2200} & \scriptsize\strut \textbf{3000} & \scriptsize\strut \textbf{4000} & \scriptsize\strut \textbf{5000} & \scriptsize\strut \textbf{6000} \\
  \hline
  \scriptsize\strut \textbf{Content} & \scriptsize\strut .518 & \scriptsize\strut .561 & \scriptsize\strut .579 & \scriptsize\strut .588 & \scriptsize\strut .595 & \scriptsize\strut .604 & \scriptsize\strut .610 \\
  \hline
  \scriptsize\strut \textbf{Reviews} & \scriptsize\strut .520 & \scriptsize\strut .578 & \scriptsize\strut .602 & \scriptsize\strut .618 & \scriptsize\strut .630 & \scriptsize\strut .639 & \scriptsize\strut .646 \\
  \hline
  \scriptsize\strut \textbf{Tags} & \scriptsize\strut \textbf{.533} & \scriptsize\strut \textbf{.600} & \scriptsize\strut \textbf{.629} & \scriptsize\strut \textbf{.647} & \scriptsize\strut \textbf{.660} & \scriptsize\strut \textbf{.669} & \scriptsize\strut \textbf{.680} \\
  \hline
  \hline
  \multicolumn{8}{|c|}{\scriptsize\strut \textbf{Second level}} \\
  \hline
  \textbf{} & \scriptsize\strut \textbf{600} & \scriptsize\strut \textbf{1400} & \scriptsize\strut \textbf{2200} & \scriptsize\strut \textbf{3000} & \scriptsize\strut \textbf{4000} & \scriptsize\strut \textbf{5000} & \scriptsize\strut \textbf{6000} \\
  \hline
  \scriptsize\strut \textbf{Content} & \scriptsize\strut .337 & \scriptsize\strut .394 & \scriptsize\strut .422 & \scriptsize\strut .437 & \scriptsize\strut .450 & \scriptsize\strut .464 & \scriptsize\strut .470 \\
  \hline
  \scriptsize\strut \textbf{Reviews} & \scriptsize\strut .349 & \scriptsize\strut .423 & \scriptsize\strut .459 & \scriptsize\strut .478 & \scriptsize\strut .497 & \scriptsize\strut .511 & \scriptsize\strut .524 \\
  \hline
  \scriptsize\strut \textbf{Tags} & \scriptsize\strut \textbf{.359} & \scriptsize\strut \textbf{.453} & \scriptsize\strut \textbf{.498} & \scriptsize\strut \textbf{.522} & \scriptsize\strut \textbf{.541} & \scriptsize\strut \textbf{.556} & \scriptsize\strut \textbf{.568} \\
  \hline
 \end{tabular}
\end{center}
\caption[Accuracy results comparing different data sources on web page classification]{Accuracy results comparing different data sources on web page classification.}
\label{tab:delicious-sources-results}
\end{table}

Table \vref{tab:delicious-sources-results} shows the results for the Delicious dataset. In this case, self-content of web pages is the worst data source out of the three we studied. Results by self-content are far below from those by reviews and tags. Likewise, social tags are clearly the best data source for the classification task. There is a clear outperformance of tags for the top level, but the difference is even larger for the second level. This strengthens one of the main motivations of this thesis, i.e., the fact that self-content is not always representative of its aboutness, and other data sources can provide more accurate definitions.

\begin{table}[htb]
\begin{center}
 \tiny
 \begin{tabular}{|l|c|c|c|c|c|c|c|}
  \hline
  \multicolumn{8}{|c|}{\scriptsize\strut \textbf{LibraryThing \& GoodReads - DDC}} \\
  \hline
  \hline
  \multicolumn{8}{|c|}{\scriptsize\strut \textbf{Top level}} \\
  \hline
  \textbf{} & \scriptsize\strut \textbf{3000} & \scriptsize\strut \textbf{6000} & \scriptsize\strut \textbf{9000} & \scriptsize\strut \textbf{12000} & \scriptsize\strut \textbf{15000} & \scriptsize\strut \textbf{18000} & \scriptsize\strut \textbf{21000} \\
  \hline
  \scriptsize\strut \textbf{Content} & \scriptsize\strut .767 & \scriptsize\strut .792 & \scriptsize\strut .802 & \scriptsize\strut .809 & \scriptsize\strut .809 & \scriptsize\strut .815 & \scriptsize\strut .817 \\
  \hline
  \scriptsize\strut \textbf{Reviews} & \scriptsize\strut .777 & \scriptsize\strut .808 & \scriptsize\strut .820 & \scriptsize\strut .831 & \scriptsize\strut .833 & \scriptsize\strut .839 & \scriptsize\strut .840 \\
  \hline
  \scriptsize\strut \textbf{Tags (LibraryThing)} & \scriptsize\strut \textbf{.861} & \scriptsize\strut \textbf{.864} & \scriptsize\strut \textbf{.864} & \scriptsize\strut \textbf{.867} & \scriptsize\strut \textbf{.869} & \scriptsize\strut \textbf{.869} & \scriptsize\strut \textbf{.868} \\
  \hline
  \scriptsize\strut \textbf{Tags (GoodReads)} & \scriptsize\strut .745 & \scriptsize\strut .747 & \scriptsize\strut .754 & \scriptsize\strut .757 & \scriptsize\strut .757 & \scriptsize\strut .757 & \scriptsize\strut .756 \\
  \hline
  \hline
  \multicolumn{8}{|c|}{\scriptsize\strut \textbf{Second level}} \\
  \hline
  \textbf{} & \scriptsize\strut \textbf{3000} & \scriptsize\strut \textbf{6000} & \scriptsize\strut \textbf{9000} & \scriptsize\strut \textbf{12000} & \scriptsize\strut \textbf{15000} & \scriptsize\strut \textbf{18000} & \scriptsize\strut \textbf{21000} \\
  \hline
  \scriptsize\strut \textbf{Content} & \scriptsize\strut .572 & \scriptsize\strut .612 & \scriptsize\strut .631 & \scriptsize\strut .643 & \scriptsize\strut .649 & \scriptsize\strut .657 & \scriptsize\strut .660 \\
  \hline
  \scriptsize\strut \textbf{Reviews} & \scriptsize\strut .582 & \scriptsize\strut .628 & \scriptsize\strut .651 & \scriptsize\strut .667 & \scriptsize\strut .678 & \scriptsize\strut .685 & \scriptsize\strut .693 \\
  \hline
  \scriptsize\strut \textbf{Tags (LibraryThing)} & \scriptsize\strut \textbf{.690} & \scriptsize\strut \textbf{.700} & \scriptsize\strut \textbf{.707} & \scriptsize\strut \textbf{.709} & \scriptsize\strut \textbf{.715} & \scriptsize\strut \textbf{.712} & \scriptsize\strut \textbf{.715} \\
  \hline
  \scriptsize\strut \textbf{Tags (GoodReads)} & \scriptsize\strut .509 & \scriptsize\strut .520 & \scriptsize\strut .528 & \scriptsize\strut .528 & \scriptsize\strut .530 & \scriptsize\strut .529 & \scriptsize\strut .530 \\
  \hline
  \hline
  \multicolumn{8}{|c|}{\scriptsize\strut \textbf{LibraryThing \& GoodReads - LCC}} \\
  \hline
  \hline
  \multicolumn{8}{|c|}{\scriptsize\strut \textbf{Top level}} \\
  \hline
  \textbf{} & \scriptsize\strut \textbf{3000} & \scriptsize\strut \textbf{6000} & \scriptsize\strut \textbf{9000} & \scriptsize\strut \textbf{12000} & \scriptsize\strut \textbf{15000} & \scriptsize\strut \textbf{18000} & \scriptsize\strut \textbf{21000} \\
  \hline
  \scriptsize\strut \textbf{Content} & \scriptsize\strut .767 & \scriptsize\strut .789 & \scriptsize\strut .798 & \scriptsize\strut .803 & \scriptsize\strut .806 & \scriptsize\strut .807 & \scriptsize\strut .810 \\
  \hline
  \scriptsize\strut \textbf{Reviews} & \scriptsize\strut .780 & \scriptsize\strut .803 & \scriptsize\strut .816 & \scriptsize\strut .823 & \scriptsize\strut .827 & \scriptsize\strut .828 & \scriptsize\strut .833 \\
  \hline
  \scriptsize\strut \textbf{Tags (LibraryThing)} & \scriptsize\strut \textbf{.853} & \scriptsize\strut \textbf{.857} & \scriptsize\strut \textbf{.856} & \scriptsize\strut \textbf{.861} & \scriptsize\strut \textbf{.861} & \scriptsize\strut \textbf{.857} & \scriptsize\strut \textbf{.861} \\
  \hline
  \scriptsize\strut \textbf{Tags (GoodReads)} & \scriptsize\strut .725 & \scriptsize\strut .731 & \scriptsize\strut .737 & \scriptsize\strut .738 & \scriptsize\strut .734 & \scriptsize\strut .731 & \scriptsize\strut .743 \\
  \hline
  \hline
  \multicolumn{8}{|c|}{\scriptsize\strut \textbf{Second level}} \\
  \hline
  \textbf{} & \scriptsize\strut \textbf{3000} & \scriptsize\strut \textbf{6000} & \scriptsize\strut \textbf{9000} & \scriptsize\strut \textbf{12000} & \scriptsize\strut \textbf{15000} & \scriptsize\strut \textbf{18000} & \scriptsize\strut \textbf{21000} \\
  \hline
  \scriptsize\strut \textbf{Content} & \scriptsize\strut .579 & \scriptsize\strut .620 & \scriptsize\strut .645 & \scriptsize\strut .658 & \scriptsize\strut .668 & \scriptsize\strut .673 & \scriptsize\strut .681 \\
  \hline
  \scriptsize\strut \textbf{Reviews} & \scriptsize\strut .581 & \scriptsize\strut .637 & \scriptsize\strut .664 & \scriptsize\strut .683 & \scriptsize\strut .698 & \scriptsize\strut .705 & \scriptsize\strut .712 \\
  \hline
  \scriptsize\strut \textbf{Tags (LibraryThing)} & \scriptsize\strut \textbf{.703} & \scriptsize\strut \textbf{.725} & \scriptsize\strut \textbf{.729} & \scriptsize\strut \textbf{.734} & \scriptsize\strut \textbf{.734} & \scriptsize\strut \textbf{.736} & \scriptsize\strut \textbf{.739} \\
  \hline
  \scriptsize\strut \textbf{Tags (GoodReads)} & \scriptsize\strut .494 & \scriptsize\strut .507 & \scriptsize\strut .510 & \scriptsize\strut .514 & \scriptsize\strut .513 & \scriptsize\strut .517 & \scriptsize\strut .519 \\
  \hline
 \end{tabular}
\end{center}
\caption[Accuracy results comparing different data sources on book classification]{Accuracy results comparing different data sources on book classification.}
\label{tab:books-sources-results}
\end{table}

Table \vref{tab:books-sources-results} shows the results for books, using tags from LibraryThing and GoodReads. In this case, we got results similar to Delicious when using tags from LibraryThing. Again, user reviews outperform the content (although we considered synopses and editorial reviews as a summary of the content of the book in this case). Moreover, social tags perform even better than user reviews, especially for the second level classification. The results are comparable for both classification schemes, DDC and LCC.

However, using tags from GoodReads is not enough to achieve results as good as using content or user reviews. GoodReads tags clearly underperform the other data sources. For this dataset, reviews are the data source scoring the best results. We believe that this happens because most GoodReads users do not provide tags when bookmarking a book\footnote{GoodReads does not encourage users to add tags as LibraryThing does, requiring a second click from the user, what brings about a large set of unannotated bookmarks, representing a ratio of more than 80\% bookmarks (see Section \vref{sec:dataset-stats})}. This way, a community providing fewer annotations gives rise to a less accurate aggregation of tags.

Summarizing, tags show to be really powerful as compared to other data sources like the content of the resource, or user reviews on it. However, large amounts of annotations are necessary in order to score outperforming results.

\section{Getting the Most Out of All Data Sources}
\label{sec:getting-the-most-out-of-all-data-sources}

Even though the tag-based representation outperforms in most cases the other two data sources, namely content and user reviews, all of them yield encouraging results and look good enough to combine them and try to improve even more the classifier's performance. The following questions arise from this statement: what if a classifier is guessing correctly while the others are making a mistake? Could we combine the predictions to get the most out of each of them?

An interesting approach to combine SVM classifiers is known as classifier committees \citep{sun_support_2004}. Classifier committees rely on the predictions of several classifiers, and combine them by means of a decision function, which serves to define the weight or relevance of each classifier in the final prediction. After applying the decision function on the predictions of all classifiers, a single unified prediction can be inferred.

An SVM classifier outputs a margin for each resource over each class in the taxonomy, meaning the reliability to belong to that class. The class with the largest positive margin for each resource is then selected as the classifier's prediction. The larger is the gap between the largest positive and the rest of margins, the more reliable can be considered the classifier's prediction. Thus, combining the predictions of SVM classifiers could be done by means of adding up their margins or reliability values for each class. Each resource will then have a new reliability value for each class, i.e., the sum of margins by different classifiers for a resource. Nonetheless, in this case, since each of the three classifiers work with different type of data, the range and scale of the margins they output differ. To solve this, we propose the normalization of the margins based on the maximum margin value outputted by each classifier, $\max (m_{i})$ (see Equation \ref{eq:margin-normalization}).

\begin{equation}
 m'_{ijc} = \frac{m_{ijc}}{\max (m_{i})}
 \label{eq:margin-normalization}
\end{equation}
	
where $m_{ijc}$ is the margin by the classifier $i$ between the resource $j$ and the hyperplane for the class $c$, and $m'_{ijc}$ is its value after normalizing it.

The class maximizing this sum of margins will be predicted by the classifier. Then, the sum of margins between the class $c$ and the resource $j$ using a committee with $n$ classifiers is defined by Equation \vref{eq:sum-margins}.

\begin{equation}
 S_{jc} = \sum_i^n m_{ijc}
 \label{eq:sum-margins}
\end{equation}

If the classifiers are working over $k$ classes, then the predicted class for the resource $j$ will be defined by Equation \vref{eq:committees-prediction}.

\begin{equation}
 C^*_{j} = \arg \max_{i = 1..k} (S_{ji})
 \label{eq:committees-prediction}
\end{equation}

As a toy example of the possible advantage of using classifier committees, Table \vref{tab:example-committees} shows the outputs in the form of margins of two classifiers for a resource in a taxonomy with 3 categories. Let this resource belong to the category \#2. The example shows that, on one hand, the classifier A has predicted the category \#1, with a margin of 1.2, but a slight gap to the category \#2 which gets a margin of 1.1. On the other hand, the classifier B says that the resource should be classified in category \#3 because of a margin of 1.2 was returned, but the gap is again slight as compared to the category \#2 with a margin of 1.0. The classifier committees would consider all the outputs by adding margins up in order to return a new margin value for the resource upon each category. As a result, committees get the largest margin value for the category \#2 with a 2.1, as compared to the 1.8 for the category \#3 and 1.7 for the category \#1. Hence, both classifiers on their own were wrong classifying this resource, but their prediction criteria were good enough to merge them with other classifiers. The fact that the actual category for the resource was predicted in second place for both classifiers gives rise to the correct classification when using committees.

\begin{table}[htb]
\begin{center}
 \tiny
 \begin{tabular}{|l|c|c|c|}
  \hline
  & \scriptsize\strut \textbf{Category \#1} & \scriptsize\strut \textbf{Category \#2} & \scriptsize\strut \textbf{Category \#3} \\
  \hline
  \hline
  \scriptsize\strut \textbf{Classifier A} & \scriptsize\strut \textbf{1.2} & \scriptsize\strut 1.1 & \scriptsize\strut 0.6 \\
  \hline
  \scriptsize\strut \textbf{Classifier B} & \scriptsize\strut 0.5 & \scriptsize\strut 1.0 & \scriptsize\strut \textbf{1.2} \\
  \hline
  \hline
  \scriptsize\strut \textbf{Classifier committees} & \scriptsize\strut 1.7 & \scriptsize\strut \textbf{2.1} & \scriptsize\strut 1.8 \\
  \hline
 \end{tabular}
\end{center}
\caption[Example of classifier committees]{Example of classifier committees, where both classifiers mispredict the category of the resource. One of them predicts category \#1, whereas the other predicts category \#3. However, it should actually be classified on category \#2, which is correctly predicted when adding margins up by using classifier committees.}
\label{tab:example-committees}
\end{table}

Next, we show the results of using classifier committees on separate tables for each dataset. Note that the tag-based approach is also included, in order to enable comparing the performance of committees to it.

\begin{table}[htb]
\begin{center}
 \tiny
 \begin{tabular}{|l|c|c|c|c|c|c|c|}
  \hline
  \multicolumn{8}{|c|}{\scriptsize\strut \textbf{Delicious - ODP}} \\
  \hline
  \hline
  \multicolumn{8}{|c|}{\scriptsize\strut \textbf{Top level}} \\
  \hline
  \textbf{} & \scriptsize\strut \textbf{600} & \scriptsize\strut \textbf{1400} & \scriptsize\strut \textbf{2200} & \scriptsize\strut \textbf{3000} & \scriptsize\strut \textbf{4000} & \scriptsize\strut \textbf{5000} & \scriptsize\strut \textbf{6000} \\
  \hline
  \scriptsize\strut \textbf{Tags} & \scriptsize\strut .533 & \scriptsize\strut .600 & \scriptsize\strut .629 & \scriptsize\strut .647 & \scriptsize\strut .660 & \scriptsize\strut .669 & \scriptsize\strut .680 \\
  \hline
  \scriptsize\strut \textbf{Content + Reviews} & \scriptsize\strut .554 & \scriptsize\strut .604 & \scriptsize\strut .627 & \scriptsize\strut .642 & \scriptsize\strut .651 & \scriptsize\strut .660 & \scriptsize\strut .670 \\
  \hline
  \scriptsize\strut \textbf{Content + Tags} & \scriptsize\strut .580 & \scriptsize\strut \textbf{.633} & \scriptsize\strut \textbf{.655} & \scriptsize\strut \textbf{.671} & \scriptsize\strut .678 & \scriptsize\strut .687 & \scriptsize\strut .696 \\
  \hline
  \scriptsize\strut \textbf{Reviews + Tags} & \scriptsize\strut .561 & \scriptsize\strut .618 & \scriptsize\strut .644 & \scriptsize\strut .662 & \scriptsize\strut .675 & \scriptsize\strut .685 & \scriptsize\strut .694 \\
  \hline
  \scriptsize\strut \textbf{Content + Reviews + Tags} & \scriptsize\strut \textbf{.581} & \scriptsize\strut .632 & \scriptsize\strut \textbf{.655} & \scriptsize\strut \textbf{.671} & \scriptsize\strut \textbf{.681} & \scriptsize\strut \textbf{.691} & \scriptsize\strut \textbf{.699} \\
  \hline
  \hline
  \multicolumn{8}{|c|}{\scriptsize\strut \textbf{Second level}} \\
  \hline
  \textbf{} & \scriptsize\strut \textbf{600} & \scriptsize\strut \textbf{1400} & \scriptsize\strut \textbf{2200} & \scriptsize\strut \textbf{3000} & \scriptsize\strut \textbf{4000} & \scriptsize\strut \textbf{5000} & \scriptsize\strut \textbf{6000} \\
  \hline
  \scriptsize\strut \textbf{Tags} & \scriptsize\strut .359 & \scriptsize\strut .453 & \scriptsize\strut .498 & \scriptsize\strut .522 & \scriptsize\strut .541 & \scriptsize\strut .556 & \scriptsize\strut .568 \\
  \hline
  \scriptsize\strut \textbf{Content + Reviews} & \scriptsize\strut .382 & \scriptsize\strut .450 & \scriptsize\strut .486 & \scriptsize\strut .505 & \scriptsize\strut .522 & \scriptsize\strut .538 & \scriptsize\strut .547 \\
  \hline
  \scriptsize\strut \textbf{Content + Tags} & \scriptsize\strut .409 & \scriptsize\strut \textbf{.488} & \scriptsize\strut \textbf{.528} & \scriptsize\strut \textbf{.547} & \scriptsize\strut \textbf{.564} & \scriptsize\strut .578 & \scriptsize\strut .587 \\
  \hline
  \scriptsize\strut \textbf{Reviews + Tags} & \scriptsize\strut .389 & \scriptsize\strut .474 & \scriptsize\strut .512 & \scriptsize\strut .534 & \scriptsize\strut .555 & \scriptsize\strut .571 & \scriptsize\strut .584 \\
  \hline
  \scriptsize\strut \textbf{Content + Reviews + Tags} & \scriptsize\strut \textbf{.412} & \scriptsize\strut \textbf{.488} & \scriptsize\strut .524 & \scriptsize\strut .545 & \scriptsize\strut \textbf{.564} & \scriptsize\strut \textbf{.579} & \scriptsize\strut \textbf{.588} \\
  \hline
 \end{tabular}
\end{center}
\caption[Accuracy results of classifier committees for web page classification]{Accuracy results of classifier committees for web page classification.}
\label{tab:delicious-committees-results}
\end{table}

Table \vref{tab:delicious-committees-results} shows the results of using classifier committees on Delicious. The effect of using committees on this dataset is really positive, because all of the combinations considering tags outperform the tag-based classifier. The committees considering only reviews and content may perform worse than tags on their own. Those committees considering the tag-based classifier are the three best. Even though tags positively combine with content and reviews separately, combining all three data sources provides a slight improvement as compared to the other two.

Reviews perform better than content on their own, but the latter performs better when combined with tags. This shows that even though content performs worse, it provides more reliable predictions than reviews, performing better on committees. Nonetheless, relying on all three data sources performs the best in most cases for both levels of the taxonomy.

\begin{table}[p]
\begin{center}
 \tiny
 \begin{tabular}{|l|c|c|c|c|c|c|c|}
  \hline
  \multicolumn{8}{|c|}{\scriptsize\strut \textbf{LibraryThing - DDC}} \\
  \hline
  \hline
  \multicolumn{8}{|c|}{\scriptsize\strut \textbf{Top level}} \\
  \hline
  \textbf{} & \scriptsize\strut \textbf{3000} & \scriptsize\strut \textbf{6000} & \scriptsize\strut \textbf{9000} & \scriptsize\strut \textbf{12000} & \scriptsize\strut \textbf{15000} & \scriptsize\strut \textbf{18000} & \scriptsize\strut \textbf{21000} \\
  \hline
  \scriptsize\strut \textbf{Tags} & \scriptsize\strut \textbf{.861} & \scriptsize\strut .864 & \scriptsize\strut .864 & \scriptsize\strut .867 & \scriptsize\strut .869 & \scriptsize\strut .869 & \scriptsize\strut .868 \\
  \hline
  \scriptsize\strut \textbf{Content + Reviews} & \scriptsize\strut .778 & \scriptsize\strut .803 & \scriptsize\strut .814 & \scriptsize\strut .821 & \scriptsize\strut .823 & \scriptsize\strut .827 & \scriptsize\strut .830 \\
  \hline
  \scriptsize\strut \textbf{Content + Tags} & \scriptsize\strut .823 & \scriptsize\strut .842 & \scriptsize\strut .845 & \scriptsize\strut .849 & \scriptsize\strut .851 & \scriptsize\strut .852 & \scriptsize\strut .852 \\
  \hline
  \scriptsize\strut \textbf{Reviews + Tags} & \scriptsize\strut .857 & \scriptsize\strut \textbf{.866} & \scriptsize\strut \textbf{.868} & \scriptsize\strut \textbf{.872} & \scriptsize\strut \textbf{.875} & \scriptsize\strut \textbf{.876} & \scriptsize\strut \textbf{.876} \\
  \hline
  \scriptsize\strut \textbf{Content + Reviews + Tags} & \scriptsize\strut .824 & \scriptsize\strut .843 & \scriptsize\strut .847 & \scriptsize\strut .852 & \scriptsize\strut .855 & \scriptsize\strut .856 & \scriptsize\strut .856 \\
  \hline
  \hline
  \multicolumn{8}{|c|}{\scriptsize\strut \textbf{Second level}} \\
  \hline
  \textbf{} & \scriptsize\strut \textbf{3000} & \scriptsize\strut \textbf{6000} & \scriptsize\strut \textbf{9000} & \scriptsize\strut \textbf{12000} & \scriptsize\strut \textbf{15000} & \scriptsize\strut \textbf{18000} & \scriptsize\strut \textbf{21000} \\
  \hline
  \scriptsize\strut \textbf{Tags} & \scriptsize\strut \textbf{.690} & \scriptsize\strut .700 & \scriptsize\strut .707 & \scriptsize\strut .709 & \scriptsize\strut .715 & \scriptsize\strut .712 & \scriptsize\strut .715 \\
  \hline
  \scriptsize\strut \textbf{Content + Reviews} & \scriptsize\strut .589 & \scriptsize\strut .631 & \scriptsize\strut .652 & \scriptsize\strut .663 & \scriptsize\strut .670 & \scriptsize\strut .679 & \scriptsize\strut .684 \\
  \hline
  \scriptsize\strut \textbf{Content + Tags} & \scriptsize\strut .645 & \scriptsize\strut .672 & \scriptsize\strut .688 & \scriptsize\strut .695 & \scriptsize\strut .700 & \scriptsize\strut .706 & \scriptsize\strut .707 \\
  \hline
  \scriptsize\strut \textbf{Reviews + Tags} & \scriptsize\strut .687 & \scriptsize\strut \textbf{.708} & \scriptsize\strut \textbf{.717} & \scriptsize\strut \textbf{.721} & \scriptsize\strut \textbf{.729} & \scriptsize\strut \textbf{.729} & \scriptsize\strut \textbf{.733} \\
  \hline
  \scriptsize\strut \textbf{Content + Reviews + Tags} & \scriptsize\strut .647 & \scriptsize\strut .677 & \scriptsize\strut .693 & \scriptsize\strut .701 & \scriptsize\strut .705 & \scriptsize\strut .713 & \scriptsize\strut .713 \\
  \hline
  \hline
  \multicolumn{8}{|c|}{\scriptsize\strut \textbf{LibraryThing - LCC}} \\
  \hline
  \hline
  \multicolumn{8}{|c|}{\scriptsize\strut \textbf{Top level}} \\
  \hline
  \textbf{} & \scriptsize\strut \textbf{3000} & \scriptsize\strut \textbf{6000} & \scriptsize\strut \textbf{9000} & \scriptsize\strut \textbf{12000} & \scriptsize\strut \textbf{15000} & \scriptsize\strut \textbf{18000} & \scriptsize\strut \textbf{21000} \\
  \hline
  \scriptsize\strut \textbf{Tags} & \scriptsize\strut \textbf{.853} & \scriptsize\strut \textbf{.857} & \scriptsize\strut \textbf{.856} & \scriptsize\strut \textbf{.861} & \scriptsize\strut \textbf{.861} & \scriptsize\strut .857 & \scriptsize\strut .861 \\
  \hline
  \scriptsize\strut \textbf{Content + Reviews} & \scriptsize\strut .777 & \scriptsize\strut .800 & \scriptsize\strut .808 & \scriptsize\strut .814 & \scriptsize\strut .818 & \scriptsize\strut .817 & \scriptsize\strut .824 \\
  \hline
  \scriptsize\strut \textbf{Content + Tags} & \scriptsize\strut .787 & \scriptsize\strut .806 & \scriptsize\strut .815 & \scriptsize\strut .819 & \scriptsize\strut .824 & \scriptsize\strut .821 & \scriptsize\strut .830 \\
  \hline
  \scriptsize\strut \textbf{Reviews + Tags} & \scriptsize\strut .831 & \scriptsize\strut .845 & \scriptsize\strut .853 & \scriptsize\strut .856 & \scriptsize\strut \textbf{.861} & \scriptsize\strut \textbf{.859} & \scriptsize\strut \textbf{.864} \\
  \hline
  \scriptsize\strut \textbf{Content + Reviews + Tags} & \scriptsize\strut .791 & \scriptsize\strut .811 & \scriptsize\strut .820 & \scriptsize\strut .826 & \scriptsize\strut .831 & \scriptsize\strut .827 & \scriptsize\strut .838 \\
  \hline
  \hline
  \multicolumn{8}{|c|}{\scriptsize\strut \textbf{Second level}} \\
  \hline
  \textbf{} & \scriptsize\strut \textbf{3000} & \scriptsize\strut \textbf{6000} & \scriptsize\strut \textbf{9000} & \scriptsize\strut \textbf{12000} & \scriptsize\strut \textbf{15000} & \scriptsize\strut \textbf{18000} & \scriptsize\strut \textbf{21000} \\
  \hline
  \scriptsize\strut \textbf{Tags} & \scriptsize\strut \textbf{.703} & \scriptsize\strut \textbf{.725} & \scriptsize\strut .729 & \scriptsize\strut .734 & \scriptsize\strut .734 & \scriptsize\strut .736 & \scriptsize\strut .739 \\
  \hline
  \scriptsize\strut \textbf{Content + Reviews} & \scriptsize\strut .600 & \scriptsize\strut .648 & \scriptsize\strut .674 & \scriptsize\strut .690 & \scriptsize\strut .705 & \scriptsize\strut .704 & \scriptsize\strut .719 \\
  \hline
  \scriptsize\strut \textbf{Content + Tags} & \scriptsize\strut .640 & \scriptsize\strut .677 & \scriptsize\strut .698 & \scriptsize\strut .709 & \scriptsize\strut .723 & \scriptsize\strut .720 & \scriptsize\strut .738 \\
  \hline
  \scriptsize\strut \textbf{Reviews + Tags} & \scriptsize\strut .688 & \scriptsize\strut .723 & \scriptsize\strut \textbf{.736} & \scriptsize\strut \textbf{.746} & \scriptsize\strut \textbf{.754} & \scriptsize\strut \textbf{.755} & \scriptsize\strut \textbf{.766} \\
  \hline
  \scriptsize\strut \textbf{Content + Reviews + Tags} & \scriptsize\strut .645 & \scriptsize\strut .685 & \scriptsize\strut .708 & \scriptsize\strut .721 & \scriptsize\strut .733 & \scriptsize\strut .732 & \scriptsize\strut .750 \\
  \hline
 \end{tabular}
\end{center}
\caption[Accuracy results of classifier committees for book classification on LibraryThing]{Accuracy results of classifier committees for book classification (LibraryThing).}
\label{tab:librarything-committees-results}
\end{table}

Table \vref{tab:librarything-committees-results} shows the results of using classifier committees on LibraryThing. These results show the great potential of tags provided by users on this social tagging system. Committees combining data sources not always outperform the sole use of tags. However, combining them with user reviews gives rise to higher performance, especially for the second level classification.

On the other hand, using content on committees yields inferior results. This shows that besides performing worse on its own, content is not good enough in this case to feed classifier committees. Probably, using synopses and editorial reviews as a summary of the content because of the unavailability of the actual content of the book makes it insufficient to get solid results.

\begin{table}[p]
\begin{center}
 \tiny
 \begin{tabular}{|l|c|c|c|c|c|c|c|}
  \hline
  \multicolumn{8}{|c|}{\scriptsize\strut \textbf{GoodReads - DDC}} \\
  \hline
  \hline
  \multicolumn{8}{|c|}{\scriptsize\strut \textbf{Top level}} \\
  \hline
  \textbf{} & \scriptsize\strut \textbf{3000} & \scriptsize\strut \textbf{6000} & \scriptsize\strut \textbf{9000} & \scriptsize\strut \textbf{12000} & \scriptsize\strut \textbf{15000} & \scriptsize\strut \textbf{18000} & \scriptsize\strut \textbf{21000} \\
  \hline
  \scriptsize\strut \textbf{Tags} & \scriptsize\strut .745 & \scriptsize\strut .747 & \scriptsize\strut .754 & \scriptsize\strut .757 & \scriptsize\strut .757 & \scriptsize\strut .757 & \scriptsize\strut .756 \\
  \hline
  \scriptsize\strut \textbf{Content + Reviews} & \scriptsize\strut .778 & \scriptsize\strut .803 & \scriptsize\strut .814 & \scriptsize\strut .821 & \scriptsize\strut .823 & \scriptsize\strut .827 & \scriptsize\strut .830 \\
  \hline
  \scriptsize\strut \textbf{Content + Tags} & \scriptsize\strut .797 & \scriptsize\strut .822 & \scriptsize\strut .831 & \scriptsize\strut .837 & \scriptsize\strut .838 & \scriptsize\strut .844 & \scriptsize\strut .845 \\
  \hline
  \scriptsize\strut \textbf{Reviews + Tags} & \scriptsize\strut \textbf{.820} & \scriptsize\strut \textbf{.847} & \scriptsize\strut \textbf{.857} & \scriptsize\strut \textbf{.865} & \scriptsize\strut \textbf{.867} & \scriptsize\strut \textbf{.872} & \scriptsize\strut \textbf{.874} \\
  \hline
  \scriptsize\strut \textbf{Content + Reviews + Tags} & \scriptsize\strut .806 & \scriptsize\strut .831 & \scriptsize\strut .842 & \scriptsize\strut .849 & \scriptsize\strut .851 & \scriptsize\strut .854 & \scriptsize\strut .857 \\
  \hline
  \hline
  \multicolumn{8}{|c|}{\scriptsize\strut \textbf{Second level}} \\
  \hline
  \textbf{} & \scriptsize\strut \textbf{3000} & \scriptsize\strut \textbf{6000} & \scriptsize\strut \textbf{9000} & \scriptsize\strut \textbf{12000} & \scriptsize\strut \textbf{15000} & \scriptsize\strut \textbf{18000} & \scriptsize\strut \textbf{21000} \\
  \hline
  \scriptsize\strut \textbf{Tags} & \scriptsize\strut .509 & \scriptsize\strut .520 & \scriptsize\strut .528 & \scriptsize\strut .528 & \scriptsize\strut .530 & \scriptsize\strut .529 & \scriptsize\strut .530 \\
  \hline
  \scriptsize\strut \textbf{Content + Reviews} & \scriptsize\strut .589 & \scriptsize\strut .631 & \scriptsize\strut .652 & \scriptsize\strut .663 & \scriptsize\strut .670 & \scriptsize\strut .679 & \scriptsize\strut .684 \\
  \hline
  \scriptsize\strut \textbf{Content + Tags} & \scriptsize\strut .594 & \scriptsize\strut .633 & \scriptsize\strut .652 & \scriptsize\strut .662 & \scriptsize\strut .671 & \scriptsize\strut .676 & \scriptsize\strut .680 \\
  \hline
  \scriptsize\strut \textbf{Reviews + Tags} & \scriptsize\strut .610 & \scriptsize\strut \textbf{.651} & \scriptsize\strut .670 & \scriptsize\strut \textbf{.683} & \scriptsize\strut \textbf{.691} & \scriptsize\strut .696 & \scriptsize\strut \textbf{.705} \\
  \hline
  \scriptsize\strut \textbf{Content + Reviews + Tags} & \scriptsize\strut \textbf{.611} & \scriptsize\strut \textbf{.651} & \scriptsize\strut \textbf{.672} & \scriptsize\strut \textbf{.683} & \scriptsize\strut .689 & \scriptsize\strut \textbf{.698} & \scriptsize\strut .702 \\
  \hline
  \hline
  \multicolumn{8}{|c|}{\scriptsize\strut \textbf{GoodReads - LCC}} \\
  \hline
  \hline
  \multicolumn{8}{|c|}{\scriptsize\strut \textbf{Top level}} \\
  \hline
  \textbf{} & \scriptsize\strut \textbf{3000} & \scriptsize\strut \textbf{6000} & \scriptsize\strut \textbf{9000} & \scriptsize\strut \textbf{12000} & \scriptsize\strut \textbf{15000} & \scriptsize\strut \textbf{18000} & \scriptsize\strut \textbf{21000} \\
  \hline
  \scriptsize\strut \textbf{Tags} & \scriptsize\strut .725 & \scriptsize\strut .731 & \scriptsize\strut .737 & \scriptsize\strut .738 & \scriptsize\strut .734 & \scriptsize\strut .731 & \scriptsize\strut .743 \\
  \hline
  \scriptsize\strut \textbf{Content + Reviews} & \scriptsize\strut .777 & \scriptsize\strut .800 & \scriptsize\strut .808 & \scriptsize\strut .814 & \scriptsize\strut .818 & \scriptsize\strut .817 & \scriptsize\strut .824 \\
  \hline
  \scriptsize\strut \textbf{Content + Tags} & \scriptsize\strut .793 & \scriptsize\strut .814 & \scriptsize\strut .823 & \scriptsize\strut .829 & \scriptsize\strut .831 & \scriptsize\strut .832 & \scriptsize\strut .836 \\
  \hline
  \scriptsize\strut \textbf{Reviews + Tags} & \scriptsize\strut \textbf{.831} & \scriptsize\strut \textbf{.836} & \scriptsize\strut \textbf{.847} & \scriptsize\strut \textbf{.853} & \scriptsize\strut \textbf{.857} & \scriptsize\strut \textbf{.857} & \scriptsize\strut \textbf{.864} \\
  \hline
  \scriptsize\strut \textbf{Content + Reviews + Tags} & \scriptsize\strut .801 & \scriptsize\strut .825 & \scriptsize\strut .833 & \scriptsize\strut .839 & \scriptsize\strut .844 & \scriptsize\strut .843 & \scriptsize\strut .850 \\
  \hline
  \hline
  \multicolumn{8}{|c|}{\scriptsize\strut \textbf{Second level}} \\
  \hline
  \textbf{} & \scriptsize\strut \textbf{3000} & \scriptsize\strut \textbf{6000} & \scriptsize\strut \textbf{9000} & \scriptsize\strut \textbf{12000} & \scriptsize\strut \textbf{15000} & \scriptsize\strut \textbf{18000} & \scriptsize\strut \textbf{21000} \\
  \hline
  \scriptsize\strut \textbf{Tags} & \scriptsize\strut .494 & \scriptsize\strut .507 & \scriptsize\strut .510 & \scriptsize\strut .514 & \scriptsize\strut .513 & \scriptsize\strut .517 & \scriptsize\strut .519 \\
  \hline
  \scriptsize\strut \textbf{Content + Reviews} & \scriptsize\strut .600 & \scriptsize\strut .648 & \scriptsize\strut .674 & \scriptsize\strut .690 & \scriptsize\strut .705 & \scriptsize\strut .704 & \scriptsize\strut .719 \\
  \hline
  \scriptsize\strut \textbf{Content + Tags} & \scriptsize\strut .608 & \scriptsize\strut .649 & \scriptsize\strut .672 & \scriptsize\strut .684 & \scriptsize\strut .692 & \scriptsize\strut .696 & \scriptsize\strut .703 \\
  \hline
  \scriptsize\strut \textbf{Reviews + Tags} & \scriptsize\strut .624 & \scriptsize\strut \textbf{.674} & \scriptsize\strut .696 & \scriptsize\strut .712 & \scriptsize\strut .725 & \scriptsize\strut \textbf{.730} & \scriptsize\strut .735 \\
  \hline
  \scriptsize\strut \textbf{Content + Reviews + Tags} & \scriptsize\strut \textbf{.626} & \scriptsize\strut \textbf{.674} & \scriptsize\strut \textbf{.699} & \scriptsize\strut \textbf{.713} & \scriptsize\strut \textbf{.728} & \scriptsize\strut .727 & \scriptsize\strut \textbf{.742} \\
  \hline
 \end{tabular}
\end{center}
\caption[Accuracy results of classifier committees for book classification on GoodReads]{Accuracy results of classifier committees for book classification (GoodReads).}
\label{tab:goodreads-committees-results}
\end{table}

Table \vref{tab:goodreads-committees-results} shows the results of using classifier committees on GoodReads. In this case, tags on their own were not strong enough to reach the results by content or user reviews. However, the committees considering tags perform the best, showing their high reliability when it comes to combining predictions.

As it happened with LibraryThing, content does not seem to be a reliable source for committees. Combining it with reviews and tags yields similar or even worse results than excluding it. Combining both reviews and tags is the best option again for the top level of the taxonomies, as for LibraryThing. Surprisingly, this combination produces results almost as good as using LibraryThing tags, which perform far better on their own. This shows that even though tags from GoodReads are not accurate enough on their own, they provide reliable margins to be considered on committees.

When comparing taxonomies, DDC and LCC, neither GoodReads nor LibraryThing shows any differences as compared to the other, proving that the conclusions are the same regardless of the classification scheme.

Summarizing, tags have shown great potential, not only as a source to classify on their own, but also to provide reliable prediction criteria to take into consideration for combining them with other data sources. Moreover, in some cases like on GoodReads, tags were not good enough on their own, but have shown to be a solid data source when used with classifier committees. Nonetheless, the data source used to combine with tags must be solid enough and provide reliable predictions to get better results. When data sources are selected appropriately, the performance improvement can be considerable. In this regard, we have seen that the synopses and reviews we chose as a summary of the content of books provide inappropriate predictions.

\section{Conclusion}
\label{sec:tag-representation-conclusion}

In this chapter, we have carried out a deep experimentation and performed a thorough analysis on the use of social tags as a source to feed resource classifiers. We have compared the performance of using social tags to that by using other data sources like the content or user reviews gathered from social media. The experiments have been applied to the three large-scale social tagging datasets introduced in Chapter \vref{c:datasets}, gathered from tagging sites with different settings and annotated resources, which allow to conclude with more generalistic thoughts. Classification experiments have been realized with annotated web pages over the ODP taxonomy, and annotated books over the DDC and LCC taxonomies. The great potential shown by social tags, for both the top and second levels of taxonomies, can be strengthened by combining the predictions with other data sources. However, not all data sources are strong enough to perform well at combining predictions, so that the selection of data sources should be done appropriately.

Parts of the research in this chapter have been published in \cite{zubiaga2009getting}, \cite{zubiaga2009clasificacion} and \cite{zubiaga2011exploiting}.

By means of these experiments, we provided an answer to the following research questions:

\begin{description}
 \item[Research Question 4] \hfill \\
 \textit{\rqrepone}
\end{description}

We have shown that it is worthwhile considering all the tags annotated on a resource instead of those in the top that were annotated most. Tags in the top are the most important, and give the main information on the aboutness of resources. However, tags in the tail are helpful to a lesser extent, providing meaningful information and improving the performance of the classifier.

Regarding the weights assigned to those tags when representing a resource, the number of users annotating each tag should be considered in order to get the best results. This is the value that has shown the best results in our experiments. It has outperformed other approaches ignoring weights or considering other data such as the total number of users annotating the resource.

Thereby, the best representation in our experiments is the one that includes all the tags with the values corresponding to the number of users annotating them.

\begin{description}
 \item[Research Question 5] \hfill \\
 \textit{\rqreptwo}
\end{description}

By means of classifier committees, which combine the predictions by different classifiers, we have shown that tags provide reliable prediction criteria to take into consideration. SVM classifiers not only predict a category, but also assign a weight to each category based on the given resource. These weights, given in the form of margin values, can be used by other classifiers which rely on different data sources. Adding up weights provided by different classifiers can help predict the correct category when a single classifier fails to categorize the resource appropriately.  Weights provided by classifiers relying on social tags are especially useful when combining them with results from other classifiers.  Nonetheless, not all data sources are helpful for combination in classifier committees, and the selected data source must be solid enough and provide reliable predictions to outperform the sole use of tags. When data sources are selected appropriately, the performance improvement can be considerable.  We have shown that this varies among datasets. For example, with the Delicious dataset, it is important to analyze all three data sources (content, reviews, and tags). However, with the LibraryThing and GoodReads datasets reviews and tags suffice.

\begin{description}
 \item[Research Question 6] \hfill \\
 \textit{\rqrepthree}
\end{description}

We have analyzed the usefulness of social tags for classification on two different levels of hierarchical taxonomies. Besides broader categories in the top level, we have also explored the classification on narrower categories in the second level. In this regard, social tags have shown to outperform the other data sources on social tagging sites that encourage users to annotate resources (Delicious and LibraryThing). Tags show clear outperformance in these cases, especially on Delicious, where the difference is even more favorable in the second level. This difference is very similar on LibraryThing. Finally, tags from GoodReads do not outperform other data sources at any level because the system does not encourage users to tag books, so that many bookmarks are not annotated.

Our findings provide a different conclusion from that by \cite{noll_exploring_2008}, where the authors pointed out the hypothesis that social tags were probably useless for deeper levels of taxonomies, and alternative data should be used instead. However, the authors performed just a statistical analysis, and did not confirm the hypothesis with real experiments.

%% file: tag-distribution.tex
\chapter{Analyzing the Distribution of Tags for Resource Classification}
\label{c:tag-distribution-classification}

\textit{``Statistics will prove anything, even the truth.''}

--- Noel Moynihan

\chaptersummary{In this chapter, we deal with the task of considering the representativity of tags for resource classification within a collection of social annotations on a social tagging system. To the best of our knowledge, no effort has been invested so far on establishing the representativity of tags when it comes to finding the aboutness of resources. In this regard, we explore how the distribution of tags across the three dimensions involved in a social tagging system (namely users, resources and bookmarks) can determine their representativity. To this end, we study and analyze the effectiveness of applying an IDF-like distribution-driven weighting scheme in search of performance improvements in a resource classification task. We define three analogous weighting schemes --IUF, IRF and IBF-- which rely on distributions of tags across users, resources and bookmarks, respectively. They have been barely used for social tagging, and their usefulness has not yet been proven.}

\chaptersummary{The chapter is organized as follows. Next, in Section \vref{sec:tag-distributions} we motivate the problem of considering tag distributions as a means to determine the representativity of tags. Then, in Section \vref{sec:tfidf} we describe the TF-IDF weighting scheme and its use on classical documents collections, and introduce analogous schemes adapted to social tagging systems in Section \vref{sec:tag-weighting-functions}. We present a set of experiments --tag-based classification, classifier committees, and correlation between weighting measures--, and analyze and study the results in Section \vref{sec:tag-distribution-experiments}. Finally, in Section \vref{sec:tag-distributions-conclusion} we conclude the chapter.}

\chaptersummary{We address the following research questions in this chapter:}

\chaptersummary{
 \begin{description}
  \item[Research Question 7] \hfill \\
  \textit{\rqdistone}
 \end{description}
}

\chaptersummary{
 \begin{description}
  \item[Research Question 8] \hfill \\
  \textit{\rqdisttwo}
 \end{description}
}

\section{Tag Distributions}
\label{sec:tag-distributions}

So far, we have explored the ways of amalgamating great deals of user annotations provided in the form of social tags, in order to find a suitable representation of a resource. We considered the weighting of a tag with respect to the resource where it was annotated, but we did not explore further into the representativity of tags within the whole collection. We have considered that two tags with the same number of users annotating it on a resource have the same representativity for the resource, because they had the same number of annotators and, therefore, they were assigned the same weight. However, they do not strictly have to represent the same representativity.

From a statistical point of view, we believe that the distribution of tags across the whole collection has much to do with the overall representativity of tags. By representativity, we refer to the weight setting how important is a certain tag when it comes to representing a resource for its classification. Accordingly, we believe that a tag that concentrates within a few resources or has been used by a few users is rather representative than a tag present in most resources or used by most users. Even if two tags have the same overall use within the collection, the way they are distributed across users, resources and bookmarks may determine whether they are focused and precise, or they are spread and imprecise instead.

To this end, a collection-aware weighting scheme like the well-known TF-IDF seems to be a good alternative. We believe it is suitable to determine the representativity of tags considering their distribution across the collection. We found that it had been hardly applied to a social structure like that by tagging systems. Its adaptation from a classical text collection, where the only dimensions are terms and documents, to a collection of bookmarks, where tags spread across users, resources and bookmarks, remains unstudied. Moreover, its usefulness for tag-based resource classification has not yet been explored.

\section{TF-IDF as a Term Weighting Function}
\label{sec:tfidf}

TF-IDF is a term weighting function that serves as a statistical measure defining the importance of a word to a document in a collection \citep{salton75vector,salton88termweighting}. When computing the TF-IDF value for the term $i$ within the document $j$ as a part of a document collection $D$, it comprises two underlying measures: (1) the term frequency (TF), i.e., the number of appearances of the term $i$ within the document $j$, and (2) the inverse document frequency (IDF), i.e., the inverse of the number of documents within the whole set of documents $D$ in which the term $i$ occurs, which refers to the general importance of the term $i$ in the collection (see Equation \vref{eq:idf}). The product of these two measures defines the TF-IDF weight of term $i$ in the document $j$ (see Equation \vref{eq:tfidf}).

\begin{equation}
 idf_{i} = \log \frac{|D|}{|\{d: t_{i} \in d\}|}
 \label{eq:idf}
\end{equation}

\begin{equation}
 tf\mbox{-}idf_{ij} = tf_{ij} \times idf_{i}
 \label{eq:tfidf}
\end{equation}

Integrating the IDF factor allows to rate lower or higher such a term depending on its distribution across the collection. This weighting function yields a higher value when the term $i$ occurs in a few documents, considering that it is of utmost representativity to those documents. On the other hand, the value will be lower when the term $i$ occurs in many documents of the collection, considering that it rather spreads across the collection instead of focusing in a few documents. In the latter case, the value becomes null when the term $i$ occurs in all the documents.

This weighting scheme has been widely used for Information Retrieval, Text Mining and Text Classification, and it is commonly used for term selection tasks. There is controversy on its appropriateness for text classification \citep{Lan2005,Forman08}, since it does not consider the relations between the terms and their appearance in the categories. However, it has shown high effectiveness in several text classification tasks \citep{joachims98text,yang99reexamination,brank2002interaction,dumais98inductive}.

There are some works that study the adaptation of TF-IDF to text and web page classification tasks. They consider the distribution of terms across categories in the training set as a value to determine the representativity of a term. For instance, in \cite{Forman08} and \cite{Lan2005} the authors compared some feature scoring metrics, including TF-IDF, in a text classification problem using a linear SVM. Each of them proposed a new term weighting function that outperformed TF-IDF in their experiments. Other works, such as \cite{debole03} and \cite{Soucy05}, propose the use of supervised weighting techniques instead of unsupervised ones for text classification tasks.

Even though alternatives to TF-IDF like those mentioned above have been proposed and successfully applied to specific tasks and collections, they have barely been used subsequently. TF-IDF continues to be the most widely used term weighting scheme, and has become a ``de facto'' standard for document representation.

In this chapter, we rely on TF-IDF as the base weighting to propose analogous schemes adapted to social tagging systems. Even though we could rely on alternatives, our main goals are (1) to perform a study on its adaptability to these structures, and (2) to find out how the settings of social tagging systems affect the resulting tag distributions and thereby the values of such weights. Thus, we will not include any category data in the calculation of the weights.

\section{Tag Weighting Functions Based on Inverse Frequencies}
\label{sec:tag-weighting-functions}

Unlike classical collections of web documents or library catalogs, where the distribution of terms across documents on the collection has been studied, social tagging systems comprise more dimensions to explore into. Besides the distribution of tags across documents or annotated resources, different users set those tags within different bookmarks. These two characteristics are new on social tagging with respect to classical text document collections. Despite this clear difference in the nature of social tagging systems, not enough attention has been paid at analyzing how each of the dimensions --resources, users and bookmarks-- affects tag distributions and, therefore, establishing tag relevances.

TF-IDF has widely been applied to text collections, and has proven to be beneficial for a large number of tasks. Text collections are mainly made up by terms written by the authors, though, and the appropriateness of using a similar approach for a collection made up by tags annotated by users other than the authors on a social environment is not clear.

Next, we introduce three tag weighting approaches, taking the classical TF-IDF approach to the social tagging scenario, and adapting it to rely on resources, users and bookmarks. These three dimensions suggest the definition of that many tag weighting functions considering inverse resource frequency (IRF), inverse user frequency (IUF), and inverse bookmark frequency (IBF) values, respectively. These three approaches follow the same function for the tag $i$ within the resource $j$ (see Equation \vref{eq:tfixf}).

\begin{equation}
 TF\mbox{-}IxF_{ij} = tf_{ij} \cdot ixf
 \label{eq:tfixf}
\end{equation}

where $tf_{ij}$ is the number of occurrences of the tag $i$ in the resource $j$, and $ixf$ is the inverse frequency function considered in each case, $irf$, $iuf$ or $ibf$, thus $x$ being $r$, $u$, or $b$.

\subsection{TF-IRF}
\label{tfirf}

This is the application of the TF-IDF approach to a social tagging system with annotated resources, considering that resources are analogous to documents in this case. Tags that are widely spread across resources are penalized with low weights and, vice versa, tags within fewer resources are considered relevant with a higher weight. Thus, the function outputs the logarithm of the total number of resources divided by the number of resources in which the tag is present (see Equation \ref{eq:irf}).

\begin{equation}
 irf_{i} =  \log \frac{|R|}{|\{r: t_{i} \in R\}|}
 \label{eq:irf}
\end{equation}

It has previously been used in a few works in the social tagging literature, even though they usually referred to this approach as TF-IDF. \cite{angelova2008} rely on this measure to infer similarity of tags by creating a tag graph, weighting the TF-IDF value of each user to a tag. \cite{shepitsen2008} and \cite{liang2010} use this measure to represent the resources in a recommendation system where resources are recommended to users. The latter concluded that although both TF-IDF and TF have identical trends, the former provides superior results in their recommendation task. Likewise, \cite{ramage_clusteringtagged_2009} compared TF-IDF and TF for clustering web pages, and showed a superiority for the former. However, they did not pay attention at the effect of tag distributions on these weightings, and they showed the usefulness of TF-IDF just for a specific case. \cite{li2008} create tag vectors using TF-IDF to compute the similarity between two documents annotated on Delicious. They assumed this weighting measure, and they did not pay attention at whether or not it was appropriate.

\subsection{TF-IUF}
\label{tfiuf}

As a new dimension present in social tagging systems, the number of users using each of the tags could also be significant to know whether a tag is representative to a collection of resources. Thus, we consider that a tag used by many users is not as representative as a tag that fewer users are utilizing (see Equation \ref{eq:iuf}).

\begin{equation}
 iuf_{i} =  \log \frac{|U|}{|\{u: t_{i} \in U\}|}
 \label{eq:iuf}
\end{equation}

This function was inferred from a previous application to a collaborative filtering system \cite{breese98empiricalanalysis}. With the aim of recommending resources to users, \cite{diederich2006} and \cite{liang2010} rely on the IUF for discovering similarities among users. The latter use both IUF and IRF to represent users and resources, respectively, but no comparison is performed among their characteristics. In \cite{abbasi2009}, TF-IUF is used along with TF-IRF over Flickr tags and user groups for finding landmark photos. They concluded that their approach was effective to find landmark photos on Flickr, but they did not study whether or not relying on those weighting measures was appropriate.

\subsection{TF-IBF}
\label{tfibf}

This is a similar inverse weighting function relying on the third dimension in which tags are distributed: bookmarks. This function considers that a tag that has been used in many bookmarks is not as relevant to represent a resource as others that have been assigned to fewer bookmarks (see Equation \ref{eq:ibf}).

\begin{equation}
 ibf_{i} =  \log \frac{|B|}{|\{b: t_{i} \in B\}|}
 \label{eq:ibf}
\end{equation}

To the best of our knowledge, this tag weighting scheme has never been used so far. Even though all three frequencies can somehow be related, there are substantial differences among them. A tag used by many users can spread across many resources, or it can just congregate in a few resources. Likewise, this factor might affect the number of bookmarks.

\section{Experiments}
\label{sec:tag-distribution-experiments}

Next, we present the classification experiments that enable (1) to analyze how each of the proposed tag weighting functions contributes to the classification of annotated resources, as well as (2) to discover whether either of the inverse tag weighting approaches outperforms the baseline relying only on the tag frequency (TF). In order to further analyze their usefulness and suitability, we also experimented on their performance when applied to classifier committees. Finally, we analyze the correlation between the different tag weighting functions.

\subsection{Tag-based Classification}
\label{ssec:tag-based-classification}

The first experiment focuses on evaluating the usefulness of tag weighting functions for a resource classification task. We perform this evaluation by comparing tag-based representations by using each of the three weighting functions --TF-IRF, TF-IUF and TF-IBF-- and the absence of distributional weighting functions (TF). Note that the latter is the same as the FTA-based weighted approach we concluded as the best representation in Chapter \vref{c:tag-representation}, and it is thus the up-to-now outperforming approach. This experiment uses an SVM with the same settings as those defined in previous Chapter (see Section \vref{sec:tag-based-classification}). We show the results for all three datasets, and 4 different representations, including the three weighting measures and TF.

\begin{table}[htb]
 \begin{center}
  \tiny
  \begin{tabular}{|l|c|c|c|c|c|c|c|}
   \hline
   \multicolumn{8}{|c|}{\scriptsize\strut \textbf{Delicious - ODP}} \\
   \hline
   \hline
   \multicolumn{8}{|c|}{\scriptsize\strut \textbf{Top level}} \\
   \hline
   & \scriptsize\strut \textbf{600} & \scriptsize\strut \textbf{1400} & \scriptsize\strut \textbf{2200} & \scriptsize\strut \textbf{3000} & \scriptsize\strut \textbf{4000} & \scriptsize\strut \textbf{5000} & \scriptsize\strut \textbf{6000} \\
   \hline
   \scriptsize\strut \textbf{TF} & \scriptsize\strut \textbf{.533} & \scriptsize\strut \textbf{.600} & \scriptsize\strut \textbf{.629} & \scriptsize\strut \textbf{.647} & \scriptsize\strut \textbf{.660} & \scriptsize\strut \textbf{.669} & \scriptsize\strut \textbf{.680} \\
   \hline
   \scriptsize\strut \textbf{TF-IRF} & \scriptsize\strut .516 & \scriptsize\strut .571 & \scriptsize\strut .593 & \scriptsize\strut .607 & \scriptsize\strut .619 & \scriptsize\strut .631 & \scriptsize\strut .639 \\
   \hline
   \scriptsize\strut \textbf{TF-IBF} & \scriptsize\strut .519 & \scriptsize\strut .573 & \scriptsize\strut .596 & \scriptsize\strut .611 & \scriptsize\strut .622 & \scriptsize\strut .633 & \scriptsize\strut .641 \\
   \hline
   \scriptsize\strut \textbf{TF-IUF} & \scriptsize\strut .528 & \scriptsize\strut .580 & \scriptsize\strut .607 & \scriptsize\strut .625 & \scriptsize\strut .636 & \scriptsize\strut .653 & \scriptsize\strut .661 \\
   \hline
   \hline
   \multicolumn{8}{|c|}{\scriptsize\strut \textbf{Second level}} \\
   \hline
   & \scriptsize\strut \textbf{600} & \scriptsize\strut \textbf{1400} & \scriptsize\strut \textbf{2200} & \scriptsize\strut \textbf{3000} & \scriptsize\strut \textbf{4000} & \scriptsize\strut \textbf{5000} & \scriptsize\strut \textbf{6000} \\
   \hline
   \scriptsize\strut \textbf{TF} & \scriptsize\strut \textbf{.359} & \scriptsize\strut \textbf{.453} & \scriptsize\strut \textbf{.498} & \scriptsize\strut \textbf{.522} & \scriptsize\strut \textbf{.541} & \scriptsize\strut \textbf{.556} & \scriptsize\strut \textbf{.568} \\
   \hline
   \scriptsize\strut \textbf{TF-IRF} & \scriptsize\strut .344 & \scriptsize\strut .424 & \scriptsize\strut .463 & \scriptsize\strut .486 & \scriptsize\strut .506 & \scriptsize\strut .518 & \scriptsize\strut .529 \\
   \hline
   \scriptsize\strut \textbf{TF-IBF} & \scriptsize\strut .348 & \scriptsize\strut .429 & \scriptsize\strut .467 & \scriptsize\strut .489 & \scriptsize\strut .509 & \scriptsize\strut .520 & \scriptsize\strut .532 \\
   \hline
   \scriptsize\strut \textbf{TF-IUF} & \scriptsize\strut .358 & \scriptsize\strut .437 & \scriptsize\strut .478 & \scriptsize\strut .502 & \scriptsize\strut .523 & \scriptsize\strut .541 & \scriptsize\strut .555 \\
   \hline
  \end{tabular}
 \end{center}
 \caption[Accuracy results of tag-based web page classification using weighting schemes]{Accuracy results of tag-based web page classification using weighting schemes.}
 \label{tab:idfs-classification-delicious}
\end{table}

Table \vref{tab:idfs-classification-delicious} shows the results of using tag weighting functions on Delicious. It can be seen that the use of inverse weighting functions is not useful in this case. In the contrary, their use harms the performance of the classifier, yielding inferior results than those obtained by the TF approach not considering weighting functions. Going further into the analysis of the performance of representations relying on weighting functions, the results show that IUF gets the best results among them, followed by IBF, and then IRF. This happens for both top and second levels of the taxonomy in a similar manner.

Our conjecture about this is that resource-based tag suggestions provided by Delicious are not helpful to this end. We have already shown in Chapter \vref{c:datasets} that such a feature alters the structure of the folksonomy on Delicious. It makes the top tags become even more popular and it alters the natural distribution of tags. Thus, such a forced distribution of tags produces weights that score lower performances. Moreover, the fact that IUF is the best weighting function in this case shows the importance of users who make their own choices instead of relying on suggestions. That is, users who are able to choose their own tags and differ from those relying on suggestion-based annotations give rise to higher weights for their seldom tags. When users rely on suggestions, it does not make any difference on the IRF values of tags, because the frequency remains unchanged. This difference is also little for IBF values. However, it makes a big difference on IUF values, because those suggestions increase the user frequencies of tags and thus reduce IUF values. Accordingly, users who make their own choices yield higher IUF values because it is likely that their tags are not being used that many times. Probably, IUF would perform better than TF if there were fewer users who rely on system suggestions, and hence more users providing their own tags instead.

\begin{table}[htb]
 \begin{center}
  \tiny
  \begin{tabular}{|l|c|c|c|c|c|c|c|}
   \hline
   \multicolumn{8}{|c|}{\scriptsize\strut \textbf{LibraryThing - DDC}} \\
   \hline
   \hline
   \multicolumn{8}{|c|}{\scriptsize\strut \textbf{Top level}} \\
   \hline
   & \scriptsize\strut \textbf{3000} & \scriptsize\strut \textbf{6000} & \scriptsize\strut \textbf{9000} & \scriptsize\strut \textbf{12000} & \scriptsize\strut \textbf{15000} & \scriptsize\strut \textbf{18000} & \scriptsize\strut \textbf{21000} \\
   \hline
   \scriptsize\strut \textbf{TF} & \scriptsize\strut .861 & \scriptsize\strut .864 & \scriptsize\strut .864 & \scriptsize\strut .867 & \scriptsize\strut .869 & \scriptsize\strut .869 & \scriptsize\strut .868 \\
   \hline
   \scriptsize\strut \textbf{TF-IRF} & \scriptsize\strut .877 & \scriptsize\strut .889 & \scriptsize\strut .894 & \scriptsize\strut \textbf{.897} & \scriptsize\strut \textbf{.900} & \scriptsize\strut .902 & \scriptsize\strut .902 \\
   \hline
   \scriptsize\strut \textbf{TF-IBF} & \scriptsize\strut .877 & \scriptsize\strut .889 & \scriptsize\strut .894 & \scriptsize\strut \textbf{.897} & \scriptsize\strut \textbf{.900} & \scriptsize\strut \textbf{.903} & \scriptsize\strut \textbf{.904} \\
   \hline
   \scriptsize\strut \textbf{TF-IUF} & \scriptsize\strut \textbf{.881} & \scriptsize\strut \textbf{.891} & \scriptsize\strut \textbf{.895} & \scriptsize\strut \textbf{.897} & \scriptsize\strut .899 & \scriptsize\strut .901 & \scriptsize\strut .900 \\
   \hline
   \hline
   \multicolumn{8}{|c|}{\scriptsize\strut \textbf{Second level}} \\
   \hline
   & \scriptsize\strut \textbf{3000} & \scriptsize\strut \textbf{6000} & \scriptsize\strut \textbf{9000} & \scriptsize\strut \textbf{12000} & \scriptsize\strut \textbf{15000} & \scriptsize\strut \textbf{18000} & \scriptsize\strut \textbf{21000} \\
   \hline
   \scriptsize\strut \textbf{TF} & \scriptsize\strut .690 & \scriptsize\strut .700 & \scriptsize\strut .707 & \scriptsize\strut .709 & \scriptsize\strut .715 & \scriptsize\strut .712 & \scriptsize\strut .715 \\
   \hline
   \scriptsize\strut \textbf{TF-IRF} & \scriptsize\strut .723 & \scriptsize\strut .750 & \scriptsize\strut .762 & \scriptsize\strut .768 & \scriptsize\strut .774 & \scriptsize\strut .777 & \scriptsize\strut .780 \\
   \hline
   \scriptsize\strut \textbf{TF-IBF} & \scriptsize\strut .723 & \scriptsize\strut \textbf{.751} & \scriptsize\strut \textbf{.763} & \scriptsize\strut \textbf{.770} & \scriptsize\strut \textbf{.775} & \scriptsize\strut \textbf{.779} & \scriptsize\strut \textbf{.781} \\
   \hline
   \scriptsize\strut \textbf{TF-IUF} & \scriptsize\strut \textbf{.729} & \scriptsize\strut \textbf{.751} & \scriptsize\strut .761 & \scriptsize\strut .766 & \scriptsize\strut .771 & \scriptsize\strut .771 & \scriptsize\strut .776 \\
   \hline
   \hline
   \multicolumn{8}{|c|}{\scriptsize\strut \textbf{LibraryThing - LCC}} \\
   \hline
   \hline
   \multicolumn{8}{|c|}{\scriptsize\strut \textbf{Top level}} \\
   \hline
   & \scriptsize\strut \textbf{3000} & \scriptsize\strut \textbf{6000} & \scriptsize\strut \textbf{9000} & \scriptsize\strut \textbf{12000} & \scriptsize\strut \textbf{15000} & \scriptsize\strut \textbf{18000} & \scriptsize\strut \textbf{21000} \\
   \hline
   \scriptsize\strut \textbf{TF} & \scriptsize\strut .853 & \scriptsize\strut .857 & \scriptsize\strut .856 & \scriptsize\strut .861 & \scriptsize\strut .861 & \scriptsize\strut .857 & \scriptsize\strut .861 \\
   \hline
   \scriptsize\strut \textbf{TF-IRF} & \scriptsize\strut .867 & \scriptsize\strut \textbf{.883} & \scriptsize\strut .887 & \scriptsize\strut \textbf{.893} & \scriptsize\strut .895 & \scriptsize\strut .894 & \scriptsize\strut .897 \\
   \hline
   \scriptsize\strut \textbf{TF-IBF} & \scriptsize\strut .867 & \scriptsize\strut \textbf{.883} & \scriptsize\strut \textbf{.888} & \scriptsize\strut \textbf{.893} & \scriptsize\strut \textbf{.896} & \scriptsize\strut \textbf{.895} & \scriptsize\strut \textbf{.898} \\
   \hline
   \scriptsize\strut \textbf{TF-IUF} & \scriptsize\strut \textbf{.871} & \scriptsize\strut .882 & \scriptsize\strut .885 & \scriptsize\strut .892 & \scriptsize\strut .893 & \scriptsize\strut .892 & \scriptsize\strut .894 \\
   \hline
   \hline
   \multicolumn{8}{|c|}{\scriptsize\strut \textbf{Second level}} \\
   \hline
   & \scriptsize\strut \textbf{3000} & \scriptsize\strut \textbf{6000} & \scriptsize\strut \textbf{9000} & \scriptsize\strut \textbf{12000} & \scriptsize\strut \textbf{15000} & \scriptsize\strut \textbf{18000} & \scriptsize\strut \textbf{21000} \\
   \hline
   \scriptsize\strut \textbf{TF} & \scriptsize\strut .703 & \scriptsize\strut .725 & \scriptsize\strut .729 & \scriptsize\strut .734 & \scriptsize\strut .734 & \scriptsize\strut .736 & \scriptsize\strut .739 \\
   \hline
   \scriptsize\strut \textbf{TF-IRF} & \scriptsize\strut .751 & \scriptsize\strut .780 & \scriptsize\strut .793 & \scriptsize\strut .803 & \scriptsize\strut .804 & \scriptsize\strut .809 & \scriptsize\strut .814 \\
   \hline
   \scriptsize\strut \textbf{TF-IBF} & \scriptsize\strut .751 & \scriptsize\strut \textbf{.781} & \scriptsize\strut \textbf{.796} & \scriptsize\strut \textbf{.805} & \scriptsize\strut \textbf{.806} & \scriptsize\strut \textbf{.811} & \scriptsize\strut \textbf{.818} \\
   \hline
   \scriptsize\strut \textbf{TF-IUF} & \scriptsize\strut \textbf{.754} & \scriptsize\strut .780 & \scriptsize\strut .790 & \scriptsize\strut .798 & \scriptsize\strut .800 & \scriptsize\strut .803 & \scriptsize\strut .807 \\
   \hline
  \end{tabular}
 \end{center}
 \caption[Accuracy results of tag-based book classification using weighting schemes on LibraryThing]{Accuracy results of tag-based book classification using weighting schemes (LibraryThing).}
 \label{tab:idfs-classification-librarything}
\end{table}

Table \vref{tab:idfs-classification-librarything} shows the results of using tag weighting functions on LibraryThing over DDC and LCC schemes. In this case, all the inverse weighting functions are clearly superior to TF, since the former always outperform the latter. Even though the outperformance is much larger for the second level, the superiority of weighting functions is clear for both levels. This shows that the studied inverse weighting functions can be really useful for folksonomies created in the absence of suggestions. Inverse tag weighting functions have successfully set suitable weights towards a definition of the representativity of tags in this case, in contrast to Delicious.

Among the tag weighting functions, all of them perform similarly, and no clear outperformances can be seen in these results. However, IBF seems to provide slightly better results than the other two approaches, followed by IRF. IUF is the worst function in this case, suggesting that the number of users choosing each tag is not the most relevant feature when there are no suggestions.

\begin{table}[htb]
 \begin{center}
  \tiny
  \begin{tabular}{|l|c|c|c|c|c|c|c|}
   \hline
   \multicolumn{8}{|c|}{\scriptsize\strut \textbf{GoodReads - DDC}} \\
   \hline
   \hline
   \multicolumn{8}{|c|}{\scriptsize\strut \textbf{Top level}} \\
   \hline
   & \scriptsize\strut \textbf{3000} & \scriptsize\strut \textbf{6000} & \scriptsize\strut \textbf{9000} & \scriptsize\strut \textbf{12000} & \scriptsize\strut \textbf{15000} & \scriptsize\strut \textbf{18000} & \scriptsize\strut \textbf{21000} \\
   \hline
   \scriptsize\strut \textbf{TF} & \scriptsize\strut .745 & \scriptsize\strut .747 & \scriptsize\strut .754 & \scriptsize\strut .757 & \scriptsize\strut .757 & \scriptsize\strut .757 & \scriptsize\strut .756 \\
   \hline
   \scriptsize\strut \textbf{TF-IRF} & \scriptsize\strut \textbf{.800} & \scriptsize\strut .808 & \scriptsize\strut .813 & \scriptsize\strut \textbf{.817} & \scriptsize\strut .816 & \scriptsize\strut .817 & \scriptsize\strut .816 \\
   \hline
   \scriptsize\strut \textbf{TF-IBF} & \scriptsize\strut \textbf{.800} & \scriptsize\strut \textbf{.809} & \scriptsize\strut \textbf{.814} & \scriptsize\strut \textbf{.817} & \scriptsize\strut \textbf{.817} & \scriptsize\strut \textbf{.818} & \scriptsize\strut \textbf{.818} \\
   \hline
   \scriptsize\strut \textbf{TF-IUF} & \scriptsize\strut .797 & \scriptsize\strut .805 & \scriptsize\strut .810 & \scriptsize\strut .814 & \scriptsize\strut .813 & \scriptsize\strut .814 & \scriptsize\strut .814 \\
   \hline
   \hline
   \multicolumn{8}{|c|}{\scriptsize\strut \textbf{Second level}} \\
   \hline
   & \scriptsize\strut \textbf{3000} & \scriptsize\strut \textbf{6000} & \scriptsize\strut \textbf{9000} & \scriptsize\strut \textbf{12000} & \scriptsize\strut \textbf{15000} & \scriptsize\strut \textbf{18000} & \scriptsize\strut \textbf{21000} \\
   \hline
   \scriptsize\strut \textbf{TF} & \scriptsize\strut .509 & \scriptsize\strut .520 & \scriptsize\strut .528 & \scriptsize\strut .528 & \scriptsize\strut .530 & \scriptsize\strut .529 & \scriptsize\strut .530 \\
   \hline
   \scriptsize\strut \textbf{TF-IRF} & \scriptsize\strut .579 & \scriptsize\strut .599 & \scriptsize\strut .609 & \scriptsize\strut .612 & \scriptsize\strut .617 & \scriptsize\strut .619 & \scriptsize\strut .621 \\
   \hline
   \scriptsize\strut \textbf{TF-IBF} & \scriptsize\strut \textbf{.583} & \scriptsize\strut \textbf{.602} & \scriptsize\strut \textbf{.614} & \scriptsize\strut \textbf{.618} & \scriptsize\strut \textbf{.624} & \scriptsize\strut \textbf{.626} & \scriptsize\strut \textbf{.628} \\
   \hline
   \scriptsize\strut \textbf{TF-IUF} & \scriptsize\strut .578 & \scriptsize\strut .598 & \scriptsize\strut .609 & \scriptsize\strut .613 & \scriptsize\strut .619 & \scriptsize\strut .620 & \scriptsize\strut .623 \\
   \hline
   \hline
   \multicolumn{8}{|c|}{\scriptsize\strut \textbf{GoodReads - LCC}} \\
   \hline
   \hline
   \multicolumn{8}{|c|}{\scriptsize\strut \textbf{Top level}} \\
   \hline
   & \scriptsize\strut \textbf{3000} & \scriptsize\strut \textbf{6000} & \scriptsize\strut \textbf{9000} & \scriptsize\strut \textbf{12000} & \scriptsize\strut \textbf{15000} & \scriptsize\strut \textbf{18000} & \scriptsize\strut \textbf{21000} \\
   \hline
   \scriptsize\strut \textbf{TF} & \scriptsize\strut .725 & \scriptsize\strut .731 & \scriptsize\strut .737 & \scriptsize\strut .738 & \scriptsize\strut .734 & \scriptsize\strut .731 & \scriptsize\strut .743 \\
   \hline
   \scriptsize\strut \textbf{TF-IRF} & \scriptsize\strut \textbf{.781} & \scriptsize\strut \textbf{.792} & \scriptsize\strut \textbf{.797} & \scriptsize\strut .801 & \scriptsize\strut \textbf{.802} & \scriptsize\strut .799 & \scriptsize\strut .804 \\
   \hline
   \scriptsize\strut \textbf{TF-IBF} & \scriptsize\strut \textbf{.781} & \scriptsize\strut \textbf{.792} & \scriptsize\strut \textbf{.797} & \scriptsize\strut \textbf{.803} & \scriptsize\strut \textbf{.802} & \scriptsize\strut \textbf{.800} & \scriptsize\strut \textbf{.805} \\
   \hline
   \scriptsize\strut \textbf{TF-IUF} & \scriptsize\strut .776 & \scriptsize\strut .788 & \scriptsize\strut .792 & \scriptsize\strut .797 & \scriptsize\strut .797 & \scriptsize\strut .794 & \scriptsize\strut .800 \\
   \hline
   \hline
   \multicolumn{8}{|c|}{\scriptsize\strut \textbf{Second level}} \\
   \hline
   & \scriptsize\strut \textbf{3000} & \scriptsize\strut \textbf{6000} & \scriptsize\strut \textbf{9000} & \scriptsize\strut \textbf{12000} & \scriptsize\strut \textbf{15000} & \scriptsize\strut \textbf{18000} & \scriptsize\strut \textbf{21000} \\
   \hline
   \scriptsize\strut \textbf{TF} & \scriptsize\strut .494 & \scriptsize\strut .507 & \scriptsize\strut .510 & \scriptsize\strut .514 & \scriptsize\strut .513 & \scriptsize\strut .517 & \scriptsize\strut .519 \\
   \hline
   \scriptsize\strut \textbf{TF-IRF} & \scriptsize\strut .578 & \scriptsize\strut .599 & \scriptsize\strut .608 & \scriptsize\strut .617 & \scriptsize\strut .618 & \scriptsize\strut .622 & \scriptsize\strut .627 \\
   \hline
   \scriptsize\strut \textbf{TF-IBF} & \scriptsize\strut \textbf{.582} & \scriptsize\strut \textbf{.605} & \scriptsize\strut \textbf{.615} & \scriptsize\strut \textbf{.625} & \scriptsize\strut \textbf{.625} & \scriptsize\strut \textbf{.628} & \scriptsize\strut \textbf{.634} \\
   \hline
   \scriptsize\strut \textbf{TF-IUF} & \scriptsize\strut .576 & \scriptsize\strut .600 & \scriptsize\strut .610 & \scriptsize\strut .619 & \scriptsize\strut .620 & \scriptsize\strut .623 & \scriptsize\strut .628 \\
   \hline
  \end{tabular}
 \end{center}
 \caption[Accuracy results of tag-based book classification using weighting schemes on GoodReads]{Accuracy results of tag-based book classification using weighting schemes (GoodReads).}
 \label{tab:idfs-classification-goodreads}
\end{table}

Table \vref{tab:idfs-classification-goodreads} shows the results of using inverse tag weighting functions over DDC and LCC schemes on GoodReads. Similar to LibraryThing, tag weighting functions clearly outperform the sole use of TF. Moreover, these outperformances are even superior than for LibraryThing. As on LibraryThing, IBF performs the best among the weighting functions, followed by IRF, and then IUF.

Even though there are also system suggestions on GoodReads, they rely on tags previously used by the user, i.e., their personomy, and thus these suggestions can only be applied to different bookmarks and resources. Thereby, those users who tend to choose new tags instead of reusing tags from their personomy are yielding more natural bookmark frequencies. This affects and helps IBF perform better, but has no impact on IUF, as it is not altered by personomy-based suggestions. This shows that the effect of personomy-based suggestions in much smaller, and it affects to a lower extent or does not almost affect the distribution of tags, because suggestions do not spread to the users. Accordingly, the studied tag weighting functions perform well when this type of suggestion exists. On both LibraryThing and GoodReads, the results for the different classification schemes, DDC and LCC, are comparable and show a similar trend.

Summarizing, results show that the studied inverse tag weighting functions can be really useful for determining the representativity of each tag within the collection. However, folksonomies can suffer from resource-based tag suggestions, transforming the structure and distributions of folksonomies. This transformation can even be harmful for the definition of tag weighting functions, and can bring about worse performance results than simply relying on TF, as happened on Delicious. Otherwise, in the absence of resource-based tag suggestions, the use of tag weighting functions contribute in a positive manner to the performance of the classifier.

Comparing the results scored by tag weighting functions, it can be seen that IBF is always slightly better than IRF. The former is more detailed than the latter, because it considers the exact number of appearances of the tag besides the number of resources it appears in. Actually, IBF is the best approach for both LibraryThing and GoodReads, where there are no suggestions, or suggestions rely on user's personomy. When these suggestions rely on tags previously annotated by others to the resource, as on Delicious, IUF performs better than the other two weighting functions, showing the relevance of the ability of users to dismiss suggestions. However, even IUF is not able to outperform TF in this case.

\subsection{Revisiting Classifier Committees}
\label{revisiting-classifier-committees}

Apart from the results scored using tag weighting functions and the comparison of their performance to that by relying on TF, it is interesting to analyze their appropriateness to combine with other data sources. As we did in Chapter \vref{c:tag-representation}, we use classifier committees to evaluate the ability of the approaches using tag weighting functions to be combined with content and/or reviews, and improve even more the performance of the classifier. This time, we rely on the best committees for each datasets, i.e., the triple combination of tags, content and reviews for Delicious, and the double combination of tags and reviews for LibraryThing and GoodReads. We run them using the 4 different weightings for tags: TF, TF-IBF, TF-IRF, and TF-IUF, i.e., those compared in the previous section as well. By using classifier committees upon these weightings, we aim at analyzing how well they perform not only on their own, but also providing their prediction criteria when combining with other data sources.

\begin{table}[htb]
 \begin{center}
  \tiny
  \begin{tabular}{|l|c|c|c|c|c|c|c|}
   \hline
   \multicolumn{8}{|c|}{\scriptsize\strut \textbf{Delicious - ODP}} \\
   \hline
   \hline
   \multicolumn{8}{|c|}{\scriptsize\strut \textbf{Top level}} \\
   \hline
   & \scriptsize\strut \textbf{600} & \scriptsize\strut \textbf{1400} & \scriptsize\strut \textbf{2200} & \scriptsize\strut \textbf{3000} & \scriptsize\strut \textbf{4000} & \scriptsize\strut \textbf{5000} & \scriptsize\strut \textbf{6000} \\
   \hline
   \scriptsize\strut \textbf{TF} & \scriptsize\strut \textbf{.581} & \scriptsize\strut \textbf{.632} & \scriptsize\strut \textbf{.655} & \scriptsize\strut .671 & \scriptsize\strut .681 & \scriptsize\strut .691 & \scriptsize\strut .699 \\
   \hline
   \scriptsize\strut \textbf{TF-IRF} & \scriptsize\strut .576 & \scriptsize\strut .629 & \scriptsize\strut .653 & \scriptsize\strut .669 & \scriptsize\strut .680 & \scriptsize\strut .690 & \scriptsize\strut .697 \\
   \hline
   \scriptsize\strut \textbf{TF-IBF} & \scriptsize\strut .576 & \scriptsize\strut .630 & \scriptsize\strut .653 & \scriptsize\strut .670 & \scriptsize\strut .680 & \scriptsize\strut .690 & \scriptsize\strut .698 \\
   \hline
   \scriptsize\strut \textbf{TF-IUF} & \scriptsize\strut .576 & \scriptsize\strut .631 & \scriptsize\strut .654 & \scriptsize\strut \textbf{.672} & \scriptsize\strut \textbf{.682} & \scriptsize\strut \textbf{.692} & \scriptsize\strut \textbf{.700} \\
   \hline
   \hline
   \multicolumn{8}{|c|}{\scriptsize\strut \textbf{Second level}} \\
   \hline
   & \scriptsize\strut \textbf{600} & \scriptsize\strut \textbf{1400} & \scriptsize\strut \textbf{2200} & \scriptsize\strut \textbf{3000} & \scriptsize\strut \textbf{4000} & \scriptsize\strut \textbf{5000} & \scriptsize\strut \textbf{6000} \\
   \hline
   \scriptsize\strut \textbf{TF} & \scriptsize\strut \textbf{.412} & \scriptsize\strut \textbf{.488} & \scriptsize\strut .524 & \scriptsize\strut .545 & \scriptsize\strut .564 & \scriptsize\strut .579 & \scriptsize\strut .588 \\
   \hline
   \scriptsize\strut \textbf{TF-IRF} & \scriptsize\strut .406 & \scriptsize\strut .485 & \scriptsize\strut .523 & \scriptsize\strut .546 & \scriptsize\strut .566 & \scriptsize\strut .580 & \scriptsize\strut .592 \\
   \hline
   \scriptsize\strut \textbf{TF-IBF} & \scriptsize\strut .407 & \scriptsize\strut .486 & \scriptsize\strut .525 & \scriptsize\strut \textbf{.548} & \scriptsize\strut .566 & \scriptsize\strut .580 & \scriptsize\strut .592 \\
   \hline
   \scriptsize\strut \textbf{TF-IUF} & \scriptsize\strut .408 & \scriptsize\strut \textbf{.488} & \scriptsize\strut \textbf{.526} & \scriptsize\strut \textbf{.548} & \scriptsize\strut \textbf{.569} & \scriptsize\strut \textbf{.584} & \scriptsize\strut \textbf{.595} \\
   \hline
  \end{tabular}
 \end{center}
 \caption[Accuracy results of classifier committees for web page classification using weighting schemes]{Accuracy results of classifier committees for web page classification using weighting schemes.}
 \label{tab:idfs-committees-delicious}
\end{table}

Table \vref{tab:idfs-committees-delicious} shows the classification results of the approaches considering inverse tag weighting functions on classifier committees for Delicious. Even though inverse tag weighting functions were not useful to improve the performance of the tag-based classifier reducing its overall accuracy, they seem to provide better decisions to be combined with other data sources. The predictions and margins outputted by all three approaches using inverse weighting functions yield slightly better results on the classification, especially when it comes to second level classification. Thereby, inverse tag weighting functions are useful for Delicious when their outputs are applied on classifier committees along with content and reviews. The unsuitability of tag weighting functions on their own gets fixed by the use of classifier committees. However, the little outperformance by tag weighting functions when using committees is almost irrelevant as compared to the TF-based committees. This outperformance is slightly clearer for largest training sets upon the second level classification.

\begin{table}[htb]
 \begin{center}
  \tiny
  \begin{tabular}{|l|c|c|c|c|c|c|c|}
   \hline
   \multicolumn{8}{|c|}{\scriptsize\strut \textbf{LibraryThing - DDC}} \\
   \hline
   \hline
   \multicolumn{8}{|c|}{\scriptsize\strut \textbf{Top level}} \\
   \hline
   & \scriptsize\strut \textbf{3000} & \scriptsize\strut \textbf{6000} & \scriptsize\strut \textbf{9000} & \scriptsize\strut \textbf{12000} & \scriptsize\strut \textbf{15000} & \scriptsize\strut \textbf{18000} & \scriptsize\strut \textbf{21000} \\
   \hline
   \scriptsize\strut \textbf{TF} & \scriptsize\strut .857 & \scriptsize\strut .866 & \scriptsize\strut .868 & \scriptsize\strut .872 & \scriptsize\strut .875 & \scriptsize\strut .876 & \scriptsize\strut .876 \\
   \hline
   \scriptsize\strut \textbf{TF-IRF} & \scriptsize\strut .864 & \scriptsize\strut .882 & \scriptsize\strut .886 & \scriptsize\strut .890 & \scriptsize\strut \textbf{.894} & \scriptsize\strut \textbf{.897} & \scriptsize\strut .897 \\
   \hline
   \scriptsize\strut \textbf{TF-IBF} & \scriptsize\strut \textbf{.865} & \scriptsize\strut \textbf{.883} & \scriptsize\strut \textbf{.887} & \scriptsize\strut \textbf{.891} & \scriptsize\strut \textbf{.894} & \scriptsize\strut \textbf{.897} & \scriptsize\strut \textbf{.898} \\
   \hline
   \scriptsize\strut \textbf{TF-IUF} & \scriptsize\strut \textbf{.865} & \scriptsize\strut \textbf{.883} & \scriptsize\strut .886 & \scriptsize\strut .889 & \scriptsize\strut .892 & \scriptsize\strut .894 & \scriptsize\strut .895 \\
   \hline
   \hline
   \multicolumn{8}{|c|}{\scriptsize\strut \textbf{Second level}} \\
   \hline
   & \scriptsize\strut \textbf{3000} & \scriptsize\strut \textbf{6000} & \scriptsize\strut \textbf{9000} & \scriptsize\strut \textbf{12000} & \scriptsize\strut \textbf{15000} & \scriptsize\strut \textbf{18000} & \scriptsize\strut \textbf{21000} \\
   \hline
   \scriptsize\strut \textbf{TF} & \scriptsize\strut .687 & \scriptsize\strut .708 & \scriptsize\strut .717 & \scriptsize\strut .721 & \scriptsize\strut .729 & \scriptsize\strut .729 & \scriptsize\strut .733 \\
   \hline
   \scriptsize\strut \textbf{TF-IRF} & \scriptsize\strut .709 & \scriptsize\strut \textbf{.742} & \scriptsize\strut .754 & \scriptsize\strut .765 & \scriptsize\strut .770 & \scriptsize\strut .773 & \scriptsize\strut .778 \\
   \hline
   \scriptsize\strut \textbf{TF-IBF} & \scriptsize\strut .710 & \scriptsize\strut \textbf{.742} & \scriptsize\strut \textbf{.756} & \scriptsize\strut \textbf{.767} & \scriptsize\strut \textbf{.772} & \scriptsize\strut \textbf{.776} & \scriptsize\strut \textbf{.780} \\
   \hline
   \scriptsize\strut \textbf{TF-IUF} & \scriptsize\strut \textbf{.712} & \scriptsize\strut .741 & \scriptsize\strut .752 & \scriptsize\strut .763 & \scriptsize\strut .767 & \scriptsize\strut .769 & \scriptsize\strut .775 \\
   \hline
   \hline
   \multicolumn{8}{|c|}{\scriptsize\strut \textbf{LibraryThing - LCC}} \\
   \hline
   \hline
   \multicolumn{8}{|c|}{\scriptsize\strut \textbf{Top level}} \\
   \hline
   & \scriptsize\strut \textbf{3000} & \scriptsize\strut \textbf{6000} & \scriptsize\strut \textbf{9000} & \scriptsize\strut \textbf{12000} & \scriptsize\strut \textbf{15000} & \scriptsize\strut \textbf{18000} & \scriptsize\strut \textbf{21000} \\
   \hline
   \scriptsize\strut \textbf{TF} & \scriptsize\strut .831 & \scriptsize\strut .845 & \scriptsize\strut .853 & \scriptsize\strut .856 & \scriptsize\strut .861 & \scriptsize\strut .859 & \scriptsize\strut .864 \\
   \hline
   \scriptsize\strut \textbf{TF-IRF} & \scriptsize\strut .849 & \scriptsize\strut .869 & \scriptsize\strut .876 & \scriptsize\strut .880 & \scriptsize\strut .887 & \scriptsize\strut .885 & \scriptsize\strut .890 \\
   \hline
   \scriptsize\strut \textbf{TF-IBF} & \scriptsize\strut .851 & \scriptsize\strut \textbf{.871} & \scriptsize\strut \textbf{.879} & \scriptsize\strut \textbf{.882} & \scriptsize\strut \textbf{.888} & \scriptsize\strut \textbf{.887} & \scriptsize\strut \textbf{.892} \\
   \hline
   \scriptsize\strut \textbf{TF-IUF} & \scriptsize\strut \textbf{.852} & \scriptsize\strut .869 & \scriptsize\strut .875 & \scriptsize\strut .880 & \scriptsize\strut .886 & \scriptsize\strut .885 & \scriptsize\strut .888 \\
   \hline
   \hline
   \multicolumn{8}{|c|}{\scriptsize\strut \textbf{Second level}} \\
   \hline
   & \scriptsize\strut \textbf{3000} & \scriptsize\strut \textbf{6000} & \scriptsize\strut \textbf{9000} & \scriptsize\strut \textbf{12000} & \scriptsize\strut \textbf{15000} & \scriptsize\strut \textbf{18000} & \scriptsize\strut \textbf{21000} \\
   \hline
   \scriptsize\strut \textbf{TF} & \scriptsize\strut .688 & \scriptsize\strut .723 & \scriptsize\strut .736 & \scriptsize\strut .746 & \scriptsize\strut .754 & \scriptsize\strut .755 & \scriptsize\strut .766 \\
   \hline
   \scriptsize\strut \textbf{TF-IRF} & \scriptsize\strut .712 & \scriptsize\strut .750 & \scriptsize\strut .770 & \scriptsize\strut .782 & \scriptsize\strut .789 & \scriptsize\strut .793 & \scriptsize\strut .803 \\
   \hline
   \scriptsize\strut \textbf{TF-IBF} & \scriptsize\strut .717 & \scriptsize\strut \textbf{.755} & \scriptsize\strut \textbf{.773} & \scriptsize\strut \textbf{.786} & \scriptsize\strut \textbf{.792} & \scriptsize\strut \textbf{.797} & \scriptsize\strut \textbf{.806} \\
   \hline
   \scriptsize\strut \textbf{TF-IUF} & \scriptsize\strut \textbf{.719} & \scriptsize\strut .754 & \scriptsize\strut .770 & \scriptsize\strut .781 & \scriptsize\strut .788 & \scriptsize\strut .792 & \scriptsize\strut .801 \\
   \hline
  \end{tabular}
 \end{center}
 \caption[Accuracy results of classifier committees for book classification using weighting schemes on LibraryThing]{Accuracy results of classifier committees for book classification using weighting schemes (LibraryThing).}
 \label{tab:idfs-committees-librarything}
\end{table}

Table \vref{tab:idfs-committees-librarything} shows the classification results of the approaches considering tag weighting functions on classifier committees for LibraryThing over DDC and LCC schemes. In this case, inverse tag weighting functions are also useful when applied to classifier committees when compared to the TF-based one. Those classifier committees including tag weighting functions produce clearly better results than the committee using TF. This performance improvement is positive for both levels, but it is larger for the second level. However, those approaches using tag-based representations with tag weighting functions perform better on their own, without considering committees (see Table \vref{tab:idfs-classification-librarything}). That is, it is better to use the classifier based on tags on their own, without including the predictions by the classifier using reviews. This means that predictions with tag weighting functions are good enough to work on their own, and it is better to ignore the other data source, i.e., reviews, which cannot catch up with the performance of tags and harm the overall performance.

\begin{table}[htb]
 \begin{center}
  \tiny
  \begin{tabular}{|l|c|c|c|c|c|c|c|}
   \hline
   \multicolumn{8}{|c|}{\scriptsize\strut \textbf{GoodReads - DDC}} \\
   \hline
   \hline
   \multicolumn{8}{|c|}{\scriptsize\strut \textbf{Top level}} \\
   \hline
   & \scriptsize\strut \textbf{3000} & \scriptsize\strut \textbf{6000} & \scriptsize\strut \textbf{9000} & \scriptsize\strut \textbf{12000} & \scriptsize\strut \textbf{15000} & \scriptsize\strut \textbf{18000} & \scriptsize\strut \textbf{21000} \\
   \hline
   \scriptsize\strut \textbf{TF} & \scriptsize\strut .820 & \scriptsize\strut .847 & \scriptsize\strut .857 & \scriptsize\strut .865 & \scriptsize\strut .867 & \scriptsize\strut .872 & \scriptsize\strut .874 \\
   \hline
   \scriptsize\strut \textbf{TF-IRF} & \scriptsize\strut .835 & \scriptsize\strut .859 & \scriptsize\strut .867 & \scriptsize\strut .874 & \scriptsize\strut .877 & \scriptsize\strut .881 & \scriptsize\strut .884 \\
   \hline
   \scriptsize\strut \textbf{TF-IBF} & \scriptsize\strut \textbf{.837} & \scriptsize\strut \textbf{.861} & \scriptsize\strut \textbf{.868} & \scriptsize\strut \textbf{.876} & \scriptsize\strut \textbf{.878} & \scriptsize\strut \textbf{.882} & \scriptsize\strut \textbf{.885} \\
   \hline
   \scriptsize\strut \textbf{TF-IUF} & \scriptsize\strut .834 & \scriptsize\strut .858 & \scriptsize\strut .866 & \scriptsize\strut .873 & \scriptsize\strut .876 & \scriptsize\strut .881 & \scriptsize\strut .883 \\
   \hline
   \hline
   \multicolumn{8}{|c|}{\scriptsize\strut \textbf{Second level}} \\
   \hline
   & \scriptsize\strut \textbf{3000} & \scriptsize\strut \textbf{6000} & \scriptsize\strut \textbf{9000} & \scriptsize\strut \textbf{12000} & \scriptsize\strut \textbf{15000} & \scriptsize\strut \textbf{18000} & \scriptsize\strut \textbf{21000} \\
   \hline
   \scriptsize\strut \textbf{TF} & \scriptsize\strut .610 & \scriptsize\strut .651 & \scriptsize\strut .670 & \scriptsize\strut .683 & \scriptsize\strut .691 & \scriptsize\strut .696 & \scriptsize\strut .705 \\
   \hline
   \scriptsize\strut \textbf{TF-IRF} & \scriptsize\strut .637 & \scriptsize\strut .676 & \scriptsize\strut .693 & \scriptsize\strut .707 & \scriptsize\strut .716 & \scriptsize\strut .719 & \scriptsize\strut .726 \\
   \hline
   \scriptsize\strut \textbf{TF-IBF} & \scriptsize\strut \textbf{.642} & \scriptsize\strut \textbf{.681} & \scriptsize\strut \textbf{.697} & \scriptsize\strut \textbf{.711} & \scriptsize\strut \textbf{.719} & \scriptsize\strut \textbf{.723} & \scriptsize\strut \textbf{.730} \\
   \hline
   \scriptsize\strut \textbf{TF-IUF} & \scriptsize\strut .638 & \scriptsize\strut .677 & \scriptsize\strut .694 & \scriptsize\strut .708 & \scriptsize\strut .717 & \scriptsize\strut .722 & \scriptsize\strut .727 \\
   \hline
   \hline
   \multicolumn{8}{|c|}{\scriptsize\strut \textbf{GoodReads - LCC}} \\
   \hline
   \hline
   \multicolumn{8}{|c|}{\scriptsize\strut \textbf{Top level}} \\
   \hline
   & \scriptsize\strut \textbf{3000} & \scriptsize\strut \textbf{6000} & \scriptsize\strut \textbf{9000} & \scriptsize\strut \textbf{12000} & \scriptsize\strut \textbf{15000} & \scriptsize\strut \textbf{18000} & \scriptsize\strut \textbf{21000} \\
   \hline
   \scriptsize\strut \textbf{TF} & \scriptsize\strut .831 & \scriptsize\strut .836 & \scriptsize\strut .847 & \scriptsize\strut .853 & \scriptsize\strut .857 & \scriptsize\strut .857 & \scriptsize\strut .864 \\
   \hline
   \scriptsize\strut \textbf{TF-IRF} & \scriptsize\strut .826 & \scriptsize\strut .846 & \scriptsize\strut .856 & \scriptsize\strut .861 & \scriptsize\strut .866 & \scriptsize\strut .864 & \scriptsize\strut .870 \\
   \hline
   \scriptsize\strut \textbf{TF-IBF} & \scriptsize\strut \textbf{.829} & \scriptsize\strut \textbf{.848} & \scriptsize\strut \textbf{.858} & \scriptsize\strut \textbf{.863} & \scriptsize\strut \textbf{.868} & \scriptsize\strut \textbf{.866} & \scriptsize\strut \textbf{.871} \\
   \hline
   \scriptsize\strut \textbf{TF-IUF} & \scriptsize\strut .826 & \scriptsize\strut .845 & \scriptsize\strut .856 & \scriptsize\strut .860 & \scriptsize\strut .866 & \scriptsize\strut .864 & \scriptsize\strut .869 \\
   \hline
   \hline
   \multicolumn{8}{|c|}{\scriptsize\strut \textbf{Second level}} \\
   \hline
   & \scriptsize\strut \textbf{3000} & \scriptsize\strut \textbf{6000} & \scriptsize\strut \textbf{9000} & \scriptsize\strut \textbf{12000} & \scriptsize\strut \textbf{15000} & \scriptsize\strut \textbf{18000} & \scriptsize\strut \textbf{21000} \\
   \hline
   \scriptsize\strut \textbf{TF} & \scriptsize\strut .624 & \scriptsize\strut .674 & \scriptsize\strut .696 & \scriptsize\strut .712 & \scriptsize\strut .725 & \scriptsize\strut .730 & \scriptsize\strut .735 \\
   \hline
   \scriptsize\strut \textbf{TF-IRF} & \scriptsize\strut .647 & \scriptsize\strut .697 & \scriptsize\strut .716 & \scriptsize\strut .732 & \scriptsize\strut .742 & \scriptsize\strut .748 & \scriptsize\strut .757 \\
   \hline
   \scriptsize\strut \textbf{TF-IBF} & \scriptsize\strut \textbf{.651} & \scriptsize\strut \textbf{.700} & \scriptsize\strut \textbf{.720} & \scriptsize\strut \textbf{.736} & \scriptsize\strut \textbf{.746} & \scriptsize\strut \textbf{.751} & \scriptsize\strut \textbf{.759} \\
   \hline
   \scriptsize\strut \textbf{TF-IUF} & \scriptsize\strut .648 & \scriptsize\strut .698 & \scriptsize\strut .718 & \scriptsize\strut .733 & \scriptsize\strut .744 & \scriptsize\strut .749 & \scriptsize\strut .757 \\
   \hline
  \end{tabular}
 \end{center}
 \caption[Accuracy results of classifier committees for book classification using weighting schemes on GoodReads]{Accuracy results of classifier committees for book classification using weighting schemes (GoodReads).}
 \label{tab:idfs-committees-goodreads}
\end{table}

Table \vref{tab:idfs-committees-goodreads} shows the classification results of the approaches considering tag weighting functions on classifier committees for GoodReads over DDC and LCC schemes. The main conclusions drawn from these results are very similar to those on LibraryThing. Again, committees relying on tag weighting functions perform clearly better than that relying on TF, especially for the second level. However, it is better to ignore the other data source, i.e., reviews, since the results by tags on their own are good enough and cannot be improved by combining them (see Table \vref{tab:idfs-classification-goodreads}).

Again, using classifier committees obtains comparable results with very similar trends for both book taxonomies, DDC and LCC.

Summarizing the results for all three datasets, the use of tag weighting functions has shown to be helpful in all cases as compared to TF when it comes to combining them with other data sources using classifier committees. However, it is better to rely only on the tag-based classifier for both LibraryThing and GoodReads, which score good results on their own, and they get harmed when combined with other data sources. In the case of Delicious, on the other hand, the use of classifier committees for approaches relying on tag weighting functions perform better results than using tags on their own, and than the committees relying on TF. Nonetheless, the latter performs just slightly worse, and their results are very similar, suggesting that any of them could be used to perform the task.

\subsection{Correlation between Tag Weighting Functions}
\label{ssec:correlation}

All three inverse tag weighting functions consider the distribution of tags across different dimensions. The values given by these three functions could correlate or not depending on the behavior of users, e.g., if many tags annotated by a large number of users congregate into the same resource, correlation between IUF and IRF would be lower than if each of the users annotate those tags in different resources. Thus, analyzing whether these three values correlate is of utmost importance.

Table \vref{tab:correlation} shows correlation values among tag weighting schemes. The correlation values between each pair of functions are shown in each row, for both the Pearson and Spearman correlation coefficients. Note that the latter considers the rank inferred from tag weights, whereas the former considers the values to compute correlations. Both correlation values range from -1 to 1. The closer is this value to 0, the less correlation exists among the compared sets and, thus, the more independent they are.

\begin{table}[htb]
 \begin{center}
  \tiny
  \begin{tabular}{|l|c|c|c|c|c|c|}
   \hline
   & \multicolumn{2}{c|}{\scriptsize\strut \textbf{Delicious}} & \multicolumn{2}{c|}{\scriptsize\strut \textbf{LibraryThing}} & \multicolumn{2}{c|}{\scriptsize\strut \textbf{GoodReads}} \\
   \hline
   & \scriptsize\strut \textbf{r} & \scriptsize\strut \textbf{$\rho$} & \scriptsize\strut \textbf{r} & \scriptsize\strut \textbf{$\rho$} & \scriptsize\strut \textbf{r} & \scriptsize\strut \textbf{$\rho$} \\
   \hline
   \scriptsize\strut \textbf{IRF-IUF} & \scriptsize\strut .763 & \scriptsize\strut .657 & \scriptsize\strut .679 & \scriptsize\strut .603 & \scriptsize\strut .529 & \scriptsize\strut .421 \\
   \hline
   \scriptsize\strut \textbf{IRF-IBF} & \scriptsize\strut .991 & \scriptsize\strut .990 & \scriptsize\strut .989 & \scriptsize\strut .981 & \scriptsize\strut .997 & \scriptsize\strut .998 \\
   \hline
   \scriptsize\strut \textbf{IUF-IBF} & \scriptsize\strut .780 & \scriptsize\strut .677 & \scriptsize\strut .720 & \scriptsize\strut .630 & \scriptsize\strut .556 & \scriptsize\strut .436 \\
   \hline
  \end{tabular}
 \end{center}
 \caption[Pearson and Spearman correlation coefficients]{Pearson (r) and Spearman ($\rho$) correlation coefficients.}
 \label{tab:correlation}
\end{table}

Correlation values show that there is a high dependence among IBF and IRF values. Both seem to be fully dependent and, thus, that is why these two approaches achieve very similar results. The correlation decreases when IUF is considered, so that it seems to be more independent to the rest. This independence is clearest for GoodReads, and intermediate for LibraryThing, but Delicious shows the highest dependence of IUF with respect to the other two values. The main reason for the clear independence of IUF on GoodReads is that users are suggested by the system with tags in their personomy, so that they easily spread tags on bookmarks and resources, keeping the user frequency unchanged. As LibraryThing and Delicious do not have this feature, IUF correlates with the others to a greater extent.

\section{Conclusion}
\label{sec:tag-distributions-conclusion}

In this chapter, we have studied and analyzed the application of tag weighting functions based on the classical IDF scheme for the resource classification task on the three large-scale datasets introduced in Chapter \vref{c:datasets}. We have considered the distributions of tags across users, resources and bookmarks to generate three variations of such weighting, namely IBF, IRF and IUF. We have performed classification experiments by considering their results on their own, and by combining them with other data sources using classifier committees. We have analyzed their results by taking into account the settings of each social tagging system, and how it affects the distribution of tags of their underlying folksonomies.

With these experiments, we have given an answer to the following research questions:

\begin{description}
 \item[Research Question 7] \hfill \\
 \textit{\rqdistone}
\end{description}

We have analyzed the suitability of IDF-like weighting functions to define the representativity of tags, which consider the distribution of tags through the whole collection of resources. Our experiments have shown that these functions helps improve performance of a resource classification task. However, we have shown that the settings of the social tagging system have an effect on those distributions. Resource-based tag suggestions have shown to influence the structure of folksonomies greatly. Suggesting tags based on previous annotations of others on the resource causes a very different tag distribution, which in turn, affects the results of the weighting function. When a system enables the resource-based tag suggestions, the use of tag weighting functions performs worse, and combining with other data sources is required to improve performance; this method can even outperform the TF-based approach.

For our classification experiments, we have found that IDF-like weighting functions clearly outperform the TF approach when resource-based tag suggestions are not enabled, i.e., on LibraryThing and GoodReads, both when used on their own, or when combined with other data sources. We found it better to consider just the tag-based approach, without combining them with other data sources, since it provides superior results, which cannot be improved by combining them with other predictions.

\begin{description}
 \item[Research Question 8] \hfill \\
 \textit{\rqdisttwo}
\end{description}

Among the studied weighting functions, the one relying on bookmark frequencies has shown to be the best when there are no resource-based tag suggestions. In these cases, IBF performs the best, followed by IRF, and IUF. All of them clearly outperform TF, when both used on their own, and combined with other data sources using classifier committees.

On the other hand, when the social tagging system suggests tags to the user relying on the resource itself, IUF performs better than the others. IUF performs better than IBF and IRF, because of the importance of the ability of users to choose their own tags without relying on suggestions from these systems. Even though IUF does not outperform TF when used on its own, combining it with other data sources produces the best approach. However, it is only slightly better than the committees relying on TF, and any of them can be used to score similar results.

%% file: tag-categorizers.tex
\chapter{Analyzing the Behavior of Users for Classification}
\label{c:analyzing-appropriateness-users}

\textit{``Always imitate the behavior of the winners when you lose.''}

--- George Meredith

\chaptersummary{In this chapter, we explore the behavior of users on social tagging systems. Earlier works have suggested and shown that users of these systems follow different goals, and they tag resources for a certain purpose. Several classifications have been proposed to discriminate user behavior. Specifically, we consider one of those classifications of behavioral purposes. Such classification splits user behavior into two goals: (a) users who aim at maintaining an organizational structure of the resources for later browsing, so-called Categorizers, and (b) users who rather provide detailed descriptions for later search, so-called Describers. These two user behaviors yield different personomy structures, i.e., they follow a different tagging pattern, which produces different tag selections from each other.}

\chaptersummary{Such a classification of users has been previously experimented, and has shown its effectiveness to discover users who rather describe resources, i.e., Describers. However, the appropriateness of Categorizers for a resource classification task has not yet been studied. Upon this, we set out the study of the suitability of users who fit such behavior, by performing a set of resource classification and descriptiveness experiments. To this end, we split the whole set of users into smaller subsets of utmost Categorizers and Describers. We explore how each subset of users better fits the classification or descriptiveness task.}

\chaptersummary{This chapter is organized as follows. Next, in Section \vref{sec:user-behavior} we briefly summarize the research work found so far on the user motivation to tagging, and motivate its interest towards our work on resource classification. In Section \vref{sec:categorizers-vs-describers} we detail in more depth one of those classifications of user behavior, which separates Categorizers from Describers. Then, in Section \vref{sec:categorizers-experiment-settings} we present the settings of our resource classification experiments enhanced by the detection of user behavior, and present their results in Section \vref{sec:user-behavior-results}. Finally, we conclude the chapter in Section \vref{sec:user-behavior-conclusion}.}

We address the following research questions in this chapter:

\chaptersummary{
 \begin{description}
  \item[Research Question 9] \hfill \\
  \textit{\rqcatone}
 \end{description}
}

\chaptersummary{
 \begin{description}
  \item[Research Question 10] \hfill \\
  \textit{\rqcattwo}
 \end{description}
}

\section{User Behavior on Social Tagging Systems}
\label{sec:user-behavior}

It has been suggested that not all the users contributing on social tagging systems are motivated by the same goal for annotating resources. Depending on their annotations, several works propose different classifications of user behavior \citep{koerner2010categorizers}. Some of them focus on detecting the types of tags provided by users. For instance, early works such as \cite{golder_structure_2006} and \cite{sen2006tagging} propose the existence of several tag types. On the other hand, others have suggested discriminating user behavior by their annotations. In this regard, works such as \cite{marlow2006ht06}, \cite{heckner2009personal}, \cite{nov2009motivational} and \cite{strohmaier2010why} propose differentiating users by their motivation for tagging resources.

As a classification of user behavior that matches our requirements, we focus on the latter by \cite{strohmaier2010why}. In this work, the authors propose differentiating two kinds of user behavior: Categorizers, who rather organize resources, and Describers, who rather define the contents of resources. It seems reasonable that users so-called Categorizers may provide annotations that better fit the resource classification task than Describers. Next, we detail in more depth these two types of users.

\section{Categorizers vs Describers}
\label{sec:categorizers-vs-describers}

The approach we consider for discriminating users by their behavior has been introduced and experimented in earlier works \citep{koerner2010stop, koerner2010categorizers, koerner2009understanding}. They consider the existence of two major tagging motivations on social tagging systems: Categorizers and Describers.

\begin{table}
 \begin{center}
 \tiny
  \begin{tabular}{|l|c|c|}
   \hline
   & \scriptsize\strut \textbf{Categorizer} & \scriptsize\strut \textbf{Describer} \\
   \hline
   \scriptsize\strut \textbf{Goal of Tagging} & \scriptsize\strut later browsing & \scriptsize\strut later retrieval \\
   \hline
   \scriptsize\strut \textbf{Change of Tag Vocabulary} & \scriptsize\strut costly & \scriptsize\strut cheap \\
   \hline
   \scriptsize\strut \textbf{Size of Tag Vocabulary} & \scriptsize\strut limited & \scriptsize\strut open \\
   \hline
   \scriptsize\strut \textbf{Tags} & \scriptsize\strut subjective & \scriptsize\strut objective \\
   \hline
  \end{tabular}
 \end{center}
 \caption[Characteristics of Categorizers and Describers]{Characteristics of Categorizers and Describers.}
 \label{tab:categorizer-vs-describer}
\end{table}

Early works such as \cite{marlow2006ht06, Hammond2005} and \cite{heckner2009personal} suggest that a distinction between at least two types of user motivations for tagging is interesting: on one hand, users can be motivated by categorization (in the following, \emph{Categorizers}). These users view tagging as a means to categorize resources according to some (shared or personal) high-level conceptualizations. They typically use a rather elaborated tag set to construct and maintain a navigational aid to the resources for later browsing.
On the other hand, users who are motivated by description (so-called \emph{Describers}) view tagging as a means to accurately and precisely detail resources. These users tag because they want to produce annotations that are useful for later search and retrieval. The development of a personal, consistent ontology to navigate across their resources is not their intuition. Table \vref{tab:categorizer-vs-describer} gives an overview of characteristics of the two different types of users, based on \cite{koerner2009understanding}.

\subsection{Measures}
\label{ssec:user-behavior-measures}

We use three different measures to differentiate users into Categorizers and Describers: Tags Per Post (TPP), Tag Resource Ratio (TRR), and Orphan Ratio (ORPHAN). Additional measures are shown in \cite{koerner2010categorizers}, but due to the high correlation with the others, we limited our efforts to the ones above. These measures rely on two features of user behavior: verbosity, which measures the number of tags a user tends to use when annotating, and diversity, which measures the extent to which users are using new tags that were not previously applied by themselves. It is worthwhile noting that these measures provide one value for each user. The measure corresponding to each user is thus computed by considering the characteristics of their bookmarks and the attached tag assignments. The resulting measures are then ranked in a list along with the rest of the users. This list makes possible inferring the extent to which a user is rather a Categorizer or a Describer.

\subsubsection{Tags per Post (TPP)}
\label{sssec:tags-per-post}

As a Describer would focus on describing their resources in a very detailed manner, the number of tags used to annotate each resource can be taken into account as an indicator to identify the motivation of the analyzed user. The \emph{tags per post} measure (short \emph{TPP}) captures this by dividing the number of all tag assignments of a user by the number of resources (see Equation \vref{eq:tpp}). $T_{ur}$ is the number of tags annotated by a user $u$ on a resource $r$, and $R_u$ is the number of resources of the user. The more tags a user utilizes to annotate the resources, the more likely they are a Describer, reflecting it in a higher TPP score.

\begin{equation}
 TPP(u)=\frac{\displaystyle\sum^r|T_{ur}|}{|R_u|}
 \label{eq:tpp}
\end{equation}

This measure relies on the verbosity of users, as it computes the average number of tags they assigned to bookmarks.

\subsubsection{Orphan Ratio (ORPHAN)}
\label{sssec:orphan-ratio}

Since Describers do not have a fixed vocabulary and freely choose tags to describe their resources in a detailed manner, they would not focus on reusing tags. This factor is analyzed in the \emph{orphan ratio} (short \emph{ORPHAN}). This measure relates the number of seldom used tags to the total number of tags. Equation \vref{eq:orphan1} shows how seldom used tags are defined by the individual tagging style of a user. In this equation, $t_{max}$ denotes the most frequent tag of the user. Equation \vref{eq:orphan2} shows the calculation of the final measure where $T^o_u$ are seldom used tags and $T_u$ are all tags of the given user. Users with more seldom tags yield a higher orphan ratio, and they are more likely to be Describers.

\begin{equation}
 n = \left\lceil \frac{|R(t_{max})|}{100} \right\rceil
 \label{eq:orphan1}
\end{equation}

\begin{equation}
 ORPHAN(u) = \frac{|T^o_u|}{|T_u|}, T^o_u = \{t| |R(t)| \le n\}
 \label{eq:orphan2}
\end{equation}

By measuring whether users frequently use the same tags or rather rely on new ones, the ORPHAN ratio considers their diversity.

\subsubsection{Tag Resource Ratio (TRR)} 
\label{ssub:tag_resource_ratio}

The \emph{tag resource ratio} (short \emph{TRR}) relates the number of tags of a user (i.e., the size of their vocabulary) to the total number of annotated resources (see Equation \ref{eq:trr}). A typical Categorizer would use a small number of tags as compared to the number of resources and would therefore score a low TRR value. 

\begin{equation}
 TRR(u)= \frac{|T_u|}{|R_u|}
 \label{eq:trr}
\end{equation}

This measure relies on both verbosity, because users who use more tags in each bookmark would usually result in a
higher \emph{TRR} value, and diversity, as those who frequently use new tags will have a larger vocabulary. Nonetheless, the latter has a higher impact in this case, since the former could be altered by verbose users who tend to reuse tags.

\section{Calculation of Measures and Experiment Settings}
\label{sec:categorizers-experiment-settings}

Users of each social tagging site have their own weights for each of the three measures above. Thus, we computed TPP, ORPHAN and TRR values for each user. This way, we are able to generate three ranked lists of users for each site. In these rankings, Categorizers rank high, whereas Describers rank low (this is arbitrary and could be inverted as well). From these lists, we can select a subset of users in the top as Categorizers, and another subset in the tail as Describers. Both sets should have the
same size in order to compare them.

Our main goal is to conclude whether these measures can discriminate Categorizers in such a way that they perform better than Describers on a resource classification task. However, we also perform experiments measuring the descriptiveness of users' tags in order to conclude whether Describers perform better to that end. With the subsets of Categorizers and Describers defined above, we perform classification and descriptiveness experiments to know how suitable they are for each of the tasks.

Table \vref{tab:cat-desc-histogram} shows the distribution of the three measures we calculated for users on the three datasets. The X axis represents quantiles of values, whereas Y axis represents the number of users belonging to each quantile. Note that the values themselves are not relevant, but just allows us to rank each user and analyze where they fall in the distribution of all weights. On one hand, the TRR measure follows a similar distribution for all three datasets. On the other hand, for the other two measures, the distributions show there are lots of extreme users for LibraryThing and GoodReads: (1) the ORPHAN measure shows both many extreme Categorizers and extreme Describers, and (2) the TPP measure discriminates a large set of extreme Categorizers, but almost no extreme Describers. These distributions change drastically on Delicious, though. For this dataset, there are many users who have middle values, and who are not that clearly discriminated as Describers or Categorizers.

\begin{table}[htbp]
 \begin{center}
  \begin{tabular}{ c c c c }
    & \scriptsize \textbf{TRR} & \scriptsize \textbf{ORPHAN} & \scriptsize \textbf{TPP} \\
    \begin{sideways}\scriptsize \textbf{Delicious}\end{sideways} & \includegraphics[width=101px]{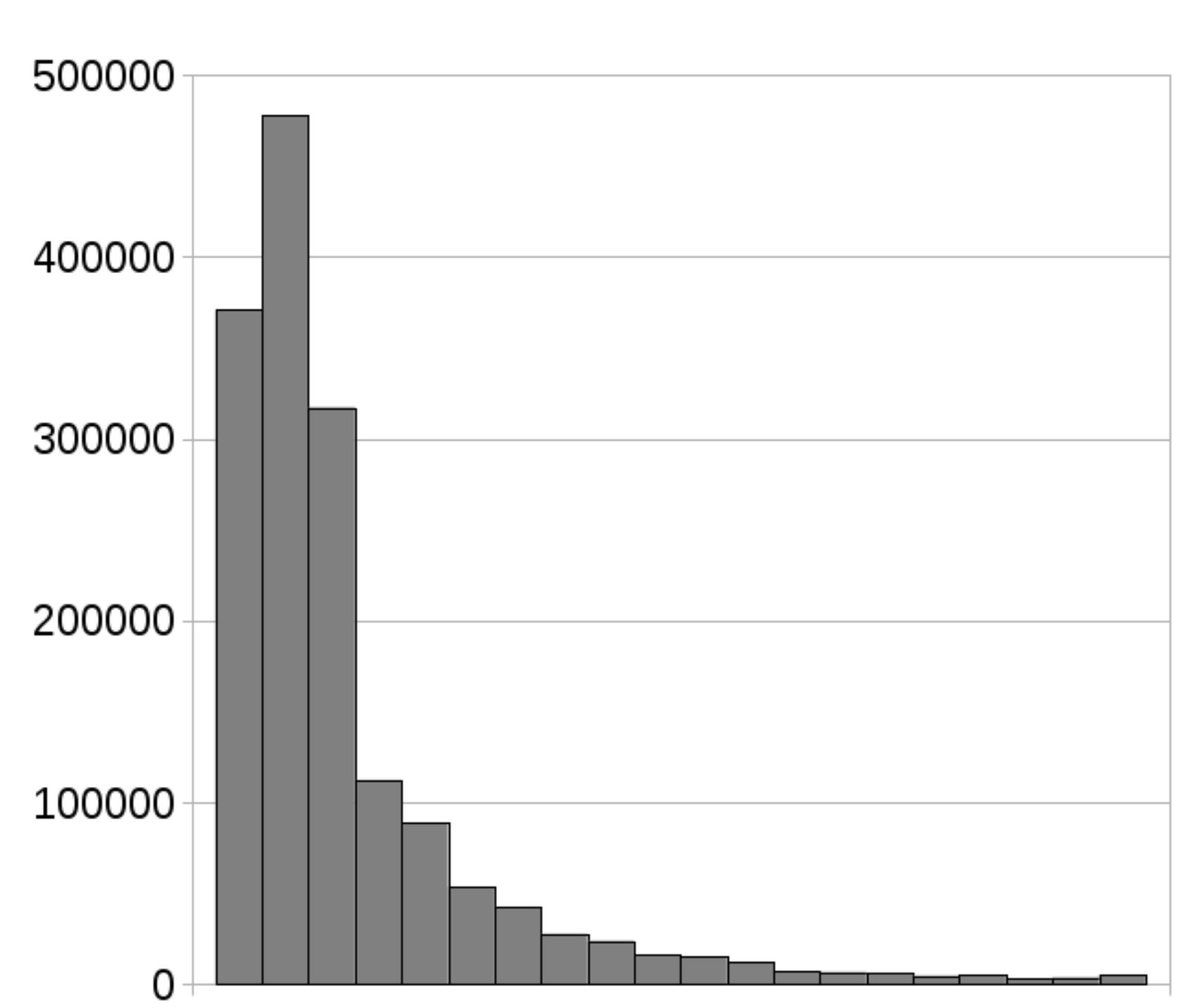} & \includegraphics[width=101px]{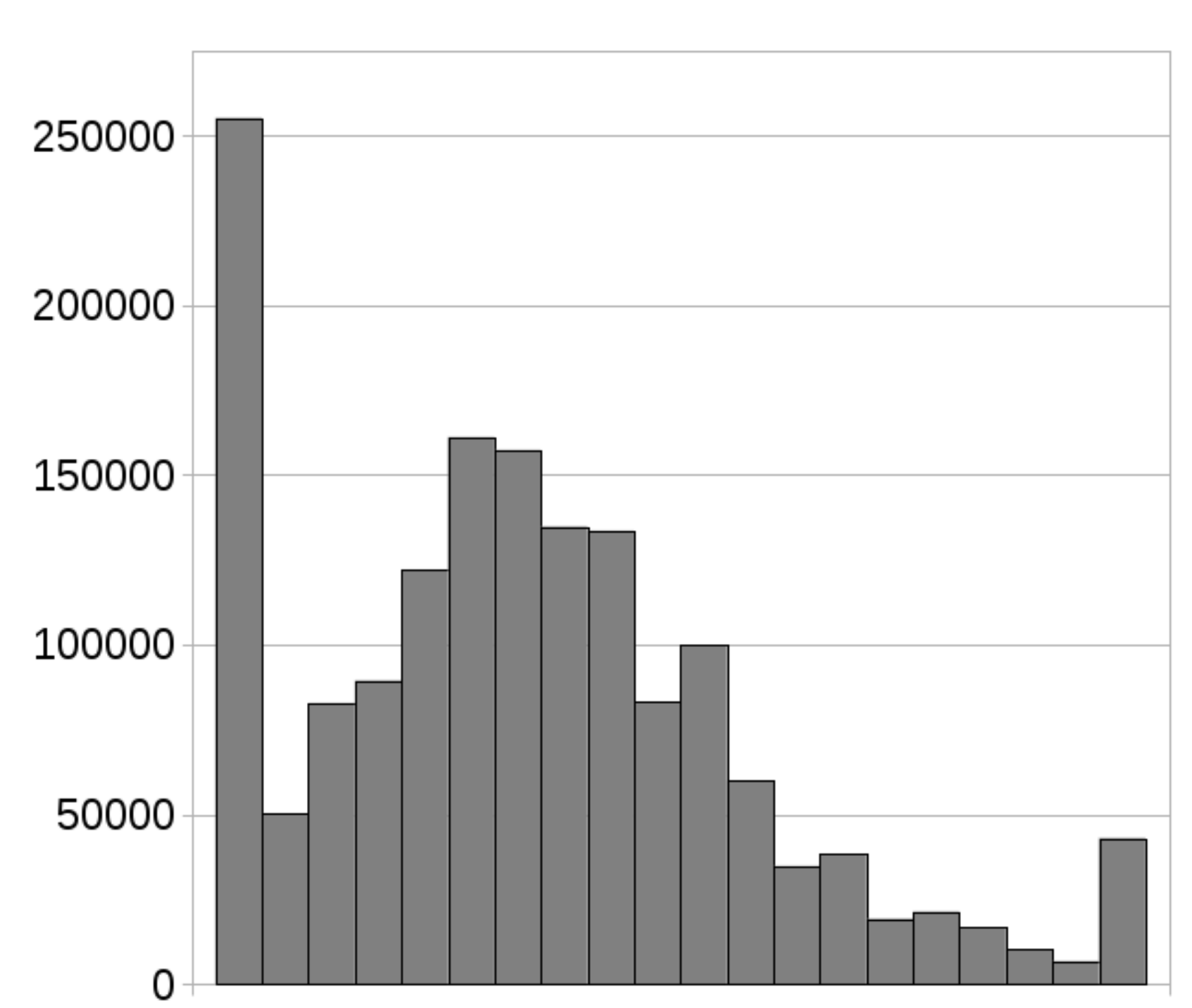} & \includegraphics[width=101px]{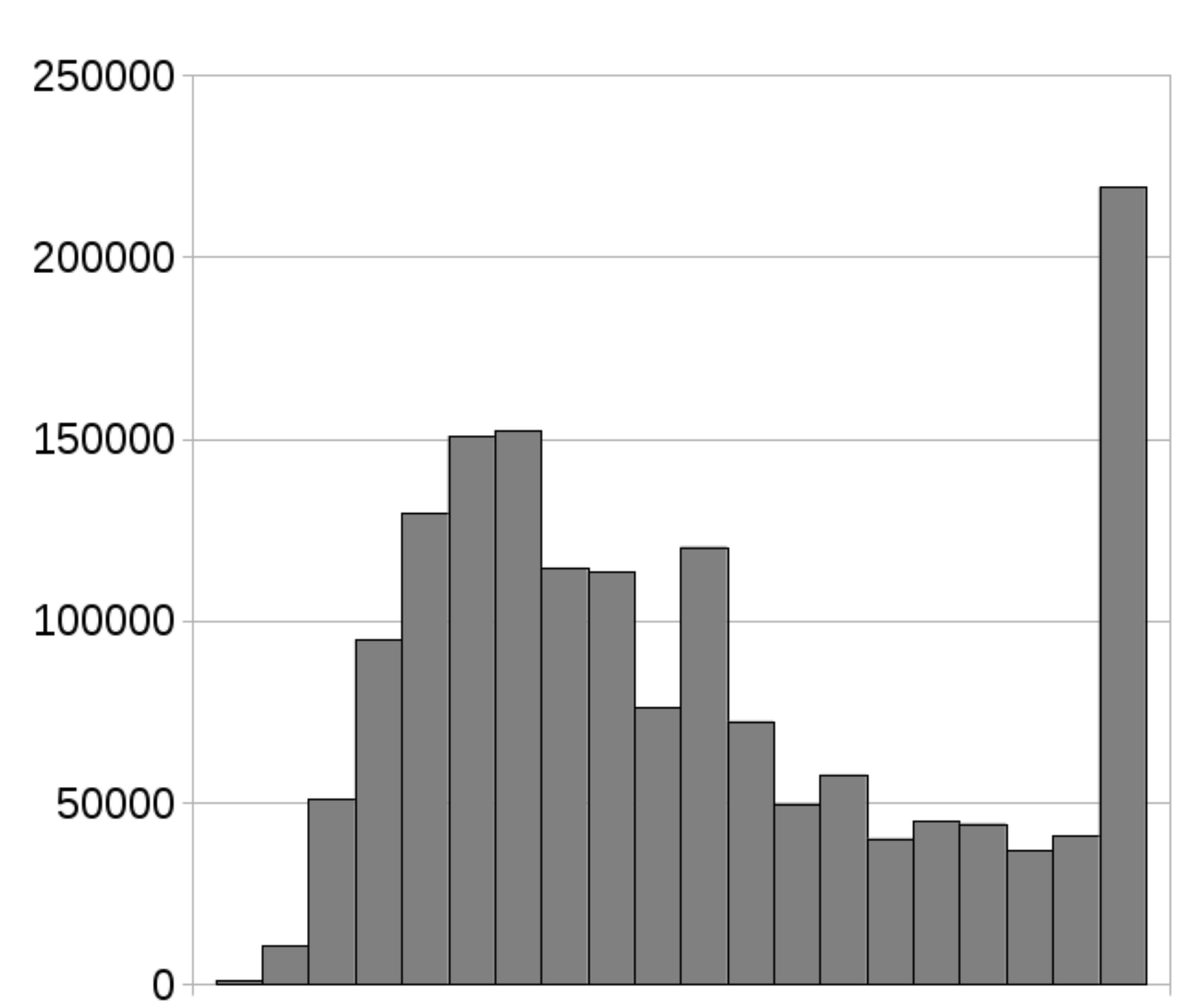} \\
    \begin{sideways}\scriptsize \textbf{LibraryThing}\end{sideways} & \includegraphics[width=101px]{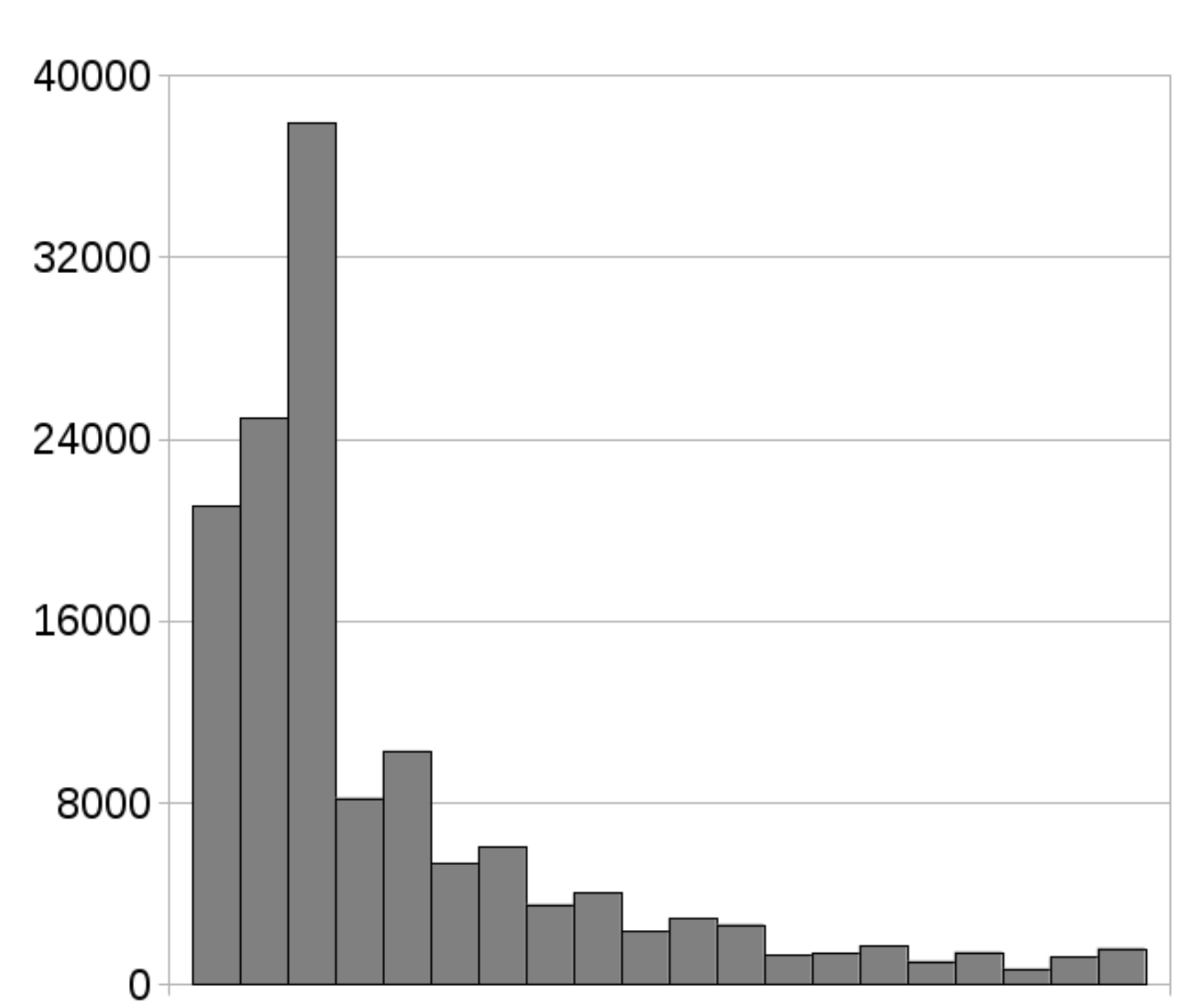} & \includegraphics[width=101px]{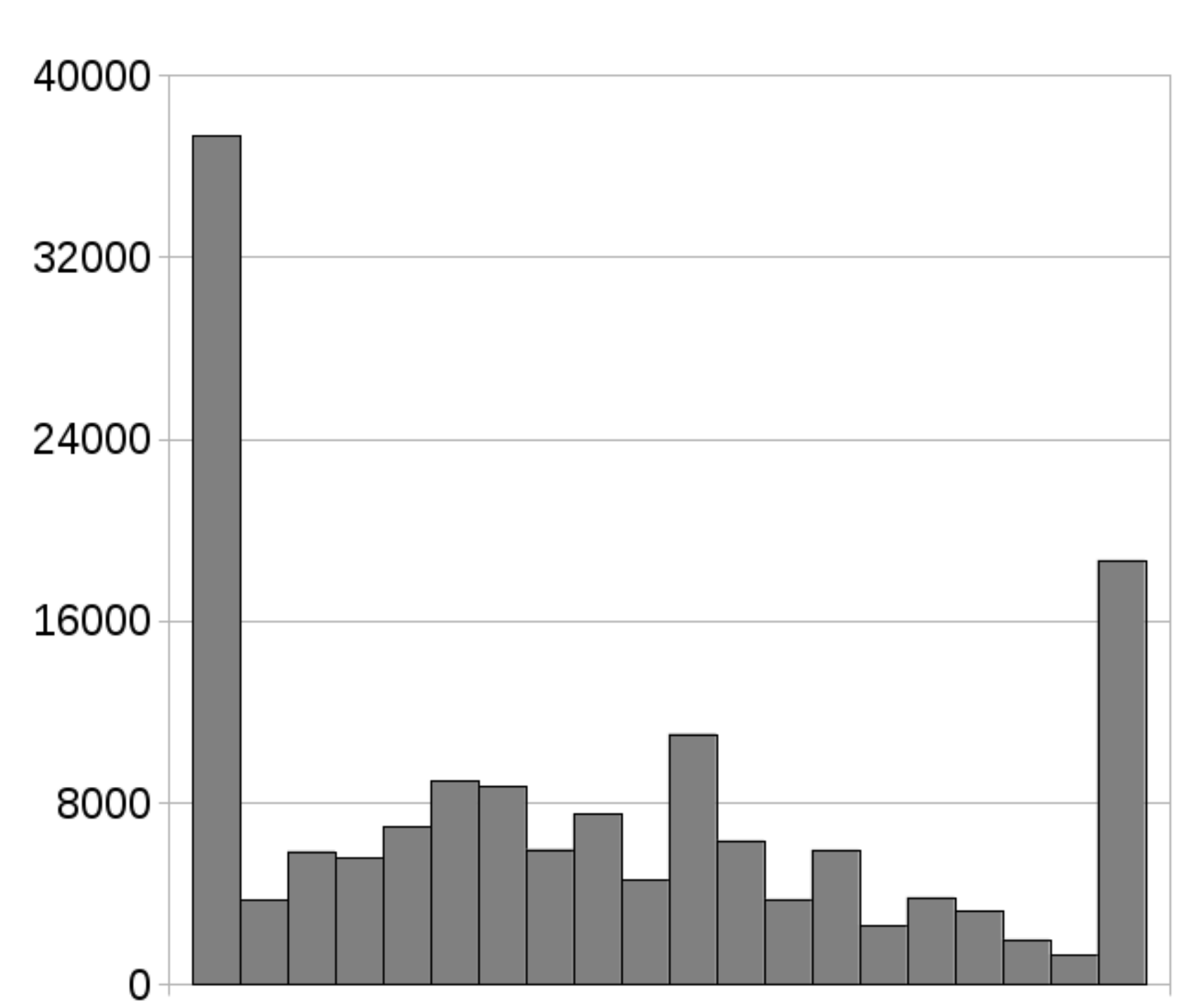} & \includegraphics[width=101px]{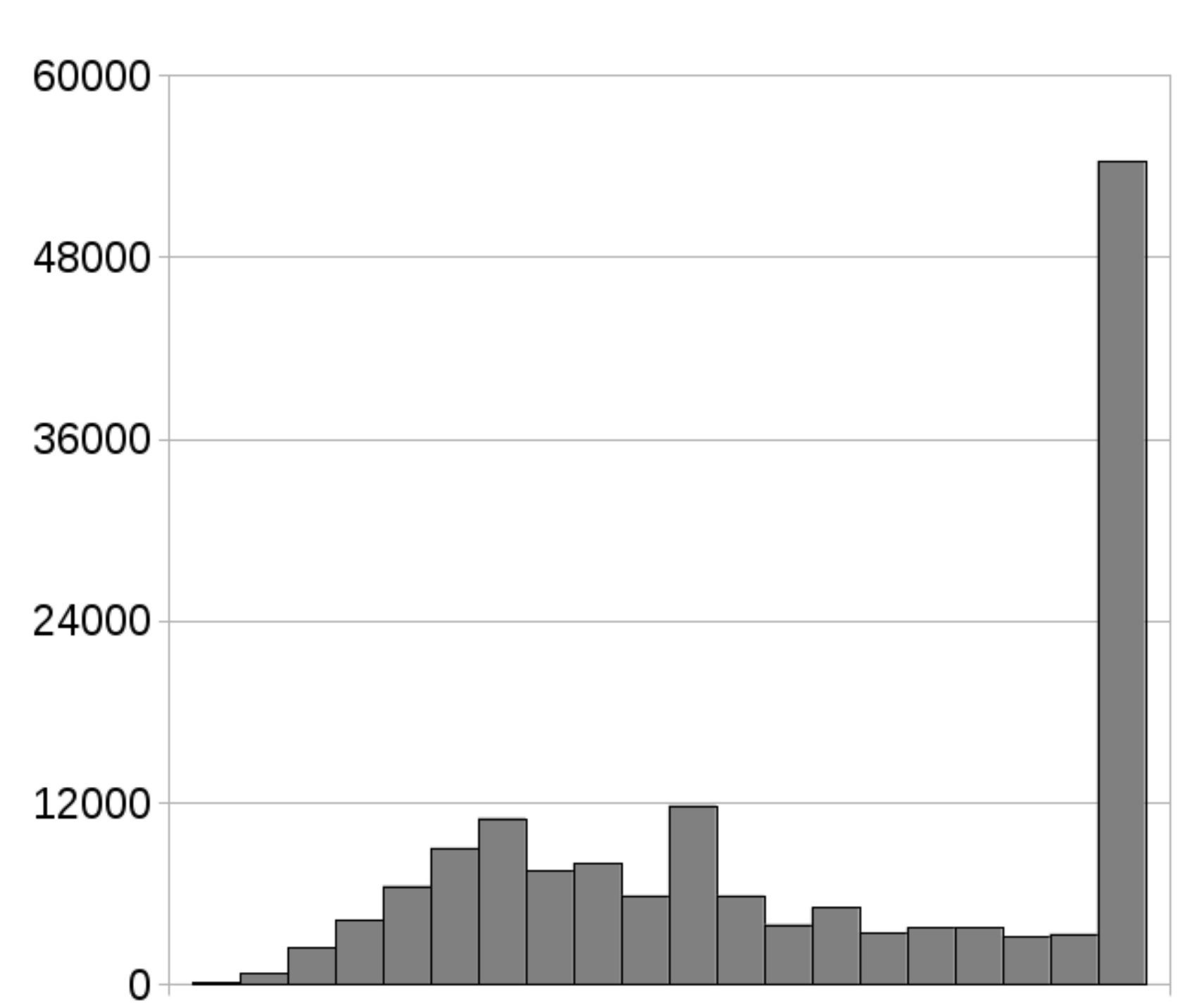} \\
    \begin{sideways}\scriptsize \textbf{GoodReads}\end{sideways} & \includegraphics[width=101px]{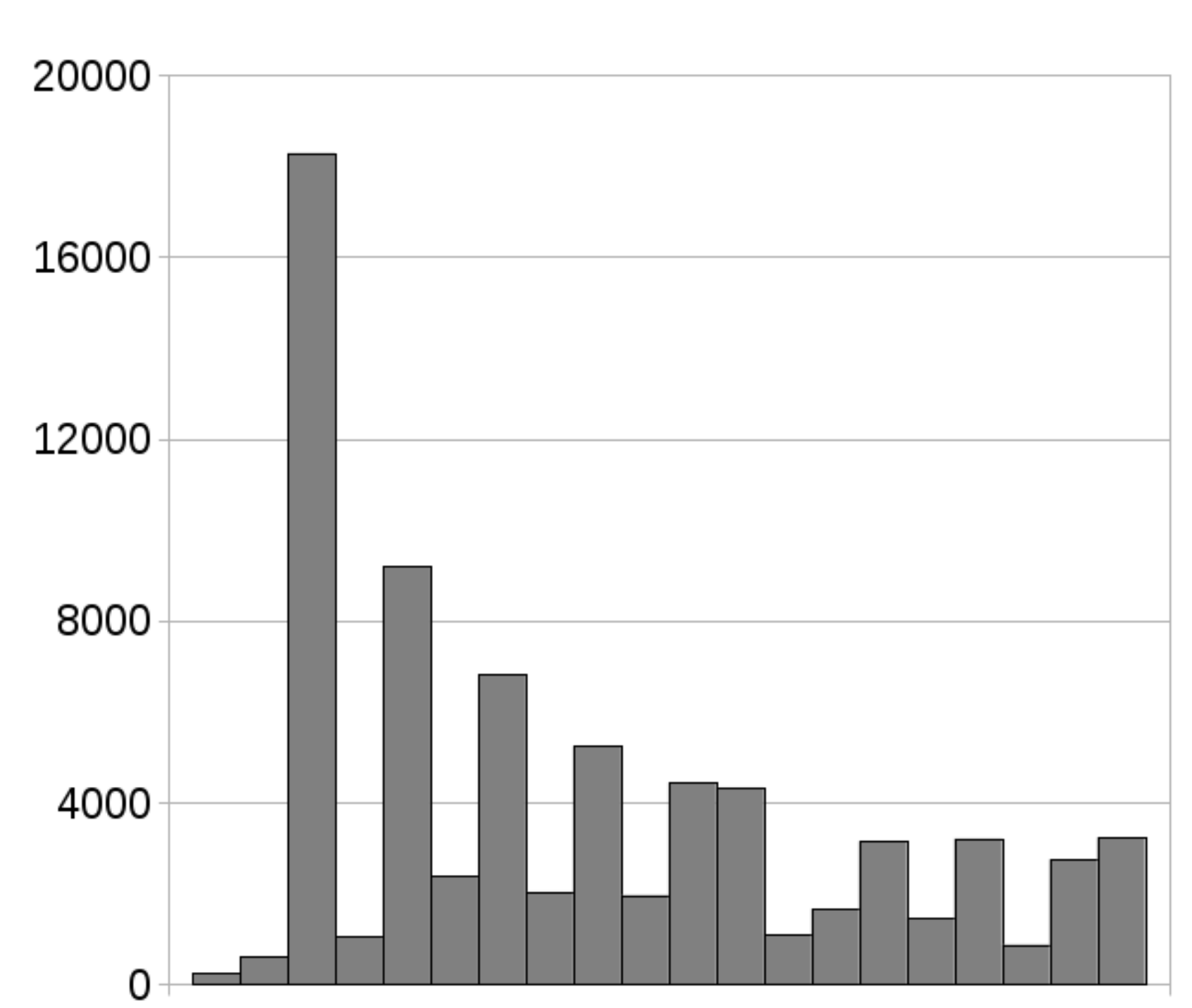} & \includegraphics[width=101px]{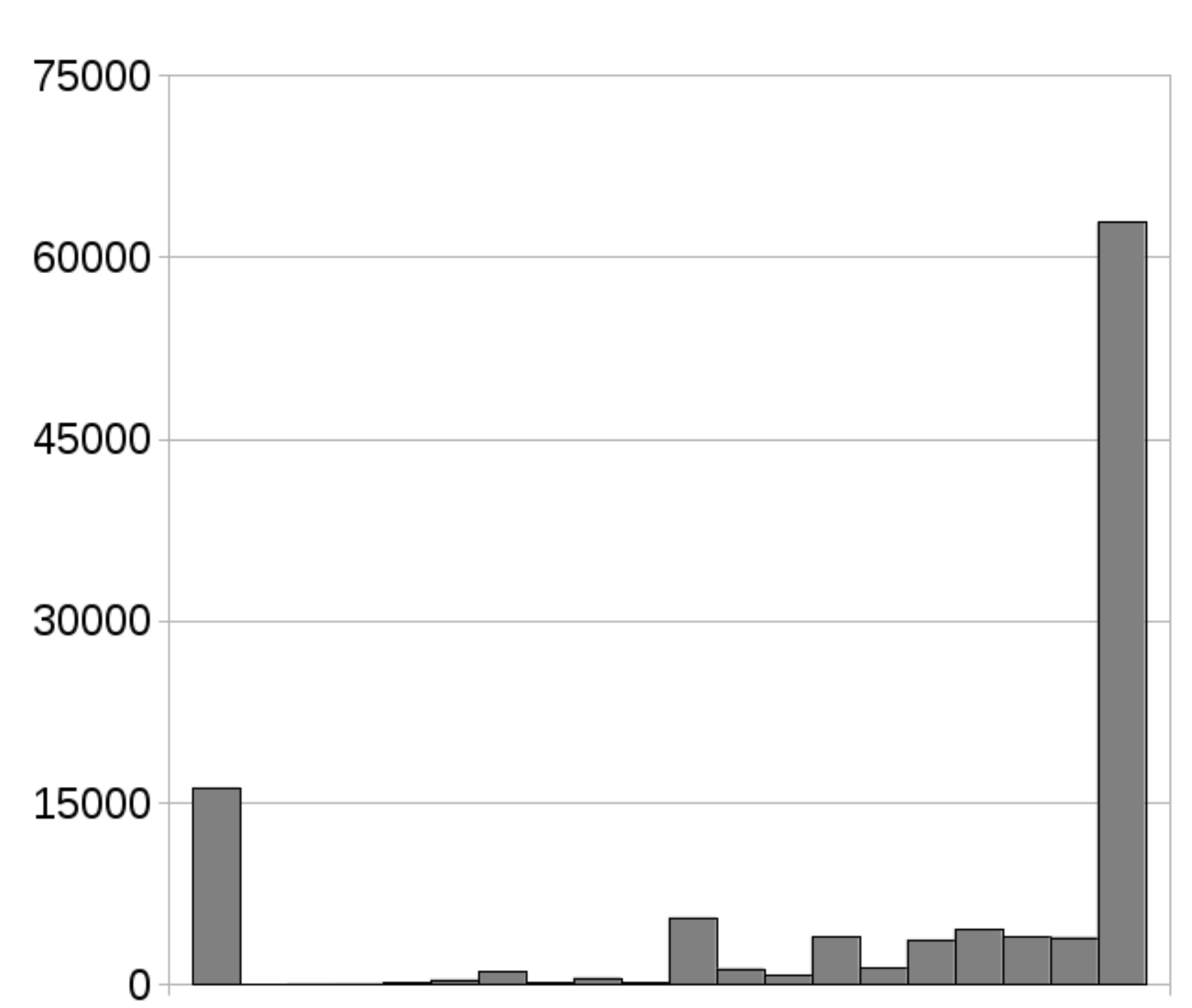} & \includegraphics[width=101px]{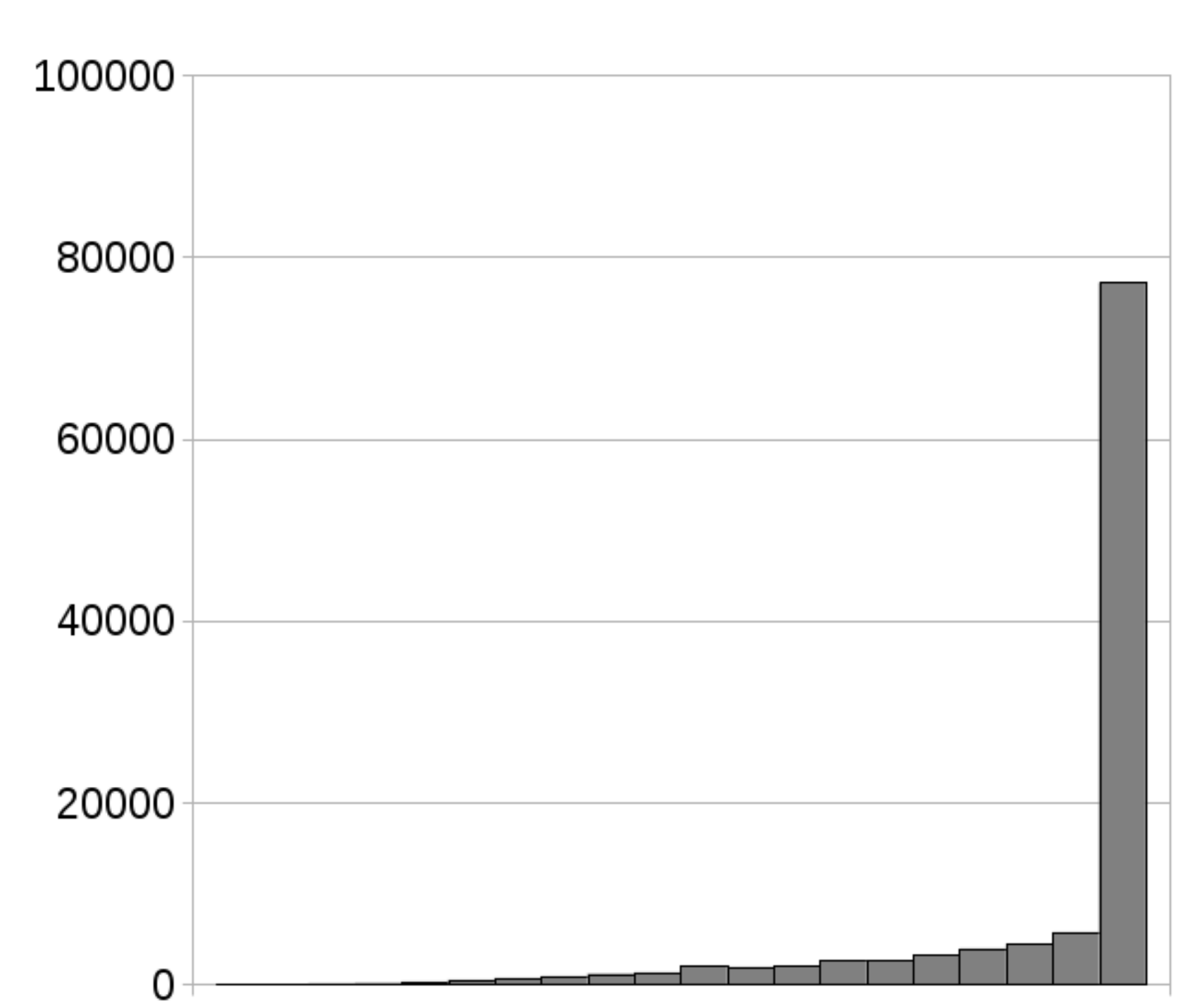} \\
  \end{tabular}
 \end{center}
 \caption[Measure distribution histograms]{Distribution histograms of the three measures (TRR, ORPHAN and TPP) for the three datasets. X axis represents the quantiles of values, whereas Y axis represents the number of users in each quantile.}
 \label{tab:cat-desc-histogram}
\end{table}

\begin{figure}[ht]
\centering
\includegraphics[width=250px]{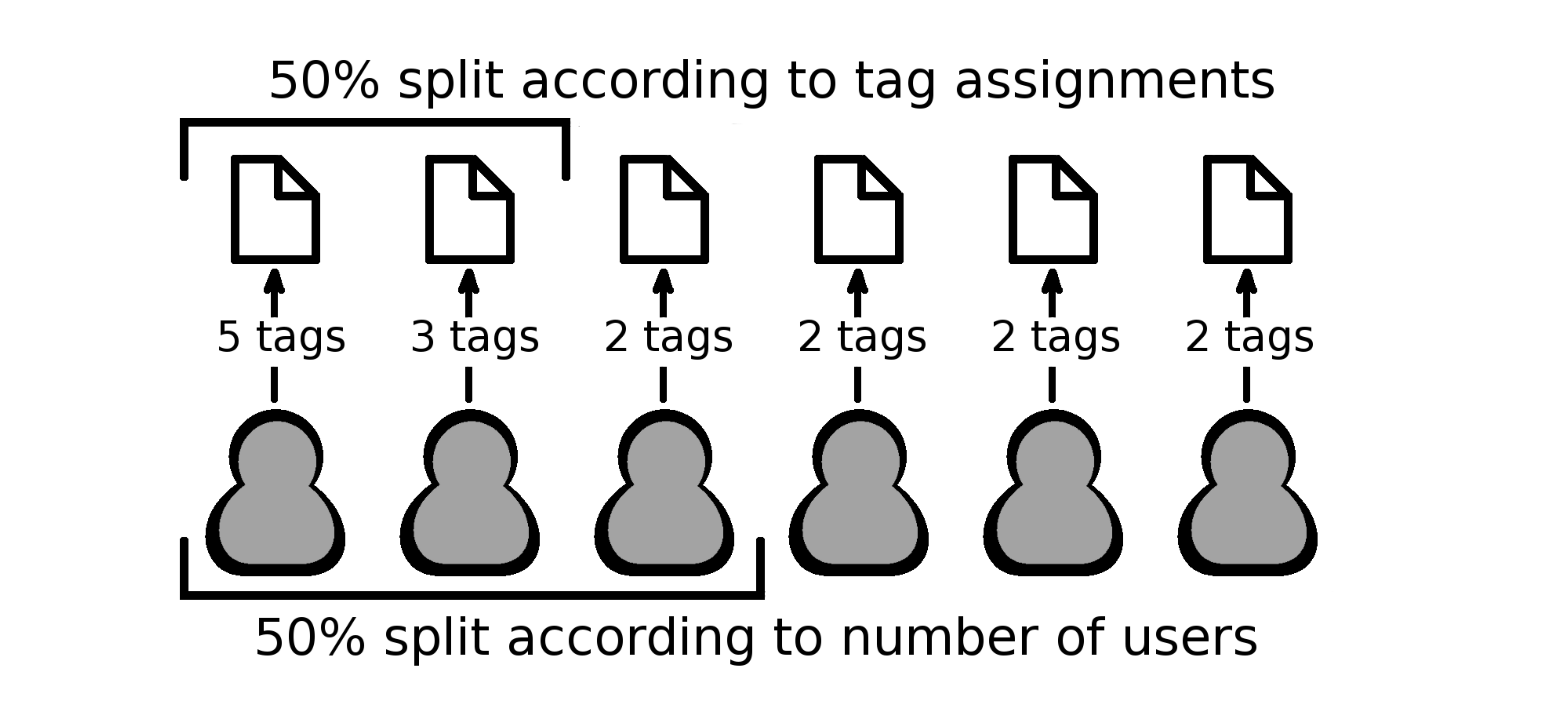}
\caption[Example of splitting based on tag assignments or number of users]{Example of a 50\% segment, selected based on tag assignments or number of users. Splitting by number of users would be unfair, since it may yield bigger amounts of data.}
\label{fig:split}
\end{figure}

To choose the sets of users with which we perform the experiments, we split the ranked lists by getting some of the top and
bottom users. Choosing fixed percents of users would be unfair, though. Some users are likely to be more verbose, by
definition of some measures, and they usually provide much more tag assignments than others. Thus, we split the users according to the percent of tag assignments they provide\footnote{We define each of the tags annotated in a bookmark as a tag assignment. Thus, a bookmark has as many tag assignments as tags has the user annotated on it.}.
This enables a fairer split of the users, with the same amount of data, e.g., a 10\% split ensures that both sets include 10\% of all tag assignments, but the number of users differs among them. Figure \vref{fig:split} shows an example of how splitting by number of tag assignments can differ from splitting by number of users. We split the user sets into smaller subsets of users ranging from 10\% to 100\%, with a step size of 10\%.

\subsection{Tag-based classification}
\label{ssec:tag-based-classification-categorizers}

For the tag-based classification, we represent the resources by aggregating annotations provided by the users within the considered subset of Categorizers or Describers. This creates reduced tagging data for each resource. With these reduced representations, we feed the multiclass SVM classifier defined in Chapter \vref{c:svm-classification}, and calculate their performance by measuring the accuracy of their predictions. This enables comparing same percents of tag assignments by Categorizers and Describers, in order to analyze whether the former outperform the latter.

\subsection{Descriptiveness of Tags}
\label{ssec:descriptiveness-of-tags}

To compute the extent to which a subset of users is providing descriptive tags, we compare their tags to the descriptive data of resource. These descriptive data include:

\begin{itemize}
 \item The textual content of the web pages, as well as user reviews for the Delicious dataset.
 \item Synopses, user reviews and editorial reviews for the book datasets, i.e., LibraryThing and GoodReads.
\end{itemize}

In the first step, we merge all these data into a single text for each resource. Accordingly, we get a single text comprising all descriptive data for each resource. After this, we compute the frequencies of each term (TF) in the texts, so that we can create a vector for each resource, where each of the dimensions in the vectors belongs to a term. On the other hand, for each selection of users, we create the vectors of tags for each resource, with the annotations of those users. This way, we have the reference descriptive vectors as well as the tag vectors we want to compare to them.

There are several measures that could compute the similarity between a tag vector ($T$) and a reference vector ($R$) for a given resource $r$. They tend to be correlated, though. Regardless of the values given by the measures, we are interested in getting comparable values towards a way to determine whether a tag set resembles to a greater or lesser extent than another set. Thus, as a well-known and robust measure for this, we compute the cosine similarity between the vectors (see Equation \vref{eq:sim1}).

\begin{eqnarray}
  \text{similarity}_r = \cos(\theta_r) = \frac{T_r \cdot R_r}{\|T_r\| \|R_r\|} = \nonumber \\
  \sum_{i=1}^{n}\frac{ {T_{ri} \times R_{ri}} }{ \sqrt{\sum_{i=1}^{n}{(T_{ri})^2}} \times \sqrt{\sum_{i=1}^{n}{(R_{ri})^2}} }
  \label{eq:sim1}
\end{eqnarray}

The above formula provides the value of similarity between the tag vector and the reference vector of a single resource. This value is the cosine of the angle between the two vectors, which could range from 0 to 1, since the term frequencies only consist of positive values. A value of 1 would mean that both vectors are exactly the same, whereas a 0 would mean they coincide in none of the terms, and so they are completely different. After getting the similarity value between each pair of vectors, we need to get the overall similarity value between users' tags and descriptive data of resources. Accordingly, the similarity between the set of $n$ reference vectors, and the set of $n$ tag vectors is computed as the average of similarities between pairs of tag and reference vectors (see Equation \vref{eq:sim2}).

\begin{equation}
  \text{similarity} = \frac{1}{n}\sum_{r=1}^n\cos(\theta_r)
  \label{eq:sim2}
\end{equation}

This similarity value shows the extent to which the tags provided by the selected set of users resembles the reference descriptive data, i.e., how descriptive are the tags by those users. The higher is the similarity value, the more descriptive are the tags provided by the users. The closer it is to 0, the more non-descriptive are the tags provided by them. Accordingly, this enables to compare same percents of tag assignments by Categorizers and Describers, and to analyze which of them provide more descriptive tags.

\section{Results}
\label{sec:user-behavior-results}

\begin{table}[htbp]
 \begin{center}
  \begin{tabular}{ c c c c }
    & \scriptsize \textbf{TRR} & \scriptsize \textbf{ORPHAN} & \scriptsize \textbf{TPP} \\
    \begin{sideways}\scriptsize \textbf{Delicious}\end{sideways} & \includegraphics[width=101px]{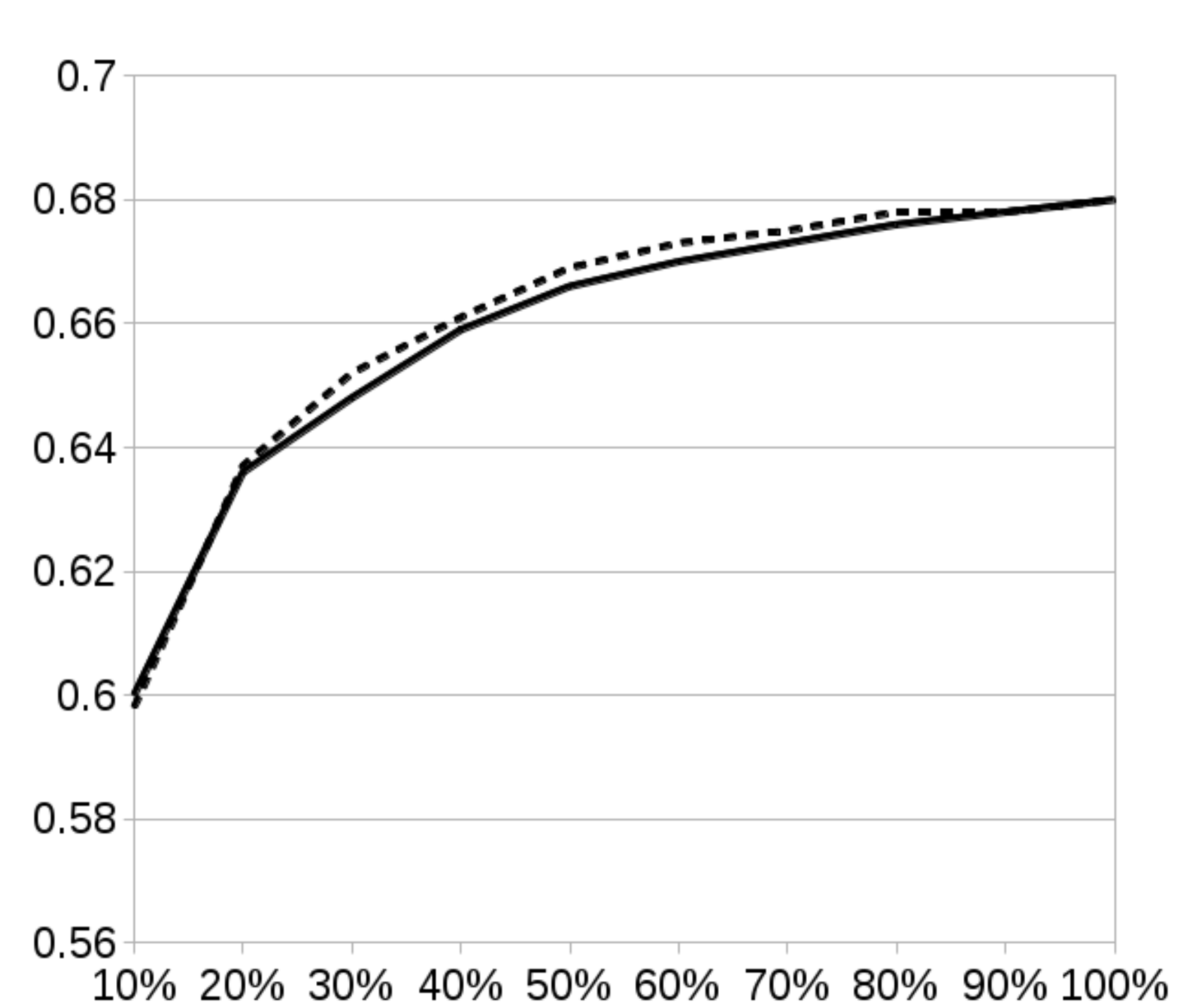} & \includegraphics[width=101px]{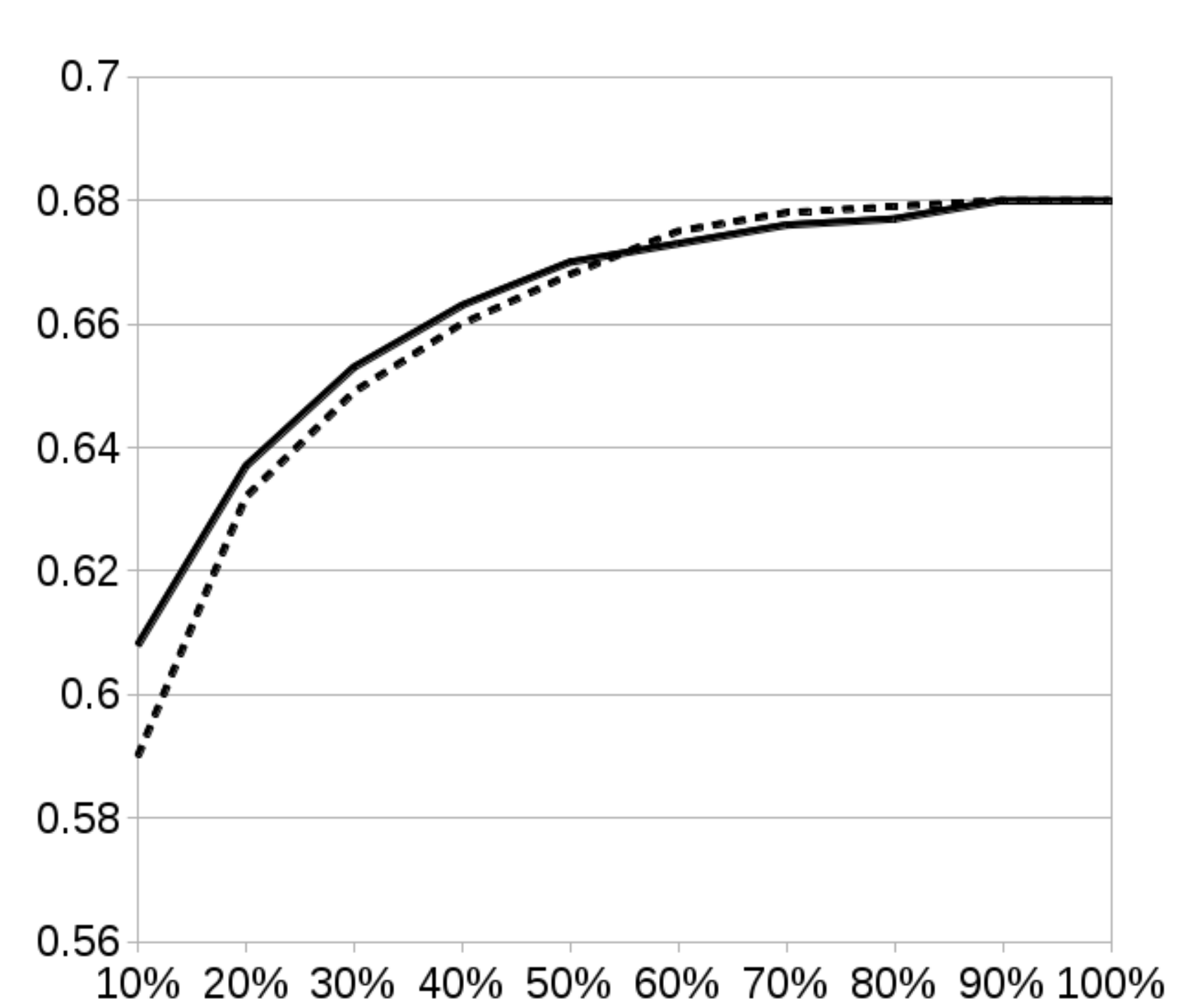} & \includegraphics[width=101px]{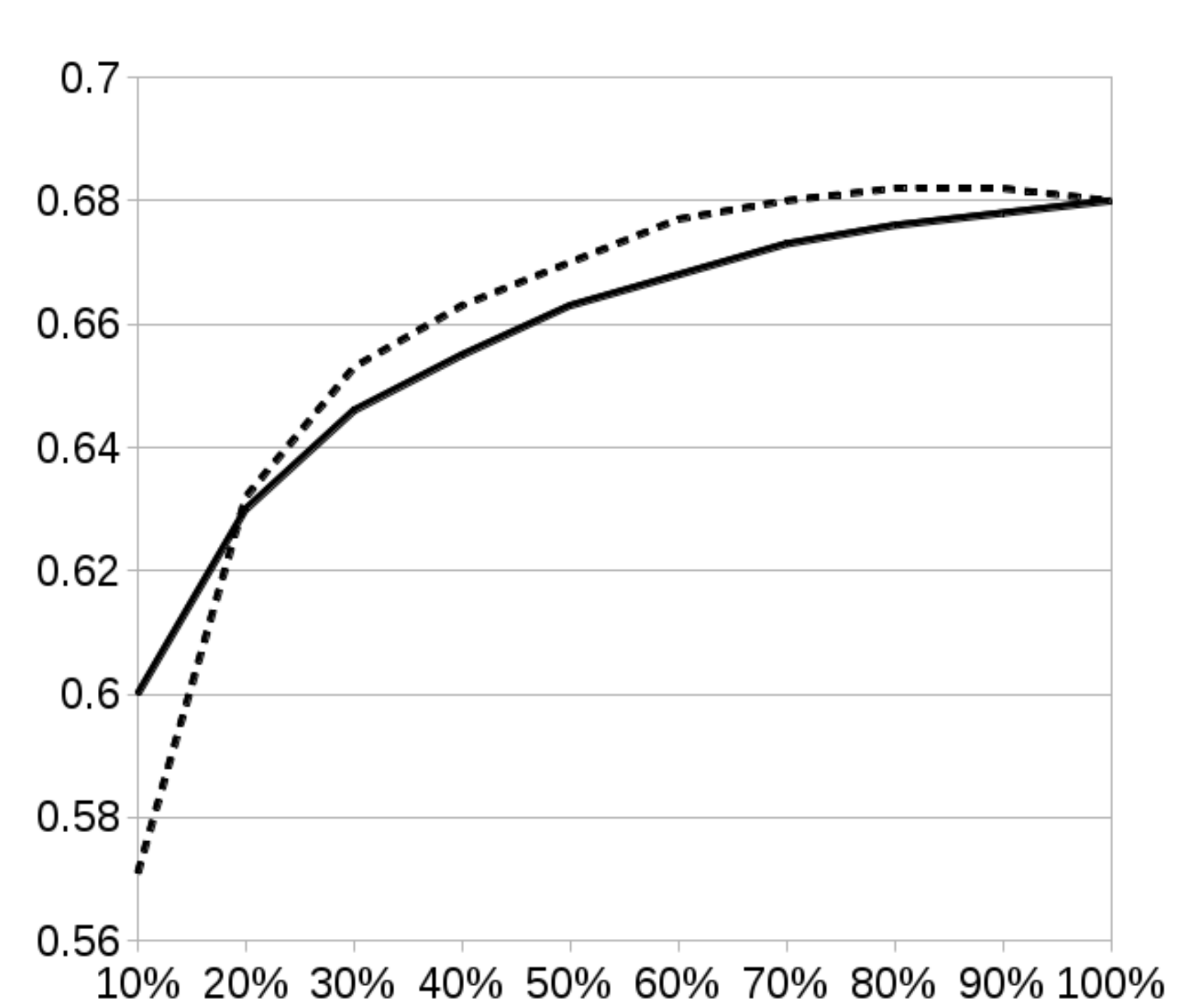} \\
    \begin{sideways}\scriptsize \textbf{LibraryThing (DDC)}\end{sideways} & \includegraphics[width=101px]{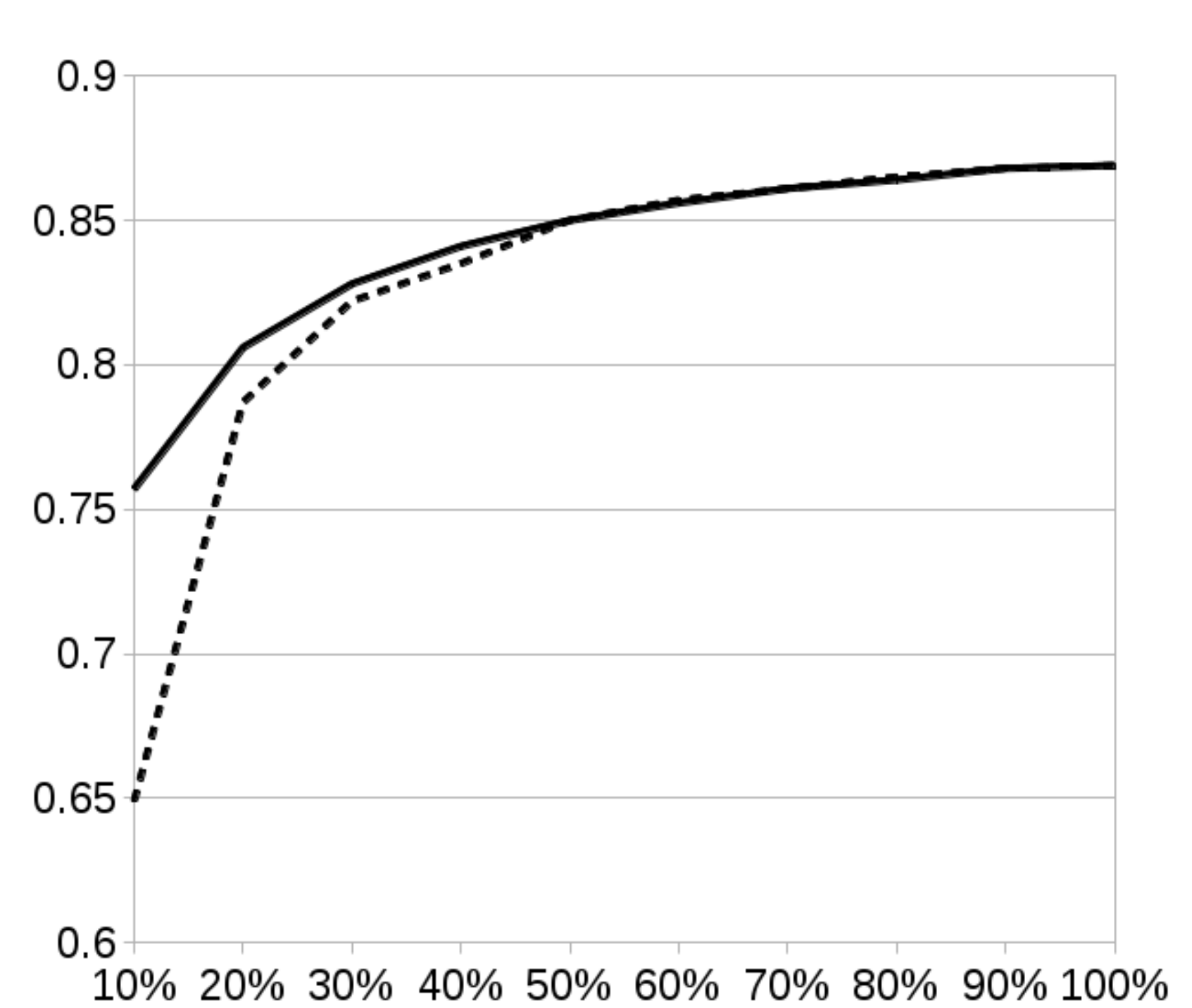} & \includegraphics[width=101px]{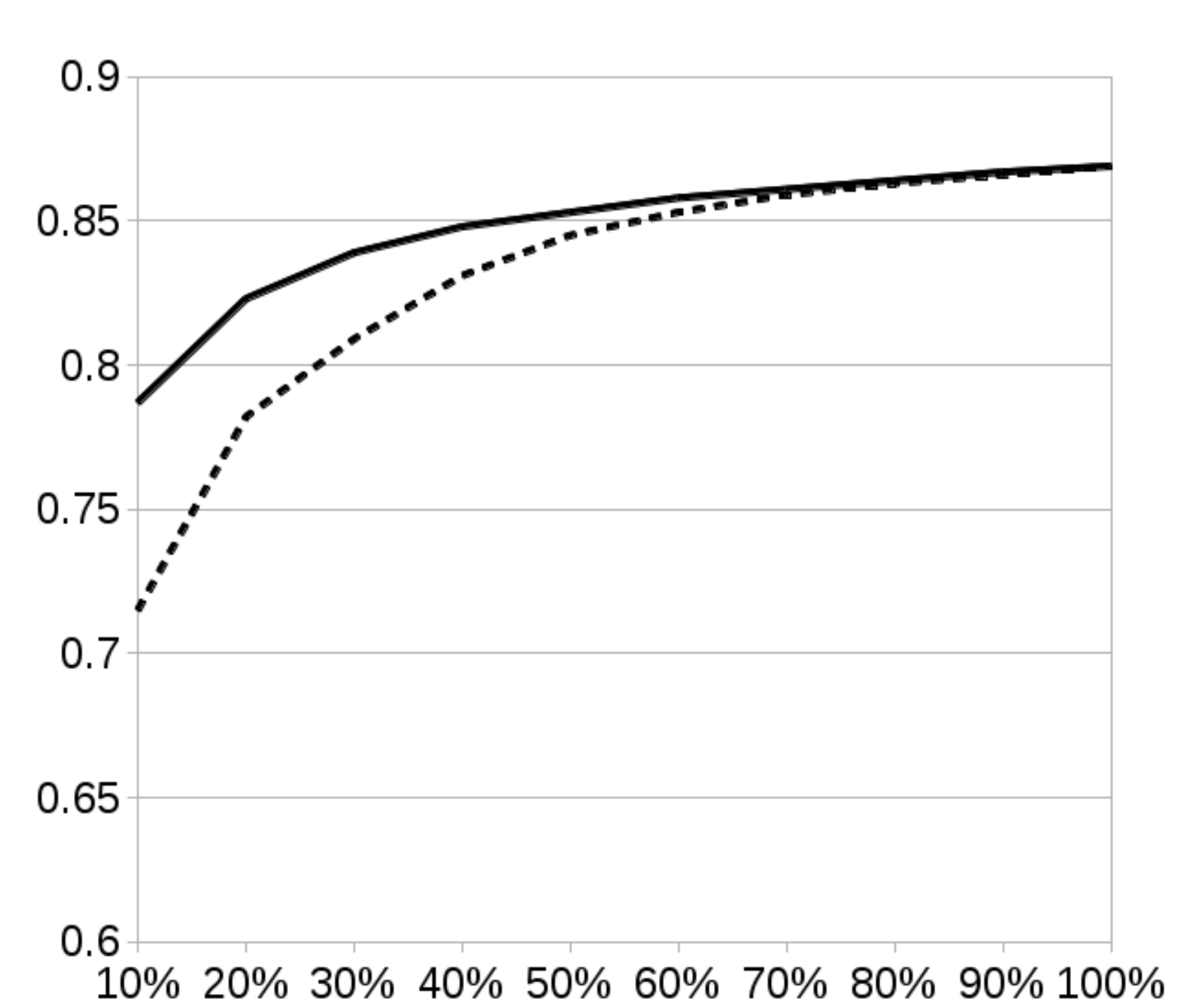} & \includegraphics[width=101px]{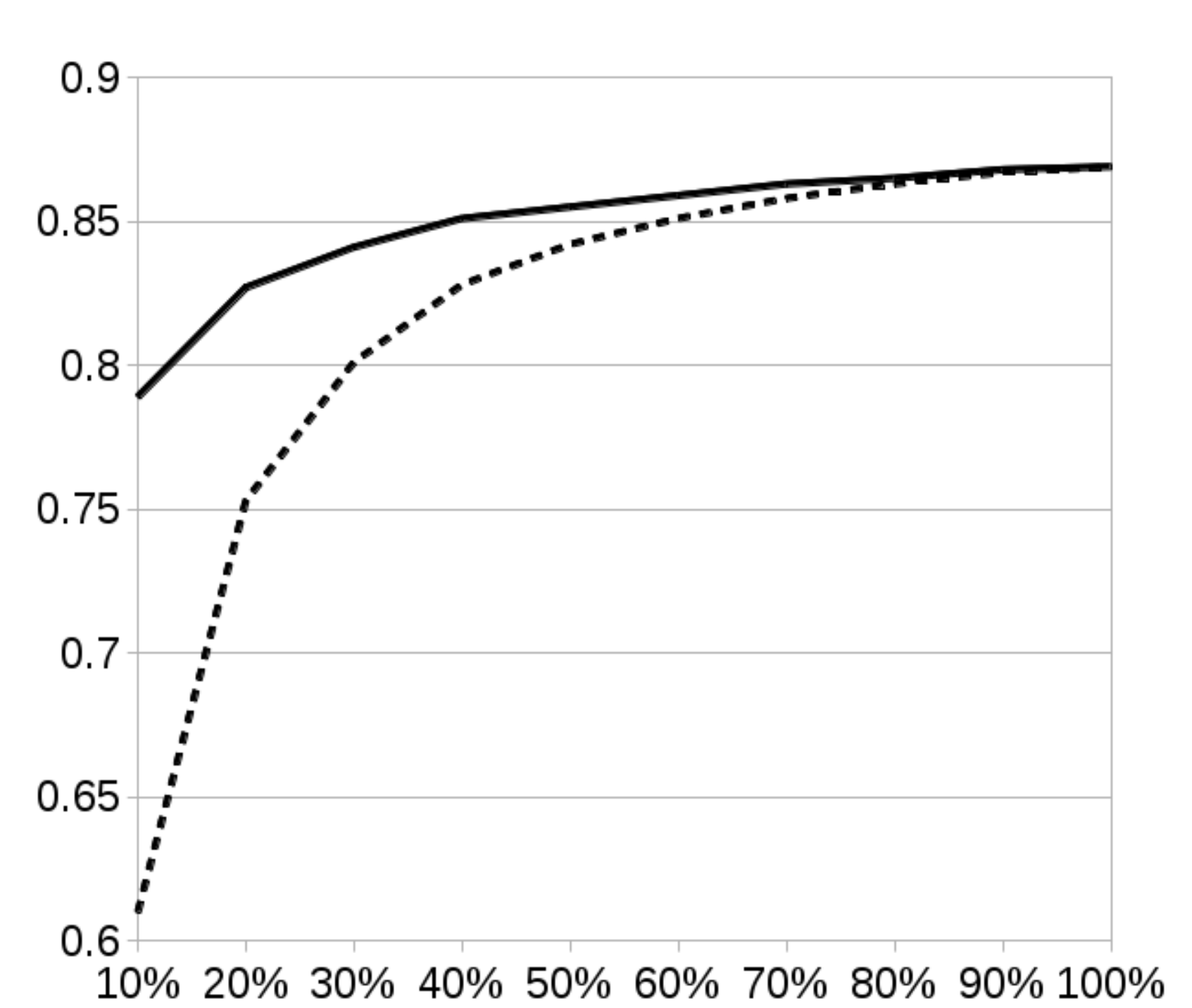} \\
    \begin{sideways}\scriptsize \textbf{LibraryThing (LCC)}\end{sideways} & \includegraphics[width=101px]{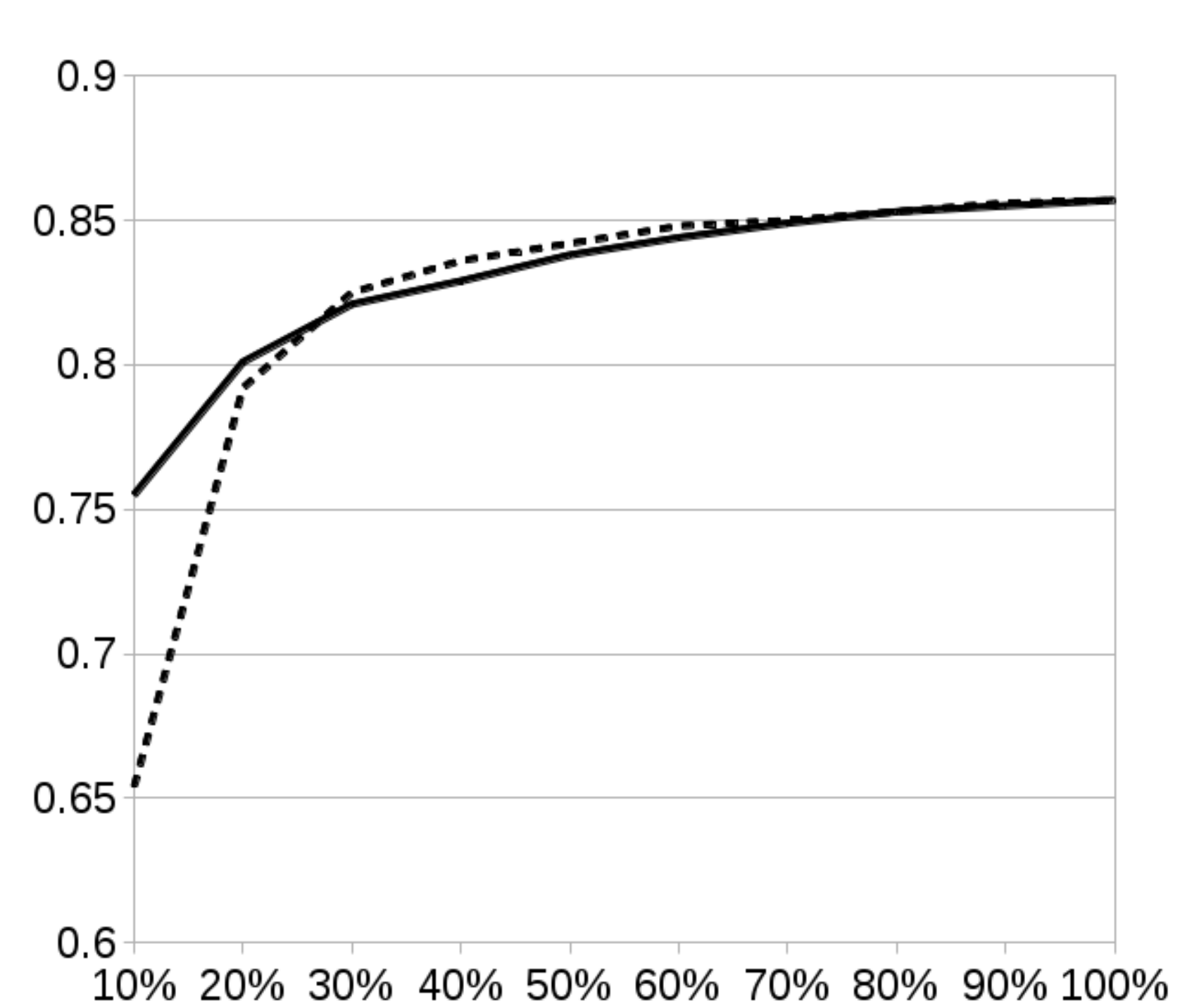} & \includegraphics[width=101px]{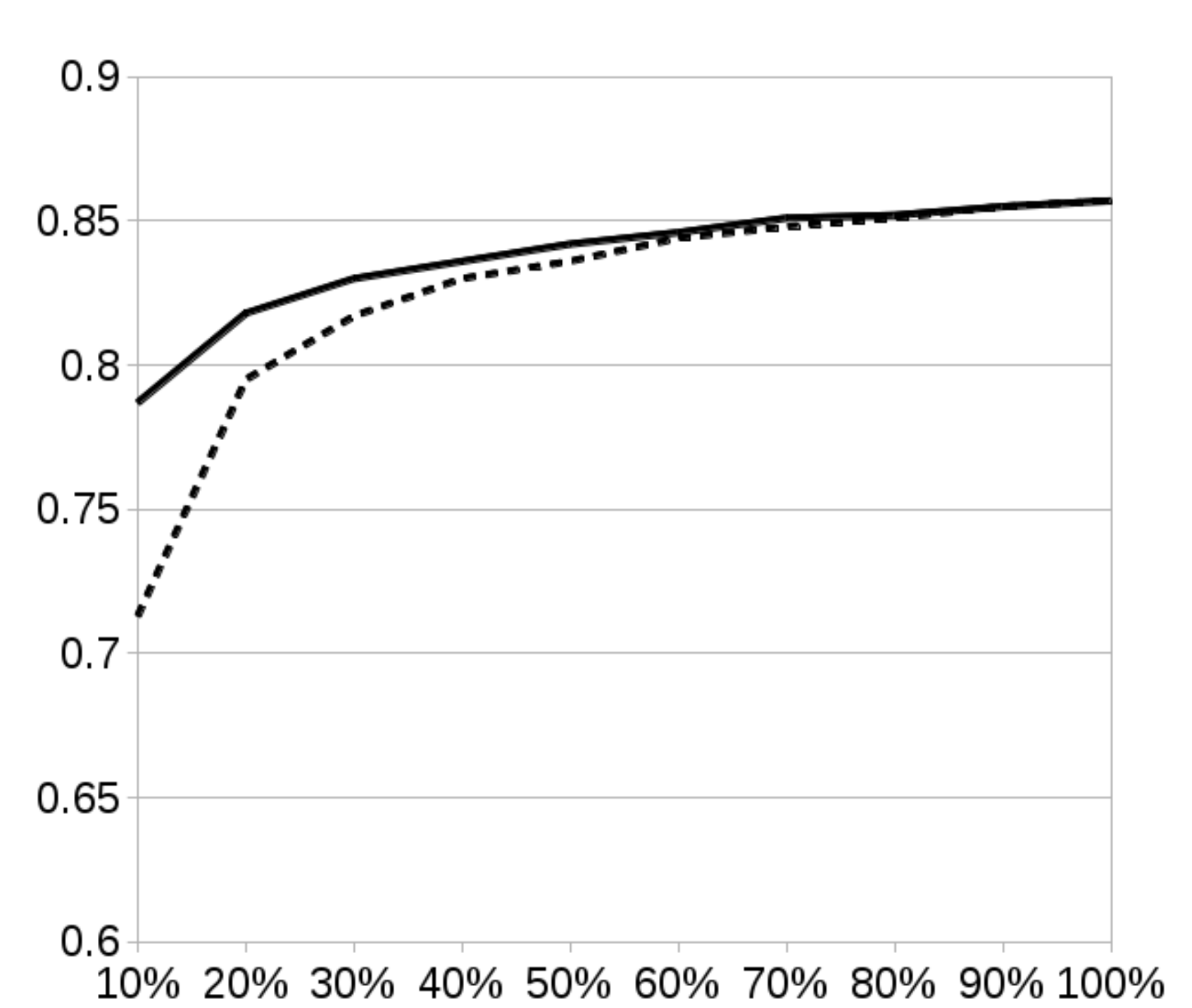} & \includegraphics[width=101px]{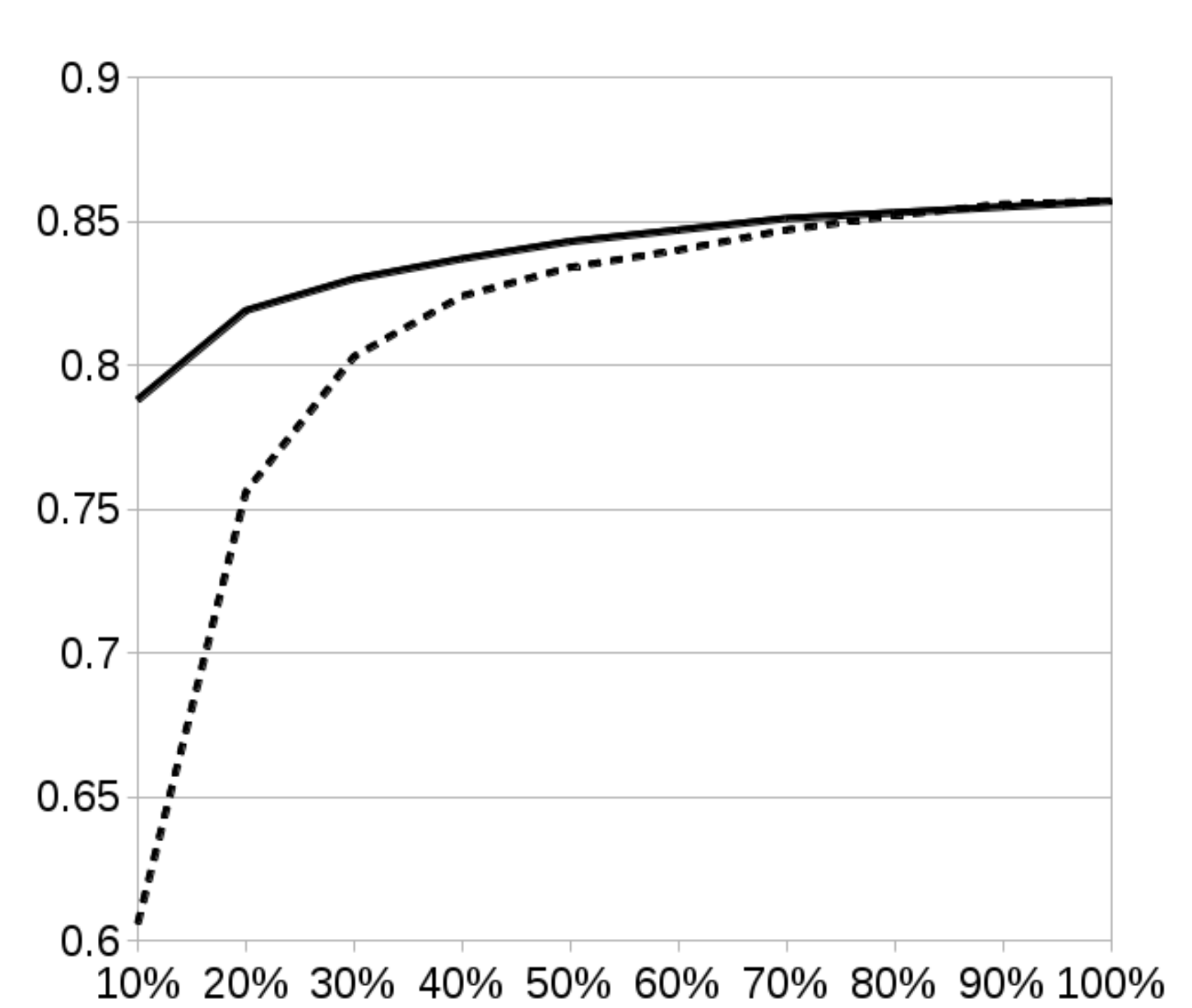} \\
    \begin{sideways}\scriptsize \textbf{GoodReads (DDC)}\end{sideways} & \includegraphics[width=101px]{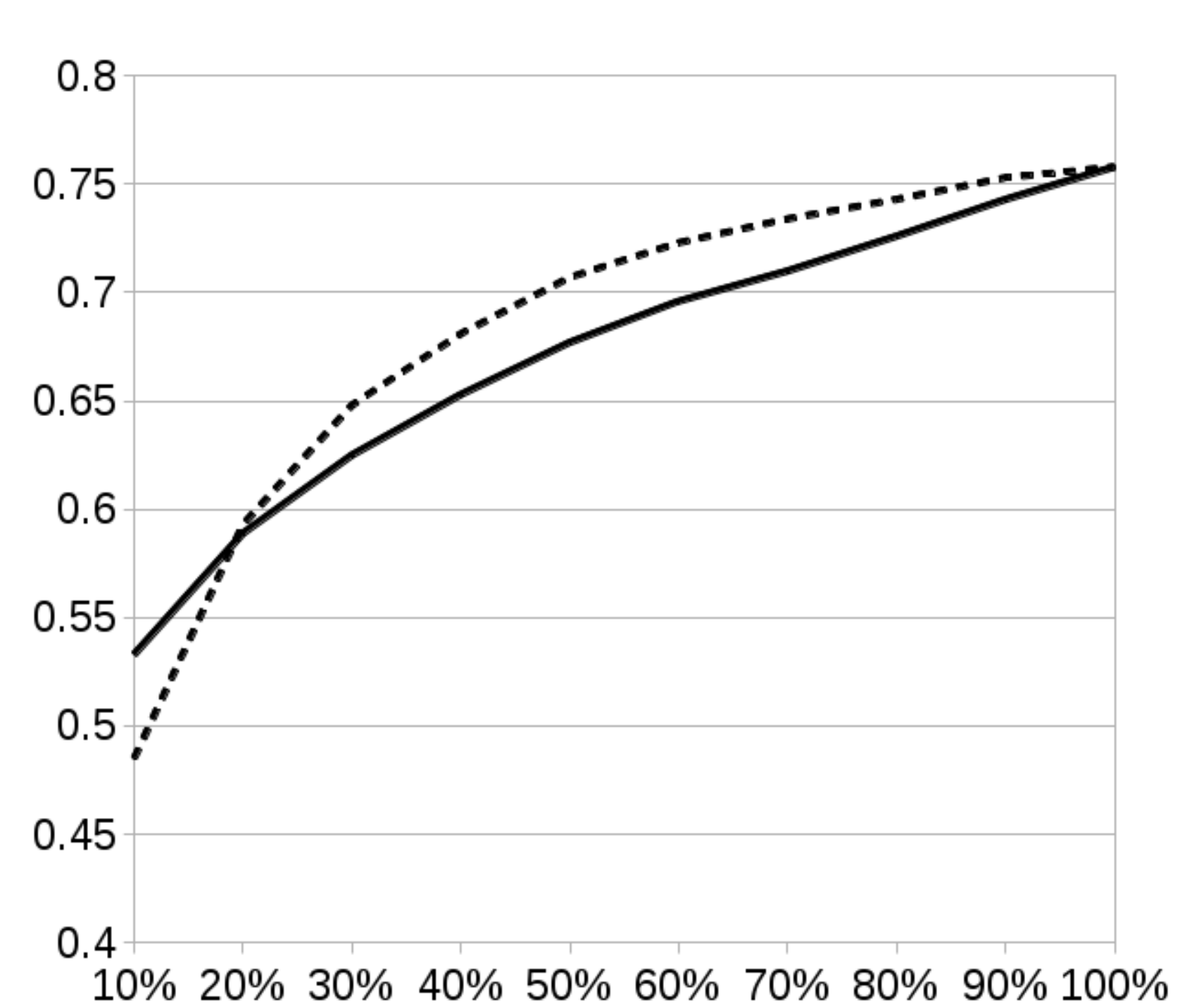} & \includegraphics[width=101px]{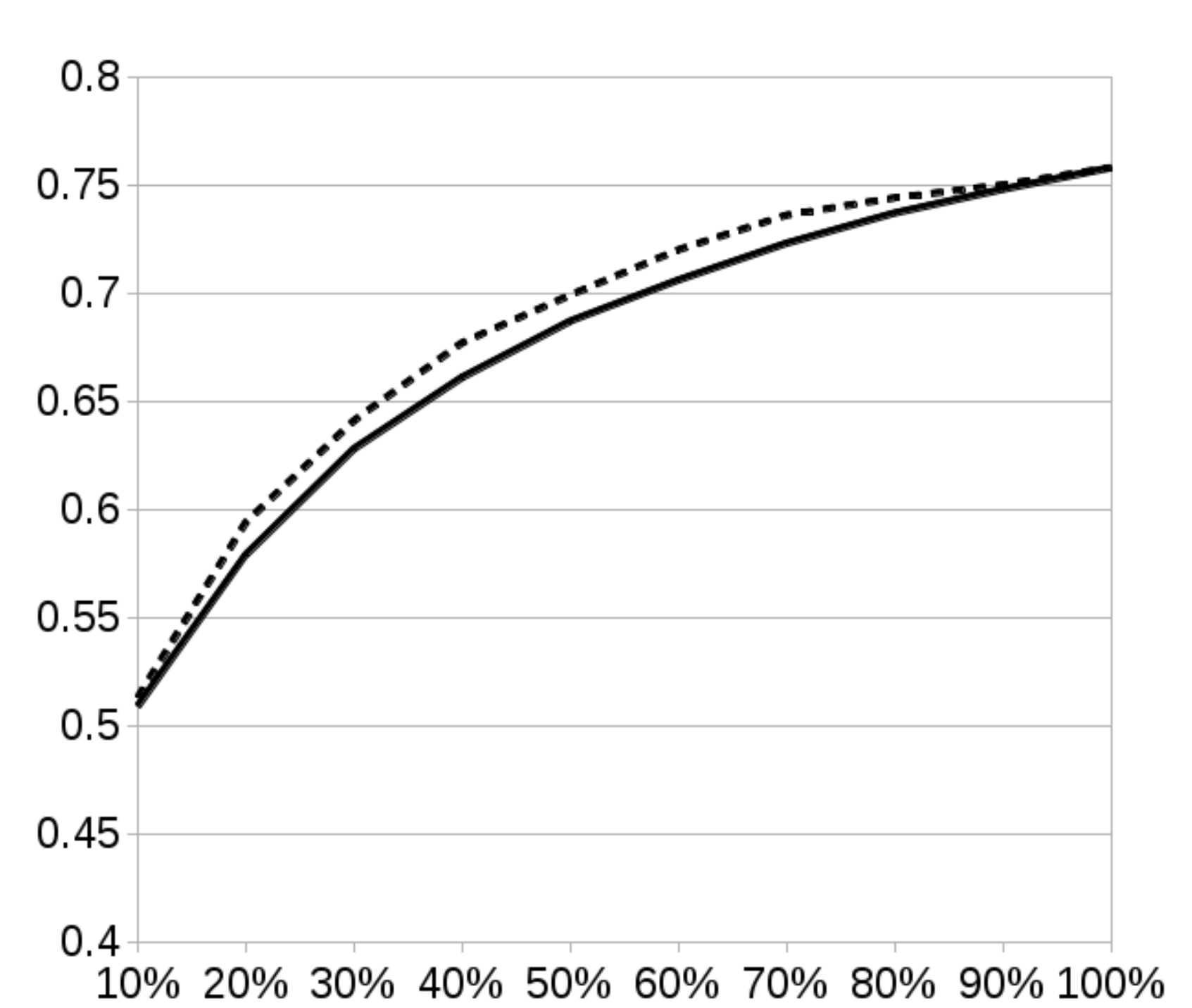} & \includegraphics[width=101px]{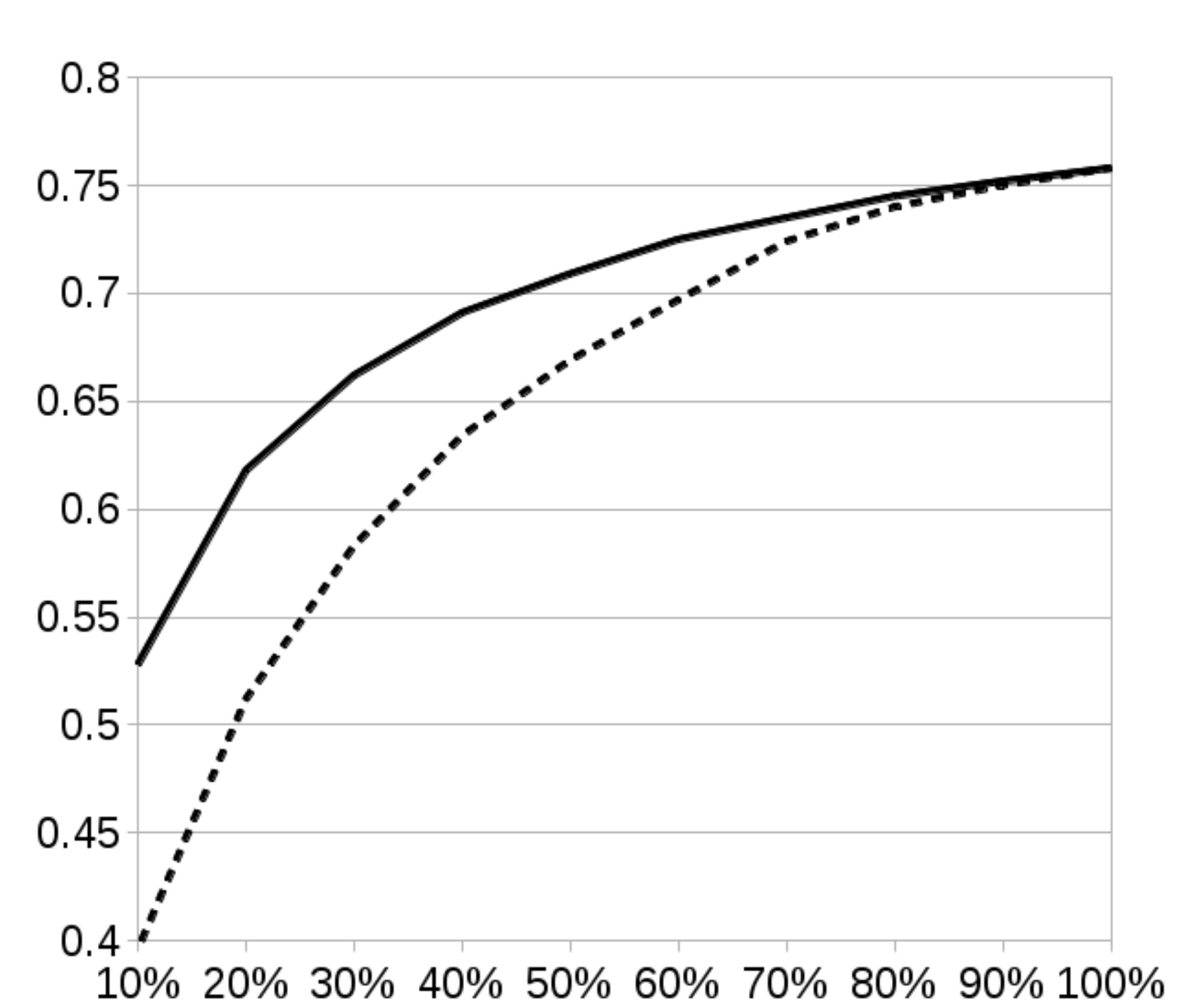} \\
    \begin{sideways}\scriptsize \textbf{GoodReads (LCC)}\end{sideways} & \includegraphics[width=101px]{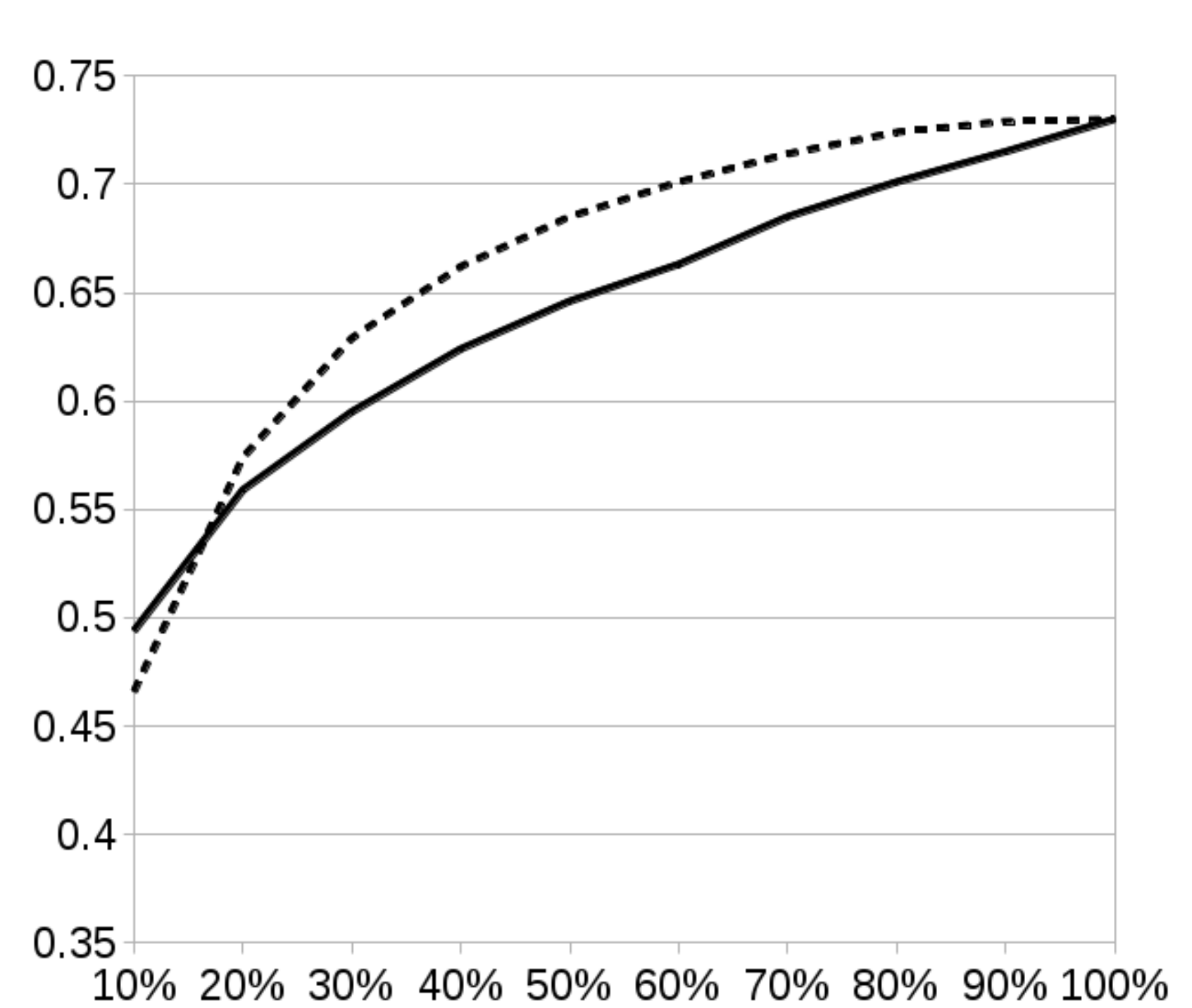} & \includegraphics[width=101px]{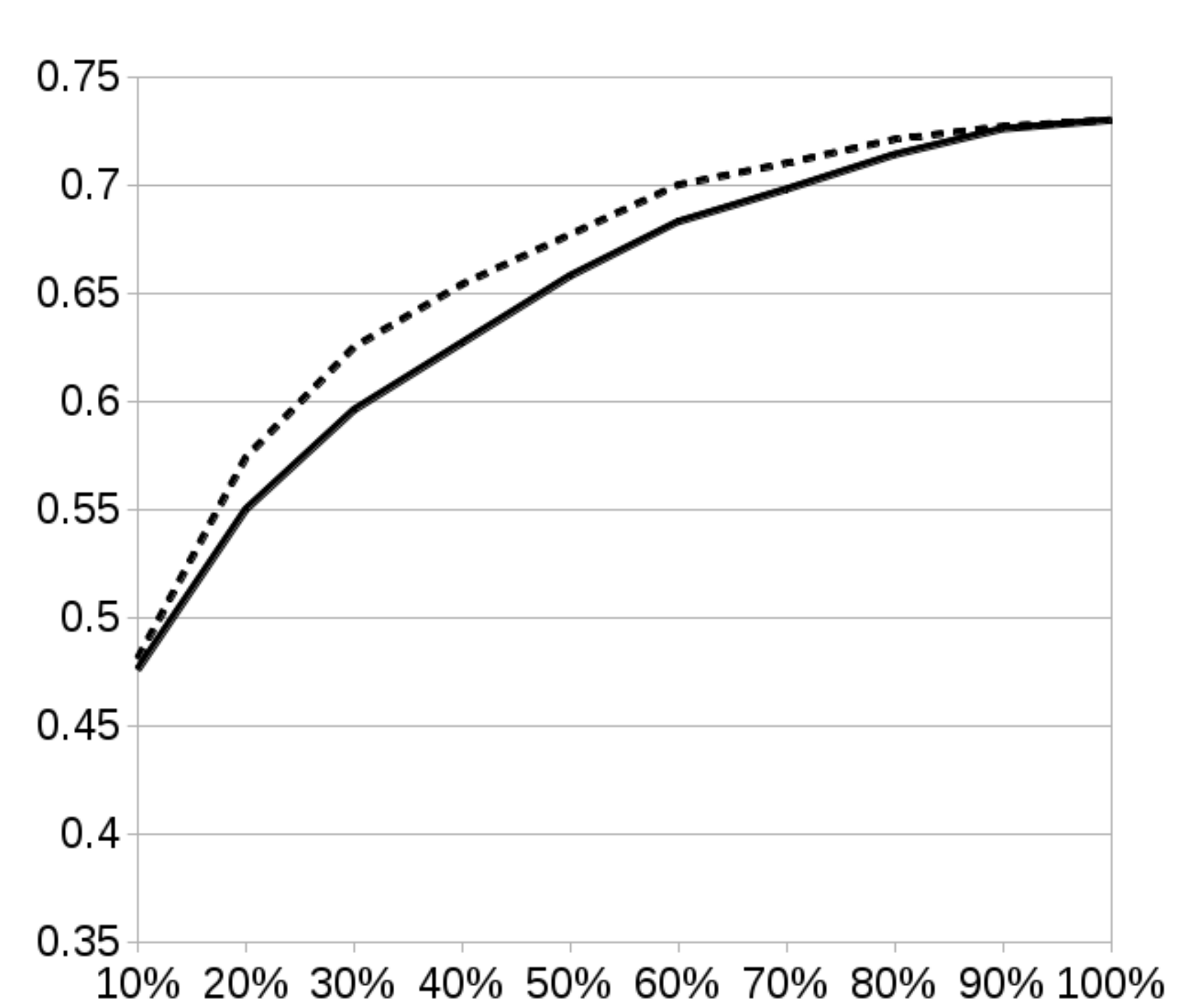} & \includegraphics[width=101px]{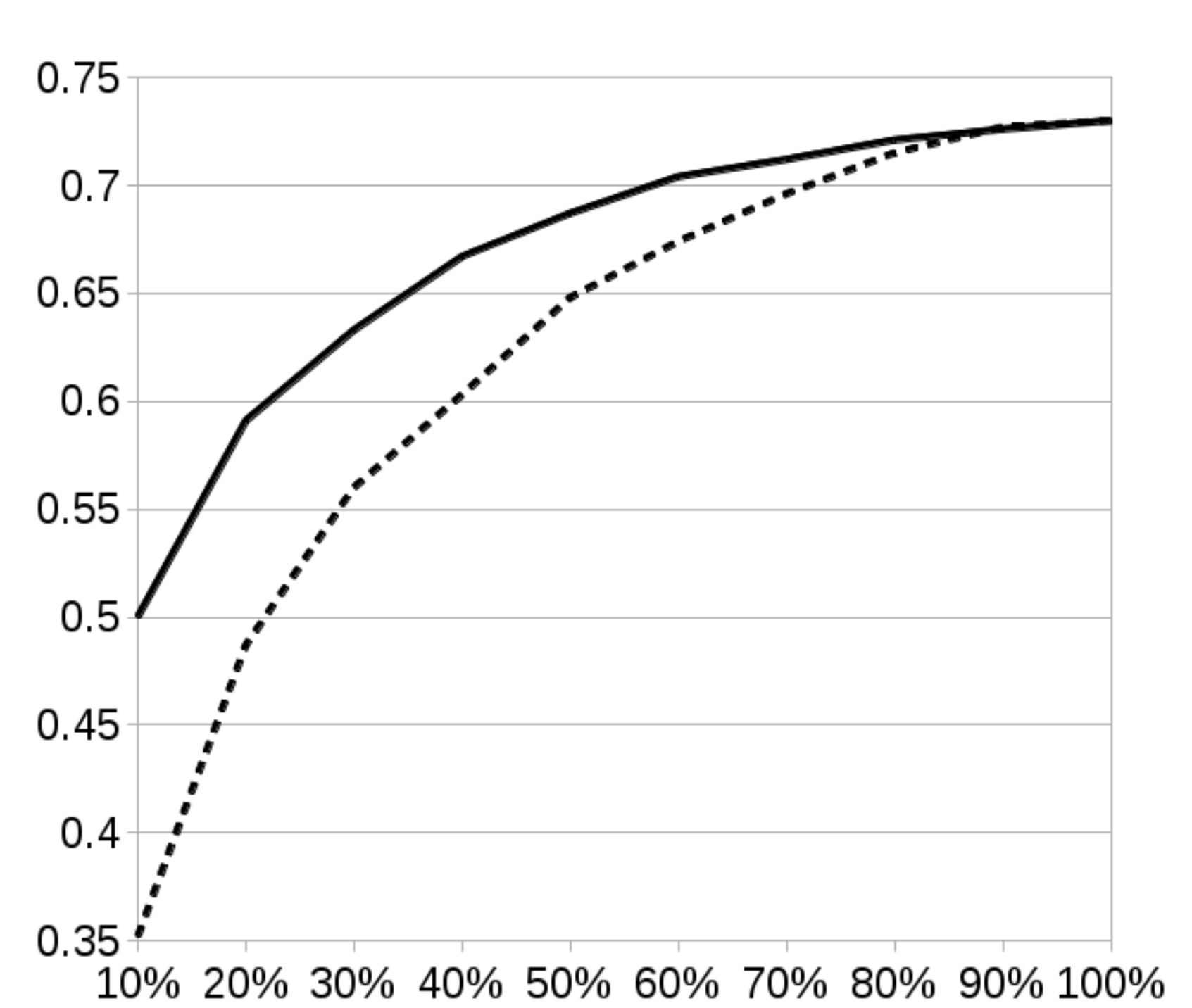} \\
  \end{tabular}
 \end{center}
 \caption[Classification results for Categorizers and Describers]{Tag-based classification accuracy results for Categorizers (continuous lines) and Describers (dashes lines) on Delicious, LibraryThing and GoodReads. The X axis represents the percents of selected top users, ranging from 10\% to 100\% with a step size of 10\%, either for Categorizers or Describers, whereas Y axis represents the accuracy.}
 \label{tab:cat-desc-classification}
\end{table}

\begin{table}[htbp]
 \begin{center}
  \begin{tabular}{ c c c c }
    & \scriptsize \textbf{TRR} & \scriptsize \textbf{ORPHAN} & \scriptsize \textbf{TPP} \\
    \begin{sideways}\scriptsize \textbf{Delicious}\end{sideways} & \includegraphics[width=101px]{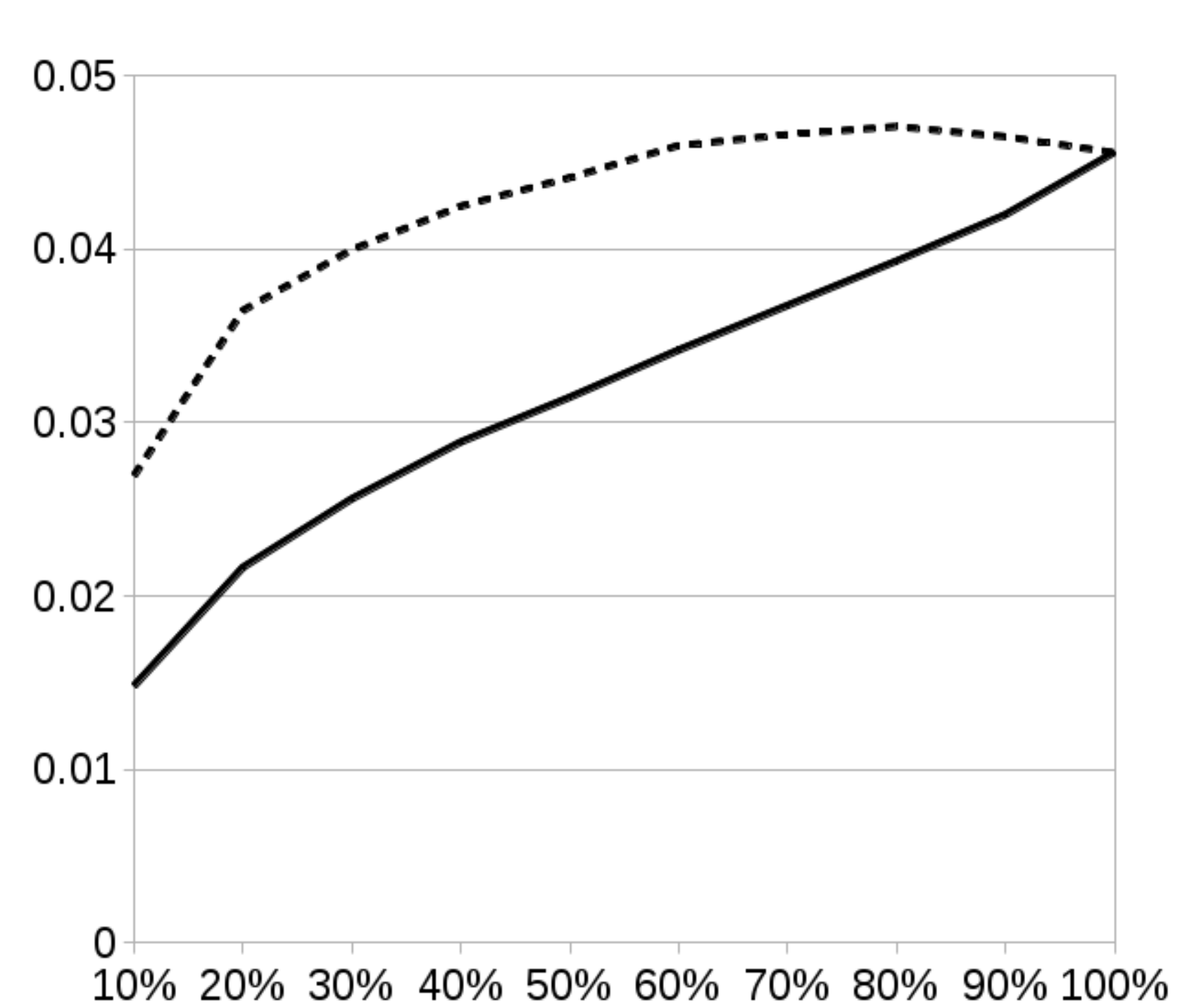} & \includegraphics[width=101px]{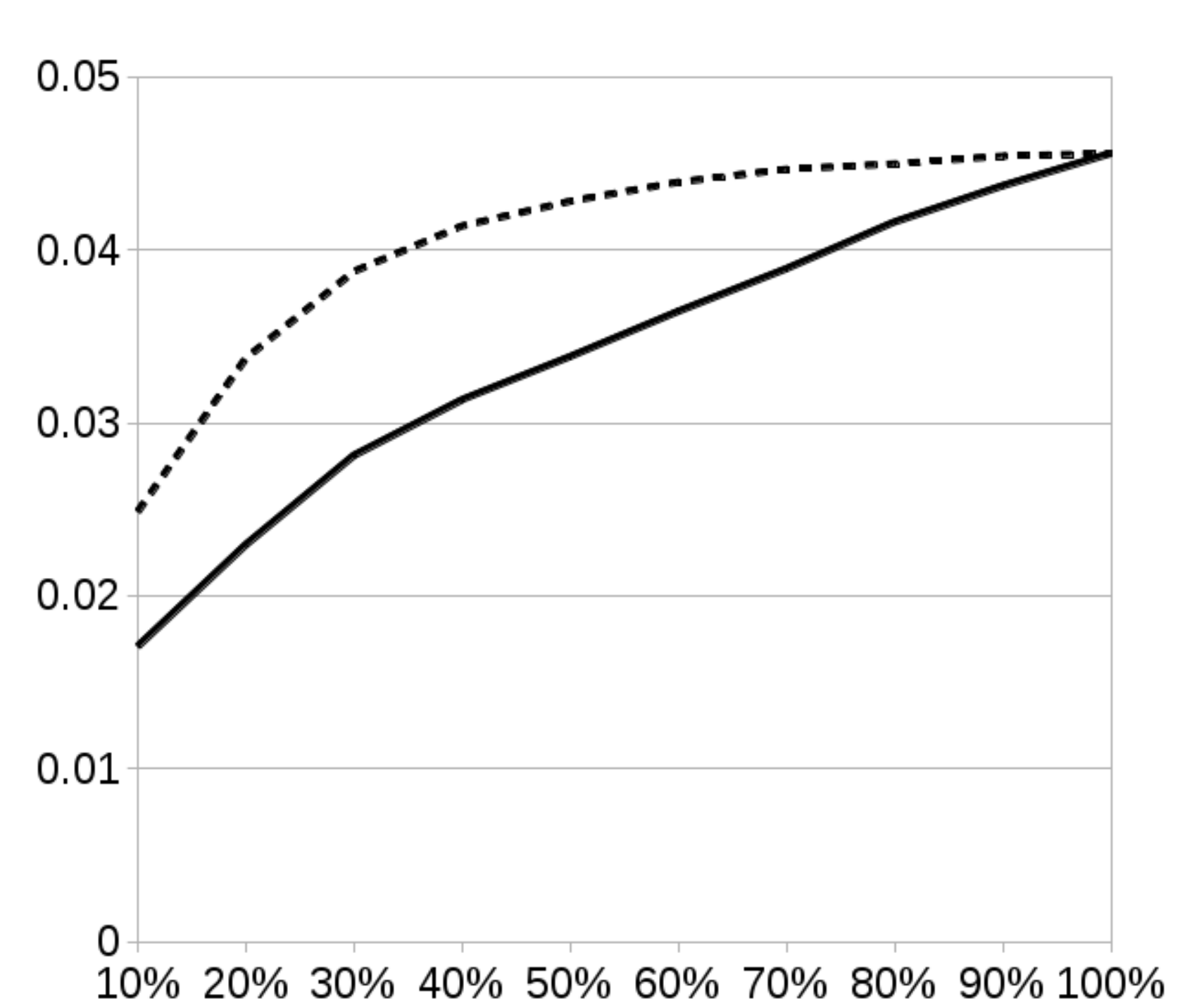} & \includegraphics[width=101px]{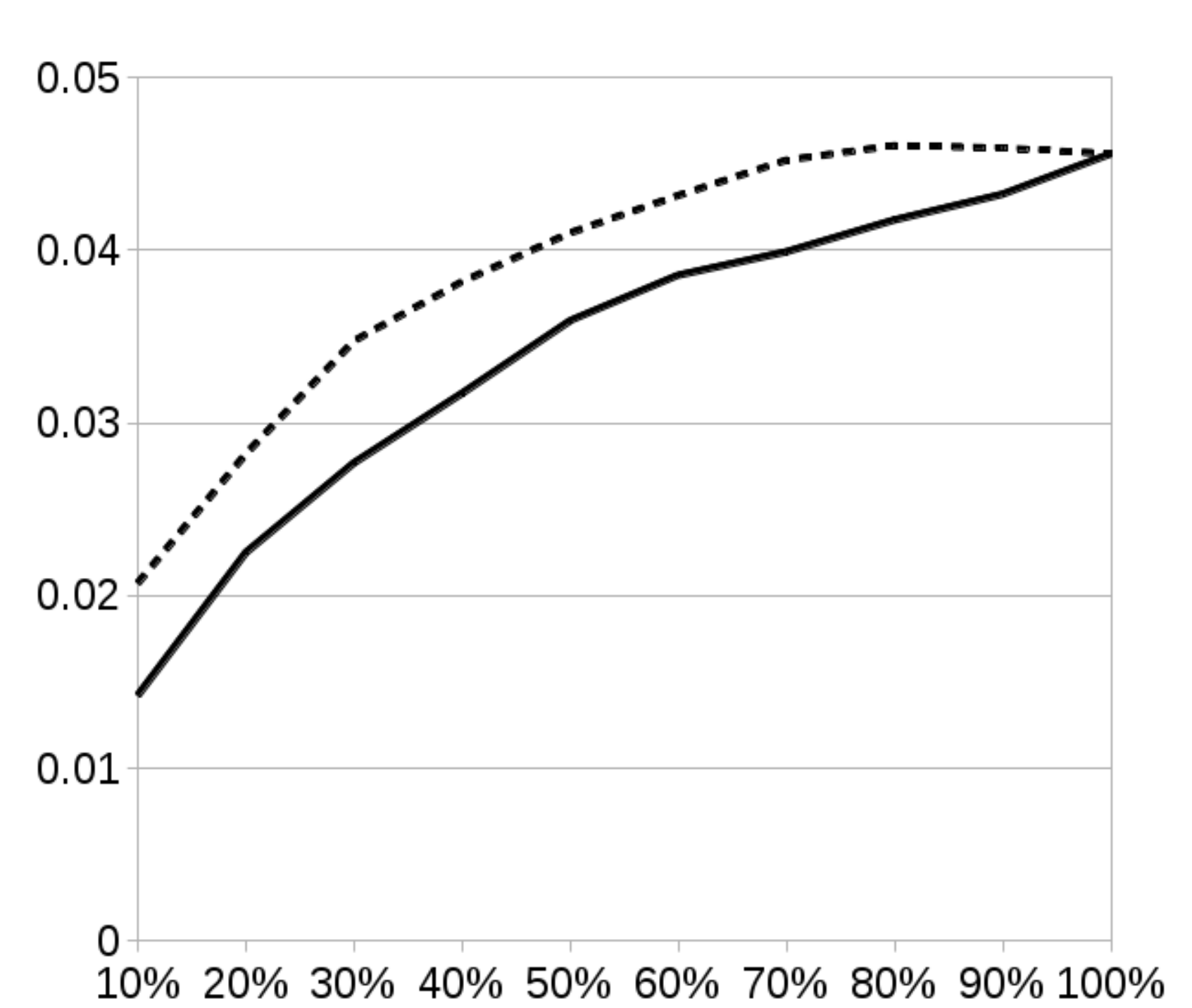} \\
    \begin{sideways}\scriptsize \textbf{LibraryThing}\end{sideways} & \includegraphics[width=101px]{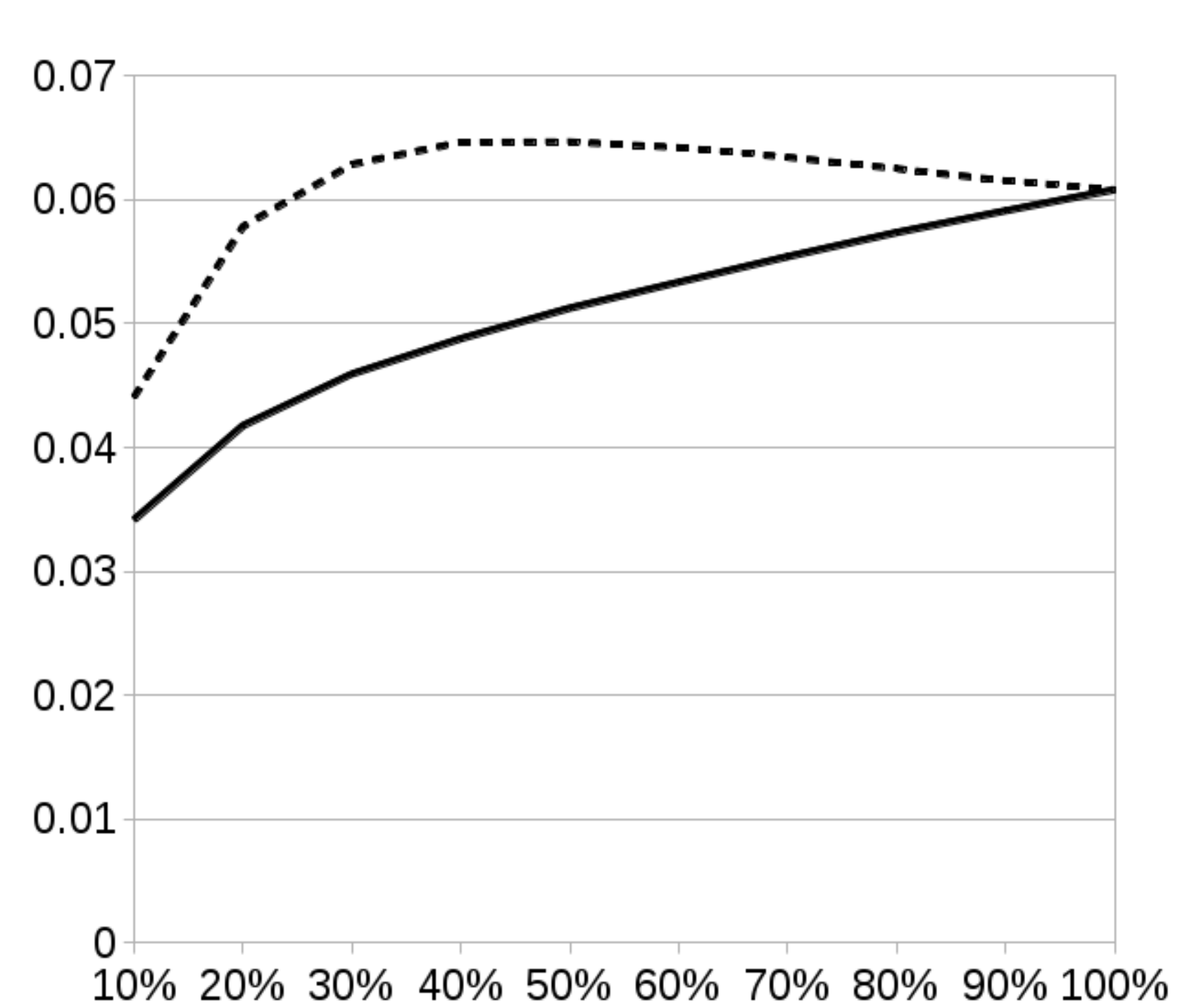} & \includegraphics[width=101px]{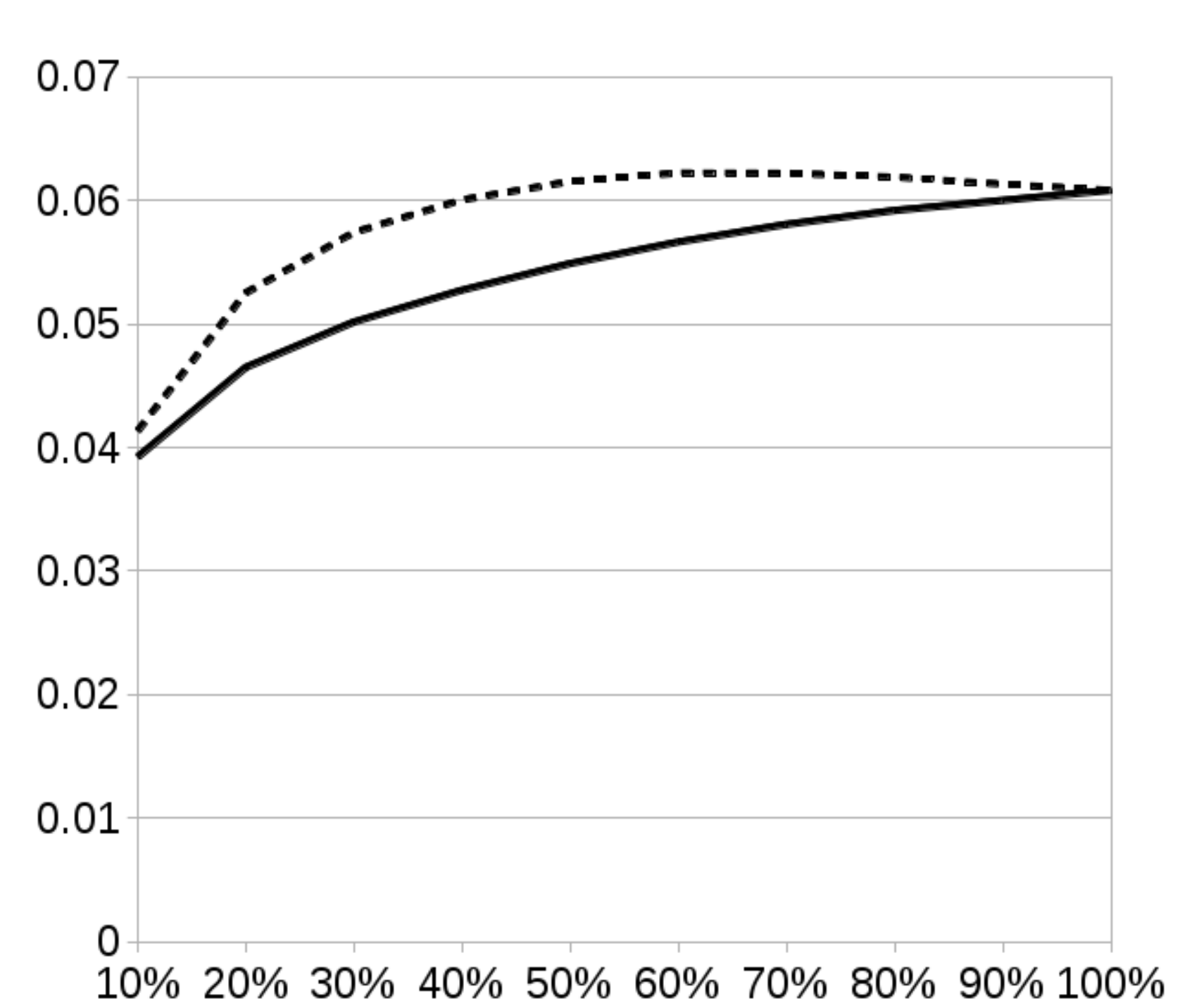} & \includegraphics[width=101px]{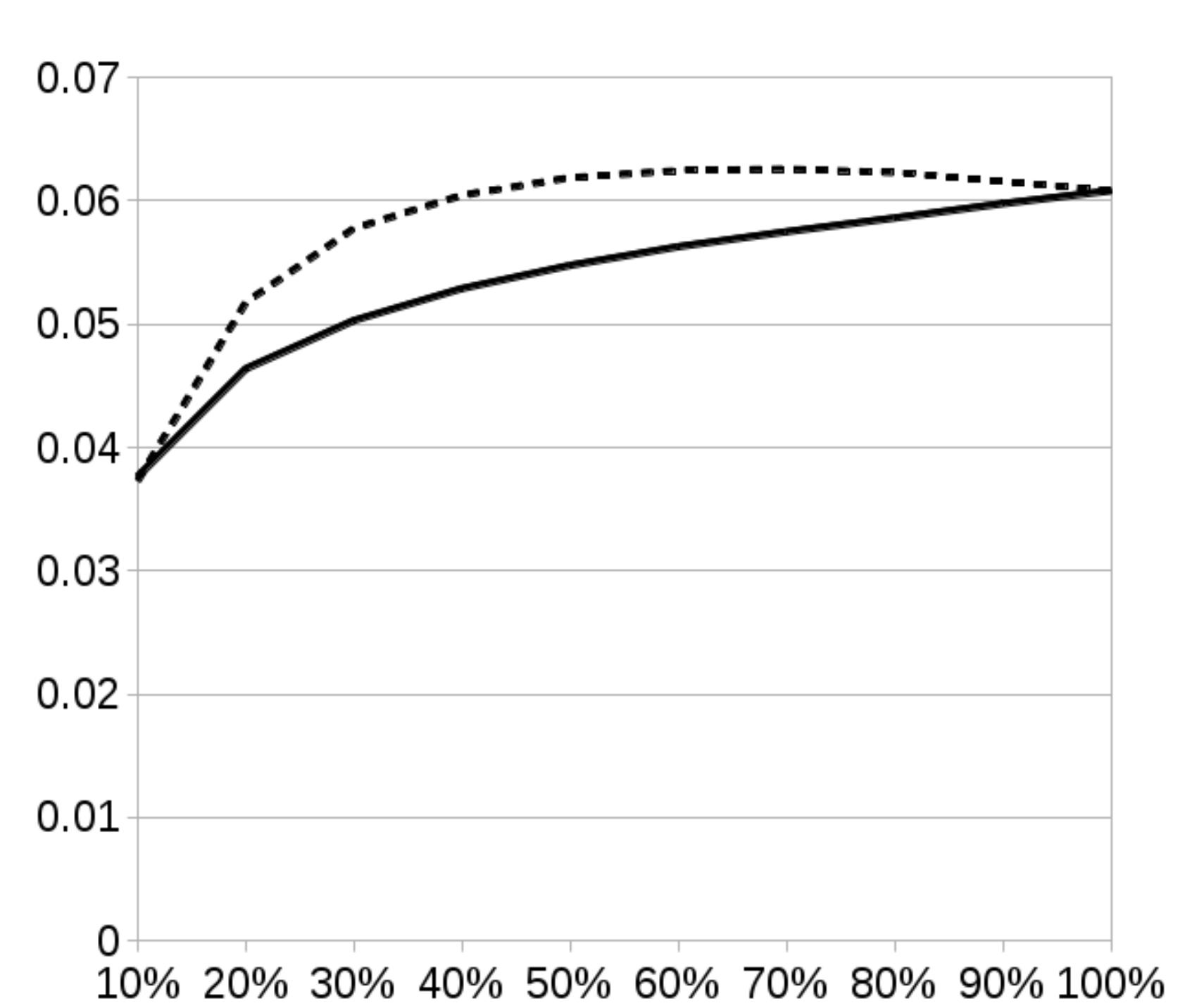} \\
    \begin{sideways}\scriptsize \textbf{GoodReads}\end{sideways} & \includegraphics[width=101px]{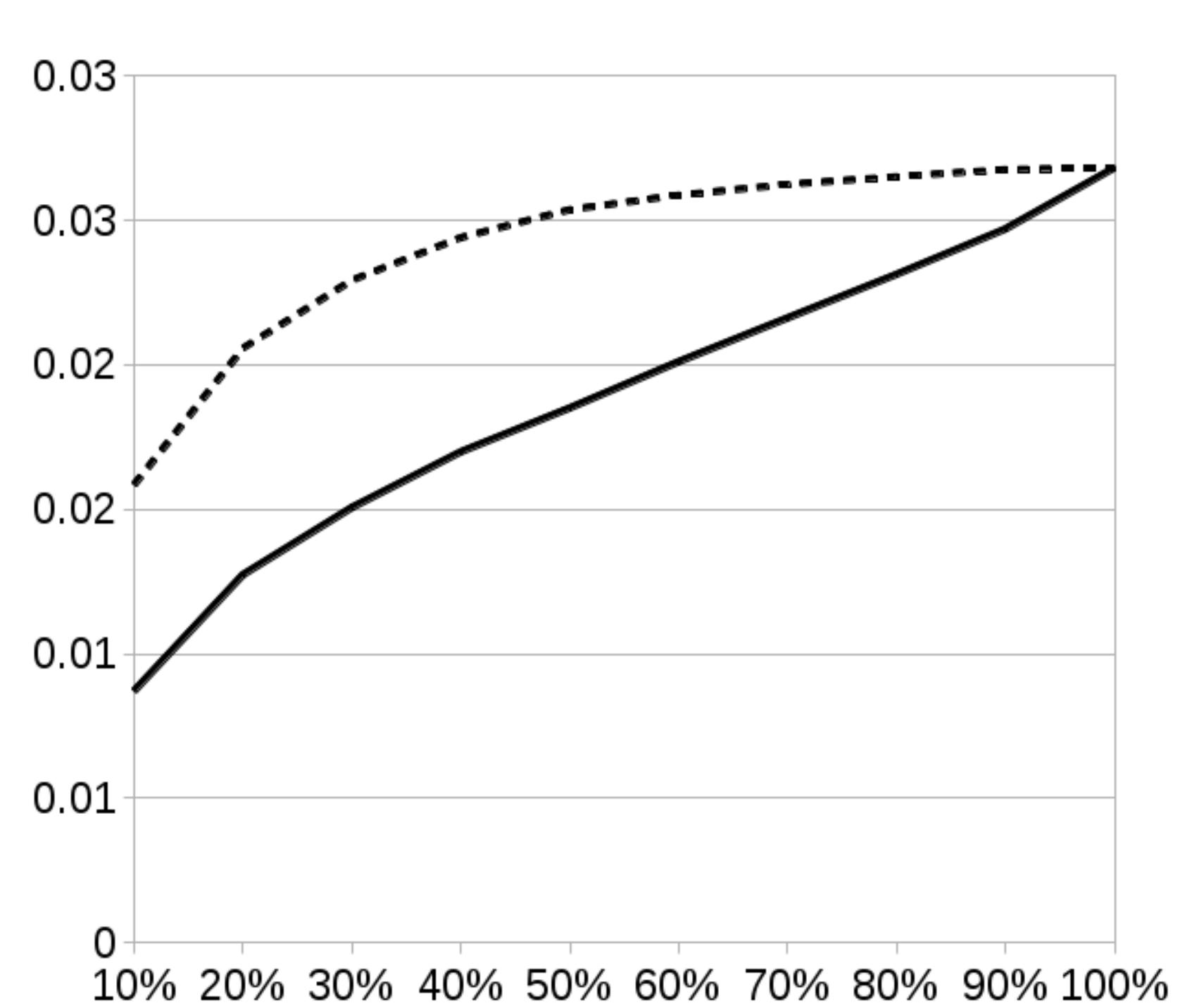} & \includegraphics[width=101px]{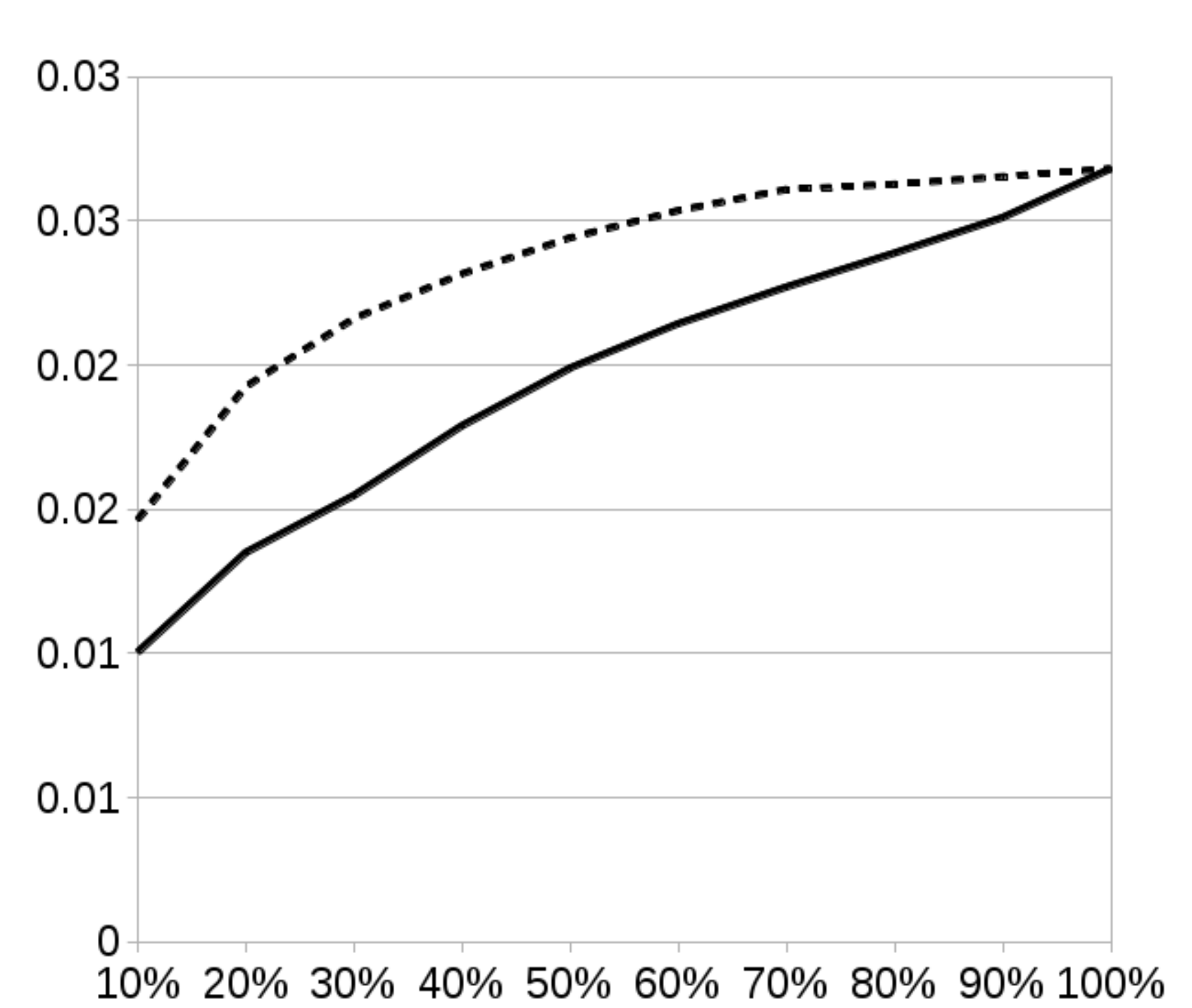} & \includegraphics[width=101px]{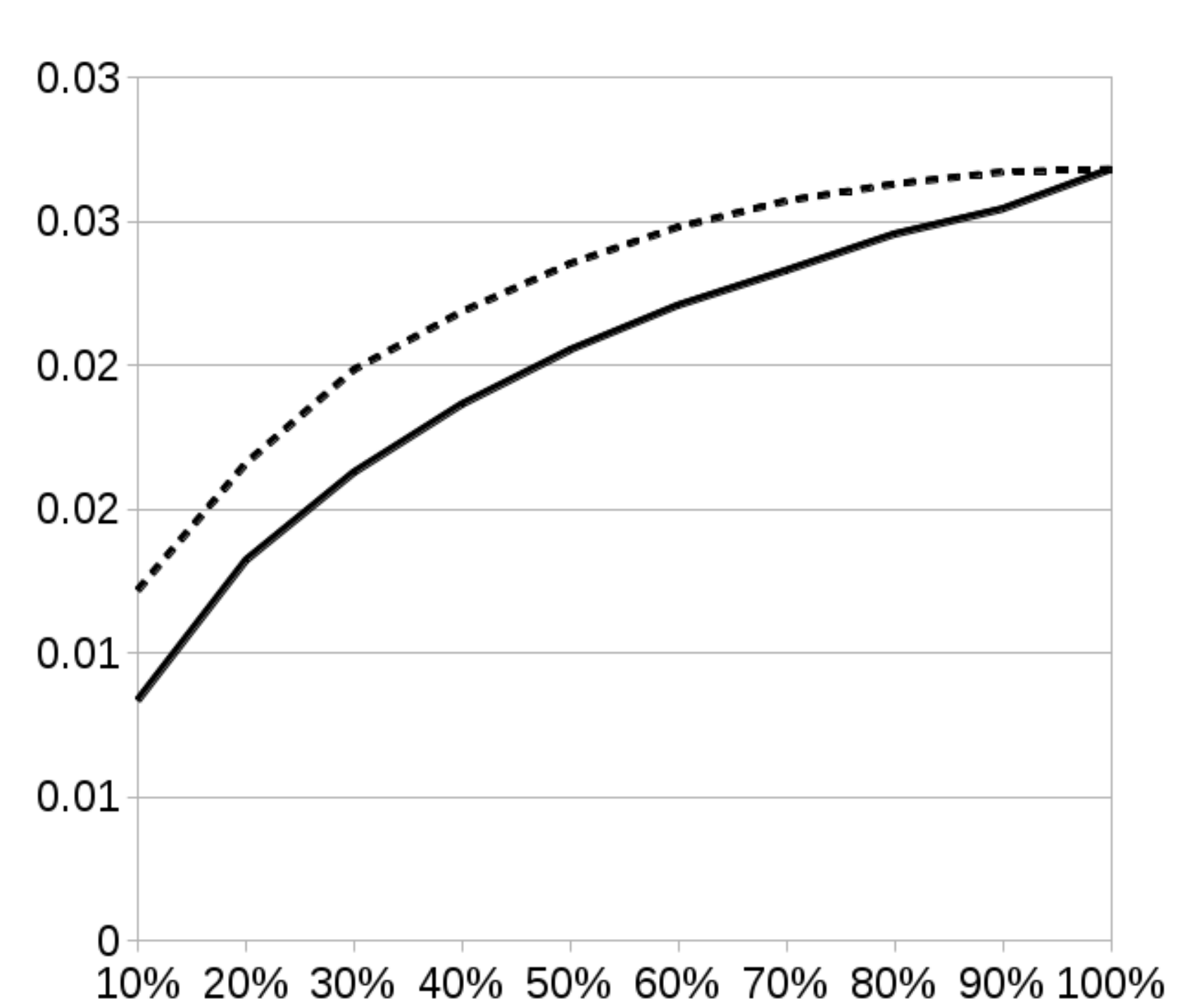} \\
  \end{tabular}
 \end{center}
 \caption[Descriptiveness results for Categorizers and Describers]{Similarity measures of the descriptiveness of tags on Delicious, LibraryThing and GoodReads. Continuous lines correspond to Categorizers, whereas dashed lines are Describers. The X axis represents the percents of selected top users, ranging from 10\% to 100\% with a step size of 10\%, either for Categorizers or Describers, whereas Y axis represents the degree of similarity (i.e., cosine value) to descriptive data.}
 \label{tab:cat-desc-descriptiveness}
\end{table}

Table \ref{tab:cat-desc-classification} shows the performance of Categorizers (continuous line) and Describers (dashed line) on the classification task, whereas Table \vref{tab:cat-desc-descriptiveness} does the same for the descriptiveness experiments. The results are presented in different graphs organized by datasets in rows --Delicious, LibraryThing and GoodReads--, and by measures in columns --TRR, ORPHAN and TPP. All of them keep the same scale and ranges for X axis, as well as for Y axis within each dataset, so that it enables an easy visual comparison of the results. When analyzing these results, we are especially interested in performance differences between Categorizers and Describers, and studying whether and why such subsets of users perform better for a certain task. Obviously, both Categorizers and Describers always yield the same performance for 100\% sets, as we are considering the whole set of users.

\subsection{Categorizers Perform Better on Classification}
\label{ssec:categorizers-better-classification}

It stands out that all three measures get positive results for both classification and descriptiveness experiments on LibraryThing. The subsets of Categorizers perform better for classification in all cases for this dataset. This means that all three measures provide a good way to discriminate Categorizers.
Among the compared measures, TPP gets the largest gap for classification, whereas TRR does it for descriptiveness.

As regards to GoodReads, results are less consistent. TPP yields especially positive results on this dataset. With the other measures, TRR and ORPHAN, Describers outperform Categorizers for classification. However, TRR works well for the 10\% subsets, suggesting that it discriminates correctly a subset of extreme Categorizers, but it fails when the subsets upsize. We speculate that the reason for this observation lies in the fact that this social tagging system is suggesting tags to users from their personomy. This encourages users to have a smaller vocabulary, and to reuse their tags frequently. It is quite easier to click on a list of tags than to type them.

In the case of Delicious, it seems that the resource-based system suggestions of its settings make it more difficult to detect Categorizers. On the one hand, TRR and ORPHAN show really slight differences between Categorizers and Describers, so that their discrimination does not seem to be performed appropriately. However, it works well with TPP for small subsets of users, where Categorizers outscore Describers. This outperformance inverts for larger subsets of users, though.

Analyzing the three datasets altogether, TPP shows the best way to discriminate Categorizers as better contributors to the resource classification task. This is clear for GoodReads and LibraryThing, but it only happens for small subsets of users on Delicious.

\subsection{Describers Perform Better on Descriptiveness}
\label{ssec:describers-better-descriptiveness}

The results of the descriptiveness experiments show that Describers are always superior to Categorizers in this regard. All three measures show to be really useful for discriminating Describers among users on social tagging systems, regardless of the settings of the site. Moreover, the measures show a similar behavior on all sites insofar as the outperformance of Describers as compared to Categorizers is fairly similar on the three datasets. However, the large gap of TRR sets it apart from the rest. Thereby, TRR gives rise to the best detection of Describers.

\subsection{Verbosity vs Diversity}
\label{ssec:verbosity-vs-diversity}

The three measures we have studied in this work rely on two different features to discriminate user behavior: verbosity and diversity. We can see a better overall performance of the TPP measure for resource classification, and the TRR measure for the descriptiveness task, we believe that: (1) verbosity can be inferred as the optimal feature for discriminating Categorizers, and (2) diversity as the feature that better discriminates Describers. In this context, we believe that Categorizers are thinking of a physical organization of resources, as librarians would do by placing books in shelves, when they annotate resources with tags. For instance, in the specific case of books, a user who thinks of the shelf where they stack their fictional books seems very likely to solely use the tag \texttt{fiction}. We could define these shelf-driven users as non-verbose. A user who adds just one tag has probably thought of the perfect tag that places it in the corresponding shelf. On the other hand, users who provide more detailed and diverse annotations rather think of describing the book instead of placing it in a specific shelf. This aspect makes the verbosity feature more powerful than the diversity feature for the detection of Categorizers. Thus, we believe that this is the feature that makes TPP so useful at discriminating Categorizers in search of an accurate resource classification as compared to TRR and ORPHAN, because it only relies on users' verbosity.

%

\subsection{Non-descriptive Tags Provide More Accurate Classification}
\label{ssec:nondescriptive-accurate-classification}

When discriminating user behavior appropriately by using a verbosity-based measure like TPP, we have shown that Categorizers better fit the classification task, whereas Describers provide annotations that further resemble the descriptive data. An interesting deduction from here is that a set of annotations that differs to a greater extent from the descriptive data produces a more accurate classification of the books. From this, we infer that Describers are using more descriptive tags, whereas Categorizers rather use non-descriptive tags. Hence, users who do not think of providing annotations in a similar way to writing reviews rely on non-descriptive tags, yielding a more accurate classification of the books.

\section{Conclusion}
\label{sec:user-behavior-conclusion}

In this chapter, we have explored the detection of user behavior on social tagging systems in search of users who rather approach to the resource classification task. To this end, we have explored the measures presented by \cite{strohmaier2010why}, which help us determine whether a user is a Categorizer rather organizing resources, or they are a Describer rather detailing the content of the resources. Specifically, we have studied the application of three different measures --TRR, ORPHAN and TPP--, which rely on two main features: verbosity and diversity of user annotations. By means of choosing different subsets of Categorizers and Describers, we have performed experiments on (1) resource classification, in order to explore whether Categorizers further resemble the classification by experts, and (2) measurement of the descriptiveness of tags, for exploring whether there is a higher similarity between tags by Describers and descriptive data of the resources.

Besides further understanding the existence of users aiming at classification on social tagging systems, i.e., Categorizers, we complemented a previous work by \cite{koerner2010stop}, where the authors showed that Describers are a good source for inferring semantic relations from folksonomies.

Parts of the research in this chapter have been published in \cite{zubiaga2011tags}.

We have answered the following research questions:

\begin{description}
 \item[Research Question 9] \hfill \\
 \textit{\rqcatone}
\end{description}

We have shown that such type of user, so-called Categorizer, actually exists. Tags assigned by Categorizers provide a more accurate classification of resources than those assigned by another set of users so-called Describers. According to our experiments, this is mostly true for systems without tag suggestions, i.e., LibraryThing, where the resource classification performed with tags by Categorizers yields clearly better results. When such suggestions exist, the detection of suitable users becomes more difficult, as we have showed happens on GoodReads and Delicious. However, the application of an appropriate measure by considering suitable features can produce a successful selection of users who fit the characteristics of a Categorizer.

\begin{description}
 \item[Research Question 10] \hfill \\
 \textit{\rqcattwo}
\end{description}

We have analyzed two features that characterize users of social tagging systems: verbosity, and diversity. We have shown that the level of verbosity helps discover Categorizers, who are better suited for the classification task. The vocabulary diversity is useful to find Describers, who tend to annotate using descriptive tags. Moreover, we have shown that users who do not rely on descriptive data provide better classification metadata than those who use descriptive tags.

%% file: conclusions.tex
\chapter{Conclusions and Future Research}
\label{c:conclusions}

\textit{``Believe those who are seeking the truth. Doubt those who find it.''}

--- Andre Gide



\chaptersummary{We conclude the thesis in this chapter. Next, we summarize the main contributions to the research field in Section \vref{sec:summary-of-contributions}. We continue by answering the formulated research questions in Section \vref{sec:answers-to-research-questions}. Finally, in Section \vref{sec:future-directions} we present an outlook on future directions of the research work in this thesis.}

\section{Summary of Contributions}
\label{sec:summary-of-contributions}

The novel idea of this work lies in the use of social annotations for carrying out a resource classification task. To the best of our knowledge, the first research work performing real classification experiments using social annotations is our first work in the field \citep{zubiaga2009getting}. Prior to that, only \cite{noll_exploring_2008} had performed a statistical analysis comparing social tags to a classification performed by experts. Taking into account the lack of work in the field, the work comprised in this thesis sheds new light on the appropriate use and representation of social tags for resource classification. More specifically, the following are the main contributions of this work:

\begin{itemize}
 \item We have created 3 large-scale social tagging datasets, including classification metadata of the annotated resources. These are among the largest datasets used so far for research and, to the best of our knowledge, the largest used for resource classification experiments. Some of these datasets, along with other smaller datasets we created, have been made publicly available for research purposes\footnote{http://nlp.uned.es/social-tagging/datasets/}. \cite{godoy2010exploiting} and \cite{strohmaier2010network}, for instance, have used some of our datasets in their recent research works. Even after we created these social tagging datasets, and made publicly available parts of them, little work has been done on creating and releasing more datasets. In \cite{koerner2010call}, the authors present a list of publicly available social tagging datasets, among which our datasets are also included. However, the authors set out the problem of the unavailability of more datasets, and encourage researchers to create and release new ones. As far as we know, no additional datasets have been released subsequently including categorization data for tagged resources.

 \item Our work is the first comparing different representations of resources based on social tags for resource classification. Moreover, it is the first work performing actual classification experiments comparing social tags to other data sources. We have shown that social tags are also useful for classification upon narrower categories in deeper levels of taxonomies. In a previous work, \cite{noll_exploring_2008} perform a statistical study concluding that social tags may not be helpful for narrower categories. In contrast to this, we have performed actual classification experiments showing a larger improvement for narrow categorization as compared to other data sources.

 \item We have analyzed the distributions of social tags in folksonomies, and performed a thorough study on how the settings of each social tagging system affect them, and therefore, a resource classification task. In this regard, we have applied a consolidated weighting scheme, TF-IDF, to the new social data structure given by folksonomies.

 \item We have shown the existence of a group of users, so-called Categorizers, whose annotations more closely resemble the classification performed by experts than social tags provided by another group of users known as Describers. The approach of differentiating Categorizers from Describers was already tested and verified in earlier works by proving the suitability of the latter for inferring semantic relations from folksonomies. Going further, we have demonstrated the suitability of Categorizers for resource classification.
\end{itemize}

The use of social annotations for the sake of resource classification tasks was a novel research line in the beginning of this thesis. However, the increasing interest of researchers on user-generated content in social media, and specifically in social tagging systems, has recently brought about more work in the field. Along with this increase, more researchers have shown their interest in the use of social annotations for resource classification tasks, and the number of works in this field has increased. \cite{godoy2010exploiting}, for instance, perform a tag-based classification study inspired by our earlier work \citep{zubiaga2009getting}. Furthermore, \cite{aliakbary_webpage_2009}, \cite{yin2009exploring}, \cite{xia2010optimizing}, and \cite{lu2010user} have recently presented their research in related matters, making use of social tags as to resource classification.

\section{Answers to Research Questions}
\label{sec:answers-to-research-questions}

At the beginning of this work, we set forth the following problem statement summarizing the main goal of the thesis:

\begin{description}
 \item[Problem Statement] \hfill \\
 \textit{\problemstatement}
\end{description}

In order to solve this problem statement, we split it into 10 research questions. Next, we list those research questions along with answers to them:

\begin{description}
 \item[Research Question 1] \hfill \\
 \textit{\rqsvmone}
\end{description}

We have shown the clear superiority of the native multiclass SVM classifiers over the other approaches combining binary classifiers. Our results show that relying on a set of binary classifiers is not a good option when it comes to multiclass taxonomies. Accordingly, native multiclass classifiers, which consider all the classes at the same time and have more knowledge of the whole task, perform much better.

\begin{description}
 \item[Research Question 2] \hfill \\
 \textit{\rqsvmtwo}
\end{description}

Semi-supervised approaches may perform better when the labeled subset is really small, but supervised approaches, which are computationally less expensive, perform similarly with more labeled documents. Therefore, we have also shown that, unlike binary tasks as shown by \cite{joachims99transductive}, a supervised approach performs very similar to a semi-supervised approach on these environments. It seems reasonable that predicting the class of uncategorized documents is much more difficult when the number of classes increases, and so the miscategorized documents are harmful for classifier's learning.

Thereby, according to these two conclusions above, we decided to use a supervised multiclass SVM approach.

\begin{description}
 \item[Research Question 3] \hfill \\
 \textit{\rqdataone}
\end{description}

To this end, we have analyzed several features that can be found in different settings of social tagging systems. Among the analyzed features, we have shown the impact of tag suggestions, which considerably alters the resulting folksonomy. In the studied social tagging sites, all of them differ on the settings regarding suggestions:

\begin{itemize}
 \item \textbf{Resource-based suggestions (Delicious):} when the system suggests tags assigned by other users to the resource at the time of bookmarking it, the likelihood of using new tags to further describe such a resource descreases. In this case, users provide less originality and tend to rely on system suggestions.
 \item \textbf{Personomy-based suggestions (GoodReads):} when the system suggests tags previously used by the user, the vocabulary in their personomy tends to be much smaller. However, users do not know how others annotated a resource, and thus they are likely to provide new tags to the resource.
 \item \textbf{Without suggestions (LibraryThing):} when the system does not suggest any tags to the user, the vocabulary in their personomy increases, as well as the diversity of tags in each resource.
\end{itemize}

\begin{description}
 \item[Research Question 4] \hfill \\
 \textit{\rqrepone}
\end{description}

We have shown that it is worthwhile considering all the tags annotated on a resource instead of those in the top that were annotated most. Tags in the top are the most important, and give the main information on the aboutness of resources. However, tags in the tail are helpful to a lesser extent, providing meaningful information and improving the performance of the classifier.

Regarding the weights assigned to those tags when representing a resource, the number of users annotating each tag should be considered in order to get the best results. This is the value that has shown the best results in our experiments. It has outperformed other approaches ignoring weights or considering other data such as the total number of users annotating the resource.

Thereby, the best representation in our experiments is the one that includes all the tags with the values corresponding to the number of users annotating them.

\begin{description}
 \item[Research Question 5] \hfill \\
 \textit{\rqreptwo}
\end{description}

By means of classifier committees, which combine the predictions by different classifiers, we have shown that tags provide reliable prediction criteria to take into consideration. SVM classifiers not only predict a category, but also assign a weight to each category based on the given resource. These weights, given in the form of margin values, can be used by other classifiers which rely on different data sources. Adding up weights provided by different classifiers can help predict the correct category when a single classifier fails to categorize the resource appropriately.  Weights provided by classifiers relying on social tags are especially useful when combining them with results from other classifiers.  Nonetheless, not all data sources are helpful for combination in classifier committees, and the selected data source must be solid enough and provide reliable predictions to outperform the sole use of tags. When data sources are selected appropriately, the performance improvement can be considerable.  We have shown that this varies among datasets. For example, with the Delicious dataset, it is important to analyze all three data sources (content, reviews, and tags). However, with the LibraryThing and GoodReads datasets reviews and tags suffice.

\begin{description}
 \item[Research Question 6] \hfill \\
 \textit{\rqrepthree}
\end{description}

We have analyzed the usefulness of social tags for classification on two different levels of hierarchical taxonomies. Besides broader categories in the top level, we have also explored the classification on narrower categories in the second level. In this regard, social tags have shown to outperform the other data sources on social tagging sites that encourage users to annotate resources (Delicious and LibraryThing). Tags show clear outperformance in these cases, especially on Delicious, where the difference is even more favorable in the second level. This difference is very similar on LibraryThing. Finally, tags from GoodReads do not outperform other data sources at any level because the system does not encourage users to tag books, so that many bookmarks are not annotated.

Our findings provide a different conclusion from that by \cite{noll_exploring_2008}, where the authors pointed out the hypothesis that social tags were probably useless for deeper levels of taxonomies, and alternative data should be used instead. However, the authors performed just a statistical analysis, and did not confirm the hypothesis with real experiments.

\begin{description}
 \item[Research Question 7] \hfill \\
 \textit{\rqdistone}
\end{description}

We have analyzed the suitability of IDF-like weighting functions to define the representativity of tags, which consider the distribution of tags through the whole collection of resources. Our experiments have shown that these functions helps improve performance of a resource classification task. However, we have shown that the settings of the social tagging system have an effect on those distributions. Resource-based tag suggestions have shown to influence the structure of folksonomies greatly. Suggesting tags based on previous annotations of others on the resource causes a very different tag distribution, which in turn, affects the results of the weighting function. When a system enables the resource-based tag suggestions, the use of tag weighting functions performs worse, and combining with other data sources is required to improve performance; this method can even outperform the TF-based approach.

For our classification experiments, we have found that IDF-like weighting functions clearly outperform the TF approach when resource-based tag suggestions are not enabled, i.e., on LibraryThing and GoodReads, both when used on their own, or when combined with other data sources. We found it better to consider just the tag-based approach, without combining them with other data sources, since it provides superior results, which cannot be improved by combining them with other predictions.

\begin{description}
 \item[Research Question 8] \hfill \\
 \textit{\rqdisttwo}
\end{description}

Among the studied weighting functions, the one relying on bookmark frequencies has shown to be the best when there are no resource-based tag suggestions. In these cases, IBF performs the best, followed by IRF, and IUF. All of them clearly outperform TF, when both used on their own, and combined with other data sources using classifier committees.

On the other hand, when the social tagging system suggests tags to the user relying on the resource itself, IUF performs better than the others. IUF performs better than IBF and IRF, because of the importance of the ability of users to choose their own tags without relying on suggestions from these systems. Even though IUF does not outperform TF when used on its own, combining it with other data sources produces the best approach. However, it is only slightly better than the committees relying on TF, and any of them can be used to score similar results.

\begin{description}
 \item[Research Question 9] \hfill \\
 \textit{\rqcatone}
\end{description}

We have shown that such type of user, so-called Categorizer, actually exists. Tags assigned by Categorizers provide a more accurate classification of resources than those assigned by another set of users so-called Describers. According to our experiments, this is mostly true for systems without tag suggestions, i.e., LibraryThing, where the resource classification performed with tags by Categorizers yields clearly better results. When such suggestions exist, the detection of suitable users becomes more difficult, as we have showed happens on GoodReads and Delicious. However, the application of an appropriate measure by considering suitable features can produce a successful selection of users who fit the characteristics of a Categorizer.

\begin{description}
 \item[Research Question 10] \hfill \\
 \textit{\rqcattwo}
\end{description}

We have analyzed two features that characterize users of social tagging systems: verbosity, and diversity. We have shown that the level of verbosity helps discover Categorizers, who are better suited for the classification task. The vocabulary diversity is useful to find Describers, who tend to annotate using descriptive tags. Moreover, we have shown that users who do not rely on descriptive data provide better classification metadata than those who use descriptive tags.

\section{Future Directions}
\label{sec:future-directions}

The use of social tags for resource classification is still a novel research field with little work done so far. The thesis has shown how social tags can be useful for the resource classification task, and provides analysis to help determine an optimal method to accurately categorize resources based on their social tags. Furthermore, this thesis paves way for future research on the utilization of social tags for resource classification.

Throughout this thesis, we have considered each tag as a different token, regardless of its semantic meaning. In this regard, future work includes analyzing the meaning of each tag trying to discover synonymous words, and relations among them. Either by using natural language processing methods or following ontology-based approaches, it could improve understanding the meaning of each tag and further exploring the knowledge provided by folksonomies.

The three weighting schemes we have used in Chapter \vref{c:tag-distribution-classification} rely on the classical TF-IDF function designed for text collections. Trying other weighting functions, as well as defining a new one that fits the structure of folksonomies would be also interesting as a future work. This would especially help for systems providing resource-based tag suggestions, like Delicious, where the tested weighting schemes did not perform well.

%% file: publications.tex
\chapter*{Publications}
\label{c:publications}

\section*{Peer-Reviewed Conferences}

\begin{itemize}
 \item Arkaitz Zubiaga, Christian K\"{o}rner, Markus Strohmaier. 2011. \emph{Tags vs Shelves: From Social Tagging to Social Classification}. In Proceedings of Hypertext 2011, the 22nd ACM Conference on Hypertext and Hypermedia, Eindhoven, Netherlands. (acceptance rate: 35/104, 34\%)
 \item Arkaitz Zubiaga, Raquel Mart\'{i}nez, V\'{i}ctor Fresno. 2009. \emph{Getting the Most Out of Social Annotations for Web Page Classification}. In Proceedings of DocEng 2009, the 9th ACM Symposium on Document Engineering, pp. 74-83, Munich, Germany. (acceptance rate: 16/54, 29.6\%)
 \item Arkaitz Zubiaga, Raquel Mart\'{i}nez, V\'{i}ctor Fresno. 2009. \emph{Clasificación de Páginas Web con Anotaciones Sociales}. In Proceedings of SEPLN 2009, XXV edici\'{o}n del Congreso Anual de la Sociedad Espa\~{n}ola para el Procesamiento del Lenguaje Natural, pp. 225-233, Donostia-San Sebasti\'{a}n. (acceptance rate: 36/72, 50\%)
 \item Arkaitz Zubiaga, Alberto P. Garc\'{i}a-Plaza, V\'{i}ctor Fresno, Raquel Mart\'{i}nez. 2009. \emph{Content-based Clustering for Tag Cloud Visualization}. In Proceedings of ASONAM 2009, International Conference on Advances in Social Networks Analysis and Mining, pp. 316-319, Athens, Greece.
\end{itemize}

\section*{Workshops}

\begin{itemize}
 \item Arkaitz Zubiaga, V\'{i}ctor Fresno, Raquel Mart\'{i}nez. 2009. \emph{Is Unlabeled Data Suitable for Multiclass SVM-based Web Page Classification?}. In Proceedings of the NAACL-HLT 2009 Workshop on Semi-supervised Learning for Natural Language Processing, pp. 28-36, Boulder, CO, United States.
\end{itemize}

\section*{Book Chapters}

\begin{itemize}
 \item Arkaitz Zubiaga, V\'{i}ctor Fresno, Raquel Mart\'{i}nez. 2011. \emph{Exploiting Social Annotations for Resource Classification}. Social Network Mining, Analysis and Research Trends: Techniques and Applications. IGI Global.
\end{itemize}

\section*{Journals}

\begin{itemize}
 \item Arkaitz Zubiaga, V\'{i}ctor Fresno, Raquel Mart\'{i}nez. 2009. Comparativa de Aproximaciones a SVM Semisupervisado Multiclase para Clasificaci\'{o}n de P\'{a}ginas Web. SEPLN, Sociedad Espa\~{n}ola para el Procesamiento del Lenguaje Natural, vol. 42, pp. 63-70.
\end{itemize}

\section*{Others}

\begin{itemize}
 \item Arkaitz Zubiaga. 2009. \emph{Enhancing Navigation on Wikipedia with Social Tags}. Wikimania 2009, Buenos Aires, Argentina.
 \item Arkaitz Zubiaga, Alberto P. García-Plaza, V\'{i}ctor Fresno, Raquel Martínez. 2009. Etiketa-lainoen Ikuskera Hobetzeko Multzokatzea. Informatikari Euskaldunen Bilkura '09, Donostia-San Sebasti\'{a}n.
\end{itemize}

%% file: topx-tags.tex
\chapter{Additional Results}
\label{c:topx-tags}

In Chapter \vref{c:tag-representation} we explored different representations of social tags in order to evaluate which of them performs better on a resource classification task. Among the approaches, we compared using all the tags annotated on each resource, and choosing just those in the top. For the latter, we focused on the top 10 tags, just to evaluate whether tags in the tail were harmful for this purpose. However, we did not show whether a selection of top 5 or 15 of tags could be a better choice. In Table \vref{tab:topx} we show the results of using different tops of tags for the FTA-based representation on the top level of the taxonomies. The results confirm that relying on all the tags performs the best, and that the selection of 5, 10 or 15 tags in the top has no impact in this regard. Going further, it also confirms that tags in the tail are far less useful, because the improvement is much smaller when low-ranked tags are included.

\begin{sidewaystable}[htb]
 \begin{center}
  \begin{tabular}{|l|c|c|c|c|c|c|c|c|c|c|c|c|c|c|c|}
   \hline
   & \scriptsize \textbf{1} & \scriptsize \textbf{2} & \scriptsize \textbf{3} & \scriptsize \textbf{4} & \scriptsize \textbf{5} & \scriptsize \textbf{6} & \scriptsize \textbf{7} & \scriptsize \textbf{8} & \scriptsize \textbf{9} & \scriptsize \textbf{10} & \scriptsize \textbf{11} & \scriptsize \textbf{12} & \scriptsize \textbf{13} & \scriptsize \textbf{14} & \scriptsize \textbf{15} \\
   \hline
   \scriptsize \textbf{Delicious} & \scriptsize .445 & \scriptsize .542 & \scriptsize .583 & \scriptsize .602 & \scriptsize .620 & \scriptsize .633 & \scriptsize .639 & \scriptsize .648 & \scriptsize .652 & \scriptsize .654 & \scriptsize .659 & \scriptsize .661 & \scriptsize .665 & \scriptsize .665 & \scriptsize .666  \\
   \hline
   \scriptsize \textbf{LibraryThing (DDC)} & \scriptsize .805 & \scriptsize .849 & \scriptsize .855 & \scriptsize .857 & \scriptsize .858 & \scriptsize .860 & \scriptsize .861 & \scriptsize .862 & \scriptsize .864 & \scriptsize .864 & \scriptsize .865 & \scriptsize .864 & \scriptsize .864 & \scriptsize .865 & \scriptsize .865 \\
   \hline
   \scriptsize \textbf{LibraryThing (LCC)} & \scriptsize .778 & \scriptsize .833 & \scriptsize .844 & \scriptsize .848 & \scriptsize .852 & \scriptsize .854 & \scriptsize .855 & \scriptsize .857 & \scriptsize .857 & \scriptsize .858 & \scriptsize .859 & \scriptsize .858 & \scriptsize .858 & \scriptsize .858 & \scriptsize .859 \\
   \hline
   \scriptsize \textbf{GoodReads (DDC)} & \scriptsize .660 & \scriptsize .714 & \scriptsize .725 & \scriptsize .731 & \scriptsize .730 & \scriptsize .730 & \scriptsize .730 & \scriptsize .730 & \scriptsize .728 & \scriptsize .730 & \scriptsize .730 & \scriptsize .729 & \scriptsize .730 & \scriptsize .730 & \scriptsize .730 \\
   \hline
   \scriptsize \textbf{GoodReads (LCC)} & \scriptsize .619 & \scriptsize .687 & \scriptsize .707 & \scriptsize .709 & \scriptsize .709 & \scriptsize .710 & \scriptsize .711 & \scriptsize .709 & \scriptsize .711 & \scriptsize .712 & \scriptsize .712 & \scriptsize .710 & \scriptsize .711 & \scriptsize .713 & \scriptsize .711 \\
   \hline
   & \scriptsize \textbf{16} & \scriptsize \textbf{17} & \scriptsize \textbf{18} & \scriptsize \textbf{19} & \scriptsize \textbf{20} & \scriptsize \textbf{21} & \scriptsize \textbf{22} & \scriptsize \textbf{23} & \scriptsize \textbf{24} & \scriptsize \textbf{25} & \scriptsize \textbf{26} & \scriptsize \textbf{27} & \scriptsize \textbf{28} & \scriptsize \textbf{29} & \scriptsize \textbf{30} \\
   \hline
   \scriptsize \textbf{Delicious} & \scriptsize .668 & \scriptsize .670 & \scriptsize .670 & \scriptsize .673 & \scriptsize .672 & \scriptsize .673 & \scriptsize .673 & \scriptsize .674 & \scriptsize .675 & \scriptsize .675 & \scriptsize .675 & \scriptsize .677 & \scriptsize .677 & \scriptsize .677 & \scriptsize .677 \\
   \hline
   \scriptsize \textbf{LibraryThing (DDC)} & \scriptsize .866 & \scriptsize .866 & \scriptsize .866 & \scriptsize .867 & \scriptsize .866 & \scriptsize .867 & \scriptsize .866 & \scriptsize .866 & \scriptsize .867 & \scriptsize .867 & \scriptsize .866 & \scriptsize .866 & \scriptsize .866 & \scriptsize .867 & \scriptsize .866 \\
   \hline
   \scriptsize \textbf{LibraryThing (LCC)} & \scriptsize .859 & \scriptsize .860 & \scriptsize .860 & \scriptsize .860 & \scriptsize .860 & \scriptsize .859 & \scriptsize .860 & \scriptsize .859 & \scriptsize .860 & \scriptsize .860 & \scriptsize .860 & \scriptsize .860 & \scriptsize .860 & \scriptsize .860 & \scriptsize .860 \\
   \hline
   \scriptsize \textbf{GoodReads (DDC)} & \scriptsize .730 & \scriptsize .729 & \scriptsize .729 & \scriptsize .730 & \scriptsize .729 & \scriptsize .730 & \scriptsize .731 & \scriptsize .730 & \scriptsize .729 & \scriptsize .728 & \scriptsize .729 & \scriptsize .730 & \scriptsize .730 & \scriptsize .729 & \scriptsize .730 \\
   \hline
   \scriptsize \textbf{GoodReads (LCC)} & \scriptsize .711 & \scriptsize .711 & \scriptsize .715 & \scriptsize .713 & \scriptsize .711 & \scriptsize .710 & \scriptsize .713 & \scriptsize .709 & \scriptsize .712 & \scriptsize .711 & \scriptsize .713 & \scriptsize .712 & \scriptsize .713 & \scriptsize .712 & \scriptsize .713 \\
   \hline
   & \scriptsize \textbf{40} & \scriptsize \textbf{50} & \scriptsize \textbf{60} & \scriptsize \textbf{70} & \scriptsize \textbf{80} & \scriptsize \textbf{90} & \scriptsize \textbf{100} & \scriptsize \textbf{150} & \scriptsize \textbf{200} & \scriptsize \textbf{250} & \scriptsize \textbf{300} & \scriptsize \textbf{350} & \scriptsize \textbf{400} & \scriptsize \textbf{450} & \scriptsize \textbf{500} \\
   \hline
   \scriptsize \textbf{Delicious} & \scriptsize .678 & \scriptsize .679 & \scriptsize .678 & \scriptsize .678 & \scriptsize .679 & \scriptsize .679 & \scriptsize .679 & \scriptsize .679 & \scriptsize .679 & \scriptsize .679 & \scriptsize .680 & \scriptsize .680 & \scriptsize .679 & \scriptsize .679 & \scriptsize .679 \\
   \hline
   \scriptsize \textbf{LibraryThing (DDC)} & \scriptsize .867 & \scriptsize .867 & \scriptsize .867 & \scriptsize .867 & \scriptsize .868 & \scriptsize .868 & \scriptsize .868 & \scriptsize .867 & \scriptsize .868 & \scriptsize .868 & \scriptsize .868 & \scriptsize .868 & \scriptsize .868 & \scriptsize .867 & \scriptsize .868 \\
   \hline
   \scriptsize \textbf{LibraryThing (LCC)} & \scriptsize .860 & \scriptsize .861 & \scriptsize .861 & \scriptsize .860 & \scriptsize .861 & \scriptsize .861 & \scriptsize .861 & \scriptsize .861 & \scriptsize .861 & \scriptsize .861 & \scriptsize .860 & \scriptsize .861 & \scriptsize .861 & \scriptsize .861 & \scriptsize .861 \\
   \hline
   \scriptsize \textbf{GoodReads (DDC)} & \scriptsize .729 & \scriptsize .729 & \scriptsize .728 & \scriptsize .729 & \scriptsize .729 & \scriptsize .728 & \scriptsize .728 & \scriptsize .729 & \scriptsize .729 & \scriptsize .729 & \scriptsize .730 & \scriptsize .730 & \scriptsize .730 & \scriptsize .730 & \scriptsize .728 \\
   \hline
   \scriptsize \textbf{GoodReads (LCC)} & \scriptsize .713 & \scriptsize .713 & \scriptsize .713 & \scriptsize .709 & \scriptsize .710 & \scriptsize .709 & \scriptsize .714 & \scriptsize .711 & \scriptsize .710 & \scriptsize .712 & \scriptsize .713 & \scriptsize .713 & \scriptsize .711 & \scriptsize .713 & \scriptsize .710 \\
   \hline
  \end{tabular}
 \end{center}
 \caption{Accuracy results of tag-based classification relying with different number of tags in the top.}
 \label{tab:topx}
\end{sidewaystable}

%% file: key-terms.tex
\chapter{Key Terms and Definitions}
\label{c:key-terms}

Next, we list and provide the definitions for some of the most relevant terms related to this thesis, which help to better understand social tagging systems:

\begin{description}
 \item[Tagging] Tagging is an open way to assign tags or keywords to resources or items (e.g., web pages, movies or books), in order to describe them. This enables the later retrieval of the resources in an easier way, using tags as resource metadata. As opposed to a classical taxonomy-based categorization system, they are usually non-hierarchical, and the vocabulary is open, so it tends to grow indefinitely. For instance, a user could tag this thesis as social-tagging, research and thesis, whereas another user could use web2.0, social-bookmarking and tagging tags to annotate it.
 \item[Social tagging] A tagging system becomes social when its tag annotations are publicly visible, and profitable for anyone. The fact of a tagging system being social implies that a user could take advantage of tags defined by others to retrieve a resource.
 \item[Social bookmarking] Delicious, StumbleUpon and Diigo, amongst others, are known as social bookmarking sites. They provide a social means to save web pages (or other online resources like images or videos) as bookmarks, in order to retrieve them later on. In contrast to saving bookmarks in user's local browser, posting them to social bookmarking sites allows the community to discover others' links and, besides, to access the bookmarks from any computer to the user itself. In these systems, bookmarks represent references to web resources, and do not attach a copy of them, but just a link. Note that social bookmarking sites do not always rely on social tags to organize resources, e.g., Reddit is a social bookmarking approach to add comments on web pages instead of tags. However, the use of social tags in social bookmarking systems is a common approach.
 \item[Social cataloging] They are quite similar to social bookmarking sites in that resources are socially shared but, in this case, offline resources like music, books or movies are saved. For instance, LibraryThing allows to save the books you like, Hulu does it for movies and TV series, and Last.fm for music-related resources. As in social bookmarking sites, tags are the most common way to annotate resources in social cataloging sites.
 \item[Folksonomy] As a result of a community tagging resources, the collection of tags defined by them creates a tag-based organization, so-called folksonomy. A folksonomy is also known as a community-based taxonomy, where the classification scheme is plain, there are no predefined tags, and therefore users can freely choose new words as tags. A folksonomy is basically known as weighted set of tags, and may refer to a whole collection/site, a resource or a user. A summary of a folksonomy is usually presented in the form of a tag cloud.
 \item[Personomy] Personomy is a neologism created from the term folksonomy, and it refers to the weighted set of tags of a single user/person. It summarizes the topics a user tags about.
 \item[Simple tagging] users describe their own resources or items, such as photos on Flickr, news on Digg or videos on Youtube, but nobody else tags another user's resources. Usually, the author of the resource is who tags it. This means no more than one user tags an item. In many cases, like in Flickr and Youtube, simple tagging systems include an attachment to the resource, and not just a reference to it.
 \item[Collaborative tagging] many users tag the same item, and every person can tag it with their own tags in their own vocabulary. The collection of tags assigned by a single user creates a smaller folksonomy, also known as personomy. As a result, several users tend to post the same item. For instance, CiteULike, LibraryThing and Delicious are based on collaborative tagging, where each resource (papers, books and URLs, respectively) could be annotated by all the users who considered it interesting.
\end{description}

%% file: acronyms.tex
\chapter{List of Acronyms}
\label{c:acronyms}

This is a list of acronyms used in this thesis:

\begin{description}
 \item[API] Application Programming Interface
 \item[DDC] Dewey Decimal Classification
 \item[FTA] Full Tagging Activity
 \item[HTML] HyperText Markup Language
 \item[IBF] Inverse Bookmark Frequency
 \item[IDF] Inverse Document Frequency
 \item[IRF] Inverse Resource Frequency
 \item[IUF] Inverse User Frequency
 \item[LCC] Library of Congress Classification
 \item[ODP] Open Directory Project
 \item[ORPHAN] Orphan Ratio
 \item[TPP] Tags Per Post
 \item[TRR] Tag Resource Ratio
 \item[URL] Uniform Resource Locator
 \item[SVM] Support Vector Machines
 \item[S$^{3}$VM] Semi-Supervised Support Vector Machines
 \item[TF] Term Frequency
 \item[VSM] Vector Space Model
\end{description}

%% file: sp-summary.tex
\chapter{Resumen (Spanish Summary)}
\label{c:sp-summary}

\selectlanguage{spanish}

\textit{``El experimentador que no sabe lo que est\'{a} buscando no
comprender\'{a} lo que encuentra.''}

--- Claude Bernard

\translatedtitle{Utilizaci\'{o}n de Folksonom\'{i}as para Clasificaci\'{o}n de
Recursos}

\chaptersummary{En esta tesis abordamos el problema de la clasificaci\'{o}n
autom\'{a}tica de recursos, una tarea cada vez m\'{a}s importante en nuestra
vida diaria. El catalogado de libros o la organizaci\'{o}n de v\'{i}deos, entre otros, 
representan algunos ejemplos de actividades para las que un proceso autom\'{a}tico 
de clasificaci\'{o}n resulta cada vez m\'{a}s necesario e importante en nuestro d\'{i}a a d\'{i}a. 
En esta tesis aprovechamos la informaci\'{o}n contenida en las anotaciones que realizan los 
usuarios de sistemas de etiquetado social, en los cuales se recogen metadatos 
que detallan el contenido de diferentes tipos de recursos, para mejorar la clasificaci\'{o}n. 
Hasta el momento, son pocos los trabajos que han explotado estos metadatos con este fin, 
y los pocos que lo han hecho se han limitado a realizar an\'{a}lisis estad\'{i}sticos. 
En esta tesis exploramos las caracter\'{i}sticas de estos sistemas de etiquetado social y de los
usuarios involucrados en ellos, as\'{i} como de las anotaciones que aportan, siempre con el
fin de sacar el m\'{a}ximo partido a estas grandes colecciones, obteniendo as\'{i} el mayor 
rendimiento posible para un clasificador autom\'{a}tico de recursos.}

\section{Motivaci\'{o}n}
\label{sec:motivacion}

Organizar recursos dentro de categor\'{i}as supone una tarea muy com\'{u}n en nuestro d\'{i}a a d\'{i}a. 
Tener recursos asignados a categor\'{i}as predefinidas siempre ayuda
a mejorar posteriores accesos a la informaci\'{o}n contenida en ellos, ya que 
este acceso puede limitarse entonces a un conjunto reducido de categor\'{i}a(s) deseada(s). 
Por ejemplo, los bibliotecarios suelen catalogar los libros por temas, de forma que quedan 
organizados por intereses similares. Las bases de datos de pel\'{i}culas, los cat\'{a}logos de m\'{u}sica 
y los sistemas de ficheros, entre otros, suelen estar organizados tambi\'{e}n por categor\'{i}as, 
de forma que se facilita su acceso futuro. Asimismo, la clasificaci\'{o}n de p\'{a}ginas web 
resulta una tarea de especial inter\'{e}s a la hora de mejorar los resultados provistos por los motores 
de b\'{u}squeda, ya que ayudan a reducir el \'{a}mbito de esta b\'{u}squeda a la categor\'{i}a deseada por el usuario. 
Directorios web como Yahoo! Directory y Open Directory Project organizan p\'{a}ginas web en categor\'{i}as, 
ofreciendo una alternativa o complemento a la b\'{u}squeda por palabra(s) clave(s).

El problema entonces est\'{a} en lo costosa y cara que resulta la clasificaci\'{o}n manual de estos recursos 
cuando la colecci\'{o}n es grande. Por ejemplo, The Library of Congress de Estados Unidos inform\'{o} en 
2002 de que el coste medio de catalogaci\'{o}n de cada registro bibliogr\'{a}fico por profesionales fue de 94,58
d\'{o}lares\footnote{http://www.loc.gov/loc/lcib/0302/collections.html}. Catalogar 291.749 registros, 
como hicieron en aquel a\~{n}o, les lleg\'{o} a costar unos 27 millones y medio de d\'{o}lares. Dado lo cara 
que resulta la categorizaci\'{o}n manual, la utilizaci\'{o}n de clasificadores autom\'{a}ticos puede ser una buena
alternativa para reducir su coste, y asimismo mantener los cat\'{a}logos al d\'{i}a con un esfuerzo humano menor.

Hasta el momento, la mayor\'{i}a de los clasificadores autom\'{a}ticos se han 
centrado en el contenido de los recursos a la hora de representarlos, 
sobre todo en tareas de clasificaci\'{o}n de p\'{a}ginas web (\cite{qi_webpage_2009}). 
No obstante, la falta de datos representativos en el contenido de muchos de ellos hace que se
complique esta tarea. Adem\'{a}s, puede resultar muy complicado obtener
suficientes datos sobre tipos de recursos como libros o pel\'{i}culas,
para los cuales puede ser m\'{a}s complicado representar el contenido o,
incluso, puede que el contenido no est\'{e} disponible en una forma que pueda ser
procesado.

Como soluci\'{o}n a este problema, los sistemas de etiquetado social proveen una
forma sencilla y barata de obtener metadatos sobre recursos.
Sistemas como Delicious\footnote{http://delicious.com},
LibraryThing\footnote{http://www.librarything.com} y
GoodReads\footnote{http://www.goodreads.com} recopilan anotaciones de usuarios
en forma de etiquetas para grandes colecciones de recursos. Estas etiquetas
provistas por usuarios dan lugar a datos significativos que describen el
contenido de los recursos \citep{heymann_can_social_2008}.

Por medio de estas etiquetas, los usuarios proveen una especie de
organizaci\'{o}n propia de los recursos. Estas etiquetas se comparten de forma
social con la comunidad, y gracias a que un gran n\'{u}mero de usuarios contribuye
en estos sistemas, son numerosas las anotaciones que se acumulan sobre cada recurso. 
Por lo tanto, 
esa acumulaci\'{o}n hace que cada una de las anotaciones sea m\'{a}s \'{u}til.
As\'{i}, la acumulaci\'{o}n de usuarios en una comunidad activa genera un gran
n\'{u}mero de marcadores, etiquetas, y por tanto, recursos anotados.

\begin{quote}
 \textit{``Cada una de las categorizaciones individuales vale menos que la
categorizaci\'{o}n de un profesional. Pero hay muchas, muchas de
aqu\'{e}llas.''}, Joshua Schachter, fundador de Delicious, en la cumbre FOWA
2006 FOWA en Londres
(Inglaterra)\footnote{http://simonwillison.net/2006/Feb/8/summit/}.
\end{quote}

Los sistemas de etiquetado social representan un medio para guardar, organizar y buscar
recursos, todo ello por medio de la anotaci\'{o}n con etiquetas escogidas por el usuario. 
Como hip\'{o}tesis principal de este trabajo, creemos que
estas grandes colecciones de anotaciones pueden mejorar de forma considerable una
tarea de clasificaci\'{o}n de recursos. Dicho de otro modo, las anotaciones provistas por usuarios
podr\'{i}an llegar a ser muy \'{u}tiles como una fuente de datos que aporta informaci\'{o}n
significativa que podr\'{i}a ayudar a inferir la categor\'{i}a de los recursos.

Dado que un gran n\'{u}mero de usuarios provee sus propias anotaciones sobre
cada recurso, nuestro objetivo entonces se centra en descubrir la manera de amalgamar
esas aportaciones en busca de una organizaci\'{o}n que se parezca a la
categorizaci\'{o}n realizada por profesionales. En este contexto, donde los
usuarios aportan grandes cantidades de metadatos, nuestro reto se centra en
sacar el m\'{a}ximo partido de ellos con el fin de mejorar el rendimiento de la
tarea de clasificaci\'{o}n de recursos.

\begin{quote}
 \textit{``Estamos en una \'{e}poca en la que los datos son baratos, pero sacar
partido de ellos no lo es''}, Danah Boyd, Investigadora sobre Social Media en
Microsoft Research New England, en el congreso WWW2010 en Raleigh, Carolina del
Norte, Estados
Unidos\footnote{http://www.danah.org/papers/talks/2010/WWW2010.html}.
\end{quote}

\subsection{Clasificaci\'{o}n de Recursos}
\label{clasificacion-recursos}

La clasificaci\'{o}n de recursos se puede definir como la tarea consistente en
la organizaci\'{o}n de recursos dentro de un conjunto de categor\'{i}as predefinidas. 
En este trabajo utilizamos las M\'{a}quinas de Vectores de Soporte (SVM,
\cite{joachims98text}), un m\'{e}todo vanguardista para clasificaci\'{o}n que ha destacado 
por sus buenos resultados desde finales de los a\~{n}os 90. 
Este algoritmo de clasificaci\'{o}n se basa en el an\'{a}lisis de un conjunto de instancias
previamente categorizadas, con lo que se alimenta el clasificador para que
adquiera el conocimiento necesario para poder clasificar posteriormente nuevos recursos.

Un problema de clasificaci\'{o}n de recursos puede definirse a partir de diferentes
caracter\'{i}sticas. Por una parte, en lo que se refiere al m\'{e}todo de
aprendizaje, puede ser \textit{supervisado}, donde todo el conjunto de
entrenamiento est\'{a} previamente categorizado, o \textit{semisupervisado}, donde
tambi\'{e}n se aprovechan instancias sin informaci\'{o}n de categor\'{i}a durante la
fase de aprendizaje. Por otra parte, considerando el n\'{u}mero de clases, la
clasificaci\'{o}n puede ser \textit{binaria}, cuando s\'{o}lo hay dos
categor\'{i}as que pueden ser asignadas a cada recurso, o \textit{multiclase},
cuando hay tres o m\'{a}s categor\'{i}as. El primer caso se utiliza habitualmente
para sistemas de filtrado, mientras que el segundo suele ser frecuente en el
caso de taxonom\'{i}as mayores, como en el caso de la clasificaci\'{o}n
tem\'{a}tica de recursos.

Para clasificaci\'{o}n tem\'{a}tica sobre grandes colecciones de recursos, como
p\'{a}ginas web en la Web o libros en bibliotecas, las taxonom\'{i}as suelen
estar definidas por m\'{a}s de dos categor\'{i}as, y el subconjunto de recursos
previamente categorizado suele ser muy peque\~{n}o. De esta manera, creemos que
se deber\'{i}a considerar y analizar la aplicaci\'{o}n de t\'{e}cnicas
semisupervisadas y multiclase para este tipo de tareas.

Por ello, en esta tesis proponemos inicialmente el an\'{a}lisis de varias t\'{e}cnicas de
clasificaci\'{o}n que utilizan SVM, con el fin de analizar su adecuaci\'{o}n a
estas tareas. Estas t\'{e}cnicas incluyen diferentes aproximaciones a la
resoluci\'{o}n de tareas multiclase, as\'{i} como algoritmos supervisados y
semisupervisados.

\subsection{Anotaciones Sociales}
\label{anotaciones-sociales}

Los sistemas de etiquetado social permiten a sus usuarios guardar y anotar sus
recursos favoritos (como por ejemplo p\'{a}ginas web, pel\'{i}culas, libros,
fotos o m\'{u}sica), comparti\'{e}ndolos a su vez con la comunidad. Los usuarios
proveen estas anotaciones normalmente en forma de etiquetas. Se conoce como
etiquetado a la forma abierta de asignar etiquetas o palabras clave a recursos,
de manera que se pueden describir y organizar. Esto posibilita la posterior
recuperaci\'{o}n de los recursos de forma m\'{a}s sencilla, aprovechando las
etiquetas como metadatos que los describen. Normalmente, no hay etiquetas
predefinidas, y por lo tanto los usuarios pueden escoger libremente las palabras
que deseen como etiquetas.

\begin{quote}
 \textit{``El etiquetado es principalmente una interfaz de usuario - una manera
para que la gente recuerde cosas, en qu\'{e} estaban pensando en el momento en
el que lo guardaron. Bastante \'{u}til para recordar, bueno para el
descubrimiento, terrible para la distribuci\'{o}n (donde los que lo publican
a\~{n}aden tantas etiquetas como pueden para incluirlo en el mayor n\'{u}mero
posible de cajas).''}, Joshua Schachter, fundador de Delicious, en la cumbre
FOWA 2006 FOWA en Londres
(Inglaterra)\footnote{http://simonwillison.net/2006/Feb/8/summit/}.
\end{quote}

Mediante este proceso se genera una
estructura de etiquetas conocida como folksonom\'{i}a, es decir, una
organizaci\'{o}n de recursos dirigida por usuarios. Folksonom\'{i}a es una
contracci\'{o}n de las palabras \textit{folk} (gente), \textit{taxis}
(clasificaci\'{o}n) y \textit{nomos} (gesti\'{o}n). Es conocida tambi\'{e}n como
una taxonom\'{i}a basada en los usuarios, en la cual la estructura no es
jer\'{a}rquica, al contrario que una clasificaci\'{o}n taxon\'{o}mica
b\'{a}sica. Por lo tanto, una folksonom\'{i}a tiene cierta relaci\'{o}n con las
taxonom\'{i}as generadas por expertos, en cuanto a que los recursos se organizan
igualmente en grupos.

Se dice que estas anotaciones pertenecen a un entorno social cuando est\'{a}n
accesibles y utilizables para cualquier usuario. Esta caracter\'{i}stica
posibilita la b\'{u}squeda de recursos aprovechando las anotaciones aportadas 
por otros. A su vez, es uno de los motivos que anima a los usuarios a
contribuir.

\begin{figure}[ht]
\begin{center}
 \includegraphics[width=100mm,clip]{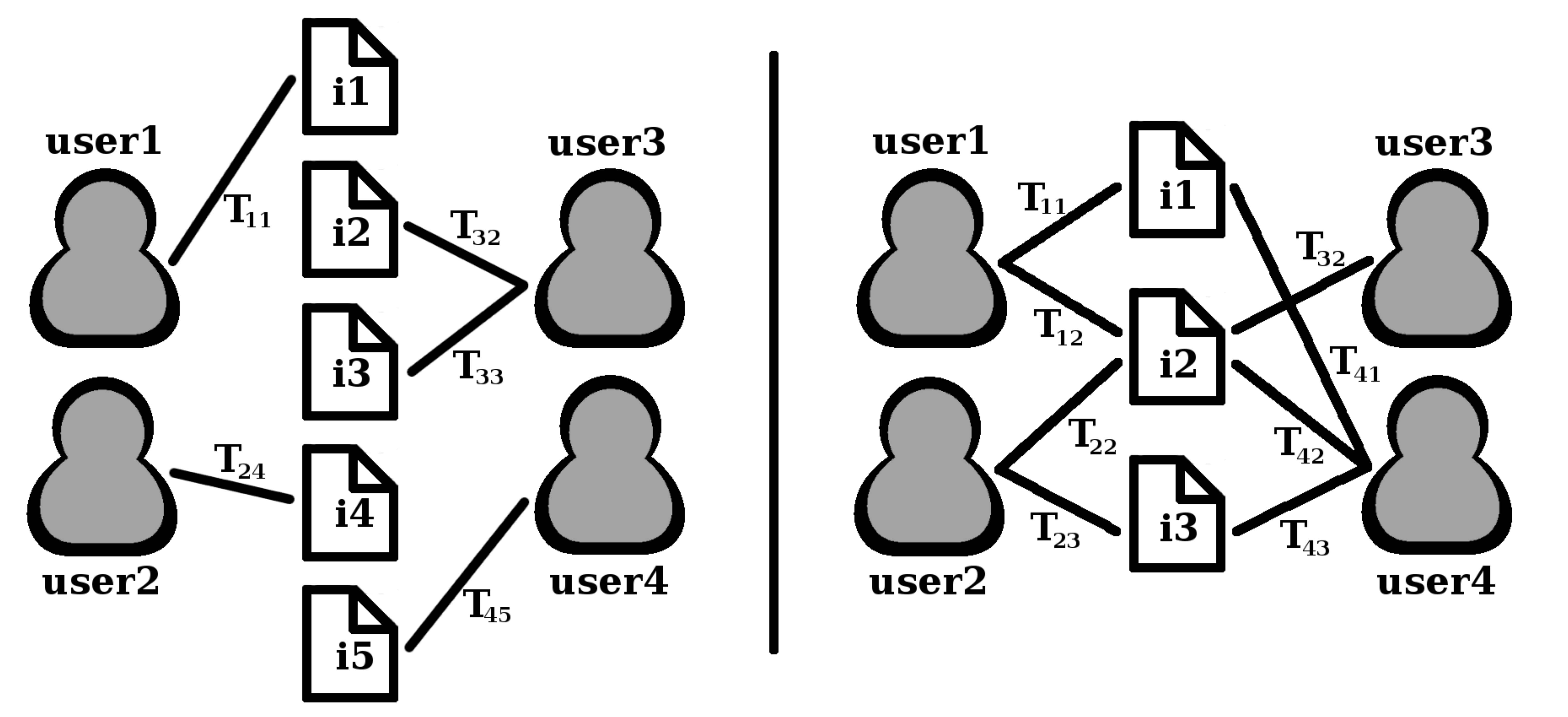}
\end{center}
\caption[Etiquetado simple y etiquetado colaborativo]{Comparaci\'{o}n de
anotaciones provistas por usuarios en sistemas de etiquetado simple y
colaborativo.}
\label{fig:simple-vs-colaborativo}
\end{figure}

No obstante, no todas las anotaciones se comparten de la misma manera. El propio
sistema de etiquetado social puede definir algunas restricciones a este
respecto, principalmente estableciendo qui\'{e}n tiene permiso para anotar cada
recurso. En este sentido, se pueden distinguir dos tipos de sistemas
\citep{smith_tagging_2008}:

\begin{itemize}
 \item \textbf{Sistemas de etiquetado simple:} los usuarios pueden describir sus
propios recursos, como es el caso del etiquetado de fotos en Flickr\footnote{http://www.flickr.com},
noticias en Digg\footnote{http://digg.com} o v\'{i}deos en
Youtube\footnote{http://www.youtube.com}, pero nadie anota los recursos de
otros. Generalmente, el autor del recurso es quien lo anota. Esto significa que
no m\'{a}s de un usuario puede etiquetar cada recurso. M\'{a}s formalmente, en un sistema de
etiquetado simple hay un conjunto de usuarios ($U$) que anota unos recursos
($R$) con unas etiquetas ($T$). Cada usuario $u_{i} \in U$ puede guardar un
recurso $r_{j} \in R$ con un conjunto de etiquetas $T_{j} =
\{t_{j1},...,t_{jp}\}$, con un n\'{u}mero $p$ variable de etiquetas. El conjunto
de etiquetas asignado a $r_{j}$ seguir\'{a} estando limitado a $T_{j}$, ya que
nadie m\'{a}s lo podr\'{a} anotar.
 
 \item \textbf{Sistemas de etiquetado colaborativo:} muchos usuarios pueden
anotar cada recurso, y todos ellos pueden etiquetarlo con su
propio vocabulario. El conjunto de etiquetas asignado por un usuario genera una
folksonom\'{i}a a menor escala, conocida como personom\'{i}a. Como resultado,
varios usuarios tienden a anotar el mismo recurso. Por ejemplo, CiteULike.org,
LibraryThing.com y Delicious se basan en anotaciones colaborativas, donde cada
recurso (art\'{i}culos, libros y URLs, respectivamente) puede ser anotado y
etiquetado por todos aquellos usuarios que lo consideren interesante. Por tanto, los
sistemas de etiquetado colaborativo son algo m\'{a}s complejos, ya que hay un
conjunto de usuarios ($U$) que guarda sus marcadores ($B$) sobre unos recursos
($R$) anot\'{a}ndolos con unas etiquetas ($T$). Cada usuario $u_{i} \in U$ puede
guardar un marcador $b_{ij} \in B$ de un recurso $r_{j} \in R$ con un conjunto
de etiquetas $T_{ij} = \{t_{ij1},...,t_{ijp}\}$, con un n\'{u}mero $p$ variable
de etiquetas. Despu\'{e}s de que $k$ usuarios guardan $r_{j}$, se describe como
un conjunto pesado de etiquetas $T_{j} = \{w_{j1} t_{j1},...,w_{jn} t_{jn}\}$,
donde $w_{j1},...,w_{jn} \leq k$ representan el n\'{u}mero de asignaciones de
cada etiqueta. Por lo tanto, cada marcador est\'{a} compuesto por la tripleta de
un usuario, un recurso y un conjunto de etiquetas: $b_{ij}: u_{i} \times r_{j}
\times T_{ij}$. As\'{i}, cada usuario guarda marcadores de diferentes recursos,
y cada recurso tiene marcadores correspondientes a diferentes usuarios. El
resultado de acumular etiquetas contenidas en los marcadores de un usuario se
conoce como la personom\'{i}a de ese usuario: $T_{i} = \{w_{i1}
t_{i1},...,w_{im} t_{im}\}$, donde $m$ es el n\'{u}mero de etiquetas diferentes
en la personom\'{i}a del usuario.

La
Figura \ref{fig:simple-vs-colaborativo} muestra un ejemplo comparativo de ambos
tipos de sistemas.
\end{itemize}

En esta tesis nos centramos en sistemas de etiquetado colaborativo. 
Generalmente, las etiquetas asociadas a un recurso tienden a coincidir entre usuarios,
haciendo de esta coincidencia algo especialmente \'{u}til en comparaci\'{o}n con las etiquetas 
que encontramos en sistemas de etiquetado simple.

En un sistema de etiquetado colaborativo, como ejemplo, un usuario podr\'{i}a
etiquetar este trabajo como \texttt{etiquetado-social},
\texttt{investigaci\'{o}n} y \texttt{tesis}, mientras que otro usuario
podr\'{i}a utilizar las etiquetas \texttt{etiquetado-social},
\texttt{marcadores-sociales}, \texttt{doctorado} y \texttt{tesis} para anotarlo.
El comportamiento de los usuarios puede diferir de forma considerable en estos
sistemas, donde la acumulaci\'{o}n de sus anotaciones se suele considerar como
consenso. Por ejemplo, el resultado de la acumulaci\'{o}n mediante suma de las
anotaciones de arriba ser\'{i}a el siguiente: \texttt{tesis} (2),
\texttt{etiquetado-social} (2), \texttt{marcadores-sociales} (1),
\texttt{doctorado} (1) e \texttt{investigaci\'{o}n} (1).


En esta tesis, analizamos y estudiamos las anotaciones provistas por usuarios en
sistemas de etiquetado social. Presentamos un estudio con el fin de sacar el
m\'{a}ximo partido de ellas, con vistas a mejorar el rendimiento de una tarea de
clasificaci\'{o}n de recursos. Concretamente, nos centramos en el an\'{a}lisis
de la utilidad de las folksonom\'{i}as generadas por usuarios como
aproximaci\'{o}n a una organizaci\'{o}n parecida a las taxonom\'{i}as creadas
por expertos. En este contexto, estudiamos diferentes representaciones basadas en el uso de
anotaciones sociales, en busca de una aproximaci\'{o}n que se parezca a la
clasificaci\'{o}n provista por expertos en la mayor media posible. Nos centramos
en obtener el m\'{a}ximo de las etiquetas sociales, tanto buscando la mejor
representaci\'{o}n, como midiendo el impacto que puede tener en este sentido la
distribuci\'{o}n de las etiquetas sobre recursos, marcadores y usuarios.
Finalmente, tambi\'{e}n estudiamos la aplicaci\'{o}n de t\'{e}cnicas
vanguardistas de an\'{a}lisis del comportamiento de los usuarios en estos
sistemas, con el fin de detectar usuarios cuyas anotaciones est\'{e}n m\'{a}s
pr\'{o}ximas a la clasificaci\'{o}n creada por expertos.

\section{Objetivos}
\label{sec:objetivos}

El objetivo principal de esta tesis se centra en aportar nuevo conocimiento
sobre el uso apropiado de la gran cantidad de datos que se pueden encontrar en
los sistemas de etiquetado social. Dado el inter\'{e}s en clasificar recursos, y
la falta de datos representativos, nos centramos en analizar en qu\'{e} medida y
de qu\'{e} manera las etiquetas sociales pueden mejorar la tarea de
clasificaci\'{o}n de recursos. Al comienzo de este trabajo comprobamos que no
hab\'{i}a investigaciones que abordaran este problema; por lo tanto, nos motiv\'{o} a
llevar a cabo esta investigaci\'{o}n. 
Con este fin, hemos definido
el siguiente planteamiento del problema, el cual resume el objetivo principal de esta tesis:

\begin{description}
 \item[Planteamiento del Problema] \hfill \\
 \textit{\problemstatementes}
\end{description}

\section{Metodolog\'{i}a}
\label{sec:metodologia}

La metodolog\'{i}a de investigaci\'{o}n seguida a lo largo del trabajo se
compone de las siguientes 6 partes:

\begin{enumerate}
 \item Revisi\'{o}n y lectura del estado del arte, as\'{i} como estudiar y
comprender detalladamente el funcionamiento de los sistemas de etiquetado
social.
 \item B\'{u}squeda de un clasificador SVM apropiado para llevar a cabo la investigaci\'{o}n.
 \item B\'{u}squeda de colecciones existentes con informaci\'{o}n extra\'{i}da de sistemas de
etiquetado social. Como no encontramos ninguno que cumpliera nuestros
requisitos, hubo que crear tres colecciones de gran escala en su lugar.
 \item Pensar y proponer aproximaciones que se ajusten a la tarea de
clasificaci\'{o}n basada en etiquetas sociales.
 \item Evaluaci\'{o}n de las aproximaciones propuestas.
 \item Realizaci\'{o}n de un riguroso an\'{a}lisis de los resultados, con el fin
de llegar a unas conclusiones s\'{o}lidas.
 \item Presentaci\'{o}n de resultados parciales en congresos y talleres
nacionales e internacionales, con el fin de obtener comentarios y sugerencias de
otros investigadores.
 \item Resumir en esta tesis la investigaci\'{o}n, aportaciones, y conclusiones
alcanzadas a lo largo de todo el trabajo.
\end{enumerate}

Del paso 4 al 6, se realiz\'{o} un proceso iterativo, realiz\'{a}ndose dichos
pasos de forma repetida varias veces.

\section{Estructura de la Tesis}
\label{sec:organizacion-tesis}

Esta tesis est\'{a} compuesta de 8 cap\'{i}tulos. A continuaci\'{o}n resumimos
brevemente el contenido de cada uno de ellos:

\begin{description}
 \item[Cap\'{i}tulo \ref{c:introduction} en la p\'{a}gina
\pageref{c:introduction}] \hfill \\
 \textbf{Introducci\'{o}n} \\
  Presentamos la motivaci\'{o}n para el estudio del uso de anotaciones
sociales para clasificaci\'{o}n de recursos. Formalizamos el problema y
motivamos la necesidad de realizar dicho estudio.
  
 \item[Cap\'{i}tulo \ref{c:related-work} en la p\'{a}gina
\pageref{c:related-work}] \hfill \\
 \textbf{Trabajo Relacionado} \\
  Ofrecemos un resumen de los trabajos previos en el campo de investigaci\'{o}n.
Resumimos los avances en campos relacionados, tanto en cuanto al uso de
anotaciones sociales, como en cuanto a la clasificaci\'{o}n de recursos.
  
 \item[Cap\'{i}tulo \ref{c:svm-classification} en la p\'{a}gina
\pageref{c:svm-classification}] \hfill \\
 \textbf{M\'{a}quinas de Vectores de Soporte para Clasificaci\'{o}n a Gran
Escala} \\
  Realizamos un estudio de diferentes aproximaciones SVM para resolver el
problema de la clasificaci\'{o}n de grandes colecciones de recursos sobre
taxonom\'{i}as multiclase. Damos con la mejor aproximaci\'{o}n SVM para estos
casos, y la utilizamos a lo largo del trabajo para realizar las tareas de
clasificaci\'{o}n.
  
 \item[Cap\'{i}tulo \ref{c:datasets} en la p\'{a}gina \pageref{c:datasets}]
\hfill \\
 \textbf{Creaci\'{o}n de Colecciones de Etiquetado Social} \\
  Describimos y analizamos en detalle las colecciones de etiquetado social utilizadas 
  en esta tesis. Detallamos el proceso de generaci\'{o}n
de dichas colecciones, y analizamos las principales caracter\'{i}sticas de sus
correspondientes folksonom\'{i}as.
  
 \item[Cap\'{i}tulo \ref{c:tag-representation} en la p\'{a}gina
\pageref{c:tag-representation}] \hfill \\
 \textbf{Representando la Acumulaci\'{o}n de Etiquetas} \\
  Proponemos y evaluamos diferentes representaciones de recursos que emplean informaci\'{o}n de
etiquetas sociales para la tarea de clasificaci\'{o}n de recursos. Estudiamos la
utilidad de las etiquetas sociales en comparaci\'{o}n a otras fuentes de datos,
y proponemos una representaci\'{o}n que saca el m\'{a}ximo partido de ellas.
Tambi\'{e}n abordamos el problema combinando las etiquetas sociales con las
otras fuentes de datos disponibles para obtener un mejor rendimiento.
  
 \item[Cap\'{i}tulo \ref{c:tag-distribution-classification} en la p\'{a}gina
\pageref{c:tag-distribution-classification}] \hfill \\
 \textbf{Analizando la Distribuci\'{o}n de Etiquetas para Clasificaci\'{o}n de
Recursos} \\
  Abordamos la idea de considerar la representatividad de las etiquetas
dentro de una colecci\'{o}n de anotaciones de un sistema de etiquetado social.
Estudiamos la aplicaci\'{o}n de funciones de pesado adaptadas a estos sistemas,
y analizamos su adecuaci\'{o}n teniendo en cuenta las configuraciones de cada
sistema.
  
 \item[Cap\'{i}tulo \ref{c:analyzing-appropriateness-users} en la p\'{a}gina
\pageref{c:analyzing-appropriateness-users}] \hfill \\
 \textbf{Analizando el Comportamiento de Usuarios para la Clasificaci\'{o}n} \\
  Exploramos el efecto que puede tener el comportamiento de usuarios en sistemas de etiquetado
social con vistas a una tarea de clasificaci\'{o}n de recursos. Bas\'{a}ndonos en
trabajos previos que sugieren la existencia de ciertos usuarios que tienden a
categorizar recursos, estudiamos si realmente se ajustan en mayor medida a la
clasificaci\'{o}n de recursos.
  
 \item[Cap\'{i}tulo \ref{c:conclusions} en la p\'{a}gina
\pageref{c:conclusions}] \hfill \\
 \textbf{Conclusiones y Trabajo Futuro} \\
  Resumimos y comentamos las principales conclusiones y aportaciones del
trabajo. Presentamos las respuestas a las preguntas de investigaci\'{o}n
formuladas al inicio, y planteamos el trabajo futuro.
\end{description}

Adem\'{a}s, la tesis contiene los siguientes ap\'{e}ndices al final, con
informaci\'{o}n adicional y res\'{u}menes en otros idiomas:

\begin{description}
 \item[Ap\'{e}ndice \ref{c:topx-tags} en la p\'{a}gina \pageref{c:topx-tags}]
\hfill \\
 \textbf{Resultados Adicionales} \\
  Presentamos algunos resultados adicionales, los cuales decidimos no incluir en
el contenido de la tesis por claridad, pero que merece la pena mostrar ya que ayudan a demostrar
y entender algunas conclusiones.
  
 \item[Ap\'{e}ndice \ref{c:key-terms} en la p\'{a}gina \pageref{c:key-terms}]
\hfill \\
 \textbf{Palabras Clave y Definiciones} \\
  Listamos los t\'{e}rminos m\'{a}s relevantes relacionados con los sistemas de
etiquetado social y proporcionamos definiciones detalladas.
  
 \item[Ap\'{e}ndice \ref{c:acronyms} en la p\'{a}gina \pageref{c:acronyms}]
\hfill \\
 \textbf{Lista de Acr\'{o}nimos} \\
  Listamos los acr\'{o}nimos utilizados a lo largo de este trabajo e indicamos a qu\'{e}
se refieren.
  
 \item[Ap\'{e}ndice \ref{c:sp-summary} en la p\'{a}gina \pageref{c:sp-summary}]
\hfill \\
 \textbf{Resumen} \\
  Resumen del contenido de este trabajo en castellano.
  
 \item[Ap\'{e}ndice \ref{c:bq-summary} en la p\'{a}gina \pageref{c:bq-summary}]
\hfill \\
 \textbf{Laburpena (Resumen en euskera)} \\
  Resumen del contenido de este trabajo en euskera.
\end{description}

\section{Preguntas de Investigaci\'{o}n Resueltas}
\label{sec:preguntas-investigacion-resueltas}

\begin{description}
 \item[Pregunta de Investigaci\'{o}n 1] \hfill \\
 \textit{\rqsvmonees}
\end{description}

Se ha demostrado una clara superioridad de los clasificadores SVM multiclase
nativos sobre las otras aproximaciones que combinan clasificadores binarios. Los
resultados muestran que basarse en un conjunto de clasificadores binarios no es
una buena opci\'{o}n cuando se trata de taxonom\'{i}as multiclase. Por lo tanto,
los clasificadores multiclase nativos, que consideran todas las clases al
mismo tiempo y tienen m\'{a}s conocimiento de la tarea completa, funcionan mejor
para estos casos.

\begin{description}
 \item[Pregunta de Investigaci\'{o}n 2] \hfill \\
 \textit{\rqsvmtwoes}
\end{description}

Los m\'{e}todos semisupervisados podr\'{i}an rendir mejor cuando el subconjunto
etiquetado es muy peque\~{n}o, pero los m\'{e}todos supervisados,
computacionalmente menos costosos, consiguen un rendimiento muy similar con unas
pocas instancias m\'{a}s etiquetadas. Por lo tanto, hemos mostrado tambi\'{e}n
que, a diferencia de las tareas de clasificaci\'{o}n binarias como ya
demostr\'{o} \cite{joachims99transductive}, un m\'{e}todo supervisado obtiene
unos resultados muy similares a los de un semisupervisado para estos casos de colecciones grandes y multiclase.
Parece razonable pensar que predecir la clase de las instancias no etiquetadas
es mucho m\'{a}s dif\'{i}cil con el incremento del n\'{u}mero de clases y, por
tanto, el incremento de errores en las predicciones se refleja tambi\'{e}n en la
fase de aprendizaje del clasificador.

Bas\'{a}ndonos en estas conclusiones, decidimos utilizar un
clasificador SVM multiclase supervisado a lo largo de esta tesis.

\begin{description}
 \item[Pregunta de Investigaci\'{o}n 3] \hfill \\
 \textit{\rqdataonees}
\end{description}

Con este fin, hemos analizado diversas caracter\'{i}sticas que se encuentran en
la configuraci\'{o}n de los sistemas de etiquetado social. Entre las
caracter\'{i}sticas analizadas, hemos mostrado el gran impacto de las
sugerencias en el etiquetado, lo cual altera de forma considerable la
folksonom\'{i}a resultante. En los sistemas de etiquetado social que hemos
estudiado, todos presentan alguna caracter\'{i}stica diferente en este aspecto:

\begin{itemize}
 \item \textbf{Sugerencias basadas en recursos (Delicious):} cuando el sistema
sugiere etiquetas asignadas por otros usuarios al recurso que se est\'{a}
guardando, se reduce la probabilidad de utilizar nuevas etiquetas que aporten nueva
informaci\'{o}n. En este caso, los usuarios dedican poco esfuerzo a pensar por
ellos mismos, y prefieren basarse en las sugerencias provistas por el sistema.
 \item \textbf{Sugerencias basadas en la personom\'{i}a (GoodReads):} cuando el
sistema sugiere etiquetas que el mismo usuario ha utilizado previamente, el
vocabulario de su personom\'{i}a tiende a ser mucho m\'{a}s reducido. No
obstante, los usuarios no saben qu\'{e} es lo que otros han anotado sobre cada
recurso, y por tanto es muy probable que aporten nuevas etiquetas que
anteriormente no se hab\'{i}an anotado sobre el recurso.
 \item \textbf{Ausencia de sugerencias (LibraryThing):} cuando el sistema no
sugiere etiquetas al usuario, el vocabulario de su personom\'{i}a tiende a ser
mayor, as\'{i} como las etiquetas asignadas a cada recurso son m\'{a}s diversas.
\end{itemize}

\begin{description}
 \item[Pregunta de Investigaci\'{o}n 4] \hfill \\
 \textit{\rqreponees}
\end{description}

Hemos demostrado que es mejor tener en cuenta todas las etiquetas anotadas sobre
un recurso que basarse s\'{o}lo en aqu\'{e}llas que han sido anotadas por
m\'{a}s usuarios. Las etiquetas m\'{a}s anotadas han demostrado ser las m\'{a}s
importantes, y aportan la informaci\'{o}n m\'{a}s relevante sobre la
tem\'{a}tica del recurso. No obstante, las etiquetas menos populares tambi\'{e}n
pueden ser \'{u}tiles en menor medida, aportando otro tipo de informaci\'{o}n \'{u}til que
mejora el rendimiento del clasificador.

En cuanto a los pesos que se asignan a las etiquetas al representar el recurso,
los mejores resultados se obtienen considerando el n\'{u}mero de usuarios que
anotan cada etiqueta. El uso de este valor ha producido los mejores resultados
en nuestros experimentos, superando a otras aproximaciones que ignoran estos
pesos, y demostrando que no hace falta considerar el n\'{u}mero total de
usuarios que anota el recurso.

Por lo tanto, a partir de nuestros experimentos, concluimos que la mejor 
representaci\'{o}n es aqu\'{e}lla que aprovecha todas las etiquetas, asignando como
peso el n\'{u}mero de usuarios que las ha anotado.

\begin{description}
 \item[Pregunta de Investigaci\'{o}n 5] \hfill \\
 \textit{\rqreptwoes}
\end{description}

Utilizando t\'{e}cnicas de combinaci\'{o}n de clasificadores, los cuales consideran las
predicciones de diferentes clasificadores, hemos demostrado que las etiquetas
aportan criterios fiables a tener en cuenta. Estos criterios son muy \'{u}tiles
para combinar dichas etiquetas con otras fuentes de datos. No obstante, no todas
las fuentes de datos son \'{u}tiles para combinar, y se deben seleccionar con
cautela las que obtienen unos resultados s\'{o}lidos y, adem\'{a}s, ofrecen unas
predicciones fiables. Cuando las fuentes de datos se escogen de manera
apropiada, la mejora de rendimiento es considerable.

\begin{description}
 \item[Pregunta de Investigaci\'{o}n 6] \hfill \\
 \textit{\rqrepthreees}
\end{description}

Hemos analizado la utilidad de las etiquetas sociales para la clasificaci\'{o}n
sobre dos niveles diferentes de las taxonom\'{i}as. Adem\'{a}s de las
categor\'{i}as de m\'{a}s alto nivel, tambi\'{e}n hemos explorado la
clasificaci\'{o}n sobre categor\'{i}as del segundo nivel, m\'{a}s precisas. En
este aspecto, los resultados usando etiquetas sociales han sido superiores a los obtenidos con otras fuentes de
datos para aquellos sistemas de etiquetado social que animan a los usuarios a
aportar anotaciones (Delicious y LibraryThing). La superioridad es muy clara en
estos casos, sobre todo para Delicious, donde la diferencia es a\'{u}n mayor
cuando se trata del segundo nivel taxon\'{o}mico. Esta diferencia es muy similar para
LibraryThing. Por \'{u}ltimo, las etiquetas de GoodReads no superan a las otras
fuentes de datos, ni siquiera para el primer nivel, ya que el sistema no anima a
los usuarios a anotar los libros, con lo que muchos de los marcadores se quedan
sin etiquetas.

Estos descubrimientos arrojan una conclusi\'{o}n diferente a la que dan
\cite{noll_exploring_2008}, donde los autores lanzan la hip\'{o}tesis de que las
etiquetas sociales podr\'{i}an no ser \'{u}tiles para niveles m\'{a}s bajos de
las taxonom\'{i}as, y que deber\'{i}an utilizarse otros tipos de datos para
estos casos.

\begin{description}
 \item[Pregunta de Investigaci\'{o}n 7] \hfill \\
 \textit{\rqdistonees}
\end{description}

A trav\'{e}s de la experimentaci\'{o}n llevada a cabo en esta tesis, hemos demostrado la
utilidad de considerar las distribuciones de etiquetas a lo largo de la colecci\'{o}n, por medio de una
funci\'{o}n de pesado inversa como la ofrecida por IDF. Estas funciones han
servido para determinar la representatividad de las etiquetas para cada
colecci\'{o}n, con el fin de mejorar el rendimiento de la tarea de
clasificaci\'{o}n de recursos. No obstante, hemos mostrado que la
configuraci\'{o}n del sistema de etiquetado social tiene mucho que ver con esas
distribuciones. Entre las caracter\'{i}sticas en la configuraci\'{o}n de los
sistemas, se ha visto que las sugerencias basadas en los
recursos influyen en gran medida la estructura de las folksonom\'{i}as
resultantes. Aquellos sistemas que sugieren etiquetas al usuario, bas\'{a}ndose
en anotaciones previas sobre el recurso, producen unas distribuciones de
etiquetas muy diferentes a aqu\'{e}llos que no sugieren etiquetas y dejan a los
usuarios que hagan su propia elecci\'{o}n. Esta caracter\'{i}stica ha sido
determinante tambi\'{e}n para la aplicaci\'{o}n con \'{e}xito de las funciones
de pesado sobre estas distribuciones.

Hemos descubierto que las funciones de pesado de etiquetas propuestas superan claramente a
la aproximaci\'{o}n basada en TF cuando el sistema no sugiere etiquetas basadas
en los recursos (es decir, en LibraryThing y GoodReads), tanto cuando se
utilizan por s\'{i} solas, como cuando se combina con otras fuentes de datos. En
realidad, es mejor considerar simplemente la aproximaci\'{o}n basada en
etiquetas que combinarla con otras fuentes de datos, ya que por s\'{i} sola
ofrece los mejores resultados, los cuales no son mejorados cuando se combinan.

No obstante, cuando el sistema sugiere etiquetas basadas en el recurso, las
folksonom\'{i}as generadas son muy diferentes al resto. Esto afecta a las
distribuciones de etiquetas en gran medida y, por lo tanto, a las funciones de
pesado que hemos estudiado. Debido a ello, el uso de funciones de pesado de
etiquetas obtiene peores resultados que no tenerlos en cuenta, y necesitan ser
combinadas con otras fuentes de datos para funcionar mejor. En este \'{u}ltimo
caso, pueden llegar a mejorar a la aproximaci\'{o}n basada en TF, gracias a las
buenas predicciones que aporta, que ayuda a alimentar de forma adecuada la
combinaci\'{o}n de clasificadores.

\begin{description}
 \item[Pregunta de Investigaci\'{o}n 8] \hfill \\
 \textit{\rqdisttwoes}
\end{description}

Entre las funciones de pesado que hemos estudiado, aqu\'{e}lla que se basa en las
frecuencias en marcadores ha demostrado ser la mejor para los sistemas sin
sugerencias de etiquetas basadas en recursos. En estos casos, IBF es la mejor
opci\'{o}n, seguida por IRF e IUF. Todos ellos superan con claridad a TF, tanto
cuando se utilizan por s\'{i} solas, como cuando se combinan con otras fuentes de
datos.

Por otro lado, cuando el sistema sugiere etiquetas basadas en el recurso es
mejor basarse en la frecuencia en usuarios. IUF funciona mejor que IBF e IRF
en estos casos, debido a la importancia de aquellos usuarios que tienden a
escoger sus propias etiquetas en lugar de basarse en las sugerencias. Aunque ni
siquiera IUF supera a TF, cuando se combina con otras fuentes de datos llega a
ser la mejor opci\'{o}n. No obstante, los resultados de este \'{u}ltimo caso son
s\'{o}lo ligeramente superiores a los obtenidos por la combinaci\'{o}n que
utiliza TF, por lo que cualquiera de ellas podr\'{i}a emplearse para llegar a
obtener unos resultados parecidos.

\begin{description}
 \item[Pregunta de Investigaci\'{o}n 9] \hfill \\
 \textit{\rqcatonees}
\end{description}

Hemos demostrado que dicho tipo de usuario, llamado Categorizador, en realidad
existe. Seg\'{u}n nuestros experimentos, esto es verdad sobre todo cuando se
trata de sistemas sin sugerencias de etiquetas como en LibraryThing, donde la
clasificaci\'{o}n de recursos realizada utilizando etiquetas de los usuarios
Categorizadores obtiene mejores resultados. Cuando las sugerencias existen, la
detecci\'{o}n de usuarios que se adec\'{u}an a la tarea se complica, como hemos
demostrado que ocurre con GoodReads y Delicious. Sin embargo, la utilizaci\'{o}n
de la medida apropiada puede producir una selecci\'{o}n exitosa de usuarios que
se ajustan a las caracter\'{i}sticas de un Categorizador.

\begin{description}
 \item[Pregunta de Investigaci\'{o}n 10] \hfill \\
 \textit{\rqcattwoes}
\end{description}

De las dos caracter\'{i}sticas que hemos considerado en este trabajo, hemos
visto que si se diferencian los usuarios por su nivel de verbosidad, se puede
encontrar un conjunto de usuarios que se ajustan m\'{a}s a la tarea de
clasificaci\'{o}n. Por otra parte, hemos visto que separando usuarios por la
diversidad de su vocabulario no se consigue una buena discriminaci\'{o}n para
este fin, sino para encontrar otro tipo de usuarios llamados Descriptores.
Adem\'{a}s de esto, hemos visto que aqu\'{e}llos usuarios que no utilizan datos
descriptivos en sus anotaciones ofrecen etiquetas que se ajustan mejor a la
clasificaci\'{o}n de recursos.

\section{Principales Contribuciones}
\label{sec:principales-contribuciones}

La idea novedosa de este trabajo de investigaci\'{o}n se basa en la
utilizaci\'{o}n de anotaciones sociales para enriquecer una tarea de
clasificaci\'{o}n de recursos. Hasta donde nosotros sabemos, el primer trabajo
de investigaci\'{o}n que llev\'{o} a cabo experimentos con tareas de
clasificaci\'{o}n reales fue nuestro primer trabajo en este campo
\citep{zubiaga2009getting}. Previamente, s\'{o}lo \cite{noll_exploring_2008}
hab\'{i}an realizado un an\'{a}lisis estad\'{i}stico que comparaba etiquetas
sociales con una clasificaci\'{o}n hecha por expertos. Teniendo en cuenta la
carencia de trabajos en el \'{a}rea, la investigaci\'{o}n recogida en esta tesis aporta
nuevo conocimiento hacia el uso y modo de representaci\'{o}n apropiados de
etiquetas sociales para la clasificaci\'{o}n de recursos. Concretamente,
nuestras aportaciones principales al \'{a}rea de investigaci\'{o}n son las
siguientes:

\begin{itemize}
 \item Hemos creado 3 colecciones de gran escala que incluyen tanto etiquetas
sociales como informaci\'{o}n de la categor\'{i}a correspondiente para una serie
de recursos. \'{E}stas pueden considerarse como unas de las mayores colecciones
utilizadas en el \'{a}rea de investigaci\'{o}n y, por lo que nosotros sabemos,
las mayores utilizadas para clasificaci\'{o}n de recursos. Algunas de estas
colecciones, junto con otras m\'{a}s peque\~{n}as que hemos creado a lo largo
del trabajo, se han hecho p\'{u}blicas para fines de
investigaci\'{o}n\footnote{http://nlp.uned.es/social-tagging/datasets/}. Entre
otros, \cite{godoy2010exploiting} y \cite{strohmaier2010network} han utilizado
alguna de nuestras colecciones para su investigaci\'{o}n.
 \item Nuestro trabajo es el primero que compara diferentes representaciones de recursos usando
etiquetas sociales. Adem\'{a}s, es el primer trabajo que realiza tareas de
clasificaci\'{o}n comparando etiquetas sociales con otros tipos de fuentes de
datos. Hemos demostrado que las etiquetas sociales son tambi\'{e}n \'{u}tiles
para categor\'{i}as m\'{a}s precisas de m\'{a}s bajo nivel. Al contrario de lo
que indican que \cite{noll_exploring_2008}, donde los autores realizan un
estudio estad\'{i}stico con el que concluyen que las etiquetas sociales
podr\'{i}an no ser \'{u}tiles para categor\'{i}as m\'{a}s precisas, hemos
demostrado que son a\'{u}n m\'{a}s \'{u}tiles que para categor\'{i}as m\'{a}s
generales.
 \item Hemos analizado las distribuciones de etiquetas sociales en
folksonom\'{i}as, y hemos realizado un riguroso estudio de c\'{o}mo la
configuraci\'{o}n de un sistema de etiquetado social afecta tales
distribuciones. En este aspecto, hemos adaptado funciones de pesado basadas en
la consolidada TF-IDF al \'{a}mbito del etiquetado social y las
folksonom\'{i}as.
 \item Hemos mostrado la existencia de un grupo de usuarios, llamados
Categorizadores, cuyas anotaciones se parecen m\'{a}s que las de otro grupo de
usuarios, llamados Descriptores, a la clasificaci\'{o}n hecha por expertos. Aunque
la aproximaci\'{o}n para diferenciar Categorizadores y Descriptores ya estaba
consolidada de previos trabajos, en \'{e}ste hemos llevado a cabo la 
tarea de demostrar que los Categorizadores se ajustan m\'{a}s a la
clasificaci\'{o}n de recursos.
\end{itemize}

La utilizaci\'{o}n de anotaciones sociales para el beneficio de tareas de
clasificaci\'{o}n de recursos era una l\'{i}nea de investigaci\'{o}n nueva al
comienzo de esta tesis. Sin embargo, el crecimiento en el inter\'{e}s de los
investigadores sobre contenidos generados por usuarios en medios de
comunicaci\'{o}n social, y concretamente en los sistemas de etiquetado social,
ha ocasionado recientemente la aparici\'{o}n de numerosos trabajos en el
\'{a}rea. Junto con este crecimiento, m\'{a}s investigadores han mostrado su
inter\'{e}s en utilizar anotaciones sociales para clasificaci\'{o}n de recursos,
y el n\'{u}mero de trabajos relacionados ha aumentado considerablemente.
\cite{godoy2010exploiting}, por ejemplo, presentan un estudio de
clasificaci\'{o}n basada en etiquetas que se inspira en un trabajo nuestro
\citep{zubiaga2009getting}.

\section{Trabajo Futuro}
\label{sec:trabajo-futuro}

La utilizaci\'{o}n de anotaciones sociales para la clasificaci\'{o}n de recursos
es un campo de investigaci\'{o}n que est\'{a} a\'{u}n en sus inicios, y se ha
realizado relativamente poco trabajo hasta el momento. El trabajo presentado en
esta tesis concluye con la manera de representar etiquetas sociales en busca de
una clasificaci\'{o}n de recursos lo m\'{a}s precisa posible. Adem\'{a}s, da
lugar al planteamiento de diversos trabajos futuros.

A lo largo de esta tesis hemos considerado cada etiqueta como un s\'{i}mbolo
diferente, sin tener en cuenta su significado sem\'{a}ntico. En este aspecto,
nuestros planes para trabajo futuro incluyen el an\'{a}lisis del significado de
las etiquetas para tratar de descubrir palabras sin\'{o}nimas y relaciones entre
ellas. Bien utilizando t\'{e}cnicas de procesamiento de lenguaje natural, o bien
mediante aproximaciones sem\'{a}nticas, esto podr\'{i}a ayudar a entender el
significado de cada etiqueta, pudiendo explorar m\'{a}s all\'{a} el conocimiento
que aportan las folksonom\'{i}as.

Las tres funciones de pesado que hemos empleado en el Cap\'{i}tulo
\ref{c:tag-distribution-classification} se basan en la conocida TF-IDF, que fue
dise\~{n}ada inicialmente para colecciones de texto. Pensamos que probar otras
funciones de pesado, as\'{i} como explorar la posible definici\'{o}n de una
nueva funci\'{o}n que se ajuste a las necesidades de estas estructuras sociales,
pueden resultar en interesantes aportaciones como trabajo futuro. Esto
ayudar\'{i}a sobre todo para sistemas que dan sugerencias de etiquetas basadas
en recursos, como pasa con Delicious, donde las funciones de pesado que hemos
experimentado no han dado buenos resultados.

%% file: bq-summary.tex
\chapter{Laburpena (Basque Summary)}
\label{c:bq-summary}

\selectlanguage{spanish}

\textit{``Hizkuntza bat ez da galtzen ez dakitenek ikasten ez dutelako, dakitenek erabiltzen ez dutelako baizik.''}

--- Joxean Artze

\translatedtitle{Baliabideen Sailkapenerako Folksonomien Ustiapena}

\chaptersummary{Tesi honetan baliabideen sailkapenaren gainean dihardugu, eguneroko bizitzan hain garrantzitsua eta ohikoa den ataza bat landuz, liburuak katalogatzea edo bideoak antolatzea izan daitekeen bezalaxe. Ataza burutzeko, etiketa sozialen sistemetan erabiltzaileek egindako anotazioez baliatzen gara. Webgune hauetan baliabide ezberdinen gainean metadatu ugari eskaini ohi dituzte erabiltzaileek. Orain arte, gutxi dira metadatu hauek helburu honetarako erabili dituzten ikerketa lanak, eta gutxi horiek analisi estatistikoak egitera mugatu dira. Tesi honetan, sistema hauen, bertako erabiltzaileen eta haien anotazioen ezaugarriak aztertzen ditugu, datu-sorta handi hauetaz ahal bezainbeste profitatu nahian, eta ahalik eta baliabideen sailkapen automatiko zehatzena lortu asmoz.}

\section{Motibazioa}
\label{sec:motibazioa}

Edozein motatako baliabideak aurrez definitutako kategoriatan sailkatzea ohiko ataza da gure eguneroko bizitzan. Baliabideei kategoriak esleitzeak ondoren berreskuratu ahal izateko erraztasunak eskaintzen ditu, bilaketa nahi den kategoriara mugatuz. Esate baterako, liburuzainek gaika antolatu ohi dituzte liburuak katalogoetan. Horrez gain, filmen datubaseak, musika katalogoak eta fitxategi sistemak, besteak beste, kategoriatan antolatu ohi dira baliabideok aurkitzea erraztuz. Era berean, \emph{Yahoo! Directory} eta \emph{Open Directory Project} bezalako web direktorioek kategoriatan antolatzen dituzte web orrialdeak. Web orrialdeak sailkatuta izateak interneteko bilatzaileen funtzionamendua hobe dezake emaitzak erabiltzailearen intereseko kategoriara mugatuz \citep{qi_webpage_2009}.

Kategorizatze lan hori eskuz egitea, ordea, oso garestia izaten da baliabide sorta handia denean. Adibide gisa, Estatu Batuetako \emph{Library of Congress} liburutegi publikoak 2002an profesionalek katalogatutako liburu bakoitzak 94,58 dolarreko kostua izan zuela adierazi zuen\footnote{http://www.loc.gov/loc/lcib/0302/collections.html}. Urte hartan katalogatu zituzten 291.749 erregistroengatik 27,5 milioitik gora ordaindu behar izan zituzten beraz. Ataza hau zein garestia den ikusita, sailkatzaile automatikoetara pasatzeak alternatiba egokia dirudi eskulana gutxitzeko, betiere katalogoak eguneratuta mantenduz.

Orain arte, sailkatzaile automatiko gehienak baliabideen edukian oinarritu dira informazio iturri gisa, web orrialdeen sailkapenari dagokionean batik bat \citep{qi_webpage_2009}. Baliabideen edukiek ez dute beti informazio esanguratsua izaten, ordea, eta horrek zaildu egiten du ataza. Gainera, batzutan ez da erraza izaten liburuak eta filmeak bezalako baliabideentzako datu nahikoa lortzea. Horrelako kasuetan zailagoa izaten da edukia errepresentatzea, eta litekeena da edukia erraz prozesatu daitekeen formatu batean ez izatea.

Arazo hauentzako soluzio posible bezala, etiketa sozialen sistemek baliabideei dagozkien metadatuak eskuratzeko modu errazago eta merkeagoa eskaintzen dute. Delicious\footnote{http://delicious.com}, LibraryThing\footnote{http://www.librarything.com} eta GoodReads\footnote{http://www.goodreads.com} bezalakoek baliabideen inguruan erabiltzaileek definitutako etiketak batzen dituzte. Erabiltzaileek sortutako etiketa hauek baliabideen edukiak deskribatzen dituzten datu esanguratsuak direla frogatu da \citep{heymann_can_social_2008}.

Etiketa hauen bitartez, baliabideen sailkapen propio baten antzekoa eskaintzen dute erabiltzaileek. Eta etiketa hauek modu sozialean elkarbanatzen dira komunitatearekin. Sistema hauetan erabiltzaile kopuru handiek parte hartzen dutenez, beraien anotazioak baliabideen gainean batu egiten dira. Ondorioz, erabiltzaile ezberdinen anotazioak baliabideetan batzeko gaitasun horrek are erabilgarriago eta baliagarriago egiten du anotazio horietako bakoitza. Komunitate aktiboetako erabiltzaileek laster-marka, etiketa eta anotatutako baliabide sorta handiak sor ditzakete.

\begin{quote}
 \textit{``Sailkapen indibidual bakoitzak profesional batek egindakoak baino gutxiago balio du. Baina ugari, mordoxka bat daude''}, Joshua Schachter, Delicious-en sortzailea, 2006ko FOWA bilkuran, Londresen (Ingalaterra)\footnote{http://simonwillison.net/2006/Feb/8/summit/}.
\end{quote}

Etiketa sozialen sistemak baliabideak gorde, antolatu eta bilatzeko tresnak dira, erabiltzaileek hautatutako etiketak baliatuz anotatzea ahalbidetzen dutenak. Gure ustez, anotazio hauek nabarmen hobe dezakete baliabideen sailkapen automatikoa. Erabiltzaileek sortutako anotazio hauek erabilgarri izan litezke baliabideen kategoriaren inguruko informazioa ematen duen informazio iturri gisa.

Baliabide bakoitzaren gainean erabiltzaile askok esleitzen dituenez anotazioak, gure helburu nagusia berauen ekarpenak batzeko modu egokia aurkitzean datza, betiere profesionalek egindako kategorizazioarekiko antzekoa den antolaketa lortuz asmoz. Erabiltzaile askok metadatu kopurua handia esleitzen duenez, gure erronka ahalik eta emaitza onena lortzean datza.

\begin{quote}
 \textit{``Garaiotan datuak eskuratzea erraza da, baina hauek zentzuz erabiltzea ea da hain erraza''}, Danah Boyd, Microsoft Research New England-eko Social Media gaineko ikertzailea, WWW2010 kongresuan, Raleigh, Ipar Karolina (Ameriketako Estatu Batuak)\footnote{http://www.danah.org/papers/talks/2010/WWW2010.html}.
\end{quote}

\subsection{Baliabideen Sailkapena}
\label{baliabideen-sailkapena}

Baliabideen sailkapena aurrez definitutako kategoria sorta batean baliabideak antolatzean datzan ataza da. Tesi honetan, Euskarri Bektoredun Makinak darabilzkigu (Support Vector Machines, SVM, \cite{joachims98text}), sailkapen metodo abangoardista. Sailkapen ataza mota hauek aurrez sailkatutako baliabide sorta batean oinarritzen dira, berau sailkatzaileak behar duen ezagutza eraikitzeko baliatzen delarik.

Baliabideen sailkapen ataza batek ezaugarri ezberdinak izan ditzake. Alde batetik, sistemaren ikasketa metodoari dagokionean, \textit{gainbegiratua} dela esaten da ikasteko erabilitako baliabide guztiak aurrez sailkatuta daudenean, eta \textit{erdi-gainbegiratua} dela, ostera, sailkatu gabeko baliabideen gainean egindako aurreikuspenak ere ikasteko erabiltzen direnean. Bestalde, kategoria kopuruari dagokionean, sailkapena \textit{bitarra} izan daiteke, bi kategoria baino ezin direnean esleitu, edo \textit{kategoria-anitza}, hiru edo kategoria gehiago daudenean. Lehena iragazte sistemetarako erabili ohi da, bigarrena taxonomia handiegoekin erabiltzen delarik, adibidez, gaikako sailkapena.

Baliabideen kolekzio handien gaikako sailkapena burutzeko, Web-eko orrialdeak edo liburutegietako liburuak izan daitezkeen bezalaxe, taxonomiak bi kategoria baino gehiagokoak izan ohi dira, eta aurrez sailkatutako baliabide kopurua oso murritza izaten da. Beraz, interesgarria deritzogu bai teknika erdi-gainbegiratuak eta bai kategoria-anitzak kontuan hartu eta aztertzea, ataza hauek burutzeko aukerarik onena zein den jakin ahal izateko.

Tesi honetan, SVM algoritmoan oinarritzen diren hainbat metodoren analisia proposatzen dugu, ataza hauekiko duten apropostasuna aztertuz. Metodo hauen artean teknika kategoria-anitz ezberdinak aztertzen ditugu, eta baita teknika gainbegiratu zein erdi-gainbegiratuak ere.

\subsection{Anotazio Sozialak}
\label{anotazio-sozialak}

Etiketa sozialen sistemek baliabide gogokoenak (web orrialdeak, filmeak, liburuak, argazkiak edo musika, besteak beste) gorde eta anotatzeko aukera eskaintzen diete erabiltzaileei, komunitatearekin elkarbanatuz. Anotazio hauek etiketa moduan eman ohi dituzte erabiltzaileek. Etiketatzea baliabideei hitz gakoak edo etiketak esleitzeari deritzo, deskribatzeko zein antolatzeko aukera emanez. Honek ondoren baliabideok bilatzea errazten du, etiketa horiek bilaketa gako bezala baliatuz. Sistema gehienetan ez daude aurrez definitutako etiketak, eta beraz nahi duten hitzak hauta ditzakete erabiltzaileek etiketa gisa.

\begin{quote}
 \textit{``Etiketatzeak interfazearekin zerikusi handia du - jendeak gauzak go\-go\-ra\-tze\-ko modu bat, gorde zuten unean zertan pentsatzen ari ziren erakusten duena. Nahiko erabilgarria gogoratzeko, ona deskubritzeko, ikaragarria hedatzeko (non argitaratzen dituztenek ahal bezainbeste etiketa definitzen dituzten kutxa gehiagotan sailkatzeko).''}, Joshua Schachter, Delicious-en sortzailea, 2006ko FOWA bilkuran, Londresen (Ingalaterra)\footnote{http://simonwillison.net/2006/Feb/8/summit/}.
\end{quote}

Etiketatze prozesu honen bitartez folksonomia deritzon egitura sortzen da, erabiltzaileek sortutako baliabideen antolaketa, alegia. Folksonomia \textit{folk} (jendea), \textit{taxis} (sailkapena) eta \textit{nomos} (kudeaketa) hitzen laburtzapena da. Komunitatean oinarritzen den taxonomia bezala ere ezagutzen da folksonomia, non sailkapen mota ez-hierarkikoa den, adituek egindako sailkapen taxonomikoetan ez bezala. Beraz, folksonomiek badute nolabaiteko zerikusia adituek egindako sailkapenekin, baliabideak taldeka sailkatzen baitira era berean.

Anotazio hauek sozialak direla esan ohi da ingurune sozial batean komunitatearekin elkarbanatuz beste guztientzako erabilgarri agertzen direnean. Honek dakarren abantaila nagusia besteek ipinitako etiketak baliatuz bilaketak egin ahal izatea da. Era berean, hauxe da erabiltzaile asko parte hartzera animatzen duen ezaugarrietako bat.

\begin{figure}[ht]
\begin{center}
 \includegraphics[width=100mm,clip]{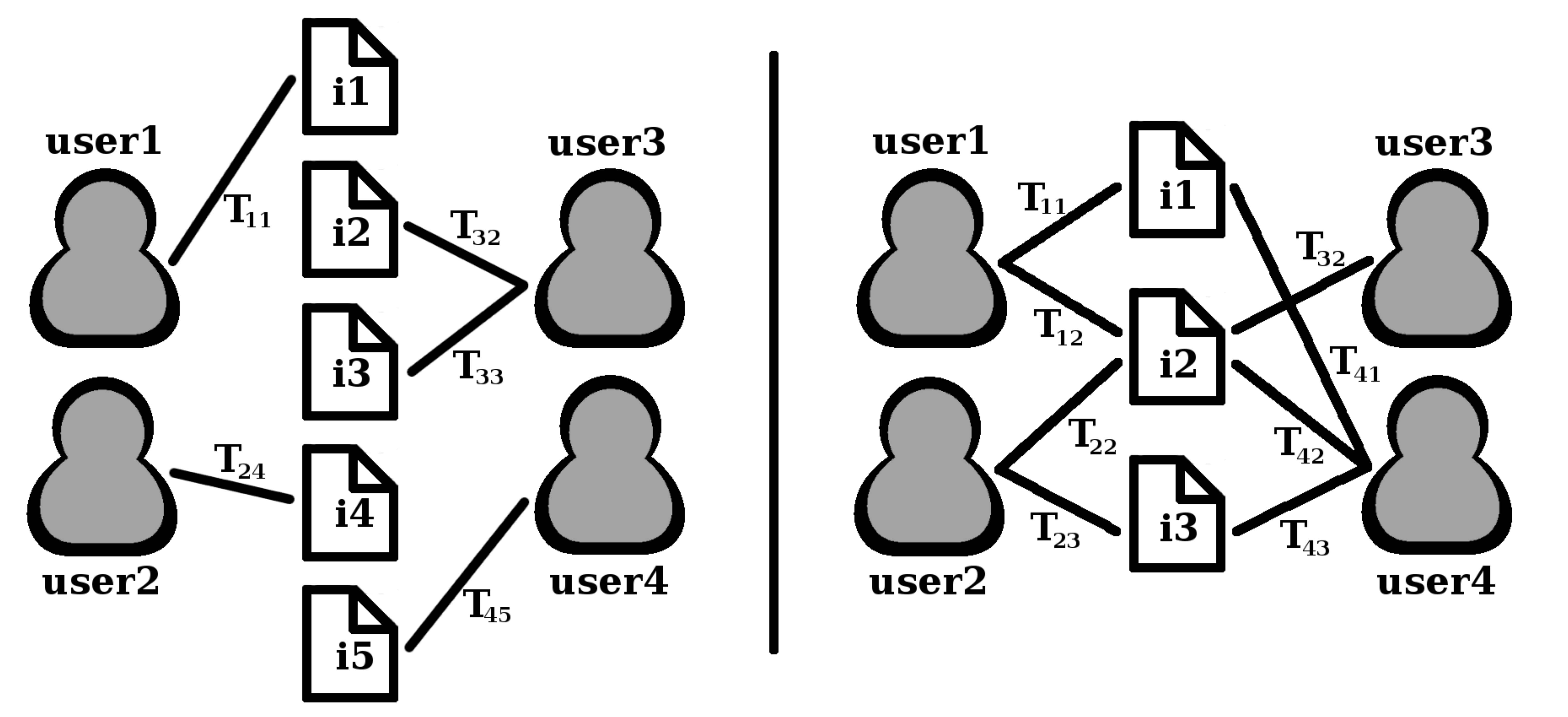}
\end{center}
\caption[Etiketatze sinplea eta etiketatzen kolaboratiboa]{Etiketatze sinple eta kolaboratiboko sistemetako anotazioen arteko konparazioa.}
\label{fig:sinplea-vs-kolaboratiboa}
\end{figure}

Anotazio guztiak ez dira modu berean elkarbanatzen, ordea. Etiketa sozialen guneak berak baldintza batzuk defini ditzake, baliabide bakoitza nork anota dezakeen mugatuz, batez ere. Honi dagokionean, bi sistema mota ezberdin di\-tza\-ke\-gu \citep{smith_tagging_2008}:

\begin{itemize}
 \item \textbf{Etiketen sistema sinpleak:} erabiltzaileek norbere baliabideak etiketa di\-tza\-ke\-te (adibidez, argazkiak Flickr-en\footnote{http://www.flickr.com}, bideoak Youtube-n\footnote{http://www.youtube.com} edo albisteak Digg-en\footnote{http://digg.com}), baina inork ezin ditu besteen baliabideak etiketatu. Normalean, baliabidearen egilea bera izaten da etiketatzen duena. Ondorioz, baliabide bakoitza erabiltzaile batek baino ez du etiketatzen. Etiketen sistema sinpleetan erabiltzaile sorta bat ($U$) izaten da, baliabide batzuen ($R$) gainean etiketa sorta bat ($T$) esleitzen duena. $u_i \in U$ erabiltzaile batek $r_j \in R$ bere baliabidea $p$ etiketa kopuru aldagarridun $T_{j} = \{t_{j1},...,t_{jp}\}$ etiketa-sortarekin anotatzen du. $r_{j}$ baliabideari esleitutako etiketa-sortak $T_{j}$ izaten jarraituko du aurrerantzean, beste inork ezingo baitu anotatu.
 
 \item \textbf{Etiketen sistema kolaboratiboak:} erabiltzaile askok anotatzen dute baliabide bera, bakoitzak etiketa ezberdin batzuk baliatuz. Erabiltzaile bakoitzak erabilitako etiketa sortak folksonomia txikiago bat sortzen du, pertsonomia deritzona. Sistema hauetan hainbat erabiltzailek etiketatu ohi du baliabide bera. Esate baterako, CiteULike.org, LibraryThing.com eta Delicious etiketatze kolaboratiboan oinarritzen dira, non baliabide bakoitza (artikuluak, liburuak eta URLak, hurrenez hurren) interesgarri deritzon erabiltzaile orok etiketa dezakeen. Etiketatze sistema kolaboratiboak sinpleak baino konplexuagoak dira. Sistema hauetan, erabiltzaile sorta bat ($U$) izaten da, baliabide batzuen ($R$) gainean laster-marka batzuk ($B$) gordetzen dabilena, etiketa sorta batekin ($T$) anotatuz. $u_i \in U$ erabiltzaile batek $r_j \in R$ baliabidearen $b_{ij} \in B$ laster-marka gorde dezake $p$ etiketa kopuru aldagarridun $T_{ij} = \{t_{ij1},...,t_{ijp}\}$ etiketa-sorta baliatuz. $k$ erabiltzailek $r_j$ baliabidea gorde eta gero, $T_{j} = \{w_{j1} t_{j1},...,w_{jn} t_{jn}\}$ pisudun etiketa-sorta bezala defini daitezke bere anotazioak, non $w_{j1},...,w_{jn} \leq k$ aldagaiek etiketa bakoitzaren esleipen kopurua adierazten duten. Ondorioz, laster-marka bakoitzak erabiltzaile, baliabide eta etiketa-sorta bana ditu bere baitan: $b_{ij}: u_{i} \times r_{j} \times T_{ij}$. Erabiltzaile bakoitzak baliabide ezberdinen laster-markak egiten ditu, eta aldi berean baliabide batek erabiltzaile ezberdinek egindako laster-markak izan ditzake. Erabiltzaile baten laster-marketako etiketak bateratzearen emaitza pertsonomia izenez ezagutzen da: $T_{i} = \{w_{i1} t_{i1},...,w_{im} t_{im}\}$, non $m$ erabiltzaileak dituen etiketa ezberdinen kopurua den.
 
 Tesi honetan etiketatze kolaboratiboko sistemekin dihardugu. \ref{fig:sinplea-vs-kolaboratiboa}. Irudiak bi sistema hauen arteko ezberdintasunak erakusten ditu adibide baten bitartez.
\end{itemize}

Baliabide bera anotatzen duten erabiltzaileen artean etiketak kointziditzeko probabilitatea altua izaten da. Ezaugarri honek bereziki interesgarri egiten du etiketatze kolaboratiboetako erabiltzaileen bateratze hau etiketatze sinpleekin alderatuz.

Etiketatze sistema kolaboratibo batean, adibidez, erabiltzaile batek lan hauxe bera anotatzeko \texttt{etiketa-sozialak}, \texttt{ikerketa} eta \texttt{tesia} etiketak erabil litzake, eta beste erabiltzaile batek \texttt{etiketa-sozialak}, \texttt{markatzaile-sozialak}, \texttt{doktoretza} eta \texttt{tesia} etiketak. Erabiltzaile bakoitzaren jarrera oso ezberdina izan daiteke sistema hauetan, eta horrexegatik izaten dira kontuan beraien guztien anotazioak bateratzeko orduan. Adibide bezala aipatutako bi horien anotazioak batuz hurrengoa lortuko genuke: \texttt{tesia} (2), \texttt{etiketa-sozialak} (2), \texttt{markatzaile-sozialak} (1), \texttt{doktoretza} (1) eta \texttt{ikerketa} (1).


Tesi honetan, erabiltzaileek etiketen sistema sozialetan egindako anotazioak aztertu eta ikertzen ditugu. Baliabiden sailkapen egoki bat lortzeko etiketa horiengandik etekina ateratzeko ikerketa lana aurkezten dugu. Konkretuki, erabiltzaileek sortutako folksonomiek adituek egindako sailkapen baten antzeko zerbait lortzeko duten erabilgarritasuna aztertzen dugu. Etiketa sozialen errepresentazio ezberdinak aztertzen ditugu bertan, betiere adituek egindako sailkapen horietara hurbildu asmoz. Bereziki, etiketa sozialei etekina ateratzea da gure helburua, bai errepresentazio egokiena bilatuz, eta baita etiketek baliabide, laster-marka eta erabiltzaile ezberdinetan aurkezten dituzten distribuzioen eraginari erreparatuz. Azkenik, sailkapen ataza batetik gertuago dauden erabiltzaileak bilatzeko ikerketa aurkezten dugu, horretarako erabiltzaileen jarrera antzemateko teknika abangoardistez baliatuz.

\section{Helburuak}
\label{sec:helburuak}

Tesi honen helburu nagusia etiketa sozialen sistemetan aurkitzen diren anotazio horiei guztiei etekina ateratzeko egin beharreko erabilpen egokiaren inguruan ezagutza berria zabaltzea da. Baliabideak sailkatzeak duen interesa jakinda, eta berauek errepresentatzeko datu esanguratsuen gabezia kontuan izanda, gure helburua baliabideen sailkapenerako etiketa sozialek lagun dezaketena aztertu, eta erabilpen aproposa egiteko modurik egokiena zein den jakitea da. Lan honen hasieran, hau ikertzen zuen lanik ez zegoela ikusi genuen. Horrek motibatu gintuen ikerketa lan hau aurrera eramatera. Helburu honekin, hurrengo planteamendua egin genuen, lanaren xede nagusia laburbilduz:

\begin{description}
 \item[Lanaren Planteamendua] \hfill \\
 \textit{\problemstatementeu}
\end{description}

\section{Metodologia}
\label{sec:metodologia-eu}

Tesi hau aurrera eramateko jarraitu den ikerketa metodologiak hurrengo 6 pausoak jarraitu ditu:

\begin{enumerate}
 \item Artearen egoeraren azterketa eta irakurketa sakona, eta baita etiketa sozialen sistemak ikertu eta funtzionamendua ondo ulertzea ere.
 \item Lana burutu ahal izateko SVM sailkatze egokia aurkitzea.
 \item Etiketen sistema sozialetan oinarrituz sortutako kolekzioak bilatzea. Gure beharrak betetzen zituenik aurkitu ez genuenez, gureak sortzea erabaki genuen, tamaina handiko hiru sortuz.
 \item Etiketa sozialetan oinarrituz, eta sailkapen atazan baliatzeko direla kontuan hartuz, lana aurrera eramateko aproposak diren hurbilketak eta errepresentazioak pentsatzea eta proposatzea.
 \item Proposatutako hurbilketa eta errepresentazioak ebaluatzea.
 \item Emaitzen azterketa sakona burutzea, ondorio sendoetara iritsi asmoz.
 \item Egindako lanaren emaitza partzialak kongresu eta tailer nazional eta internazionaletan aurkeztu, beste ikertzaileen iritzi eta gomendioak jaso ahal izateko.
 \item Egindako ikerketa lana, ekarpen nagusiak, eta lortutako ondorioak tesi honetan batu eta laburbildu.
\end{enumerate}

4. pausotik 6.era, behin eta berriz errepikatu zen prozesua, pauso horiek behin baino gehiagotan burutuz.

\section{Tesiaren Egitura}
\label{sec:tesiaren-antolaketa}

Tesi honek 8 kapitulu ditu. Jarraian azaltzen da labur-labur kapitulu hauetako bakoitzaren edukia zein den.

\begin{description}
 \item[\ref{c:introduction}. kapitulua \pageref{c:introduction}. orrialdean] \hfill \\
 \textbf{Sarrera} \\
  Baliabideen sailkapenerako etiketa sozialak ikertu nahi izateko motibazioa aurkezten dugu. Ataza formalki azaldu, eta ikerketa burutzeko beharra motibatzen dugu.
  
 \item[\ref{c:related-work}. kapitulua \pageref{c:related-work}. orrialdean] \hfill \\
 \textbf{Erlazionatutako Lana} \\
  Arlo honetan eta erlazionatutakoetan lehenago egindako lanak laburbiltzen ditugu, bai etiketa sozialen erabilpenean, eta baita baliabideen sailkapenean ere.
  
 \item[\ref{c:svm-classification}. kapitulua \pageref{c:svm-classification}. orrialdean] \hfill \\
 \textbf{Euskarri Bektoredun Makinak Neurri Handiko Sailkapenerako} \\
  Taxonomia kategoria-anitzetan baliabide kolekzio handien sailkapenerako SVM hurbilketa ezberdinen analisia aurkezten dugu. Tesian zehar erabiltzeko aproposena den SVM hurbilketa zein den jakitea ahalbidetzen digu ikerketa honek.
  
 \item[\ref{c:datasets}. kapitulua \pageref{c:datasets}. orrialdean] \hfill \\
 \textbf{Etiketa Sozialen Datu-Sorten Sorkuntza} \\
  Lan honetan guztian zehar erabiltzeko sortu genituen etiketa sozialen kolekzioak zehatz-mehatz deskribatu eta aztertzen ditugu. Kolekzioak sortzeko jarraitutako prozesua azaldu, eta folksonomien ezaugarri nagusiak aztertzen ditugu.
  
 \item[\ref{c:tag-representation}. kapitulua \pageref{c:tag-representation}. orrialdean] \hfill \\
 \textbf{Etiketen Gehikuntzaren Errepresentazioa} \\
  Baliabideen sailkapenerako etiketa sozialen errepresentazio ezberdinak proposatu eta ebaluatzen ditugu. Etiketa sozialek beste datu iturri batzuekin alderatuta baliabideen sailkapenerako duten errendimendua ikertzen dugu, eta ataza burutzeko errepresentazio egokiena zein izan daitekeen aztertzen dugu. Horrez gain, etiketa sozialak beste datu iturriekin nahasten ditugu errendimendua hobetu ahal izateko.
  
 \item[\ref{c:tag-distribution-classification}. kapitulua \pageref{c:tag-distribution-classification}. orrialdean] \hfill \\
 \textbf{Baliabideen Sailkapenerako Etiketen Distribuzioaren Azterketa} \\
  Etiketa sozialen sistemetako etiketa bakoitzak baliabideen sailkapenerako duen adierazgarritasuna aztertzen dugu. Horretarako, sistema hauetarako egokitutako pisu-funtzioak erabiltzen ditugu. Gainera, funtzio hauek zenbaterainoko aproposak diren aztertzen dugu, betiere sistema bakoitzaren ezarpenei erreparatuz.
  
 \item[\ref{c:analyzing-appropriateness-users}. kapitulua \pageref{c:analyzing-appropriateness-users}. orrialdean] \hfill \\
 \textbf{Sailkapenerako Erabiltzaileen Jarreraren Analisia} \\
  Etiketa sozialen sistemetako erabiltzaileen jarrerak baliabideen sailkapenean izan dezakeen eragina aztertzen dugu. Sailkatzea helburu duten erabiltzaileak existitzen direla dioten aurreko lanetan oinarrituz, erabiltzaile horiek baliabideen sailkapenerako egokiagoak diren aztertzen dugu.
  
 \item[\ref{c:conclusions}. kapitulua \pageref{c:conclusions}. orrialdean] \hfill \\
 \textbf{Ondorioak eta Etorkizunerako Ildoak} \\
  Lanaren ondorio eta ekarpen nagusiak laburbiltzen ditugu. Horrez gain, lanaren hasieran formulatutako galderei erantzun, eta etorkizunerako ildoak aurkezten ditugu.
\end{description}

Horrez gain, tesi honek jarraian azaltzen diren eranskin hauek ere baditu, informazio gehigarria eta beste hizkuntza batzuetako laburpenekin:

\begin{description}
 \item[\ref{c:topx-tags}. eranskina \pageref{c:topx-tags}. orrialdean] \hfill \\
 \textbf{Emaitza Gehigarriak} \\
  Gehigarri gisa, emaitza lagungarri batzuk aurkezten ditugu, tesiaren parte moduan sartu ez baditugu ere, ondorio batzuk frogatu eta ulertzeko balio dutenak.
  
 \item[\ref{c:key-terms}. eranskina \pageref{c:key-terms}. orrialdean] \hfill \\
 \textbf{Hitz Nagusiak eta Definizioak} \\
  Etiketa sozialen sistemekin zerikusia duten hainbat hitzen definizioa ematen dugu.
  
 \item[\ref{c:acronyms}. eranskina \pageref{c:acronyms}. orrialdean] \hfill \\
 \textbf{Akronimoen Zerrenda} \\
  Lanean zehar erabilitako akronimoen eta berauen esanahien zerrenda aurkezten da.
  
 \item[\ref{c:sp-summary}. eranskina \pageref{c:sp-summary}. orrialdean] \hfill \\
 \textbf{Resumen (Gaztelerazko Laburpena)} \\
  Lan honen edukiaren gaztelerazko laburpena.
  
 \item[\ref{c:bq-summary}. eranskina \pageref{c:bq-summary}. orrialdean] \hfill \\
 \textbf{Laburpena} \\
  Lan honen edukiaren euskarazko laburpena.
\end{description}

\section{Ebatzitako Ikerketa Galderak}
\label{sec:ebatzitako-ikerketa-galderak}

\begin{description}
 \item[1. Ikerketa Galdera] \hfill \\
 \textit{\rqsvmoneeu}
\end{description}

Jatorrizko kategoria-anitzeko SVM sailkatzaile bat erabiltzea sailkatzaile bitarra bateratzea baino askoz aproposagoa dela erakutsi dugu. Gure emaitzek argi eta garbi erakutsi dute kategoria-anitzeko taxonomien kasuan ez dela aukera aproposa sailkatzaile bitarretan oinarritzea. Ondorioz, jatorrizko kategoria-anitzeko sailkatzaileak, kategoria guztiak aldi berean kontuan hartuz ataza osoa hobeto ezagutzen dutenak, egokiagoak dira errendimendu hobea lortzeko.

\begin{description}
 \item[2. Ikerketa Galdera] \hfill \\
 \textit{\rqsvmtwoeu}
\end{description}

Teknika erdi-gainbegiratuek emaitza hobeak eskura ditzakete aurrez sailkatutako baliabide sorta oso-oso txikia denean, baina teknika gainbegiratuek antzeko errendimendua lortzen dute baliabide gehixeago kontuan hartuz. Horrez gain, teknika gainbegiratuek konputazio aldetik gutxiago exijitzen dute. Beraz, \cite{joachims99transductive} egileak sailkapen bitarrerako erakutsitakoaren aurkakoa erakutsi dugu, ataza kategoria-anitzetarako teknika gainbegiratu eta erdi-gainbegiratuak oso antzerakoak direla, alegia. Zentzuzkoa dirudi kategoria kopurua handitu ahala zailagoa izatea teknika erdi-gainbegiratuen ikasketa behar bezala burutzea, gaizki sailkatutako baliabideek zarata gehitzen baitute ikasketa prozesuan.

Ondorio hauei eutsiz, tesian zehar kategoria-anitzeko SVM gainbegiratua erabiltzea erabaki genuen.

\begin{description}
 \item[3. Ikerketa Galdera] \hfill \\
 \textit{\rqdataoneeu}
\end{description}

Hau jakiteko, etiketa sozialen sistemetako ezarpenen ezaugarri ezberdinak aztertu ditugu. Aztertutako ezaugarrien artean, etiketak gomendatzeak duen garrantzia nabaritu dugu, folksonomien egituran nabarmen eragiten baitu. Aztertutako etiketa sozialen sistemek ezarpen ezberdinak dituzte gomendioei dagokienean:

\begin{itemize}
 \item \textbf{Baliabidean oinarritutako gomendioak (Delicious):} baliabide bat e\-ti\-ke\-ta\-tze\-ra\-ko orduan, sistemak baliabide horretan beste erabiltzaile batzuek definitutako etiketak gomendatzen dituenean, erabiltzaileak etiketa berriak definitzeko probabilitatea izugarri jaisten da, gehienetan gomendioetan oinarritzen baitira. Kasu honetan, erabiltzaileek esfortzu txikia egiten dute etiketa berriak pentsatzen, eta gomendioei kasu egitea nahiago izaten dute.
 \item \textbf{Pertsonomian oinarritutako gomendioak (GoodReads):} baliabide bat etiketatzerako orduan, sistemak erabiltzaile horrek aurrez beste baliabide ba\-tzu\-e\-tan ipinitako etiketak gomendatzen dituenean, erabiltzailearen etiketa kopurua izugarri murrizten da. Erabiltzailearen berbategia askoz txikiagoa izaten da beraz. Hala eta guztiz ere, erabiltzaileek ez dakite beste batzuek zein etiketa ipini dizkioten baliabideari, eta baliabidearekiko berriak diren etiketak definitzeko probabilitatea mantendu egiten da.
 \item \textbf{Gomendiorik gabe (LibraryThing):} baliabide bat etiketatzerako orduan, sistemak etiketarik gomendatzen ez duenean, erabiltzailearen berbategia hazi egiten da, eta baliabide bakoitzean etiketa berriak definitzeko probabilitatea mantendu egiten da.
\end{itemize}

\begin{description}
 \item[4. Ikerketa Galdera] \hfill \\
 \textit{\rqreponeeu}
\end{description}

Erabiltzaile gehienek anotatu dituzten etiketa gutxi batzuetan oinarritu baino, etiketa guzti-guztiak kontuan hartzea merezi duela erakutsi dugu. Gehien anotatutakoak dira garrantzitsuenak, eta baliabidea zeren ingurukoa den gehien adierazten dutenak dira. Hala eta guztiz ere, erabiltzaile gutxik anotatutakoek ere badute nolabaiteko adierazgarritasuna, beste neurri batean bada ere, eta sail\-ka\-tzai\-le\-a\-ren\-tza\-ko baliagarria den informazioa eskaintzen dute.

Baliabideen errepresentazioa egiterakoan etiketei emandako pisuei dagokionean, etiketa bakoitza definitu duen erabiltzaile kopurua pisu bezala erabiltzearena da emaitza onenak ematen dituena. Erabiltzaile kopurua alde batera utzi, edo baliabidea anotatu duen erabiltzaile guztien kopurua kontuan izatea bezalako beste hurbilketa batzuk gainditu ditu aurrekoak.

Laburbilduz, aurkitu dugun errepresentazio egokiena etiketa guztiak erabili, eta etiketa bakoitza erabiltzaile kopuruaren arabera pisatzearena da.

\begin{description}
 \item[5. Ikerketa Galdera] \hfill \\
 \textit{\rqreptwoeu}
\end{description}

Sailkatzaile ezberdinen aurreikuspenak elkartzen dituzten sailkatzaile bateratuetan oinarrituz, etiketek kontuan hartu beharreko iritziak ematen dituztela erakutsi dugu. Iritzi hauek oso baliagarriak dira etiketa sozialak beste datu iturriekin bateratzeko. Edonola ere, datu iturri guztiak ez dira lagungarriak bateratzerako orduan. Aukeratutako datu iturriak nahikoa sendoak izan behar dira, aurreikuspen iritzi aproposak eman ditzaten. Datu iturriak ondo aukeratzen direnean, baina, errendimendua nabarmen hobe daiteke.

\begin{description}
 \item[6. Ikerketa Galdera] \hfill \\
 \textit{\rqrepthreeeu}
\end{description}

Baliabideen sailkapenerako etiketa sozialean erabilgarritasuna taxonomien bi mailatan aztertu dugu. Goi-mailako kategoriez gainera, bigarren mailako kategoria zehatzagoekin ere egin dugu ikerketa. Guneak erabiltzaileak anotatzera animatzen dituenean (Delicious eta LibraryThing-en), etiketa sozialek beste datu iturriek baino emaitza hobeak lortzen dituzte. Etiketek askogatik gainditzen dituzte beste datu iturriak kasu hauetan, Delicious-en batez ere, bigarren mailako emaitzek abantaila askoz garbiagoa erakusten baitute. Ezberdintasun hau LibraryThing-en ere gertatzen da. Azkenik, GoodReads-eko etiketek ez dituzte beste datu iturriak gainditzen, ezta goi-mailako kategorietan ere, sistemak ez dituelako erabiltzaileak etiketatzera animatzen, eta horrela anotazio gutxiago egiten direlako.

Gure emaitza hauek \cite{noll_exploring_2008} egileen hipotesia deusezten dute. Beraiek egindako analisi estatistikoan, etiketek kategoria zehatzagoetan sailkapenak egiteko balioko ez zutela uste zuten, eta horretarako beste datu iturri batzuk erabili beharko liratekeela.

\begin{description}
 \item[7. Ikerketa Galdera] \hfill \\
 \textit{\rqdistoneeu}
\end{description}

Etiketen distribuzioak kontuan hartzea, IDFn oinarritutako pisu-funtzio batean oinarrituz, baliabideen sailkapenerako etiketen adierazgarritasuna zehazteko interesgarria dela erakutsi dugu. Etiketa sozialen sistemaren ezarpenek, ordea, zerikusi handia dute distribuzio hauekin. Guneen ezarpenen artean baliabideetan oinarritutako gomendioek garrantzia handia dutela ikusi dugu, folksonomiaren egitura erabat aldatzen baitute. Gomendio hauek dituzten sistemek oso distribuzio ezberdinak aurkezten dituzte. Honen arabera, pisu-funtzioen erabilgarritasuna jakin daiteke.

Gure sailkapen esperimentuetan ikusi ahal izan dugu pisu-funtzioek TF gainditzen dutela baliabideetan oinarritutako gomendioak existitzen ez direnean, hau da, LibraryThing eta GoodReads-en, bai bakarrik erabilita, eta baita beste datu iturri batzuekin elkartzerakoan ere. Halaber, aproposagoa da berauek bakarrik erabiltzea, beste datu iturriekin elkartu gabe, emaitza hobeak lortzen baitira horrela.

Baliabideetan oinarritutako gomendioak ematen direnean, ordea, folksonomien egitura oso ezberdina da, honek distribuzioetan eragiten du eta, ondorioz, baita pisu-funtzioetan ere. Hau dela-eta, pisu-funtzioak erabiltzerakoan emaitza txarragoak lortzen dira, eta beste datu iturriekin elkartu beharra dago hobetu ahal izateko. Elkartzerakoan, baina, TFk baino emaitza hobeak lortzen ditu, sailkatzailearen iritzi egokiei esker.

\begin{description}
 \item[8. Ikerketa Galdera] \hfill \\
 \textit{\rqdisttwoeu}
\end{description}

Ikertutako pisu-funtzioen artean, laster-marka frekuentzietan oinarritzen denak lortzen ditu emaitza onenak sistemak baliabideetan oinarritutako gomendioak ematen ez dituenean. Kasu hauetan, IBF da onena, IRF eta IUFk jarraituta. Horiek guztiek argi eta garbi gainditzen dute TFren errendimendua, bai bakarrik erabilita, eta baita sailkatzaile bateratuen bitartez beste datu iturri batzuekin elkartzerakoan ere.

Bestalde, guneak baliabideetan oinarritutako gomendioak ematen dituenean, erabiltzaileen frekuentziak emaitza hobeak ematen ditu. IUFren errendimendua IBF eta IRFrena baino hobea da, mota honetako guneetako gomendioetan oinarritu beharrean bere etiketa propioak definitzen dituzten erabiltzaileek duten garrantzia dela-eta. Bakarrik erabilita IUFk kasu honetan TF gainditzen ez badu ere, beste datu iturri batzuekin elkartzean emaitzarik onenak lortzen ditu. Hala ere, gutxigatik gainditzen du TFn oinarritutako sailkatzaile bateratuen emaitza, eta bietako edozein erabil liteke emaitza antzekoak eskuratuz.

\begin{description}
 \item[9. Ikerketa Galdera] \hfill \\
 \textit{\rqcatoneeu}
\end{description}

Erabiltzaile mota hori, Sailkatzaile izenekoa, existitzen dela frogatu dugu. Gure esperimentuen arabera, hau egia da, batez ere, etiketen gomendioak ez dituzten guneetan, hau da, LibraryThing-en. Gune honetan Sailkatzaileek definitutako etiketek emaitza hobeak lortzen dituzte sailkapenerako. Gomendioak ematen direnean, ordea, erabiltzaile hauek antzematea zailagoa da, GoodReads eta Delicious-ekin gertatzen den bezala. Hala ere, erabiltzaileak antzemateko neurri aproposa erabiltzeak Sailkatzaileak antzematea ahalbidetzen du, kasu hauetan ere bai.

\begin{description}
 \item[10. Ikerketa Galdera] \hfill \\
 \textit{\rqcattwoeu}
\end{description}

Aztertutako bi ezaugarrien artean, sailkapen atazarako aproposenak diren Sailkatzaileak antzemateko ezaugarri interesgarriena erabiltzailearen hiztuntasuna dela erakutsi dugu. Erabiltzaile batek etiketa gutxiago edo gehiago definitzeko duen ohitura adierazten du hiztuntasunak. Ezaugarri hau baliatuz, posible da sailkapen atazatik gertuago dauden erabiltzaileak aurkitzea. Bestalde, erabiltzaileen berbategiaren aniztasunaren arabera Deskribatzaileak diren erabiltzaileak antzeman daitezke. Honez gain, etiketa deskribatzaileak erabiltzen ez dituztenek sailkapen hobea sortzen dutela deskubritu dugu.

\section{Ekarpen Nagusiak}
\label{sec:ekarpen-nagusiak}

Lan honen ideia berritzailea baliabideen sailkapenerako etiketa sozialak ba\-li\-a\-tze\-an datza. Guk dakigula, etiketa sozialak baliatuz egiazko sailkapen esperimentuak burutzen dituen lehen lana guk aurkeztutako lehena da \citep{zubiaga2009getting}. Horren aurretik, \cite{noll_exploring_2008} egileek etiketa sozialak eta adituen sailkapenak alderatu zituzten analisi estatistikoa eginez. Arlo honetako lanen gabezia kontuan hartuz, tesi honetan aurkezten dugun lanak baliabideen sailkapenerako etiketa sozialen erabilpen eta errepresentazio aproposerako argipenak ematen dira. Konkretuki, hurrengo ekarpen nagusiak aurkeztu ditugu lan honetan:

\begin{itemize}
 \item Tamaina handiko 3 kolekzio sortu ditugu etiketa sozialen sistemetan oinarrituz, kontuan hartutako baliabideei adituek esleitutako sailkapen datuekin batera. 3 hauek ikerketan erabilitako datu-sorta handienen artean daudela esan genezake eta, guk dakigula, baliabideen sailkapenerako erabilitako handienak dira. Datu-sorta hauetako batzuk, beste txikiago batzuekin batera, publikoki eskuragarri utzi ditugu beste ikertzaile batzuek erabili ahal izan dezaten\footnote{http://nlp.uned.es/social-tagging/datasets/}. Datu-sorta hauek, besteak beste, \cite{godoy2010exploiting} eta \cite{strohmaier2010network} egileek baliatu dituzte beraien ikerketa lanetarako.
 \item Gure lana etiketa sozialen errepresentazio ezberdinak alderatzen dituen lehena da. Gainera, etiketa sozialak eta beste datu iturri batzuk alderatuz egiazko sailkapen esperimentuak egiten dituen lehen ikerketa lana da. Etiketa sozialak goi-mailako kategoriatarako baizik, maila baxuagoko kategoria zehatzagoetan sailkatzeko ere baliagarriak direla erakutsi dugu. \cite{noll_exploring_2008} egileek ondorioztatuko hipotesia ezeztatzen dugu honenbestez. Lan horretako analisi estatistikoaren arabera, kategoria zehatzagoetarako etiketen erabilgarritasuna oso txikia izan zitekeela diote egileek.
 \item Etiketa sozialek folksonomiatan dituzten distribuzioak aztertu ditugu, eta sistema bakoitzaren ezarpenek zentzu honetan duten eragina ikertu dugu. Horretarako, pisu-funtzio ezagun baten oinarritu gara, TF-IDF, folksonomien egitura hauetara egokituz.
 \item Sailkatzaile bezala definitu ditugun erabiltzaileek osatutako multzoa existitzen dela erakutsi dugu. Erabiltzaile hauen anotazioak gertuago daude adituen sailkapen taxonomikoetatik, Deskribatzaile deitu ditugun bere erabiltzaile batzuen anotazioetatik baino. Sailkatzaile eta Deskribatzaileak ezberdintzeko hurbilketak lehendik ere frogatu baziren, hauxe da Sailkatzaileak baliabideen sailkapenerako aproposagoak direla erakusten duen lehen lana.
\end{itemize}

Etiketa sozialak baliabideen sailkapenerako erabiltzea ikerketa lerro berria zen tesi honekin hasi ginenean. Hala ere, azkenaldian sare sozialetan, eta bereziki etiketa sozialen sistemetan, erabiltzaileek sortutako edukien gainean ikertzeko sortu den interesa dela-eta, lan berri ugari ekarri du. Hazkunde honekin batera, ikertzaile gehiagok erakutsi du etiketa sozialak sailkapenerako erabiltzeko interesa, eta arlo honetan egindako ikerketa lanen kopuruak nabarmen egin du gora. \cite{godoy2010exploiting} egileek, esate baterako, etiketak erabiltzen dituzte sailkapenerako, gure aurreko lan baten oinarrituz \citep{zubiaga2009getting}.

\section{Etorkizunerako Ildoak}
\label{sec:etorkizunerako-ildoak}

Etiketa sozialak baliabideen sailkapenerako erabiltzea oraindik ere ikerketa arlo berria da, eta lan gutxi egin da honen inguruan. Tesi honetan aurkezten den lanak ahalik eta baliabideen sailkapen zehatzena lortzera bidean etiketa sozialak errepresentatzeko modu egokia zein den argitzen du. Horrez gain, etorkizunerako ildo berriak ireki ditu.

Tesi honetan guztian zehar, etiketa bakoitza ikur ezberdin bat bezala hartu dugu kontuan, izan dezakeen esanahi semantikoa aztertu gabe. Zentzu honetan, etorkizunerako lan interesgarria litzateke analisi semantikoa egitea etiketen artean dauden sinonimoak eta erlazio ezberdinak antzemateko. Lengoaia naturalen prozesamendurako teknika baliatuz, edo hurbilketa semantikoetara joz, etiketen inguruko ezagutza areagotzea lortu liteke, folksonomien azterketa sakonagoa ahalbidetuz.

\ref{c:tag-distribution-classification}. Kapituluan erabili ditugun pisu-funtzioak testu kolekzioetan erabiltzeko pentsatutako TF-IDF funtzioan oinarritzen dira. Etorkizunerako interesgarria izan liteke beste funtzio batzuk probatzea, eta baita folksonomien egitura hauetara egokitu daitekeen beste funtzio batzuk proposatzea ere. Honek asko lagunduko luke gomendioak ematen dituzten sistemetarako, Delicious-en esate baterako, izan ere probatu ditugun funtzioek ez baitute behar bezala funtzionatu sistema honetan.